\documentclass[twoside,12pt]{article}
\usepackage{epsfig}

\def\Journal#1#2#3#4{{#1} {#2} (#4) #3 }

\def\NPA{{\em Nucl. Phys.} A}

\def\PLB{{\em Phys. Lett.} B}

\def\PRL{\em Phys. Rev. Lett.}

\def\PREP{\em Phys. Rep.}

\def\PRD{{\em Phys. Rev.} D}
\def\PRC{{\em Phys. Rev.} C}

\newcommand{\be}{\begin{equation}}
\newcommand{\ee}{\end{equation}}
\newcommand{\bea}{\begin{eqnarray}}
\newcommand{\eea}{\end{eqnarray}}

\def\eq#1{{Eq.~(\ref{#1})}}
\def\fig#1{{Fig.~\ref{#1}}}

\newcommand{\slpartial}{\raise.15ex\hbox{$/$}\kern-.53em\hbox{$\partial$}}
\newcommand{\slA}{\raise.15ex\hbox{$/$}\kern-.73em\hbox{$A$}}
\newcommand{\slp}{\raise.15ex\hbox{$/$}\kern-.57em\hbox{$p$}}
\newcommand{\slk}{\raise.15ex\hbox{$/$}\kern-.57em\hbox{$k$}}
\newcommand{\slq}{\raise.15ex\hbox{$/$}\kern-.57em\hbox{$q$}}
\newcommand{\slepsilon}{\raise.15ex\hbox{$/$}\kern-.53em\hbox{$\epsilon$}}
\newcommand{\slvarepsilon}{\raise.15ex\hbox{$/$}\kern-.53em\hbox{$\varepsilon$}}
\newcommand{\slC}{\raise.15ex\hbox{$/$}\kern-.53em\hbox{$C$}}

\newcommand{\ben}{\begin{eqnarray*}}
\newcommand{\een}{\end{eqnarray*}}
\newcommand{\stackeven}[2]{{{}_{\displaystyle{#1}}\atop\displaystyle{#2}}}
\newcommand{\lsim}{\stackeven{<}{\sim}}
\newcommand{\gsim}{\stackeven{>}{\sim}}
\newcommand{\bas}{\overline{\alpha}_s}
\newcommand{\un}[1]{\underline{#1}}

\newcommand{\as}{\alpha_s}

\begin{document}

\title{Saturation Physics and Deuteron--Gold Collisions at RHIC}
\author{Jamal Jalilian-Marian$^1$\footnote{Email: jamal@phys.washington.edu} \
and Yuri V. Kovchegov$^2$\footnote{Email: yuri@mps.ohio-state.edu} \\~~\\
$^1$ Institute for Nuclear Theory, Box 351550 \\
University of Washington \\
Seattle, WA 98195-1550, USA \\~~\\
$^2$ Department of Physics \\
The Ohio State University \\
191 West Woodruff Avenue \\
Columbus, OH 43210, USA}

\maketitle

\begin{abstract}
We present a review of parton saturation/Color Glass Condensate
physics in the context of deuteron-gold ($d+Au$) collisions at
RHIC. Color Glass Condensate physics is a universal description of all
high energy hadronic and nuclear interactions. It comprises classical
(McLerran-Venugopalan model and Glauber-Mueller rescatterings) and
quantum evolution (JIMWLK and BK equations) effects both in small-$x$
hadronic and nuclear wave functions and in the high energy scattering
processes. Proton-nucleus (or $d+A$) collisions present a unique
opportunity to study Color Glass Condensate predictions, since many
relevant observables in proton-nucleus collisions are reasonably
well-understood theoretically in the Color Glass Condensate approach. In this
article we review the basics of saturation/Color Glass Condensate
physics and reproduce derivations of many important observables in proton
(deuteron)--nucleus collisions. We compare the predictions of Color
Glass physics to the data generated by $d+Au$ experiments at RHIC and
observe an agreement between the data and the theory, indicating that
Color Glass Condensate has probably been discovered at RHIC. We point
out further experimental measurements which need to be carried out to
test the discovery.
\end{abstract}

\vspace*{1cm}

\begin{center}
{\sl Prepared for publication in Prog. Part. Nucl. Phys.}
\end{center}

\newpage

\tableofcontents

\newpage

\section{Introduction}

Saturation/Color Glass Condensate physics \cite{GLR}-\cite{yuri_bk} is
a rapidly developing field of strong interactions at high
energy. Color Glass Condensate (CGC) physics describes high parton
densities inside the hadronic and nuclear wave functions at small
values of Bjorken $x$ variable. It demonstrates how the gluon fields
in the hadronic and nuclear wave functions reach their maximum
allowable values in quantum chromodynamics (QCD) corresponding to
$A_\mu \sim 1/g$ \cite{MV,yuri1,yuri2,jkmw}. The CGC formalism is also
successfully applied to calculation of total, elastic and diffractive
cross sections of high energy hadronic and nuclear scattering. There,
by resumming the strong gluon field dynamics, it resolves such
long-standing questions as unitarity of the scattering $S$-matrix
\cite{froi} and infrared (IR) safety \cite{Bartels}, which are known 
to be violated by the Balitsky-Fadin-Kuraev-Lipatov (BFKL) evolution
equation \cite{BFKL} corresponding to the weak field limit of
CGC. Saturation/CGC approach allows one to calculate particle
production in hadronic and nuclear scattering
\cite{KMW1}-\cite{Lappi}. The resulting inclusive particle production
cross sections are infrared-safe, which is a significant theoretical
improvement over the IR-divergent perturbative QCD results
\cite{GB,Lipatov1} making the small coupling approach to particle
production self-consistent. One of the interesting application of
particle production in CGC framework is understanding the initial
conditions for the evolution of the quark-gluon system produced in
heavy ion collisions toward the possible thermalization leading to
formation of quark-gluon plasma (QGP) \cite{yuriaa,Ian04,KV,Lappi}.

Perturbative QCD (pQCD) has been extremely successful in describing
the particle production data over a large kinematic window
\cite{amueller}. Applications of pQCD to particle production in high
energy hadronic or nuclear collisions are based on the use of
collinear factorization theorems. The essence of a collinearly
factorized cross section is the idea of incoherence. In other words, a
hadronic cross section can be written as a convolution of parton
distributions and fragmentation functions, which are universal
non-perturbative objects that are also subject to perturbative
evolution (DGLAP), with a hard scattering cross section involving
partons, which is perturbatively calculable but is process
dependent. Parton distribution functions are typically measured in
Deep Inelastic Scattering experiments such as the ones performed at
HERA, where it has been observed that the gluon and sea quark
distributions grow very fast with decreasing Bjorken $x$. This fast
growth can be understood in pQCD as driven by radiation of gluons with
small Bjorken $x$ via DGLAP evolution equations. Collinear
factorization theorems are not exact and are violated by effects that
are typically suppressed by inverse of the hard momentum transfer but
can be enhanced by energy (or $\ln \, 1/x$) or $A$ dependent factors,
which may be large at high energy and/or for large nuclei. This
necessitates construction of a new formalism that does not rely on
collinear factorization and can include these potentially large
effects.  The hint for this new formalism comes from pQCD itself,
noticing that the rise of parton distribution functions can not
continue for ever since it would lead to growth of hadronic cross
sections at a rate which would violate unitarity. A weak coupling
mechanism which can tame this fast growth is gluon recombination and
saturation. Color Glass Condensate formalism is the natural
generalization of pQCD in order to make it applicable to dense
partonic systems.

The extensive theoretical progress in the field of saturation/Color
Glass has been summarized in several review articles, mostly
concentrating on the issues of non-linear small-$x$ evolution
\cite{ILM,IV,Heribert}. Our article here deals with saturation/Color
Glass Condensate physics putting more emphasis on particle production
in proton (deuteron)--nucleus collisions ($p(d)A$) and in deep
inelastic scattering (DIS). Many of the relevant particle production
observables in $pA$ and DIS have been well-understood theoretically in
the saturation/Color Glass approach, at least at the partonic level
\cite{KMW1}-\cite{Gelis:2002nn}. It is therefore, very important to be
able to verify our theoretical understanding by comparing the
predictions of CGC physics for particle production in $p(d)A$ to the
experimental data produced by deuteron-gold ($d+Au$) scattering
program at Relativistic Heavy Ion Collider (RHIC) at Brookhaven
National Laboratory (BNL) \cite{phenix}-\cite{stardA}. Below we will
review both the CGC predictions \cite{KLM}-\cite{KLN} and experimental
data reported by RHIC experiments \cite{phobos_mult}-\cite{stardA}. We
will point out the apparent agreement between the two indicating a
possible discovery of Color Glass Condensate at RHIC \cite{GM}. We
will also review future experimental test which can be carried out to
test the CGC discovery both by $d+Au$ program at RHIC and by $pA$
scattering program at Large Hadron Collider (LHC) at CERN.

The paper is structured as follows. We begin in Sect. \ref{satrev}
with a general review of saturation/CGC physics. This review is by no
means all-inclusive: we will concentrate on the material that we will
need later in our discussion of particle production. We refer the
interested reader who wants to learn more about various aspects of
small-$x$ evolution to the dedicated reviews in
\cite{ILM,IV,Heribert,genya}. In Sect. \ref{ed} we discuss the BFKL 
evolution equation \cite{BFKL} along with its problems, such as
violation of unitarity \cite{froi} and diffusion into infrared
\cite{Bartels}. We also review the Gribov-Levin-Ryskin--Mueller-Qiu 
(GLR-MQ) evolution equation \cite{GLR,MQ}. We proceed in
Sect. \ref{QCA} by discussing quasi-classical approximation in
small-$x$ physics. We review Glauber-Mueller multiple rescatterings in
DIS \cite{GlaMue,Glauber} and McLerran-Venugopalan model of small-$x$
wave functions of large nuclei
\cite{MV,yuri1,yuri2,jkmw,LM}. Quasi-classical regime at small-$x$ is
valid when $x$ is small enough, so that coherent interactions of
nucleons in the nucleus with the projectile are possible. This
translates into the requirement for parton coherence length $l_{coh}$
to be larger than the nuclear radius $R$ \cite{KS},
\be\label{clcond}
l_{coh} \, = \, \frac{1}{2 \, m_N \, x_{Bj}} \, > \, R,
\ee
with $m_N$ the nucleon mass and $x_{Bj}$ the Bjorken $x$ variable of a
parton. Defining the {\sl rapidity} variable $Y = \ln 1/x_{Bj}$ we
recast the condition (\ref{clcond}) as $Y \, > \, \ln A$, with $A$ the
atomic number of the nucleus. On the other hand, when $x_{Bj}$ becomes
too small, BFKL evolution effects become important, breaking down the
quasi-classical approximation. BFKL evolution brings in powers of $\as
\, \ln (1/x_{Bj}) \, \sim \, \as \, Y$. Requiring for such effects to
be small, $\as \, Y < 1$, we obtain an upper bound on the allowable
rapidity range. The applicability window for the quasi-classical
approximation is then
\be\label{clcond3}
\ln A \, \le \, Y \, \le \, \frac{1}{\as}. 
\ee
We also discuss in Sect. \ref{QCA} how the {\sl saturation} scale
$Q_s$ arises in the quasi-classical limit.

We continue our review of saturation/Color Glass physics by
re-deriving the
Jalilian-Marian--Iancu--McLerran--Weigert--Leonidov--Kovner (JIMWLK)
\cite{jklw1}-\cite{Jalilian-Marian:1998cb},
\cite{Iancu:2000hn}-\cite{Ferreiro:2001qy} and Balitsky-Kovchegov (BK) 
\cite{balitsky,yuri_bk} non-linear evolution equations in Sect. \ref{QE}. 
Quantum small-$x$ evolution corrections become important when $\as \,
Y \, \gsim \, 1$ \cite{BFKL}, such that
\be\label{qcond}
Y \, \ge \, \frac{1}{\as}.
\ee
\eq{qcond} gives a lower bound on the region of applicability of JIMWLK 
and BK evolution equations. 

We conclude the review of CGC by solving the non-linear evolution
equations in Sect. \ref{see}. There we discuss the solution of linear
(BFKL) evolution equation outside of the saturation region,
demonstrate an interesting property of the solution of JIMWLK and BK
known as {\sl geometric scaling} inside the saturation region
\cite{geom,LT} and reproduce the derivation of {\sl extended geometric
scaling} outside of that region \cite{IIM1}. We demonstrate how
saturation scale $Q_s$ grows with energy once the quantum evolution
effects are included \cite{MT,Mueller3,MP}. We explain how JIMWLK and
BK evolution equations resolve the problems of the BFKL evolution by
unitarizing the corresponding total cross sections of DIS and by
prohibiting diffusion into the infrared, making the small-coupling
approximation self-consistent \cite{Braun1}-\cite{AAMSW}.

We continue in Sect. \ref{pppa} by deriving expressions for a number
of particle production observables in $pA$ collisions in the
saturation/CGC framework. We start in Sect. \ref{clpa} by calculating
inclusive gluon production cross section in $pA$ in the
quasi-classical approximation \cite{KM,KTS,DM,KW2}. We show that
quasi-classical multiple rescatterings lead to Cronin enhancement
\cite{Cronin} of gluon production in $pA$ \cite{BKW,KKT,KNST}-\cite{ktbroadening3}, 
as can be seen from \fig{cron}. We then proceed in Sect. \ref{qepa} by
including the effects of quantum BK evolution in the expression for
gluon production cross section in $pA$ \cite{KT,Braun2,Braun3}. As one
can see from \fig{toy}, the effect of small-$x$ evolution is to
flatten the Cronin maximum introducing suppression of particle
production at all transverse momenta $p_T$ (see \fig{toy})
\cite{KLM}-\cite{BKW,AAMSW,IIT}. In Sect. \ref{clquark} we calculate
valence quark production cross section in $pA$ both in the
quasi-classical limit and including small-$x$ evolution
\cite{Dumitru:2002qt}-\cite{Gelis:2002nn}. We move on to
electromagnetic probes in Sect. \ref{ep}, where we calculate prompt
photon and dilepton production cross sections in $pA$ collisions
\cite{Gelis:2002ki}-\cite{BGD}.  Finally, in Sect.\ref{2pi}
we analyze two-particle correlations
\cite{KLM2}-\cite{Fujii:2005vj}. We rederive production cross section
in $pA$ for two gluons at mid-rapidity \cite{JMK}, for a quark and a
gluon at forward rapidity \cite{JMK} and for $q\bar q$ pair at
mid-rapidity \cite{Kharzeev:2003sk}-\cite{Fujii:2005vj}.

We review some of the data generated by $d+Au$ scattering program at
RHIC \cite{phenix}-\cite{stardA} in Sect.~\ref{data}. We show that the
data reported by BRAHMS \cite{brahms-2}, PHENIX
\cite{phenix}, PHOBOS \cite{phobos} and STAR \cite{star} experiments show 
Cronin-like enhancement of particle production at mid-rapidity. We
discuss how this result, combined with suppression of produced hadrons
in $Au+Au$ collisions \cite{brahmsaa}-\cite{starAA}, serves as a
control experiment for the signals of quark-gluon plasma formation in
heavy ion collisions at RHIC \cite{Bj}-\cite{GM1}. We then review
BRAHMS data at forward rapidity \cite{brahms-1,brahms-2}, indicating
suppression predicted by saturation/CGC approach
\cite{KLM,KKT,AAKSW}. This data is confirmed by preliminary results
from PHENIX \cite{phenixdA}, PHOBOS \cite{phobosdA} and STAR
\cite{stardA}, demonstrating that Color Glass Condensate has probably
been discovered at RHIC \cite{GM1}. We conclude Sect. \ref{data} by
listing future experimental tests
\cite{Jalilian-Marian:2004er,KKT2,JamalH} which need to be carried out
to verify the discovery of CGC at RHIC \cite{GM1}.

We conclude in Sect. \ref{conc} by discussing some of the issues which
we inevitably had to leave out in this review, including recent
progress in our understanding of pomeron loop corrections to small-$x$
evolution \cite{loop1}-\cite{loop5} and some exclusive processes
\cite{Hebecker}-\cite{yuridiff}, which can not be measured at RHIC.

Throughout the paper we will use the following notation for the
4-vectors: for a 4-vector $v_\mu = (v_0, v_1, v_2, v_3)$ we define the
light cone components by $v^\pm = (v_0 \pm v_3)/\sqrt{2}$ and combine
the transverse components into a two-dimensional vector ${\un v} =
(v_1, v_2)$. The metric tensor is chosen in such a way that $v_\mu
v^\mu = 2 \, v^+ \, v^- - {\un v}^2$.

%%%%%%%%%%%%%%%%%%%%%%%%%%%%%%%%%%%%%%%%%%%%%%%%%%%%%%%%%%%%%%%%%%%%%%%%%%%%%%%%%%%%%%%%%

\section{Overview of Saturation/Color Glass Condensate Physics}
\label{satrev}

In this Section we will review the developments and advances of high
energy QCD which led to our modern understanding of the
saturation/Color Glass Condensate physics.

\subsection{Early Developments}
\label{ed}

\subsubsection{The BFKL Equation}
\label{BFKLeq}

A milestone in the development of small-$x$ physics was the
derivation by Balitsky, Fadin, Kuraev, and Lipatov of what has become
known as the BFKL equation \cite{BFKL}. The BFKL equation describes
the behavior of scattering amplitudes and gluon distribution functions
at asymptotically high energies. It does it by resumming leading
logarithms of energy, i.e., powers of the parameter $\as \ln s$, where
in the perturbative QCD regime the coupling constant is small, $\as
\ll 1$, and, in Regge kinematics, the center of mass energy $s$ of the
scattering process is much larger than any other momentum scale
involved.

Derivation of the BFKL equation is rather complicated. For derivations
in the transverse momentum space we refer the interested reader to
\cite{BFKL,Jar}, as well as to Sect. \ref{JIMWLKsect} below. Derivation in 
transverse coordinate space can be found in \cite{dip1,dip2,dip3,CM},
as well as in Sect. \ref{BKsect} below. Here we are going to present a
simple physical picture of the BFKL evolution following Mueller in
\cite{MuellerCarg}. \\

\begin{figure}
\begin{center}
\epsfxsize=8cm
\leavevmode
\hbox{\epsffile{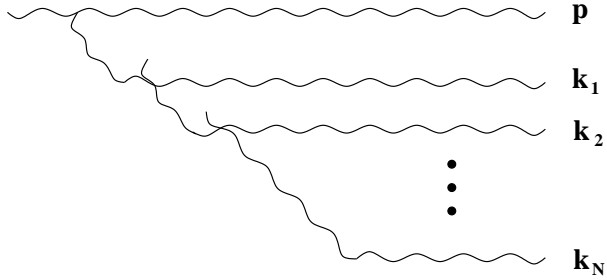}}
\end{center}
\caption{Gluon cascade leading to the BFKL evolution. All gluon emissions 
are ordered in light cone ``+'' momenta, and, correspondingly, in
light cone time. }
\label{bfkl_phys}
\end{figure}

{\sl Physical Picture} \\
~\\
Consider an ultrarelativistic gluon scattering on a target at
rest. Before scattering on the target, the gluon can emit one or more
extra gluons. The emissions are illustrated in \fig{bfkl_phys}. In the
infinite momentum frame considered here, the gluon moving in the light
cone ``+'' direction has a very large typical light cone ``+''
component of its momentum, which we denote by $p^+$. The original
gluon emits another gluon with a much smaller light cone momentum
$k_1^+ \ll p^+$. A simple calculation shows that the emission
probability is given by
\be\label{P1}
d P_1 \, = \, \frac{\as \, N_c}{\pi} \, \frac{d^2 k_1}{{\un k}_1^2} \,
\frac{d k_1^+}{k_1^+},
\ee
where $N_c$ is the number of colors (and the $SU(N_c)$ Casimir
operator in the adjoint representation) and ${\un k}_1$ is the
two-component transverse momentum vector of the gluon $\# 1$. \eq{P1}
can be obtained from the usual formula for photon Bremsstrahlung by
multiplying it with the color Casimir $N_c$ and replacing $\alpha_{EM}
\rightarrow \as$. Following \cite{MuellerCarg} we assume that all 
emitted gluons have fixed transverse momentum of the order of some
scale $Q$. This allows us to simplify the problem by replacing the
transverse momentum integral $d^2 k_1 / {\un k}_1^2$ by a constant
$c$. Defining the {\it rapidity} variable
\be
y_1 \, = \, \ln \frac{p^+}{k_1^+} 
\ee
we rewrite \eq{P1} as
\be\label{P11}
d P_1 \, = \, c \, \frac{\as \, N_c}{\pi} \, d y_1.
\ee
Using \eq{P11} we conclude that given the rapidity interval
\be
\Delta y_1 \, = \, \frac{\pi}{\as \, c \, N_c}
\ee
the original gluon would split into two gluons with probability 1. 

To generalize the picture to many gluon emissions, we have to
understand the space-time picture of the process. First we note that
the typical light cone coherence time of the gluon $\# 1$ is 
\be\label{lctime}
\tau_1 \, \equiv \, x^+_1 \, = \, \frac{2 \, k_1^+}{{\un k}_1^2} \, \approx
\, \frac{2 \, k_1^+}{Q^2}. 
\ee
Similarly, for the $i$th gluon in the gluon cascade, the light cone
time is of the order of 
\be
\tau_i \, \approx \, \frac{2 \, k_i^+}{Q^2}.
\ee
Thus, if the original gluon emits a cascade of gluons with their
longitudinal momenta being progressively smaller
\be\label{kord}
p^+ \gg k_1^+ \gg k_2^+ \gg \ldots \gg k_N^+
\ee
than their light cone times would also be ordered
\be
\tau_1 \gg \tau_2 \gg \ldots \gg \tau_N.
\ee
%%%%%%%%%%%%%%%%%%%%%%%%%%%%%
%%%%%%%%%%%%%%%%%%%%%%%%%%%%%%
This cascade is shown in \fig{bfkl_phys}. There the gluons are ordered
in time, such that the typical coherence time of each emitted gluon is
much shorter than the coherence times of all the preexisting (harder)
gluons. Therefore, for the $i$th gluon, all the gluons $1, \ldots,
i-1$ appear frozen in time. The gluon can be emitted off any of these
preexisting gluons, which is shown by disconnected gluon lines in
\fig{bfkl_phys}. Indeed, in the transverse direction the gluons are 
coherent only over short distances of the order of $\Delta x_\perp
\sim 1/Q$, such that each gluon is emitted only by some fraction $c'$
of the preexisting gluons. Since the colors of all these gluons are
random, the $i$th gluon ``sees'' an effective color charge
\be
g_i \, = \, \sqrt{c' \, i} \, g
\ee
which results from a random walk in color space of $i$ preexisting
gluons. The probability of $i$th gluon emission is
\be\label{Pi}
d P_i \, = \, c \, c' \, i \, \frac{\as \, N_c}{\pi} \, d y_i.
\ee
The rapidity interval required for $i$th gluon emission is
\be
\Delta y_i \, = \, \frac{\pi}{c \, c' \, i \, \as \, N_c}.
\ee
Thus, the rapidity needed to emit $N$ gluons is given by
\be
Y_N \, = \, \sum_{i=1}^N \Delta y_i \, = \, \frac{\pi}{c \, c' \, \as
\, N_c} \, \sum_{i=1}^N \, \frac{1}{i} \, \approx \, \frac{\pi}{c \, c' 
\, \as \, N_c} \, \ln N,
\ee
so that the total number of gluons emitted in rapidity interval $Y$ is
given by
\be\label{expglue}
N(Y) \, = \, e^{c \, c' \, \frac{\as \, N_c}{\pi} \, Y}.
\ee
This simple physical picture that we borrowed from \cite{MuellerCarg}
gives the right qualitative behavior of the BFKL evolution. The exact
BFKL evolution also leads to the exponentially increasing number of
gluons, just like we obtained in \eq{expglue}. The successive time
ordered emissions discussed above and shown in \fig{bfkl_phys} resum
the leading logarithms of energy, just like the exact BFKL equation:
each power of $\as$ gets enhanced by a power of rapidity $Y$ (which is
equivalent to the logarithm of center of mass energy $s$), such that
\eq{expglue} resums powers of $\as Y$. Even the kinematics of successive 
emissions considered above (see \eq{kord}) is the same as in the exact
BFKL evolution, where extracting the leading logarithmic contribution
requires ordering of the gluons' longitudinal momenta.~\\

{\sl The BFKL Equation}~\\

\begin{figure}
\begin{center}
\epsfxsize=5cm
\leavevmode
\hbox{\epsffile{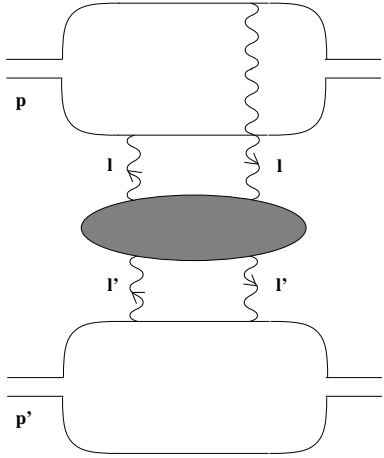}}
\end{center}
\caption{Scattering of two onia with the interaction mediated by the 
BFKL evolution.}
\label{bfkl_stand}
\end{figure}

Let us now consider a scattering of a bound heavy quark-antiquark
state (quarkonium, or simply onium) on another quarkonium. The
interaction between the onia is shown in \fig{bfkl_stand}.
%%%%%%%%%%%%%%%%%%%%%%%%%%%%%
%%%%%%%%%%%%%%%%%%%%%%%%%%%%%%
In general, we can write down the total onium-onium scattering cross
section as a convolution of the onia light cone wave functions
\cite{BL,BPP} with the imaginary part of the forward scattering amplitude 
of two quark-antiquark pairs (the imaginary part is denoted by $F$)
\begin{equation}\label{onium_tot}
\sigma_{tot} \, = \, 2 \, \int d^2 x \, \int_0^1 dz \, \int d^2 x' 
\, \int_0^1 dz' \, \Phi_{q\bar q} (\underline{x}, z) \, \Phi_{q\bar q}
 (\underline{x}', z') \, F ({\un x}, {\un x}', Y),
\end{equation}
where $\Phi_{q\bar q} (\underline{x}, z)$ and $\Phi_{q\bar q}
(\underline{x}', z')$ are light cone wave functions of the onia with
transverse sizes ${\un x}$ and ${\un x}'$. In the center of mass
frame, each of the onia carries a large light cone momentum, $p^+$ and
$p'^-$ correspondingly. For high energy eikonal scattering considered
here, these two momenta ($p^+$ and $p'^-$) are much larger than any
transverse momentum scales in the problem. In each of the onia the
quark carries the fraction $z$ (or $z'$) of the light cone component
of the total momentum of the quark-antiquark state. $Y = \ln (s/4
M^2)$ is the rapidity variable, defined as a logarithm of the ratio of
the center of mass energy of the onium-onium scattering $s$ over some
typical transverse momentum scale, like the onium mass $M$ used here.

At the lowest order the interaction in \fig{bfkl_stand} is mediated by
the two-gluon exchange. The corresponding imaginary part of the
forward amplitude, which we denote by $F=F^{(0)}$, is given by
\begin{equation}\label{twogl}
F^{(0)} (\underline{x},\underline{x}' ) \, = \, \frac{\as^2 \, C_F}{N_c} \, 
\int \, \frac{d^2 l}{[\underline{l}^2]^2} \, (2 - e^{- i
\underline{l} \cdot \underline{x}} - e^{ i \underline{l} \cdot
\underline{x}} ) \, (2 - e^{- i \underline{l} \cdot \underline{x}'} - e^{
i \underline{l} \cdot \underline{x}'} ),
\end{equation}
where $C_F = (N_c^2 - 1)/2 N_c$ is the $SU(N_c)$ Casimir operator in
the fundamental representation and ${\un l}$ is the transverse
momentum of each of the gluons. 

The BFKL evolution allows one to calculate quantum corrections to
\eq{twogl} that bring in powers of $\as Y$. Apparently, as was shown in 
\cite{BFKL}, such corrections preserve the two-gluon exchange structure 
of the interactions and can be summarized by the blob in
\fig{bfkl_stand}. Corresponding generalization of the amplitude (\ref{twogl}) 
reads
\be\label{forw}
F ({\un x}, {\un x}', Y)  \, = \, \frac{\as^2 \, C_F}{N_c} \, 
\int \, \frac{d^2 l \, d^2 l'}{\underline{l}^2 \, {\un l}'^2} \, (2 - e^{- i
\underline{l} \cdot \underline{x}} - e^{ i \underline{l} \cdot
\underline{x}} ) \, (2 - e^{- i \underline{l}' \cdot \underline{x}'} - e^{
i \underline{l}' \cdot \underline{x}'} ) \, f ({\un l}, {\un l}', Y)
\ee
with ${\un l}$ and ${\un l}'$ the gluons' transverse momenta on both
sides of the blob as shown in \fig{bfkl_stand}. At the lowest
(two-gluon) order the amplitude $f ({\un l}, {\un l}', Y)$ is given by
a delta-function
\be\label{bfkl_init}
f^{(0)} ({\un l}, {\un l}') \, = \, f ({\un l}, {\un l}', Y=Y_0) \, =
\, \delta^2 ({\un l} - {\un l}'),
\ee
which, after substitution into \eq{forw} readily gives
\eq{twogl}. ($Y_0$ is some initial rapidity, corresponding to the 
center of mass energy where the two-gluon exchange dominates the
interaction.)

The BFKL equation for the amplitude $f ({\un l}, {\un l}', Y)$ with
the initial condition (\ref{bfkl_init}) reads \cite{BFKL}
\be\label{bfkleq}
\frac{\partial f ({\un l}, {\un l}', Y)}{\partial Y} \, = \, 
\frac{\as \, N_c}{\pi^2} \int \! \frac{d^2
k}{(\underline{k} - \underline{l})^2} \left[ f
(\underline{k},\underline{l}', Y) - \frac{{\un l}^2 \, f
(\underline{l},\underline{l}', Y)}{{\un k}^2 + (\underline{k} -
\underline{l})^2} \right].
\ee
%%%%%%%%%%%%%%%%%%%%%%%%%%%%%
\begin{figure}
\begin{center}
\epsfxsize=18cm
\leavevmode
\hbox{\epsffile{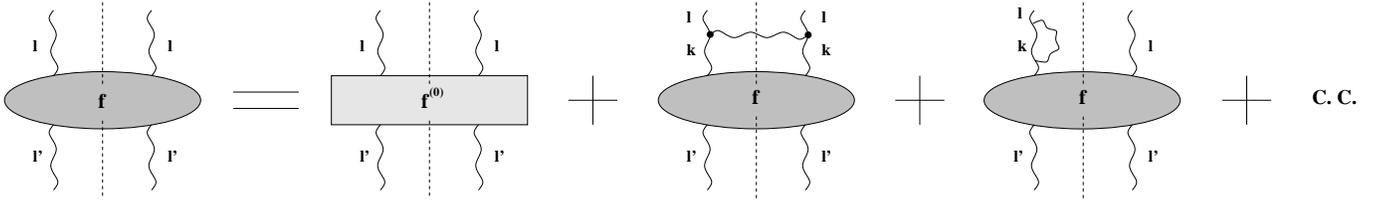}}
\end{center}
\caption{Schematic representation of a single rung of the BFKL evolution.}
\label{bfkl_iter}
\end{figure}
%%%%%%%%%%%%%%%%%%%%%%%%%%%%%%
The equation is illustrated in \fig{bfkl_iter}, where the dashed
vertical line denotes the cut. The equation in \fig{bfkl_iter} states
that the blob can either consist of just a two gluon exchange without
any evolution (the first term on the right hand side of
\fig{bfkl_iter}), corresponding to the initial condition in
\eq{bfkl_init}, or the blob may have small-$x$ evolution corrections
included. The evolution corrections can be real (the second term on
the right in \fig{bfkl_iter}) and virtual (the third and fourth terms
on the right of
\fig{bfkl_iter}). The real term contains a gluon in the final state 
(crossing the cut) and corresponds to the first term on the right hand
side of \eq{bfkleq}. The gluon is emitted off the $t$-channel gluons
via the effective Lipatov vertices \cite{BFKL}, denoted by thick dots
in \fig{bfkl_iter}, which represent the sum of all possible
emissions. The virtual terms in \fig{bfkl_iter} contain no gluon in
the final state and correspond to the last term on the right of
\eq{bfkleq}.
%%%%%%%%%%%%%%%%%%%%%%%%%%%%%
\begin{figure}[hb]
\begin{center}
\epsfxsize=5cm
\leavevmode
\hbox{\epsffile{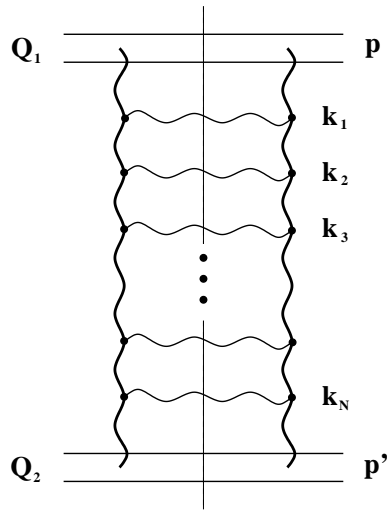}}
\end{center}
\caption{BFKL evolution as a ladder diagram.}
\label{ladder}
\end{figure}
%%%%%%%%%%%%%%%%%%%%%%%%%%%%%%

The solution of the BFKL equation should contain all iterations of the
kernel depicted in \fig{bfkl_iter}. Iterating the BFKL kernel leads
to the ladder diagrams, like the one shown in \fig{ladder}. It depicts
interaction of two onia characterized by typical transverse momentum
scales $Q_1$ and $Q_2$ interacting via a BFKL-evolved
amplitude. \fig{ladder} shows the structure of the blob in
\fig{bfkl_stand} in more detail. Again, the triple gluon vertices in
\fig{ladder} are not the usual QCD vertices: they are effective
Lipatov vertices responsible for the real part of the BFKL kernel
\cite{BFKL}. Similarly, the $t$-channel gluon lines do not correspond
to the usual QCD gluon propagators: they give the so-called reggeized
gluon propagators \cite{BFKL}.

\begin{figure}[h]
\begin{center}
\epsfxsize=18cm
\leavevmode
\hbox{\epsffile{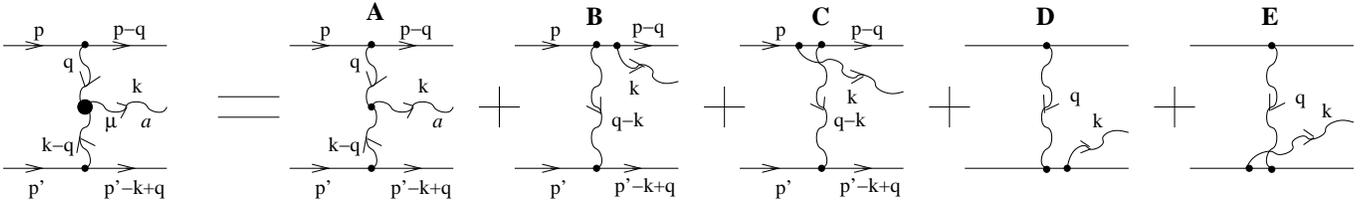}}
\end{center}
\caption{The diagrams contributing to effective Lipatov vertex, which is 
denoted by a very thick dot. }
\label{lvert}
\end{figure}

The emissions resummed by a Lipatov vertex are shown in
\fig{lvert}. There we consider scattering of two ultrarelativistic quarks 
leading to production of a soft gluon with momentum $k^+ \ll p^+$ and
$k^- \ll p'^-$ \cite{BFKL,GB,KMW1,KMW2,KR,GM}. The correct emission
amplitude, which is obtained by summing diagrams A-E in \fig{lvert},
can be written as the first diagram in \fig{lvert} with the effective
vertex triple gluon given by \cite{BFKL,Lipatov1}
\be\label{lipa}
C_\mu^a (k, \underline{q}) \, = \, g \, T^a \, \left( \frac{
\underline{q}^2 }{k^-} - k^+ \, , \, - \frac{
(\underline{k} - \underline{q})^2 }{k^+} + k^- \, , \,
2\underline{q} - \underline{k}
\right)
\ee
in the $(+,-,\perp)$ form with $T^a$ the $SU(N_c)$ color matrix in the
adjoint representation.
%%%%%%%%%%%%%%%%%%%%%%%%%%%%%
%%%%%%%%%%%%%%%%%%%%%%%%%%%%%%

To construct an effective reggeized gluon propagator one has to resum
all leading logarithmic corrections to a single $t$-channel gluon
exchange keeping the exchange amplitude in a color octet state. The
details of this sophisticated resummation procedure can be found in
\cite{FR}. Here we will only give the final result: the leading logarithmic 
corrections to the $t$-channel gluon propagator exponentiate,
modifying the gluon propagator
\be\label{gregg}
\frac{D_{\mu\nu}^{ab} (q)}{q^2} \, \longrightarrow \, \frac{D_{\mu\nu}^{ab} 
(q)}{q^2} \, e^{- \omega ({\un q}^2) \, \Delta y}
\ee
where $\omega ({\un q}^2)$ is the gluon Regge trajectory
\cite{BFKL,Lipatov1}
\be
\omega ({\un q}^2) \, = \, \frac{\as \, N_c}{(2 \pi )^2} \, \int \, 
\frac{d^2 k  \, {\un q}^2}{{\un k}^2 ({\un k} - {\un q})^2}
\ee
and $\Delta y$ is the rapidity interval spanned by a given $t$-channel
gluon.

Eqs. (\ref{lipa}) and (\ref{gregg}) give us the rules for vertices and
$t$-channel propagators necessary to construct the ladder diagram in
\fig{ladder}. There the leading logarithmic contribution is given by 
the multi-Regge kinematics of the produced $s$-channel gluons
\be
p^+ \, \gg \, k_1^+ \, \gg \, k_2^+ \, \gg \, \ldots \, \gg \, k_N^+ \,
\gg \, p'^+,
\ee
\be
p^- \, \ll \, k_1^- \, \ll \, k_2^- \, \ll \, \ldots \, \ll \, k_N^-
\, \ll \, p'^-,
\ee
and
\be\label{regge3}
k_{1}^\perp \, \sim \, k_{2}^\perp \, \sim \, \ldots \, \sim \,
k_N^\perp,
\ee
where $k_i^\perp = |{\un k}_i|$. \\

{\sl Solution of the BFKL Equation} \\

To solve \eq{bfkleq} we first have to find the eigenfunctions of its
integral kernel. The kernel of BFKL equation is conformally
invariant. It is easy to verify that a complete and orthogonal set of
eigenfunctions of \eq{bfkleq} is formed by the functions
\cite{BFKL,Lipatov2}
\be\label{eigf}
({\un l}^2)^{-\frac{1}{2} + i \nu} \, e^{i \, n \, \phi},
\ee
where $\phi$ is the angle between vector ${\un l}$ and some chosen
axes and $n$ is integer. The eigenvalues of the eigenfunctions in
\eq{eigf} are \cite{BFKL,Lipatov2}
\be\label{eigv}
\frac{2 \, \as \, N_c}{\pi} \, \chi (n, \nu)
\ee
where
\be\label{chi}
\chi (n, \nu) \, = \, \psi (1) - \frac{1}{2} \psi \left( \frac{1 + |n|}{2} 
+ i \nu \right) - \frac{1}{2} \psi \left( \frac{1 + |n|}{2} 
- i \nu \right)
\ee
with $\psi (z) = d \ln \Gamma (z) / dz$.  Denoting $l \equiv |{\un
l}|$ and $l' \equiv |{\un l}'|$ we write the solution of \eq{bfkleq}
as
\be\label{ans1}
f ({\un l}, {\un l}', Y) \, = \, \sum_{n=-\infty}^\infty \,
\int_{-\infty}^\infty \, d \nu \, C_{n,\nu} (Y) \, l^{-1 + 2 i \nu} \, 
l'^{-1 - 2 i \nu} \, \, e^{i \, n \, (\phi - \phi')}.
\ee
Substituting \eq{ans1} into \eq{bfkleq} and using the eigenvalues from
\eq{eigv} we find
\be\label{Cnn}
C_{n,\nu} (Y) \, = \, C_{n,\nu}^{(0)} \ e^{\frac{2 \, \as \,
N_c}{\pi} \, \chi (n, \nu) \, Y},
\ee
where the coefficient $C_{n,\nu}^{(0)}$ is fixed by the initial
conditions (\ref{bfkl_init}) giving
\be\label{C0}
C_{n,\nu}^{(0)} \, = \, \frac{1}{2 \pi^2}.
\ee
Combining Eqs. (\ref{Cnn}), (\ref{C0}) and (\ref{ans1}) yields
\be\label{bfklsol} 
f ({\un l}, {\un l}', Y) \, = \, \frac{1}{2 \pi^2} \, \sum_{n=-\infty}^\infty \,
\int_{-\infty}^\infty \, d \nu \,  e^{\frac{2 \, \as \,
N_c}{\pi} \, \chi (n, \nu) \, Y} \, l^{-1 + 2 i \nu} \, l'^{-1 - 2 i
\nu} \, \, e^{i \, n \, (\phi - \phi')}.
\ee
\eq{bfklsol} provides us the solution of \eq{bfkleq} with the initial 
conditions given by \eq{bfkl_init}. As one can see already from
\eq{bfklsol}, the BFKL equation generates amplitudes which grow exponentially 
with rapidity $Y$. Remembering that $Y \sim \ln s$, this translates
into a power of energy growth.

Let us evaluate the amplitude in \eq{bfklsol} a little
further. Consider the case when $l \sim l'$, i.e., the two momentum
scales involved in the problem are not very much different from each
other. A simple analysis of the function $\chi (n, \nu)$ allows one to
conclude that the dominant contribution to the amplitude is given by
the $n=0$ term in the sum in \eq{bfklsol}. Expanding $\chi (n=0, \nu)$
around the saddle point at $\nu = 0$ we get
\be\label{sp}
\chi (0, \nu) \, \approx \, 2 \ln 2 - 7 \, \zeta (3) \, \nu^2,
\ee
where $\zeta (z)$ is the Riemann zeta-function. Using \eq{sp} in
\eq{bfklsol} we can perform the $\nu$-integration obtaining \cite{BFKL}
\be\label{spa}
f ({\un l}, {\un l}', Y) \, \approx \, \frac{1}{2 \, \pi^2 \, l \, l'}
\, \sqrt{\frac{\pi}{14 \, \zeta (3) \, \bas \, Y}} \ \exp \left[ 
(\alpha_P - 1) \, Y - \frac{\ln^2 l/l'}{14 \, \zeta (3) \, \bas \, Y}
\right],
\ee
where we have defined the {\sl intercept} of the perturbative pomeron
\be\label{bfkl_int}
\alpha_P - 1 \, = \, \frac{4 \, \as \, N_c}{\pi} \, \ln 2
\ee
and
\be
\bas \, \equiv \, \frac{\as \, N_c}{\pi}.
\ee
The essential feature of \eq{spa} is that it shows that cross sections
mediated by the BFKL exchange grow as a power of energy 
\be\label{poe}
\sigma \, \sim \, e^{(\alpha_P - 1) \, Y} \, \sim \, s^{\alpha_P - 1}.
\ee 
This behavior is reminiscent of the Pomeranchuk singularity in the
reggeon calculus, and is, therefore, sometimes referred to as the pomeron
or the hard pomeron (to distinguish it from the soft non-perturbative
interaction phenomenon). \\

%%%%%%%%%%%%%%%%%%%%%%%%%%%%%
\begin{figure}[b]
\begin{center}
\epsfxsize=7cm
\leavevmode
\hbox{\epsffile{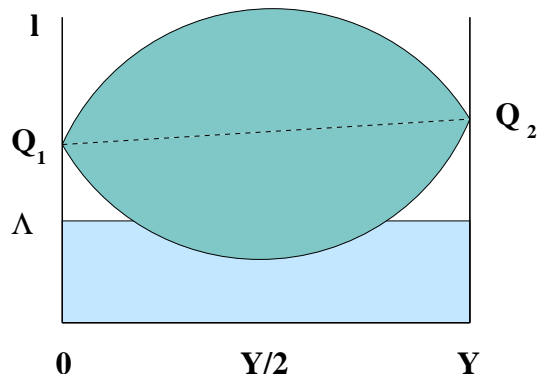}}
\end{center}
\caption{Diffusion of the gluon transverse momentum in the BFKL ladder. 
Non-perturbative region is denoted by the scale $\Lambda$.}
\label{cone}
\end{figure}
%%%%%%%%%%%%%%%%%%%%%%%%%%%%%%

{\sl Problems of the BFKL Evolution} \\

The BFKL equation poses some important questions even in the case of
heavy onium-onium scattering and at the leading order in $\as$ in the
kernel.

{\bf (i)} The power of energy growth of the total cross section
(\eq{poe}) violates Froissart unitarity bound, which states that the
growth of the total cross sections with energy at asymptotically high
energies is bounded by \cite{froi}
\be\label{fro}
\sigma \, \le \, \frac{const}{m_\pi^2} \, \ln^2 s
\ee
with $m_\pi$ the pion's mass. (For a good pedagogical derivation of
the Froissart bound we refer the readers to \cite{genya}.) This
implies that some new physical effects should modify the BFKL equation
at very large $s$ making the resulting amplitude unitary.

{\bf (ii)} The solution in \eq{bfklsol} includes a diffusion term,
which is the last term in its exponent. To see the potential danger of
this term, let us consider a half of the ladder of \fig{ladder},
stretching from one of the onia (the top one) to some intermediate
gluon in the middle of the ladder carrying transverse momentum
$l_\perp$ and having rapidity $Y/2$. Applying \eq{spa} to that
half-ladder we see that it includes a term
\be\label{diff}
\exp \left[-{\ln^2 (l/Q_1) \over 14 \, \zeta (3) \, \bas \, \, (Y/2)}\right].
\ee
This term is responsible for diffusion of the transverse momenta from
the initial perturbative scale $Q_1$ both to high and low momenta,
i.e., into infrared and ultraviolet. It implies that the distribution
of gluons' transverse momentum in the ladder, while still centered
around $Q_1$, may have significant fluctuations towards high and low
momenta $l$ as shown in \fig{cone}, where we plot the typical range of
transverse momentum $l$ in the ladder of \fig{ladder} as a function of
rapidity of gluons in the ladder. The width of the diffusion grows
with rapidity $Y$ allowing for larger fluctuations at higher
energies. Thus, no matter how large the starting scale $Q_1$ is, at
certain very high energy the momentum of some gluons in the middle of
the ladder would become of the order of $\Lambda_{QCD}$ leading to the
coupling constant $\as (\Lambda_{QCD}) \sim 1$ and thus invalidating
further application of perturbative QCD ($\as \ll 1$) and,
consequently, of the BFKL evolution \cite{Bartels}. The diffusion
starts from the scale $Q_1$ at one end of the ladder and from the
scale $Q_2$ at the other end (see
\fig{cone}). At high energy the allowed momentum range broadens
towards the middle and touches the non-perturbative region of low
momenta at mid-rapidity. The plot of allowed momenta as a function of
rapidity shown here in \fig{cone} is sometimes referred to as Bartels
cone \cite{Bartels}. Again it hints that the BFKL equation should be
modified at higher energies to avoid the problem of running into the
non-perturbative region. Alternatively, if such modification is not
found within perturbation theory, we would be forced to admit that
high energy asymptotics is dominated by non-perturbative
physics. Fortunately this is not the case, as will be shown below.

\subsubsection{The GLR-MQ Equation}

\begin{figure}[ht]
\begin{center}
\epsfxsize=5cm
\leavevmode
\hbox{\epsffile{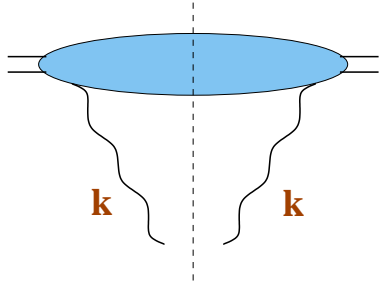}}
\end{center}
\caption{Unintegrated gluon distribution of an onium.}
\label{gluelo}
\end{figure}

As we have seen above, the BFKL evolution leads to exponential growth
of total cross sections with energy, violating the Froissart bound
\cite{froi}. It also leads to exponential growth of the density of partons 
in the onium (or hadron) wave function. To see this let us define an
unintegrated gluon distribution of an onium by
\be\label{oniglue}
\phi (x_{Bj}, {\un k}^2) \, = \, \frac{\as \, C_F}{\pi} \, \int \, 
\frac{d^2 l}{\underline{l}^2} \, (2 - e^{- i
\underline{l} \cdot \underline{x}} - e^{ i \underline{l} \cdot
\underline{x}} ) \, f ({\un l}, {\un k}, Y = \ln 1/x_{Bj}) \, d^2 x \, dz \, 
\Phi_{q\bar q} ({\un x}, z).
\ee
The definition (\ref{oniglue}) is illustrated in \fig{gluelo}. To
obtain it we have, essentially, truncated the lower onium in
\fig{bfkl_stand}, leaving two disconnected gluon lines. The diagram in
\fig{gluelo} has to be calculated in $A^+ = 0$ light cone gauge to
give the gluon distribution function.  The unintegrated gluon
distribution function in \eq{oniglue} gives us the number of gluons in
the onium wave function having transverse momentum ${\un k}$ and
carrying the fraction $x_{Bj}$ of the onium ``+'' component of the
momentum (Bjorken, or Feynman $x$).
%%%%%%%%%%%%%%%%%%%%%%%%%%%%%
%%%%%%%%%%%%%%%%%%%%%%%%%%%%%%

Using the BFKL solution from \eq{spa} in \eq{oniglue} one can easily
see that the gluon distribution grows as
\be
\phi (x_{Bj}, {\un k}^2) \, \sim \, \left( \frac{1}{x_{Bj}} 
\right)^{\alpha_P - 1} .
\ee
Therefore, the number of gluons rises sharply at small $x$ / high
energy, in agreement with the semi-qualitative estimate of
\eq{expglue}. This feature is illustrated qualitatively by the
gluon cascade representation of the BFKL evolution presented in the
Sect. \ref{BFKLeq}. There, the gluons are produced in a multi-Regge
kinematics with comparable transverse momenta (see \eq{regge3}). That
means that the typical transverse sizes of the gluons, given by
$r_i^\perp \sim 1/k_i^\perp$ are also of the same order for all the
gluons
\be
r_1^\perp \, \sim \, r_2^\perp \, \sim \, \ldots \, \sim \, r_N^\perp.
\ee
Therefore, the BFKL cascade produces many gluons in the onium or
hadron wave function, with roughly the same transverse size. As energy
increases, more and more gluons are produced in the cascade. The
gluons overlap in the transverse plane, creating areas of high gluon
density. Thus, not only the {\it number} of gluons, but their {\it
density} in the transverse plane increase with energy.
%%%%%%%%%%%%%%%%%%%%%%%%%%%%%
\begin{figure}[ht]
\begin{center}
\epsfxsize=15cm
\leavevmode
\hbox{\epsffile{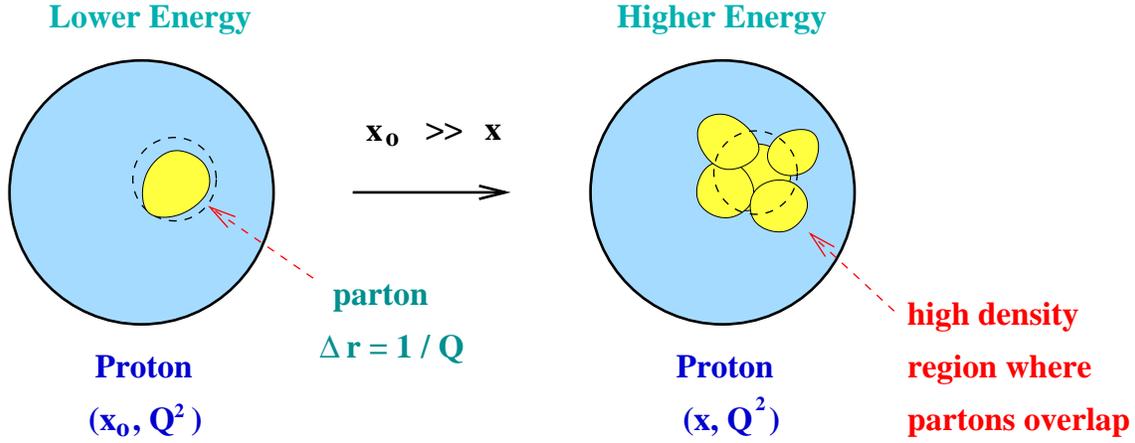}}
\end{center}
\caption{BFKL equation as a high density machine.}
\label{hdens}
\end{figure}
%%%%%%%%%%%%%%%%%%%%%%%%%%%%%%
This is illustrated in \fig{hdens} for a wave function of a proton. At
some initial value of Bjorken $x$, corresponding to lower energy, the
proton's wave function may have one parton with transverse size $\sim
1/Q$ in it. As we go to smaller $x$, BFKL evolution would generate
many more partons of comparable size, creating a region of high gluon
density in the wave function
\cite{MuellerCarg}.

However, the gluon density can not rise forever. It is known that in
QCD the gluon fields can not be stronger than $A_\mu \sim 1/g$ for
small coupling $g$. Therefore, when the gluon field reaches the
density corresponding to field strength 
\be
\frac{F_{\mu \nu}}{Q^2} \, \sim \, \frac{1}{g} 
\ee
some new, possibly non-linear effects should become important slowing
down the density growth \cite{MuellerCarg}. This strong field
constraint may be intimately connected to the problem of unitarization
of cross sections at fixed impact parameter, i.e., to the black disk
limit (see below).

\begin{figure}[ht]
\begin{center}
\epsfxsize=10cm
\leavevmode
\hbox{\epsffile{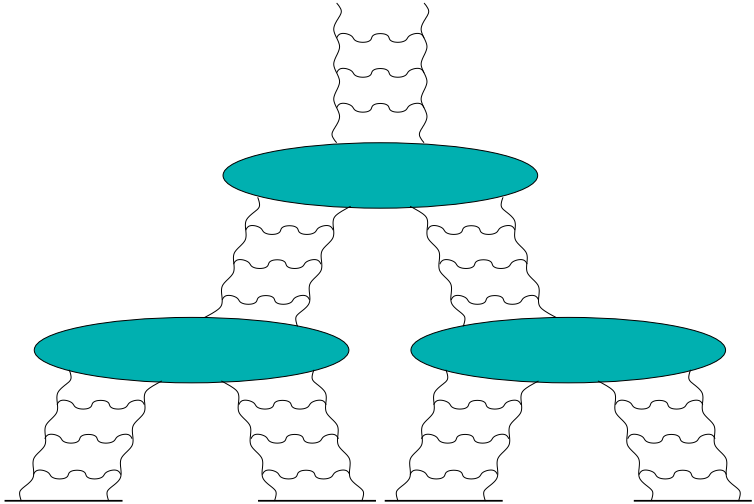}}
\end{center}
\caption{An example of fan diagram resummed by the GLR-MQ equation.}
\label{fan}
\end{figure}

To understand how the growth of the gluon distribution can be tamed,
Gribov, Levin and Ryskin (GLR) \cite{GLR,GLR2} considered distribution
functions of a ``dense'' proton or a nucleus. By ``dense'' proton we
imply a model of a proton filled with sources of color charge --- sea
quarks and gluons, which were pre-created in the proton's wave
function by some non-perturbative mechanism (see, e.g.,
\cite{Chevy}). Gribov, Levin and Ryskin  \cite{GLR,GLR2} argued that 
for such systems multiple ladder exchanges may become important. Since
one is interested in gluon distribution, which is a correlation
function for two gluonic fields, these multiple ladders should come in
as the so-called ``fan'' diagrams. An example of a fan diagram is
shown in \fig{fan}.
%%%%%%%%%%%%%%%%%%%%%%%%%%%%%
%%%%%%%%%%%%%%%%%%%%%%%%%%%%%%
There multiple BFKL ladders start from different quarks and gluons in
the proton or nucleus, shown by straight lines at the bottom of
\fig{fan}. Due to high density of gluon fields, the ladders 
can not stay independent forever. As the energy increases so does the
gluon density, eventually leading to {\it recombination} of the
ladders, as shown in \fig{fan}. Ladder recombination is described by
effective ladder merger vertices, denoted by blobs in \fig{fan}. These
vertices are sometimes called the triple pomeron vertices, since they
connect three different ladders (BFKL pomerons). For their calculation
we refer the reader to \cite{BW} and references therein.

Gribov, Levin and Ryskin \cite{GLR,GLR2} suggested that, before the
energy gets sufficiently high for all nonlinear effects to become
important, there could be an intermediate energy region where the
physics of gluon distributions is dominated by $2 \rightarrow 1$
ladder recombination only. This recombination brought in a quadratic
correction to the linear BFKL equation, leading to the GLR evolution equation
\cite{GLR,GLR2}
\be\label{glreq}
\frac{\partial \, \phi (x, {\un k}^2)}{\partial \ln (1/x)} \, = \, 
\frac{\as \, N_c}{\pi^2} \int \! \frac{d^2
l}{(\underline{k} - \underline{l})^2} \left[ \phi
(x, \underline{l}^2) - \frac{{\un k}^2 \, \phi
(x, \underline{k}^2)}{{\un l}^2 + (\underline{k} -
\underline{l})^2} \right] - \frac{\as^2 \, \pi}{S_\perp} \, \left[ \phi
(x, \underline{k}^2) \right]^2,
\ee
where, for simplicity, we assumed that the proton or nucleus has a
shape of a cylinder oriented along the beam axis with the cross
sectional area $S_\perp = \pi R^2$. As expected, the linear term in
\eq{glreq} is equivalent to the BFKL equation (\ref{bfkleq}), while the 
quadratic term, responsible for ladder mergers, introduces damping,
thus slowing down the growth of the gluon distributions with
energy. The growth of gluon distributions with energy given by
\eq{glreq}  should slow down and {\sl saturate} at very high 
energies. This phenomenon became known as the {\sl saturation} of
parton distributions.

The ansatz (\ref{glreq}) of GLR \cite{GLR} was proven by Mueller and
Qiu \cite{MQ} in the double leading logarithmic approximation (DLA)
for the (integrated) gluon distribution functions, which are defined
as
\cite{MQ}
\be\label{xg}
x G (x, Q^2) \, = \, \int^{Q^2} \, d k^2 \, \phi (x, k^2).
\ee
The double logarithmic approximation is a resummation of the powers of
the parameter
\be
\as \ln (1/x) \ln (Q^2/\Lambda^2).
\ee 
The BFKL equation \cite{BFKL} was derived in the leading logarithmic
approximation (LLA), corresponding to resummation of the parameter
$\as \ln 1/x$, or, equivalently, $\as Y$ (see Sect. \ref{BFKLeq}). In
the limiting case of large $Q^2$ in the distribution functions (or
large ${\un k}^2$ in the unintegrated distribution functions) another
large logarithm becomes important: $\ln Q^2/\Lambda^2$, where
$\Lambda$ is the non-perturbative QCD scale. It becomes possible to
define a new resummation parameter, $\as \ln (1/x) \ln
(Q^2/\Lambda^2)$. In principle, the LLA is much broader than DLA: it
resums powers of $\as
\ln 1/x$ with {\it any} dependence of the obtained terms on $Q^2$, not 
restricting it to the leading logarithmic regime. Leading powers of
$\as \ln (Q^2/\Lambda^2)$ are, of course, resummed by the Dokshitzer,
Gribov, Lipatov, Altarelli, Parisi (DGLAP) equation
\cite{DGLAP}. Indeed, the DLA limits of the BFKL and DGLAP equations 
are identical, since they are resumming the same parameter, $\as \ln
(1/x) \ln (Q^2/\Lambda^2)$.

Employing the DLA and analyzing diagrams with two merging DGLAP
ladders, Mueller and Qiu arrived at the following evolution equation
\cite{MQ} (again written here for a cylindrical nucleus)
\begin{eqnarray}\label{mq}
  \frac{\partial^2 \, xG (x, Q^2) }{\partial \ln (1/x)\, \partial \ln
  (Q^2 / \Lambda^2) } \, = \, \frac{\as \, N_c}{\pi} \, xG (x, Q^2) -
  \frac{\as^2 \, \pi}{S_\perp} \, \frac{1}{Q^2} \, [xG (x, Q^2)]^2,
\end{eqnarray}
which is known as GLR-MQ equation. \eq{mq} is in agreement with
\eq{glreq}, and could be be obtained from the latter by taking the DLA 
limit and using \eq{xg}. Thus Mueller and Qiu \cite{MQ} proved
\eq{glreq} in the double logarithmic limit.

\eq{mq} allows one to estimate at which $Q^2$ the non-linear saturation 
effects become important. To do that we have to equate the linear and
quadratic terms on the right hand side of \eq{mq}. The corresponding
value of $Q^2$ is called the {\sl saturation scale} and is denoted by
$Q_s^2$. It is determined by
\be
Q_s^2 \, = \, \frac{\as \, \pi^2}{S_\perp \, N_c} \, xG (x, Q_s^2).
\ee
The non-linear saturation effects are important for all $Q \, \lsim \,
Q_s$, which is known as the {\sl saturation region}.

The quadratic damping term in both \eq{glreq} and \eq{mq} was believed
to be important only near the border of the saturation region, for $Q
\sim Q_s$, where the non-linear effects were only starting to become
important \cite{GLR,MQ}. It was expected that higher order non-linear
corrections would show up as one goes deeper into the saturation
region towards $Q < Q_s$ (see, e.g., \cite{AGL}). In the next Section
we will talk about the model where all such corrections could be
resummed in a particular quasi-classical approximation, where the BFKL
evolution can be neglected.

%%%%%%%%%%%%%%%%%%%%%%%%%%%%%%%%%%%%%%%%%%%%%%%%%%%%%%%%%%%%%%%%%%%%%%%%%%%%%%%%%%%%%%%%%%%%%%

\subsection{Quasi-Classical Approximation}
\label{QCA}

As was suggested by the GLR equation, the effects of high gluonic
field strengths come in as multiple exchanges. To better understand
the effect of multiple rescatterings one can consider a specific model
of hadronic or nuclear wave functions. In this section we will
consider the case of deep inelastic scattering on a large dilute
nucleus, where multiple rescatterings take place on individual
nucleons and, as we will demonstrate, can be described by the
Glauber-Mueller formula \cite{GlaMue}. In the infinite momentum frame
this large nucleus can be thought of as a Lorentz-contracted
``pancake'' in the direction transverse to its velocity. Due to a
large number of nucleons in the nucleus, the parton color charge
density fluctuations in this ``pancake'' are large, described by a
hard scale referred to as the {\sl saturation scale} $Q_s$. For very
large nuclei, the saturation scale is large $Q_s \gg \Lambda_{QCD}$
making the strong coupling constant small $\as (Q_s) \ll 1$. At weak
coupling the dynamics of field theories becomes classical: therefore,
as was concluded by McLerran and Venugopalan \cite{MV}, the gluon
field of such a large ultrarelativistic nucleus should be given by the
solution of the classical Yang-Mills equations of motion \cite{YM}
with the nucleus providing the source current. The resulting small-$x$
nuclear wave function systematically includes all multiple
rescatterings and exhibits the effects of saturation.

\subsubsection{Glauber-Mueller Rescatterings}
\label{GM}

%%%%%%%%%%%%%%%%%%%%%%%%%%%%%
\begin{figure}[b]
\begin{center}
\epsfxsize=7cm
\leavevmode
\hbox{\epsffile{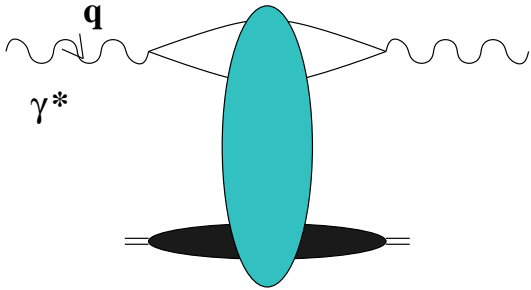}}
\end{center}
\caption{Deep inelastic scattering in the rest frame of the target.}
\label{dis}
\end{figure}
%%%%%%%%%%%%%%%%%%%%%%%%%%%%%%
Let us consider deep inelastic scattering (DIS) on a large nucleus. In
DIS the incoming electron emits a virtual photon, which in turn
interacts with the proton or nucleus. In the rest frame of the
nucleus, the interaction can be thought of as virtual photon splitting
into a quark-antiquark pair, which then interacts with the nucleus
(see \fig{dis}). Since the light cone lifetime of the $q\bar q$ pair
is much longer than the size of the target nucleus, the total cross
section for the virtual photon--nucleus scattering can be written as a
convolution of the virtual photon's light cone wave function (the
probability for it to split into a $q\bar q$ pair) with the forward
scattering amplitude of a $q\bar q$ pair interacting with the nucleus
\be\label{sigN}
\sigma_{tot}^{\gamma* A} (Q^2, x_{Bj}) \, = \,  \int \frac{d^2 x \, d z }{2 \pi} \, 
[\Phi_T ({\un x}, z) + \Phi_L ({\un x}, z) ] \ d^2 b \ N({\un x}, {\un
b} , Y)
\ee
with the help of the light-cone perturbation theory \cite{BL,BPP}.
Here the incoming photon with virtuality $Q$ splits into a
quark--antiquark pair with the transverse separation ${\un x}$ and the
impact parameter (transverse position of the center of mass of the
$q\bar q$ pair) ${\un b}$. $Y$ is the rapidity variable given by $Y =
\ln (s \, x_T^2) \approx \ln 1/x_{Bj}$. The square of the light cone 
wave function of $q \overline{q}$ fluctuations of a virtual photon is
denoted by $\Phi_T ({\un x}, z)$ and $\Phi_L ({\un x}, z)$ for
transverse and longitudinal photons correspondingly, with $z$ being
the fraction of the photon's longitudinal momentum carried by the
quark. At the lowest order in electromagnetic coupling ($\alpha_{EM}$)
$\Phi_T ({\un x}, z)$ and $\Phi_L ({\un x}, z)$ are given by
\cite{NZ,KMc}
\begin{equation}
  \Phi_T ({\un x}, z) = \frac{2 N_c}{\pi} \, \sum_f \alpha^f_{EM} \,
  \left\{ a_f^2 \ K_1^2 (x_T a_f) \ [z^2 + (1 - z)^2] + m_f^2 K_0 (x_T
  a_f)^2 \right\},
\end{equation}
\begin{equation}
  \Phi_L ({\un x}, z) = \frac{2 N_c}{\pi} \, \sum_f \alpha^f_{EM} \, \
  4 \, Q^2 \, z^2 (1 - z)^2 \ K_0^2 (x_T a_f),
\end{equation}
with $a_f^2 = Q^2 z (1 - z) + m_f^2$, $x_T = |{\bf x}|$ and $\sum_f$
denoting the sum over all relevant quark flavors with quark masses
denoted by $m_f$. $\alpha^f_{EM} = e_f^2 /4 \pi$ with $e_f$ the
electric charge of a quark with flavor $f$.

\begin{figure}
\begin{center}
\epsfxsize=12cm
\leavevmode
\hbox{\epsffile{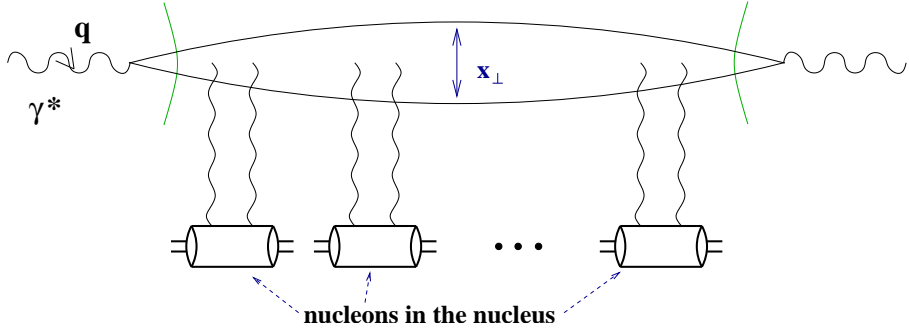}}
\end{center}
\caption{Deep inelastic scattering in the quasi-classical 
Glauber-Mueller approximation in $\partial_\mu A^\mu =0$ gauge.}
\label{qqbar}
\end{figure}

Our goal is to calculate the forward scattering amplitude of a
quark--anti-quark dipole interacting with the nucleus, which is
denoted by $N({\un x}, {\un b} , Y)$ in \eq{sigN}, including all
multiple rescatterings of the dipole on the nucleons in the
nucleus. To do this we need to construct a model of the target
nucleus. Following Mueller \cite{GlaMue} we assume that the nucleons
are dilutely distributed in the nucleus. Let us chose to perform
calculations in the $\partial_\mu A^\mu =0$ covariant gauge. There we
can represent the dipole-nucleus interaction as a sequence of
successive dipole-nucleon interactions, as shown in \fig{qqbar}. Since
each nucleon is a color singlet, the lowest order dipole-nucleon
interaction in the forward amplitude from \fig{qqbar} is a two-gluon
exchange. The exchanged gluon lines in \fig{qqbar} are disconnected at
the top: this denotes a summation over all possible connections of
these gluon lines either to the quark or to the anti-quark lines in
the incoming dipole.

%%%%%%%%%%%%%%%%%%%%%%%%%%%%%
%%%%%%%%%%%%%%%%%%%%%%%%%%%%%%

%%%%%%%%%%%%%%%%%%%%%%%%%%%%%
\begin{figure}[ht]
\begin{center}
\epsfxsize=5cm
\leavevmode
\hbox{\epsffile{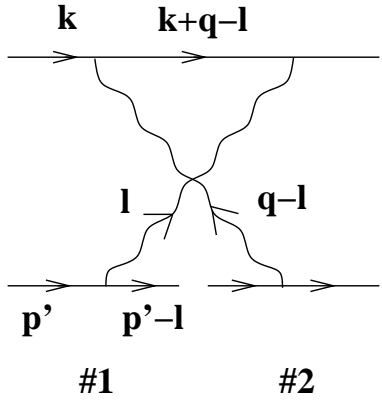}}
\end{center}
\caption{An example of a diagram which does not contribute to the 
total dipole-nucleus cross section due to incorrect ordering of
interactions with nucleons ($x_2^+ > x_1^+$).}
\label{cross}
\end{figure}
%%%%%%%%%%%%%%%%%%%%%%%%%%%%%%
To demonstrate that the ordered gluon exchanges of \fig{qqbar} are
indeed the only possible interactions, let us consider a diagram where
such ordering is broken, as shown in \fig{cross}. There, for
simplicity, we consider scattering of a single quark on a pair of
nucleons, denoted as $\# 1$ and $\# 2$, with interchanged ordering of
quark-nucleon rescatterings. To show that such diagrams are suppressed
in the dilute nucleus case, let us consider the scattering in the
center of mass frame, where the projectile quark is moving in the
light cone ``$+$'' direction and the nucleus is moving in the light cone
``$-$'' direction. In a dilute nucleus all the nucleons are spatially
separated. For a nucleus moving in the light cone ``$-$'' direction this
translates into a separation between the nucleons along the ``$+$'' axis:
specifically, let us denote $x_1^+$ and $x_2^+$ the light cone
coordinates of the nucleons $\# 1$ and $\# 2$ in
\fig{cross}, choosing $x_2^+ > x_1^+$ for certainty. (As we will see, 
the opposite ordering will be indeed allowed.) We are interested in
the integration over the $l$-momentum in the diagram of
\fig{cross}. More specifically, we are interested in integrating over
the ``$-$'' component of this momentum. In the high energy scattering
case considered here, the incoming projectile quark carries large
momentum $k^+$, and the nucleons in the nucleus carry a large ``$-$''
component of momentum, e.g., $p'^-$ carried by a quark in the nucleon
$\# 1$. In such eikonal approximation the quark-gluon vertices do not
depend on $l$, so that all the $l$-dependence is given by the gluon
propagators for the $l$ and $q-l$ lines, and by the quark propagator
of the $k+q-l$ line. By requiring that the outgoing quark in the
nucleon $\# 1$ is on mass-shell, $(p'-l)^2 =0$ for massless quarks, we
obtain $0=(p'-l)^2 \approx -2 p'^- l^+$, such that with the eikonal
accuracy $l^+ = 0$. This means that the propagator of the $l$-line,
given in general by $-i/l^2$, becomes $i/{\un l}^2$ and does not
depend on $l^-$ anymore. Similarly one can show that the propagator
for the $q-l$-line is $i/({\un q} - {\un l})^2$ and is also
independent of $l^-$. Thus all the $l^-$-dependence of the diagram in
\fig{cross} is given by the denominator of the propagator of the $k+q-l$
quark line, since the numerator also becomes $l^-$-independent in the
eikonal limit.  The relevant $l^-$ integral can be written as
\be\label{qp1}
\int_{-\infty}^\infty \frac{d l^-}{(k+q-l)^2 + i \epsilon} \, 
e^{- i \, l^- \, (x_2^+ - x_1^+)}.
\ee
Since $k^+$ is very large, we rewrite \eq{qp1} as
\be\label{qp2}
\int_{-\infty}^\infty \frac{d l^-}{2 \, k^+ \, (k^- + q^- - l^-) 
- ({\un k} + {\un q} - {\un l})^2 + i \epsilon} \, 
e^{- i \, l^- \, (x_2^+ - x_1^+)}.
\ee
Since $k^+ >0$ the integrand in \eq{qp2} has a pole in the upper half
plane. However, as $x_2^+ > x_1^+$, the exponent in \eq{qp2} allows us
to close the integration contour in the lower half plane
only. Therefore expression in \eq{qp2} is zero, which means that the
diagram in \fig{cross} can be neglected in the eikonal limit. (In
fact, \eq{qp2} dictates which ordering of the nucleons is allowed: if
$x_2^+ < x_1^+$ the expression in \eq{qp2} becomes non-zero, bringing
us back to ordered picture of interactions shown in \fig{qqbar}.)
%%%%%%%%%%%%%%%%%%%%%%%%%%%%%
\begin{figure}
\begin{center}
\epsfxsize=11cm
\leavevmode
\hbox{\epsffile{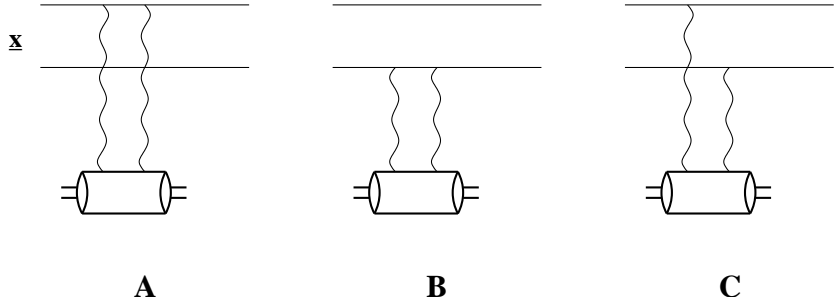}}
\end{center}
\caption{Diagrams contributing to dipole-nucleon scattering at the two-gluon order.}
\label{prop}
\end{figure}
%%%%%%%%%%%%%%%%%%%%%%%%%%%%%%

To resum the diagrams like the one shown in \fig{qqbar} we start by
calculating the graphs contributing to the scattering of a dipole on a
single nucleon, as shown in \fig{prop}. Remembering that we are
working in the $\partial_\mu A^\mu =0$ gauge and assuming for the
moment that all scattering in the nucleon happens on one of its
valence quarks we write for the sum of the diagrams in \fig{prop}
(similarly to \eq{twogl})
\be\label{1sc1}
\frac{\as^2 \, C_F}{N_c} \, \int \frac{d^2 l}{[{\un l}^2]^2} \, (2 - 
e^{- i {\un l} \cdot {\un x}} - e^{i {\un l} \cdot {\un x}}),
\ee
where ${\un x}$ is the transverse separation between the quark and the
anti-quark in the incoming dipole, as shown in \fig{prop}. Instead of
performing the integration in \eq{1sc1} exactly, let us note that
gluon are correct degrees of freedom responsible for the interactions
only if the dipole is much smaller than the typical nucleon size,
$x_\perp \ll 1/\Lambda$, with $\Lambda$ some non-perturbative infrared
cutoff. Expanding the argument of the integral in
\eq{1sc1} and integrating over ${\un l}$ we obtain (see also \cite{BDMPS1})
\be\label{1sc2}
\frac{\as^2 \, \pi \, C_F}{N_c} \, x_\perp^2 \, \int_\Lambda^{1/x_\perp} 
\frac{d l_T}{l_T} \, = \, \frac{\as^2 \, \pi \, C_F}{N_c} \, x_\perp^2 \, 
\ln \frac{1}{x_\perp \, \Lambda}.
\ee
As one can show by explicitly calculating the diagram in \fig{gluelo}
without the evolution (or by using \eq{bfkl_init} in \eq{oniglue},
replacing $q\bar q$ dipole by a valence quark in the latter, and using
the resulting unintegrated gluon distribution in \eq{xg}), at the
lowest two-gluon order the gluon distribution function of a valence
quark is given by
\be\label{xglo}
x_{Bj} G (x_{Bj}, Q^2) \, = \, \frac{\as \, C_F}{\pi} \, \ln
\frac{Q^2}{\Lambda^2}.
\ee
Using \eq{xglo} in \eq{1sc2} yields
\be\label{1sc3}
\frac{\as \, \pi^2}{2 \, N_c} \,  x_\perp^2 \, x_{Bj} G (x_{Bj}, 1/x_\perp^2).
\ee
\eq{1sc3} can be easily generalized to the case of a nucleon by using the 
gluon distribution of a nucleon $xG_N$ in it instead of that of a
valence quark
\be\label{1sc4}
\frac{\sigma^{q{\bar q}N}}{2} \, = \, 
\frac{\as \, \pi^2}{2 \, N_c} \,  x_\perp^2 \, x_{Bj} G_N (x_{Bj}, 1/x_\perp^2).
\ee
\eq{1sc4} gives us the scattering cross section of a quark dipole on 
a single nucleon. The factor of $2$ is due to the definition of total
cross section in terms of the forward scattering amplitude, as shown
in \eq{onium_tot}.

To obtain the multiple rescattering amplitude of \fig{qqbar} we have
to resum all multiple rescatterings of \fig{prop} (or, equivalently,
\eq{1sc4}). Defining the $S$-matrix of a quark--anti-quark pair scattering 
on a nucleus at rest by $S(z, {\un x})$, where $z$ is the longitudinal
coordinate of the pair as it propagates through nuclear matter with
$z=0$ at the edge of the nucleus, we can write the following equation
\be\label{glaprop}
\frac{\partial}{\partial z} S(z, {\un x}) \, = \, - 
\frac{\rho \, \sigma^{q{\bar q}N}}{2} S(z, {\un x})
\ee
with $\rho$ the density of nucleons in the nucleus ($\rho = A / [(4/3)
\pi R^3]$ for a spherical nucleus of radius $R$ with atomic number $A$.) 
\eq{glaprop} has the following physical meaning: as the dipole propagates 
through the nucleus it may encounter nucleons, with the probability of
interaction per unit path length $dz$ given by the product of nucleon
density $\rho$ and the interaction probability $\sigma^{q{\bar q}N}/2$
from \eq{1sc4}. (In the literature devoted to jet quenching in the
medium this quantity is referred to as opacity \cite{EL,EL2}.) The
initial condition for \eq{glaprop} is given by a freely propagating
dipole without interactions, such that $S (z=0, {\un x}) = 1$. We
refer the interested reader to \cite{BDMPS1,KM,BDMPS} for a more
detailed derivation of \eq{glaprop}. Solving \eq{glaprop} with $S
(z=0, {\un x}) = 1$ initial condition yields
\be\label{glaS}
S(z, {\un x}) \, = \, \exp \left\{ - \frac{\rho \,
\sigma^{q{\bar q}N}}{2} \, z \right\}.
\ee
To construct the forward scattering amplitude $N$ of a dipole
scattering on a nucleus we need to know the $S$-matrix for the dipole
which went through the whole nucleus. To obtain it we need to put $z =
T({\un b})$ in \eq{glaS}, where $T({\un b})$ is the nuclear profile
function equal to the length of the nuclear medium at a given impact
parameter $\un b$, such that $T({\un b}) = 2 \, \sqrt{R^2 - {\un
b}^2}$ for a spherical nucleus of radius $R$. Identifying $N({\un x},
{\un b} , Y)$ with the scattering $T$-matrix, such that $N = 1-S$ we
finally write
\cite{GlaMue}
\be\label{glaN}
N({\un x}, {\un b} , Y=0) \, = \, 1 - \exp \left\{ - \frac{\rho \,
\sigma^{q{\bar q}N}}{2} \, T({\un b}) \right\} \, = \, 1 - 
\exp \left\{ - \frac{\as \, \pi^2}{2 \, N_c} \,  \rho \, T({\un b}) \,
x_\perp^2 \, x_{Bj} G_N (x_{Bj}, 1/x_\perp^2) \right\}.
\ee
This expression is known as Glauber-Mueller formula
\cite{Glauber,GlaMue}, since it presents a generalization of the
Glauber model of independent multiple rescatterings in the nucleus
\cite{Glauber} to the case of QCD \cite{GlaMue}.

We put $Y=0$ in the argument of $N$ in \eq{glaN} to underline that
this expression does not include any small-$x$ evolution which would
bring in the rapidity dependence. Indeed, \eq{glaN} was derived in the
approximation where the interaction of the dipole with each of the
nucleons is limited to a two-gluon exchange. At this order the gluon
distribution $x_{Bj} G_N$ is given by \eq{xglo} and is, therefore,
$x_{Bj}$-independent, leading to rapidity-independence of the whole
expression in \eq{glaN}.

\eq{glaN} allows us to determine the parameter corresponding to the 
resummation of the diagrams like the one shown in \fig{qqbar}. Using
the gluon distribution from \eq{xglo} in \eq{glaN}, and noting that
for large nuclei the profile function scales as $T({\un b}) \sim
A^{1/3}$ and the nucleon density scales as $\rho \sim A^0$, we
conclude that the resummation parameter of multiple rescatterings is
\cite{yuri2}
\be\label{mrp}
\as^2 \, A^{1/3}.
\ee
The physical meaning of the parameter $\as^2 \, A^{1/3}$ is rather
straightforward: at a given impact parameter the dipole interacts with
$\sim A^{1/3}$ nucleons exchanging two gluons with each. Since the
two-gluon exchange is parametrically of the order $\as^2$ we obtain
$\as^2 \, A^{1/3}$ as the resummation parameter.
%%%%%%%%%%%%%%%%%%%%%%%%%%%%%
\begin{figure}
\begin{center}
\epsfxsize=9cm
\leavevmode
\hbox{\epsffile{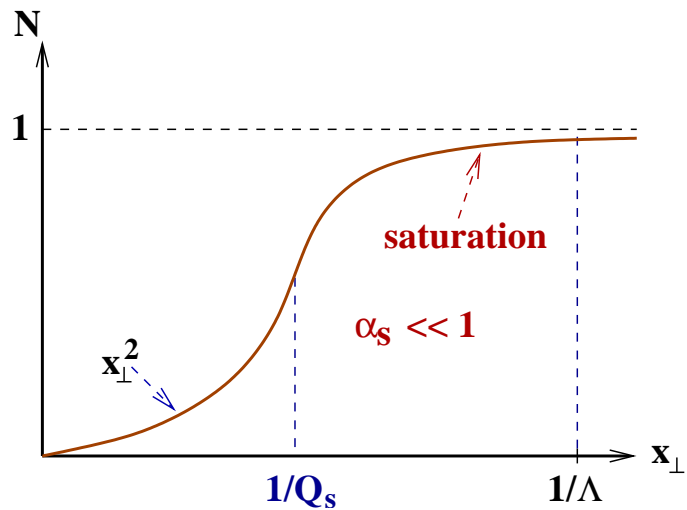}}
\end{center}
\caption{The forward amplitude of the dipole--nucleus scattering $N$ 
plotted as a function of the transverse separation between the quark
and the anti-quark in a dipole $x_\perp$ using \eq{glaN2}. }
\label{glamue}
\end{figure}
%%%%%%%%%%%%%%%%%%%%%%%%%%%%%%

Defining the quark {\sl saturation} scale
\be\label{qsmv}
Q_s^2 ({\un b}) \, \equiv \, \frac{4 \, \pi \, \as^2 \, C_F}{N_c} \, \rho \, T({\un b})
\ee
we rewrite \eq{glaN} as
\be\label{glaN2}
N({\un x}, {\un b} , Y=0) \, = \, 1 - \exp \left\{ - \frac{x_\perp^2
\, Q_s^2 ({\un b}) \, \ln (1/x_T \, \Lambda)}{4} \,  \right\}.
\ee
The dipole amplitude $N$ from \eq{glaN2} is plotted (schematically) in
\fig{glamue} as a function of $x_\perp$. One can see that, at small 
$x_\perp$, $x_\perp \ll 1/Q_s$, we have $N \sim x_\perp^2$ and the
amplitude is a rising function of $x_\perp$. However, at large dipole
sizes $x_\perp \gsim 1/Q_s$, the growth stops and the amplitude levels
off ({\sl saturates}) at $N = 1$. This regime corresponds to the black
disk limit for the dipole-nucleus scattering: for large dipoles the
nucleus appears as a black disk. To understand that $N = 1$ regime
corresponds to the black disk limit let us note that the total
dipole-nucleus scattering cross section is given by
\be\label{sigtot}
\sigma^{q{\bar q}A}_{tot} \, = \, 2 \, \int d^2 b \, N({\un x}, {\un b} , Y)
\ee
where the integration goes over the cross sectional area of the
nucleus. If $N=1$ at all impact parameters $\un b$ inside the nucleus,
\eq{sigtot} gives for a spherical nucleus of radius $R$
\be\label{black}
\sigma^{q{\bar q}A}_{tot} \, = \, 2 \, \pi \, R^2,
\ee
which is a well-known formula for the cross section for a particle
scattering on a black sphere \cite{LL}.

The transition between the $N \sim x_\perp^2$ to $N=1$ behaviors in
\fig{glamue} happens at around $x_\perp \sim 1/Q_s$. For dipole sizes 
$x_\perp \gsim 1/Q_s$ the amplitude $N$ {\sl saturates} to a
constant. This translates into saturation of quark distribution
functions in the nucleus, as was shown in \cite{GlaMue} (as $xq +
x\bar q \sim F_2 \sim \sigma_{tot}^{\gamma* A}$), and thus can be
identified with parton saturation, justifying the name of the {\sl
saturation scale}. We will show this connection between saturation for
the amplitude $N$ and for the partonic wave functions using gluons as
an example in Sect. \ref{MV} below.

Before we proceed let us finally note that, since $T({\un b}) \sim
A^{1/3}$, the saturation scale in \eq{qsmv} scales as \cite{MV,GlaMue}
\be\label{qsmvsc}
Q_s^2 \, \sim \, A^{1/3}
\ee
with the nuclear atomic number. \eq{qsmvsc} implies that for a very
large nucleus the saturation scale would become very large, much
larger than $\Lambda_{QCD}$. If $Q_s \gg \Lambda_{QCD}$, the
transition to the black disk limit in \fig{glamue} happens at the
momentum scales (corresponding to inverse dipole sizes) where the
physics is perturbative and gluons are correct degrees of
freedom. Therefore, \eq{qsmvsc} is of paramount importance, since it
justifies the approximation we made throughout this Section that
dipole-nucleon interactions can be described by perturbative gluon
exchanges instead of some non-perturbative mechanisms.

\newpage

\subsubsection{The McLerran-Venugopalan Model}
\label{MV}
{\sl \ \ \ \ \ Point Charges Approach} \\

\begin{figure}[h]
\begin{center}
\epsfxsize=15cm
\leavevmode
\hbox{\epsffile{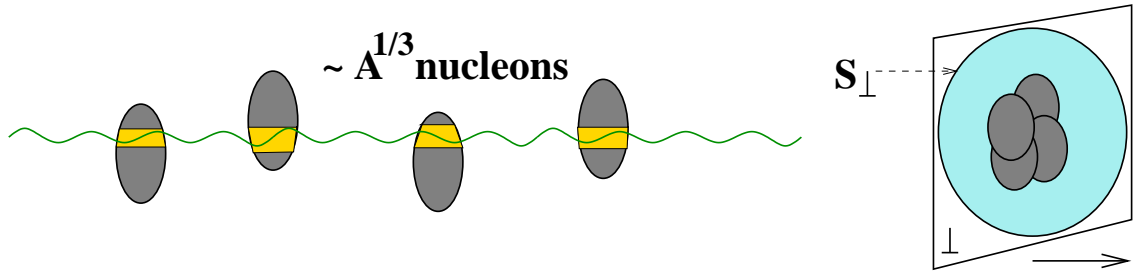}}
\end{center}
\caption{Small-$x$ gluon sees the whole nucleus coherently in the 
longitudinal direction and interacts with several different nucleons
in it. The effective color charge seen by the gluon is a result of a
random walk in the color space (see text).  }
\label{rv}
\end{figure}

Let us consider a large ultrarelativistic nucleus in the infinite
momentum frame. We are interested in the small-$x$ tail of the gluon
wave function in the nucleus. In the rest frame of the nucleus the
small-$x$ gluons have coherence length of the order of \cite{KS}
\be
l_{coh} \, \sim \, \frac{1}{2 \, m_N \, x_{Bj}}
\ee
where $m_N$ is the mass of a nucleon. If $x_{Bj}$ is sufficiently
small, the coherence length may become very long, much longer than the
size of the nucleus. Such small-$x$ gluons would be produced by the
whole nucleus coherently (only in the longitudinal direction). An
example of such interaction is shown in \fig{rv}. There the small-$x$
gluon (wavy line) interacts coherently with several Lorentz-contracted
nucleons. Indeed the nucleons, and the nucleus as a whole, are
color-neutral and one may worry that a coherent gluon simply would not
``see'' them. However, the gluon is coherent only in the longitudinal
direction: in the transverse direction the gluon is localized on the
scale $x_\perp \sim 1/k_T$ with $k_T$ the transverse momentum of the
gluon. If $k_T \gg \Lambda_{QCD}$, which is a necessary condition for
using gluon degrees of freedom, the transverse extent of the gluon
would be much smaller than the sizes of the nucleons. Because of that
the gluon would interact only with a part of each nucleon in the
transverse direction, as shown in \fig{rv}. The color charge in that
segment of the nucleon that a gluon is traversing does not have to be
zero: the gluon may run into, say, a single valence quark there. As a
result of such interactions, the gluon would ``feel'' some effective
color charge of all the nucleons' segments that it would traverse
through. In the spirit of Glauber approximation we may assume that all
nucleons are independent of each other, so that interactions with
parts of different nucleons are similar to a random walk in color
space. If each individual nucleons' segment has a typical color charge
$g$, than, due to the random walk nature of the process, the total
color charge seen by the gluon at a fixed impact parameter would be $g
\, \sqrt{n}$, where $n \sim A^{1/3}$ is the number of nucleons at this 
impact parameter.

%%%%%%%%%%%%%%%%%%%%%%%%%%%%%
%%%%%%%%%%%%%%%%%%%%%%%%%%%%%%

In the infinite momentum frame, due to Lorentz contraction, all the
nucleons appear to be squeezed into a thin ``pancake'' of
Lorentz-contracted nucleus, as shown on the right of \fig{rv}. One may
then define an effective color charge density seen by a gluon in the
transverse plane of the nucleus \cite{MV,jkmw,yuri1,LM}. The typical
magnitude of these color charge density fluctuations is given by the
color charge squared divided by the transverse area of the nucleus,
$(g \, \sqrt{n})^2 / S_\perp \, = \, g^2 \, n / S_\perp$. The number
of color charge sources in the whole nucleus is proportional to the
number of nucleons in the nucleus, $n \sim A$.  The typical color
charge density fluctuations are, therefore, characterized by the
momentum scale
\be\label{mu1}
\mu^2 \, \sim \, \frac{g^2 \, A}{S_\perp} \, \sim \, A^{1/3}.
\ee
It is important to notice that the momentum scale in \eq{mu1} grows
with $A$ as $A^{1/3}$, similar to the saturation scale in
\eq{qsmvsc}. The important conclusion we can draw from \eq{mu1} is that 
for sufficiently large nuclei their small-$x$ wave functions are
characterized by a hard momentum scale $\mu$, which is much larger
than $\Lambda_{QCD}$ allowing for a small coupling $\as$ description
of the process. Field theories with small coupling are usually
dominated by classical fields, with quantum corrections bringing in
extra powers of the small coupling constant $\as$. Therefore the
dominant small-$x$ gluon field of a large nucleus is {\sl classical}.
It can be found by solving the classical Yang-Mills equations of
motion \cite{YM}
\be\label{YMeq}
{\cal D}_\mu \, F^{\mu\nu} \, = \, J^\nu
\ee
with the nucleus providing the source current $J^\nu$, such that, in
the infinite momentum frame 
\be\label{jmv}
J^\nu \, = \, \delta^{\, \nu +} \, \rho ({\un x}, x^-)
\ee 
with $\rho ({\un x}, x^-)$ the color charge density. This conclusion
is originally due to McLerran and Venugopalan
\cite{MV}. Another way of understanding why the classical field
dominates is by arguing that large color charge density $\mu$ implies
high occupation numbers for the color charges in the nuclei: high
occupation numbers lead to classical description of the relevant
physics.

We need to find the gluon field of the nucleus in order to construct
its gluon distribution function. The latter is most easily related to
the gluon field in the light cone gauge of the nucleus. However, the
classical gluon field of a nucleus is easier to find in the covariant
$\partial_\mu A^\mu =0$ gauge. To do this we will, for simplicity,
assume that all the relevant color charge in the nucleus is carried by
the valence quarks. Furthermore, we will specifically chose to
consider a nucleus with ``mesonic'' nucleons made out of $q\bar q$
pairs instead of three valence quarks \cite{yuri1}. (This latter
assumption would only simplify the calculations, with the conclusions
being easily generalizable to the case of real nuclei.) Considering
the nucleus moving ultrarelativistically in the light cone $+$
direction, we label the ``valence'' quark and anti-quark coordinates
by ${\un x}_i, x^-_i$ and ${\un x'}_i, x'^-_i$. In the recoilless
eikonal approximation considered here neither one of these coordinates
changes due to ``emission'' of the gluon fields \cite{MV}. In
covariant gauge the color charge density of such ultrarelativistic
nucleus made of valence quarks is given by
\begin{eqnarray}\label{rho}
 {\rho}^a (\underline {x}, x^{-})= g \, \sum_{i=1}^A \, (t_i^a) \,
 [\delta(x^{-}-x^-_i) \, \delta({\underline{x}}-{\underline{x}}_i ) -
\delta(x^{-}-x'^-_i) \, \delta({\underline{x}}-{\underline{x}}'_i ) ]
\end{eqnarray}
with $(t_i^a)$ the color charge matrix of the quark and the anti-quark
in the $i$th ``nucleon'' in fundamental representation. The classical
electromagnetic field of an ultrarelativistic point charge $e$ at
${\un x}={\un 0}, x^-=0$ in $\partial_\mu A^\mu =0$ gauge can be
easily found \cite{Muesol,yuri1}
\be\label{emcl}
A'^{+} (x) = - {e \over {2 \pi}} \delta (x^{-}) \ln (|\underline {x}|
\Lambda)
\ee
with all other field components being zero and $\Lambda$ some infrared
cutoff.  Since this field is localized by the delta-function along the
light cone, the fields of a ensemble of charges located at different
longitudinal coordinates $x^-_i$ do not overlap. Therefore, if we want
to generalize \eq{emcl} to the non-Abelian case we need not worry
about the non-Abelian effects due to the overlap of the fields from
different charges, since those do not take place. The non-Abelian
generalization can be easily accomplished by replacing the electric
charge by its QCD analogue, $e \rightarrow g (t^a)$, such that the
gluon field of the nucleus in covariant gauge is given by \cite{yuri1}
\begin{eqnarray}
A'^{+} = {-} { g \over {2 \pi }} \sum_{a=1}^{N_c^2 -1} \sum_{i=1}^A
 t^a (t_i^a) \left[ \delta(x^{-}-x^-_i) \ln
 (|{\underline{x}}-{\underline{x}}_i| \Lambda ) 
 - \delta(x^{-}-x'^-_i) \ln
 (|{\underline{x}}-{\underline{x}}'_i| \Lambda ) \right] , \underline
 A' ={\un 0} , A'^{-}=0 .
\label{covfd}
\end{eqnarray}
As one can explicitly verify, \eq{covfd} satisfies \eq{YMeq} with the
source current given by \eq{jmv} with the color charge density from
\eq{rho}. We now have to perform a gauge transformation of this field
into the light-cone gauge. The field in a new gauge is
\begin{eqnarray}
A_{\mu} = S A'_{\mu} S^{-1} - { i \over g } (\partial _{\mu} S)
S^{-1}.
\end{eqnarray}
Requiring the new gauge to be the light-cone gauge, $A^+ =0 $, we
obtain
\begin{eqnarray}
S(\underline {x}, x^{-} ) = \mbox{P} \exp 
\left( { -ig \int_{-\infty}^{x^{-}} dx'^{-} A'^{+} 
(\underline {x}, x'^{-})} \right).
\label{smatr}
\end{eqnarray}
The field in $A^+ =0$ light cone gauge is
\begin{eqnarray}\label{lcfield}
{\underline {A}}(\underline {x}, x^{-} ) = \int_{-\infty}^{x^{-}}
dx'^{-} F^{+\perp}(\underline {x}, x'^{-} ) = \int_{-\infty}^{x^{-}}
dx'^{-} S(\underline {x}, x'^{-} ) F'^{+\perp}(\underline {x}, x'^{-}
) S^{-1} (\underline {x}, x'^{-} ).
\end{eqnarray}
Using the explicit expression for the field in covariant gauge from
\eq{covfd} in Eqs. (\ref{lcfield}) and (\ref{smatr}) we obtain \cite{yuri1}
\begin{eqnarray}
{\underline {A}} (\underline {x}, x^{-} ) = { g \over {2 \pi }}
\sum_{a=1}^{N_c^2 -1} \sum_{i=1}^A (t_i^a) \left( S(\underline {x}, x^-_i )
t^a S^{-1} (\underline {x}, x^-_i) \, {{\underline{x}}-{\underline{x}}_i
\over |{\underline{x}}-{\underline{x}}_i|^2} \, \theta (x^{-} - x^-_i)
\right. \nonumber
\end{eqnarray}
\begin{eqnarray}
 - \left. S(\underline {x}, x'^-_i ) t^a S^{-1} (\underline {x},
x'^-_i) \, {{\underline{x}}-{\underline{x}'}_i \over
|{\underline{x}}-{\underline{x}'}_i|^2} \, \theta (x^{-} -
x'^-_i) \right)
\label{clsol}
\end{eqnarray}
with
\begin{eqnarray}
S(\underline {x}, x^{-} ) = \prod_{i=1}^A 
\exp \left[ {i g^2 \over {2 \pi }}
\sum_{a=1}^{N_c^2 -1} t^a (t_i^a) \ln \left({|\underline {x} - {\underline
{x}}_i| \over { |\underline {x} - {\underline {x}}'_i|}} \right)
\theta (x^{-} - x^-_i) \right].
\label{cutoff}
\end{eqnarray}
Eqs. (\ref{clsol}) and (\ref{cutoff}) give us the classical field of a
large ultrarelativistic nucleus moving in the light cone $+$ direction
calculated in the light cone gauge of the nucleus $A^+ =0$. We will
also refer to this field as a non-Abelian Weizs\"{a}cker-Williams
field, since it is a non-Abelian analogue of the well-known
Weizs\"{a}cker-Williams field in electrodynamics \cite{WW}. Below we
will rederive this result in the approach where the nucleus is
described by a continuous color charge density $\rho ({\un x})$ and
will finally calculate the correlator of two of the fields in
\eq{clsol} to obtain the classical gluon distribution function of a
nucleus.
%%%%%%%%%%%%%%%%%%%%%%%%%%%%%
\begin{figure}
\begin{center}
\epsfxsize=10cm
\leavevmode
\hbox{\epsffile{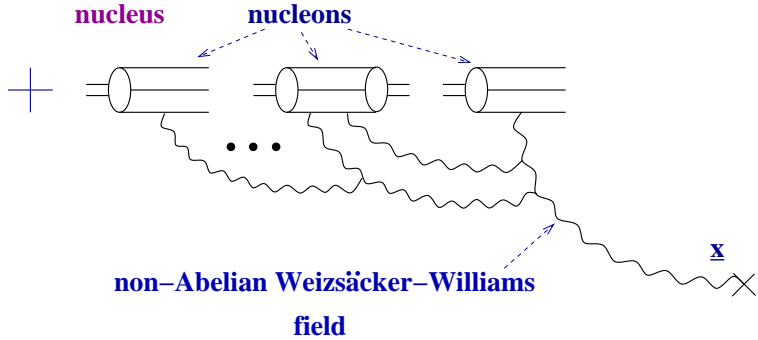}}
\end{center}
\caption{Classical gluon field of an ultrarelativistic large nucleus: 
the non-Abelian Weizs\"{a}cker-Williams field. }
\label{ww}
\end{figure}
%%%%%%%%%%%%%%%%%%%%%%%%%%%%%%

Let us first identify which Feynman diagrams correspond to the
non-Abelian Weizs\"{a}cker-Williams field of \eq{clsol}. Since a
detailed analysis of the problem can be found in \cite{yuri2}, here we
will only quote the answer. An example of the diagram contributing to
the classical field of \eq{clsol} is shown in \fig{ww}. Classical
fields are usually given by tree diagrams (graphs without loops): this
is indeed the case in \fig{ww}, where the gluons from different
sources (valence quarks in nucleons) keep merging with each other
until they form a gluon field at the point where we ``measure'' it,
which is denoted by a cross in \fig{ww}. Each nucleon can only emit
one (inelastic) or two (elastic) gluons: as discussed in more detail
in \cite{yuri2} emission of more than two gluons per nucleon allows
for diagrams with quantum loops to be of the same order as the
classical fields. Such graphs has to be discarded because the
classical fields does not give the dominant contribution at that
order.  The two-gluon per nucleon limit is the same as the one
considered in Sect. \ref{GM} giving rise to the resummation parameter
for multiple rescatterings from \eq{mrp}. As we will see below
classical field from \eq{clsol} resums powers of the same parameter
$\as^2 A^{1/3}$. The discarded diagrams having more than two gluons
per nucleon are suppressed by extra powers of $\as$ brought in by the
extra gluons, which are not enhanced by powers of $A$. \\

{\sl Effective Action Approach} \\

In order to investigate high gluon density effects, McLerran and
Venugopalan proposed an effective action for QCD at small $x$ and/or
for large nuclei (the MV model) \cite{MV}. To illustrate the approach,
it is easier to consider a nucleus (this can be easily generalized to
a proton) in a frame where it is traveling with the speed of light so
that it has a large $P^+$ (light cone momentum). In this frame, the
longitudinal size of nucleus is Lorentz contracted ($2R \rightarrow
2R/\gamma$) so that all the valence quarks are localized in a
longitudinal ``pancake'' of width $2R/\gamma$. Furthermore, consider
the transverse area of the nucleus as a rectangular grid of size $r
\ll 1/\Lambda_{QCD}$ so that we are not sensitive to details of
confinement. Since the nucleus is taken to be large, there is a large
number of valence quarks (and large $x$ gluons) in this rectangular
``pancake''. As we discussed above, the small $x$ gluons have large
longitudinal wavelengths and therefore can not resolve individual
valence quarks in the nucleus. Rather, they couple coherently to the
effective color charge generated by the valence quarks and large $x$
gluons, denoted by $\rho(x^-,{\un x})$. This color charge is taken to be
independent of the light cone time since the small-$x$ gluons, to
which it couples, have very short life times ($\Delta x^+ \sim
1/xP^+$) and see the effective color charge as frozen in time.

Based on the above physical picture, one can write the following
effective action for QCD at small $x$ and in the light cone gauge
$A^+=0$ \cite{MV,jkmw} (for an alternative but equivalent form of the
action see \cite{JMJV})
\begin{eqnarray}\label{eq:mv_action}
S \equiv -{1\over 4} \int d^4 x \, F_{\mu\nu}^{a \, 2} (x) + {i\over
N_c} \int d^2 x dx^- \, Tr \rho (x^-,{\un x}) W_{-\infty,\infty}[A^-]
(x^-,{\un x}) + i \int d^2 x \, dx^- F[\rho({\un x},x^-)]
\end{eqnarray}
where 
\be
W_{-\infty,\infty}[A^-] (x^-,{\un x}) \equiv P
exp\bigg[-ig\int^{\infty}_{-\infty} dx^+ A^-_a (x^+,x^-,{\un x})\,
T_a\bigg]
\ee
and $\exp \left[i \left\{i \int d^2 x dx^- F[\rho({\un
x},x^-)]\right\}\right]$ can be thought as the statistical weight of a
given color charge configuration. In the original MV model, $F$ was
taken to be a Gaussian, given by
\cite{MV,yuri1,LM}
\be
\int dx^- \, F[\rho({\un x},x^-)] \, = \, {{1\over 2 \, \mu^2} \, \rho^2({\un x})}
\ee 
while the interaction term was taken to be $\delta (x^-) \, \rho ({\un x})
A^-$. It turns out that as long as one is interested in the evolution
of observables with $x$, it is valid to take $\rho({\un x},x^-)
\rightarrow \rho({\un x}) \delta(x^-)$ with logarithmic accuracy. However,
if one is interested in calculating observables at the same point in
$x^-$, one needs to take the structure of the color charge density
$\rho$ in the longitudinal direction $x^-$ into account.

In order to calculate a physical observable, one first solves the
classical equations of motion and then averages over the color charge
configurations $\rho$ with the weight function $F[\rho]$.
\begin{eqnarray}
<O(A)> \equiv { \int [D\rho] D[A] \exp\{-F[\rho]\} \, O(A) e^{i S[\rho,A]} \over
 \int [D\rho] D[A] \exp\{-F[\rho]\}  e^{i S[\rho,A]}} 
\end{eqnarray}

As we have just mentioned, in the original MV model, the weight
function $F$ is taken to be a Gaussian. This is a good approximation
as long as the number of color sources is large and if the color
sources are not correlated, such as in a large nucleus
\cite{yuri1}. However, evolution in $x$ will change the weight
function and it will not be a Gaussian in general.

The classical equation of motion involving the non-zero component of
the current ($J^+$) is
\be\label{eom1}
\partial_i \partial^+ A^i + - i g [A_i, \partial^+A^i] = g J^+
\ee
and $F_{ij} = 0$. The solution has the form $A^i = \theta (x^-)
\alpha^i ({\un x})$ where $\alpha_i ({\un x})$ satisfies $\partial_i \alpha_i
({\un x}) = g \rho ({\un x})$. In the original work of McLerran and
Venugopalan \cite{MV}, it was argued that the commutator term is zero
since it involves the commutator of the fields at the same point in
$x^-$. This is not quite correct \cite{yuri1}, however, and leads to
infrared singular solutions. The reason is that this term is very
singular due to the presence of the $\delta (x^-)$ so that one needs
to understand the structure of the field across this singularity,
which in turn means that one needs to know the structure of the color
source distribution $\rho$ in $x^-$.

In \cite{jkmw,yuri1} the structure of the color sources in $x^-$ was
taken into account. Following \cite{jkmw} the equation of motion now
becomes
\be
D_i {\partial \over \partial y} A^i ({\un x},y) = g \rho (y,{\un x})
\label{eq:eom2}
\ee
where we have introduced the space-time rapidity variable $y \sim \ln
1/x^-$. Furthermore, the weight function is also modified in order to
take the rapidity dependence of the sources into account and is given
by
\begin{eqnarray}
\int D[\rho] \, \exp \Bigg\{ 
- \int_0^{\infty} dy \int d^2x {\rho^2 (y,{\un x}) \over \mu^2 (y)}\Bigg\}.
\end{eqnarray}

To find the classical solution, we introduce the path ordered Wilson
line, given by
\be
U(y,{\un x}) \equiv \hat{P} exp \bigg [ i  \int_y^{\infty} dy^{\prime}
\Lambda (y^{\prime},{\un x})\bigg].
\ee
Since $F_{ij} = 0$, the classical field $A_i$ must be a pure gauge in
two dimensions so that it can be written as
\be
A^i (y,{\un x}) = {i \over g} \, U(y,{\un x}) \, \partial^i \, U^{\dagger}(y,{\un x}),
\ee
which, substituted into (\ref{eq:eom2}) yields the following equation for
$\Lambda$: ${\un \partial}^2 \Lambda = -g^2 U^{\dagger}
\rho \, U$. The classical solution can be written as
\begin{eqnarray}\label{clsol2}
A^i (y,{\un x}) = {1\over g} \int_y^{\infty} dy^{\prime} \, U_{\infty,y^{\prime}}({\un x}) \,
[\partial^i \Lambda (y^{\prime},{\un x})] \, U_{y^{\prime},\infty}({\un x}).
\end{eqnarray}
Furthermore, since the transformation between $\rho$ and $\Lambda$
does not depend on $\rho$, one can perform the color averaging using
$\Lambda$. In other words, for any operator $O$, we have the following
relation
\begin{eqnarray}
\int D[\rho] \, O(\rho) \, exp \Bigg[ - \int_0^{\infty} dy \int d^2x 
{\rho^2(y,{\un x}) \over \mu^2(y)} \Bigg] =
\int D[\Lambda] \, O(\Lambda) \, exp \Bigg[ - \int_0^{\infty} \int d^2x 
{[\partial_T^2\Lambda(y,{\un x})]^2 \over g^4 \mu^2(y)}\Bigg].
\end{eqnarray}
Comparing \eq{clsol2} with \eq{lcfield} we can identify matrices $U$
and $S$ and the function $\Lambda/x^-$ with the gluon field $A'^+$ in
covariant gauge, which, in the case of point charges approach, is
given by \eq{covfd}. Therefore, the solution from \eq{clsol2} is
equivalent to the solution found previously in \eq{clsol}. \\

{\sl Classical Gluon Distribution} \\

Now we can compute the correlator of two gluon fields 
\begin{eqnarray}\label{gcorr}
G_{ij}^{ab} (y,{\un x},{\un y}) \, \equiv \, 
\langle A_i^a(y,{\un x})\, A_j^b(y,{\un y}) \rangle
\end{eqnarray}
which is related to the unintegrated gluon distribution via
\begin{eqnarray}\label{ww1}
\phi^{WW} (x_{Bj}, \un k^2) \, = \, \frac{1}{2 \, \pi^2} \, \int d^2 b \, 
d^2 r \, e^{- i {\un k} \cdot {\un r}} \ \mbox{Tr} \langle {\un A}
({\un 0}) \cdot {\un A} ({\un r}) \rangle ,
%\, = \, \frac{1}{(2 \, \pi)^2} \, \int d^2 b \, 
%d^2 r \, e^{- i {\un k} \cdot {\un r}} \ G_{ij}^{ab} (y, {\un 0}, {\un r}),
\end{eqnarray}
where the index $WW$ denotes the quasi-classical
Weizs\"{a}cker-Williams distribution function and ${\un A} ({\un x}) =
\lim_{y \rightarrow -\infty} {\un A} (y, {\un x})$.

The easiest way to compute the correlator (\ref{gcorr}) is by
expanding the path ordered exponentials, performing the color
averaging and exponentiating the result. The first term of the
expansion gives
\begin{eqnarray}
G_{ij,0}^{ab} (y,{\un x};y,{\un y}) = {1 \over g^2} \,  \delta_{ab} \int^{\infty}_{y} 
dy^{\prime} g^4 \, \mu^2 (y^{\prime})
\partial_i\partial_j {1\over \partial_T^4} ({\un x},{\un y}).
\end{eqnarray}
The inverse of the operator $\partial_T^4$ is infrared singular and
must be regulated. A natural cutoff is the QCD confinement scale
$\Lambda_{QCD}$. A more refined treatment shows that the saturation
scale $Q_s$ provides a natural infrared cutoff for the operator $1/
\partial_T^4$. Nevertheless, we define
\begin{eqnarray}
{1\over \partial_T^4} ({\un x}) \equiv \gamma({\un x}) = {1\over 8\pi} x_T^2\,
\ln [x_T^2 \Lambda^2_{QCD}] + \gamma(0)
\end{eqnarray}
so that  the next term in the expansion gives
\begin{eqnarray}
G_{ij,1}^{ab}(y,{\un x},{\un y}) = {\delta_{ab} \over g^2} {N_c\over 2} \bigg[ g^4
\int_{y}^{\infty} \mu^2(y^{\prime})\bigg]^2 [\partial_i\partial_j
\gamma ({\un x} - {\un y})] [\gamma({\un x} - {\un y}) - \gamma(0)].
\end{eqnarray}
Similarly, the $n^{th}$ term in the expansion gives
\begin{eqnarray}
G_{ij,n}^{ab} = {\delta_{ab} \over g^2} {N_c^n \over (n+1)!} \bigg[ g^4
\int_{y}^{\infty} \mu^2(y^{\prime})\bigg]^{n+1}
\partial_i\partial_j \gamma({\un x} - {\un y}) [\gamma({\un x} - {\un y}) - \gamma (0)]^n.
\end{eqnarray}
We can now sum the series and get the following expression for the
gluon distribution function
\begin{eqnarray}
G_{ii}^{aa}(y,x_T) %= {dN \over dy\,d^2 x} 
= {4 (N_c^2 -1) \over g^2 N_c}\,{1\over x_T^2} 
\Bigg[ 1 - [x_T^2 \Lambda^2_{QCD}]^{{g^4 N_c \over 8\pi} x_T^2\,\chi(y)}\Bigg]
\label{eq:G_ii}
\end{eqnarray}
where we have defined $\chi(y) \equiv \int_y^{\infty} dy^{\prime}
\mu^2(y^{\prime})$. Using \eq{rho} to calculate a correlator of two 
$\rho$'s in light cone gauge we can identify $\chi (y) = \rho \, T
({\un b}) / 4 N_c$ \cite{yuri1}. Alternatively the correlator in
\eq{gcorr} can be re-calculated using the gluon field from \eq{clsol}
\cite{KM}. In the end one can rewrite \eq{eq:G_ii} as
\be\label{gla2}
G_{ii}^{aa}(y, x_T) \, = \, \frac{2 \, C_F}{\as \, \pi} \,
\frac{1}{x_T^2} \, \left[ 1 - \exp \left( - \frac{x_T^2 \, Q_s^2 \, 
\ln (1/x_T \Lambda_{QCD})}{4} \right) \right],
\ee
which looks very similar to the Glauber-Mueller formula (\ref{glaN2}),
except that the saturation scale in \eq{gla2} is now for gluons,
\be\label{qsmvg}
Q_s^2 ({\un b}) \, \equiv \, 4 \, \pi \, \as^2 \, \rho \, T({\un b}),
\ee
and is different from the quark saturation scale (\ref{qsmv}) by
replacing Casimir operators, $C_F \rightarrow N_c$.
%%%%%%%%%%%%%%%%%%%%%%%%%%%%%
\begin{figure}[t]
\begin{center}
\epsfxsize=8cm
\leavevmode
\hbox{\epsffile{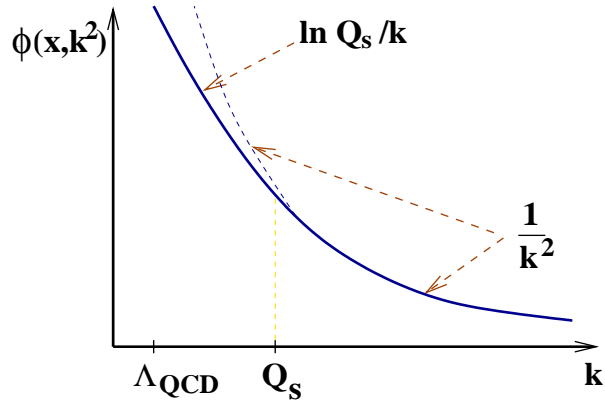}}
\end{center}
\caption{Unintegrated gluon distribution in the quasi-classical approximation 
as a function of gluons' transverse momentum (see \eq{gla2}).}
\label{mv1}
\end{figure}
%%%%%%%%%%%%%%%%%%%%%%%%%%%%%%

Combining Eqs. (\ref{ww1}), (\ref{gcorr}) and (\ref{gla2}) we obtain
the following gluon distribution function
\be\label{ww2}
\phi^{WW} (x_{Bj}, {\un k}^2) \, = \, \frac{C_F}{\as \, 2 \, \pi^3} \, 
\int d^2 b \, d^2 r \, e^{- i \, {\un k} \cdot {\un r}} \, \frac{1}{r_\perp^2} \, 
N_G ({\un r}, {\un b}, Y= 0)
\ee
expressed in terms of the gluon dipole-nucleus forward scattering
amplitude
\be\label{NGgm}
N_G ({\un r}, {\un b}, Y= 0) \, = \,  1 - \exp \left( - \frac{r_T^2 \, Q_s^2 \, 
\ln (1/r_T \Lambda_{QCD})}{4} \right).
\ee
Note that the unintegrated gluon distribution in \eq{ww2} is
proportional to $1/\as$ when $N_G \sim 1$: this corresponds to gluon
fields being $A_i \sim 1/g$ characteristic of all classical solutions.

It should be noted that (\ref{eq:G_ii}) is an all twist result for the
gluon distribution function, valid in the classical regime, which
resums multiple scattering of gluons from the target hadron or
nucleus, in the spirit of the Glauber-Mueller formalism and can be
thought of as the initial condition for an evolution equation which
would take gluon emission into account, such as the JIMWLK equation
which we will consider below.

It is instructive to look at the different limits of (\ref{eq:G_ii}),
or, equivalently, \eq{ww2}. In the limit where $x_T \rightarrow 0$
(perturbative QCD limit), we get $G(x_T) \sim \ln (x_T)$ so that in
momentum space using \eq{ww1} we have $\phi (x_{Bj}, {\un k}^2) \sim
1/k_T^2$ in agreement with pQCD.  On the other hand, in the limit
where $x_T$ is large (but is smaller than $1/\Lambda_{QCD}$), we get
$G(x_T) \sim 1/x_T^2$ so that in momentum space we have $\phi (x_{Bj},
{\un k}^2) \sim \ln (k_T)^2$. The momentum distribution of gluons is
shown in Fig. \ref{mv1} where the slowing down of the infrared
divergence of the gluon distribution function is evident.
%%%%%%%%%%%%%%%%%%%%%%%%%%%%%
\begin{figure}
\begin{center}
\epsfxsize=11cm
\leavevmode
\hbox{\epsffile{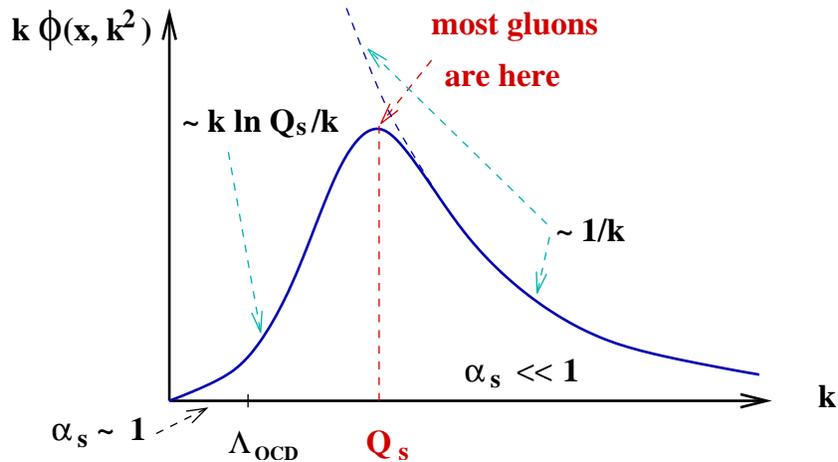}}
\end{center}
\caption{Phase space distribution of gluons in the transverse momentum space.}
\label{mv2}
\end{figure}
%%%%%%%%%%%%%%%%%%%%%%%%%%%%%%

It is important to notice that the gluon distribution function (solid
line in \fig{mv1}) is now only logarithmically infrared divergent, in
contrast to pQCD (the first term in expansion of \eq{gla2} in powers
of $Q_s$ shown by dashed line in \fig{mv1}), which would lead to power
divergence. In other words, the non-linearities of the classical gluon
field have (almost) regulated the infrared divergence present in
leading twist pQCD, making the residual singularity integrable over
${\un k}$.  This phenomenon is usually referred to as {\it gluon
saturation}: the increasing gluon distribution slows down its growth
in the infrared from the power-law scaling to the logarithmic one. It
is interesting to note that the saturation scale characterizing
unitarization of the dipole-nucleus forward scattering amplitude $N$
in Sect. \ref{GM} turns out to be the same saturation scale (modulo
the Casimir factors) as the one governing the saturation of gluon
distribution functions in
\fig{mv1}!

The phase space distribution of gluons is shown in Fig. \ref{mv2},
where we multiplied the unintegrated gluon distribution $\phi (x_{Bj},
{\un k}^2)$ from \fig{mv1} by the phase space factor $k_T$. In
contrast with leading twist pQCD (dashed line), the full classical
line in \fig{mv2} has a peak, which is due to the fact that the
infrared divergence is now regularized. The location of the peak
determines the typical momentum of the gluons in the small-$x$
hadron/nucleus wavefunction and is given by the saturation momentum
since it is the only scale in the problem. The fact that most gluons
in the wave function of a hadron or nucleus reside in a state with a
finite momentum $Q_s$ is the reason why this state is also called a
{\it condensate}. It allows us to treat the gluon wave function of a
high energy hadron or nucleus in the small coupling approximation.

%%%%%%%%%%%%%%%%%%%%%%%%%%%%%%%%%%%%%%%%%%%%%%%%%%%%%%%%%%%%%%%%%%%%%%%%%%%%%%%%%%%%%%%%%%%%%

\subsection{Quantum Evolution}
\label{QE}

It is known that quantum corrections to the classical results
presented in the previous section are potentially large. These large
corrections arise from emission of gluons, both real and virtual. If
the available energy is sufficiently high, it leads to a large phase
space for gluon emission which gives rise to large logs of energy (or
of $1/x$) which need to be resumed. In the CGC formalism, the presence
of these large corrections to the classical approximation was proved
in \cite{ajmv1,ajmv2}. A Wilson renormalization group formalism was
developed in \cite{jklw1,jklw2} which allows for resummation of these
large logs. The resulting equation can be written as an evolution
equation for correlator of any number of Wilson lines and is known as
the JIMWLK equation. This (functional) equation was later shown to be
equivalent to a set of coupled equations derived by Balitsky
\cite{balitsky}. In the large-$N_c$ limit, these equations can be
written in a closed form, as a single integral equation, independently
derived by one of the authors for the correlator of two Wilson lines
\cite{yuri_bk}, known as the BK equation.

\subsubsection{The JIMWLK Equation}
\label{JIMWLKsect}

The starting point is the action given by (\ref{eq:mv_action}). We
introduce the following decomposition of the full field $A_{\mu}$,
\begin{eqnarray}
A_{\mu} = b_{\mu} + \delta A_{\mu} + a_{\mu}
\label{eq:decomp}
\end{eqnarray}
where $b_{\mu}$ is the solution to the classical equation of motion
considered in the previous section, $\delta A_{\mu}$ denotes the
quantum fluctuations having longitudinal momenta $q^+$ between $P^+_n$
and $P^+_{n-1}$, and $a_{\mu}$ is a soft field having longitudinal
momentum $k^+$ where $k^+ < P^+_n$. The effective action is calculated
for field $a_{\mu}$, integrating out hard fluctuations $\delta
A_{\mu}$, assuming that these fluctuations are much smaller than the
classical field $b_{\mu}$.  This procedure reproduces the form of the
action given in (\ref{eq:mv_action}) with a modified functional
$F^{\prime}$, due to inclusion of the hard fluctuations into the color
sources. The change of the functional $F$ with $log 1/x$ leads to an
evolution equation for the statistical weight functional, known as the
JIMWLK equation, which can then be used to derive evolution equations
for any $n$ point function of the effective theory.

Expanding the action around the classical solution and keeping first
and second order terms in hard fluctuations $\delta A_{\mu}$, we get
\begin{eqnarray}
S= -{1\over 4}G(a)^2-{{1}\over{2}} 
\delta A_{\mu} [{\rm D}^{-1} (\rho)]^{\mu\nu} \delta A_{\nu} + 
ga^- \rho^\prime+O((a^-)^2)
+ iF[\rho]
\label{eq:efaction}
\end{eqnarray}
where
\begin{eqnarray}
\rho^\prime =\rho+ \delta \rho_1 + \delta \rho_2
\label{eq:rhoprime}
\end{eqnarray}
with
\begin{eqnarray}
\delta \rho^a_1({\un x},x^+, x^-) &=& -2 f^{abc} \alpha^{b}_{i} ({\un x}) 
\delta A^{c}_{i}({\un x}, x^+, x^-=0)
\label{eq:rho1}
\\ 
&-& {{g}\over{2}} f^{abc} \rho^{b}
({\un x}) \int dy^+ \Bigg[\theta (y^+ - x^+) - \theta (x^+ - y^+) \Bigg] 
\delta A^{-c}({\un x}, y^+, x^-=0)\nonumber
\end{eqnarray}
and
\begin{eqnarray}
\delta \rho^a_2({\un x}, x^+,x^-) &=& -f^{abc} [\partial^+ \delta 
A^{b}_{i}({\un x},x^+,x^-) ]\delta A^{c}_{i}({\un x},x^+,x^-) 
\nonumber\\
&-& {{g^2}\over{N_c}} \rho^{b}({\un x}) \int dy^+ \delta
A^{-c}({\un x},y^+, x^-=0) \int dz^+ \delta A^{-d}({\un x},y^+,x^-=0) \nonumber
\\ &\times &\Bigg[\theta (z^+ -y^+) \theta (y^+ -x^+) {\rm tr} T^a T^c
T^d T^b \nonumber \\ &+&\theta (x^+ -z^+) \theta (z^+ -y^+) {\rm tr}
T^a T^b T^c T^d \nonumber \\ &+&\theta (z^+ -x^+) \theta (x^+ -y^+)
{\rm tr} T^a T^d T^b T^c \Bigg].
\label{eq:rho2}
\end{eqnarray}
The first term in both $\delta\rho_1^a$ and $\delta \rho^a_2$ is
coming from expansion of $G^2$ in the action while the rest of the
terms proportional to $\rho$ are from the expansion of the Wilson line
term.  The three terms correspond to different time ordering of the
fields. Since the longitudinal momentum of $a^-$ is much lower than of
$\delta A$, we have only kept the eikonal coupling which gives the
leading contribution in this kinematics.  The higher order terms in
$a^-$ do not affect the derivation here and can be reconstructed later
using the uniqueness of the action.

At first stage, we integrate over $\delta A_{\mu}$ while keeping both
$\rho$ and $\delta \rho$ fixed. We note that it is sufficient to keep
only the first and second order terms in correlation functions of
$\delta \rho$, all higher order terms being suppressed by powers of
$\alpha_s$. We therefore define
\begin{eqnarray}
<\delta\rho^a({\un x})>_{\delta A}& = & \alpha_s \, y\, \sigma^a({\un x})\nonumber \\
<\delta\rho^a({\un x},x^+)\delta\rho^b({\un y},x^+)>_{\delta A}& = &
\alpha_s \, y \,\chi^{ab}({\un x},{\un y})
\label{eq:chisig}
\end{eqnarray}
with $y=\ln (1/x)$. Integrating out the hard fluctuations $\delta A_{\mu}$ then leads to
\begin{eqnarray}
\lefteqn{
\int D[\rho,\rho^\prime]
[ {\rm Det}(\chi)]^{-1/2}\exp\left(-F[\rho]\right) 
} 
\nonumber \\ && \times 
\exp\left(-{1\over 2\,\alpha_s\,y\,}
  \left[\rho^\prime_x-\rho_x-
    \alpha_s\,y\, \sigma_x\right][\chi^{-1}_{x y}]\left[
    \rho^\prime_y-\rho_y-\alpha_s\,y\,\sigma_y\right]\right)
\nonumber\\ & \equiv & 
\int D[\rho,\rho^\prime]\exp\{-U[\rho,\rho^\prime]\}
\label{eq:intoutdA}
\end{eqnarray}
where an integration (summation) over repeated indices is implied. In
the next step, we integrate out $\rho$ at fixed $\rho^{\prime}$. This
can be done using the steepest decent method since the integrand is a
steep function of $\rho$, peaked around $\rho^{\prime}$. The steepest
decent equation is given by
\begin{eqnarray}
\rho^{\prime }_x-\rho_x-\alpha_s\,y\,\sigma_x=\alpha_s\,y\,
\chi_{x u}\left[{\delta F\over \delta\rho_u}+{1\over 2}{\rm
    tr}(\chi^{-1} 
{\delta\chi\over\delta\rho_u})\right ].
\label{eq:sd}
\end{eqnarray}
Substitution of this into Eq.(\ref{eq:intoutdA}) gives
\begin{eqnarray}
U & = & F+{1\over 2}\rm{tr}\ln(\chi)
\\ &&
+{\alpha_s\,y\, \over 2}\left[{\delta F\over \delta\rho_u}+{1\over 2}
{\rm tr}(\chi^{-1}
{\delta\chi\over\delta\rho_u})\right ]\chi_{u v}
\left[{\delta F\over \delta\rho_v}+{1\over 2}{ \rm tr}(\chi^{-1}
{\delta\chi\over\delta\rho_v})\right ].\nonumber
\label{eq:dF}
\end{eqnarray}
In the above expression all the functionals are taken at $\rho^0$
which is the solution of the steepest descent equation (\ref{eq:sd}).
We also need the correction due to quadratic fluctuations of $\rho$
around the steepest decent solution $\rho_0$. It is given by
\begin{eqnarray}
{\delta^2 U\over\delta\rho_x\delta\rho_y} & = &
{1\over \eta}\chi^{-1}_{x y}+
{\delta^2 \over\delta\rho_x\delta\rho_y}\left[F+{1\over 2}
\rm{tr}\ln(\chi)\right] \\
&& + \chi^{-1}_{x u}{\delta \sigma_u\over\delta\rho_y}
+{\delta \sigma_u\over\delta\rho_x}\chi^{-1}_{u y}\nonumber\\
&& + \left[\chi^{-1}_{x u}{\delta \chi_{u v}\over\delta\rho_y}
+\chi^{-1}_{y u}{\delta \chi_{u v}\over\delta\rho_x} \right]
\left[{\delta F\over\delta\rho_v}+{1\over 2}{\rm tr}\left(\chi^{-1}
{\delta \chi\over\delta\rho_v}\right)\right].\nonumber
\label{eq:deltau}
\end{eqnarray}
Furthermore, there are contributions from the third and the fourth
derivatives of $U$, explicit expressions for which are long and can be
found in \cite{jklw2}. Putting everything together, we recover the
form of the action given in (\ref{eq:mv_action}) with the new
functional $F^{\prime}$ given by
\begin{eqnarray}
F^{\prime} &\equiv& F +  
{\alpha_s\, y\over 2}\left[\chi_{u v}
{\delta^2 \over\delta\rho_u\delta\rho_v}F
-{\delta^2\chi_{u v} \over\delta\rho_u\delta\rho_v}
-{\delta F\over \delta\rho_u}
\chi_{u v}
{\delta F\over \delta\rho_v}
\right. \nonumber \\ && \left. \hspace{1cm}
+2{\delta F\over \delta\rho_u}{\delta\chi_{u v}\over
\delta\rho_v}+2{\delta \sigma_u\over\delta\rho_u}
-2{\delta F\over\delta\rho_u}\sigma_u\right]
\label{eq:F_prime}
\end{eqnarray}
which can be rewritten as an evolution equation for the functional $F$ in terms
of the one and two point fluctuations $\sigma$ and $\chi$  as
\begin{eqnarray}
{d\over d y}F
& = &{\alpha_s\over 2}\left[\chi_{u v}
{\delta^2 \over\delta\rho_u\delta\rho_v}F
-{\delta^2\chi_{u v} \over\delta\rho_u\delta\rho_v}
-{\delta F\over \delta\rho_u}
\chi_{u v}
{\delta F\over \delta\rho_v}
\right. \nonumber\\ && \left. \hspace{1cm}
+
2{\delta F\over \delta\rho_u}{\delta\chi_{u v}\over
\delta\rho_v}
+2{\delta \sigma_u\over\delta\rho_u}
-2{\delta F\over\delta\rho_u}\sigma_u\right].
\label{eq:rg_F}
\end{eqnarray}
If we define the statistical weight functional $Z \equiv e^{-F}$, the
above equation takes a very simple form if written as an evolution
equation for $Z$ known as the JIMWLK equation
\begin{eqnarray}
{d\over d \,y}\, Z = \alpha_s 
\left\{{1\over 2}{\delta^2
    \over\delta\rho_u\delta\rho_v}\left[Z\chi_{u v}
\right]
-{\delta \over\delta\rho_u}\left[Z\sigma_u\right]\right\}
\label{eq:rg_Z}
\end{eqnarray}

The equation (\ref{eq:rg_Z}) can be used to derive evolution equations
for any number of correlators of the color charge density $\rho$. As
an example, we consider the two point function. Multiplying both sides
of (\ref{eq:rg_Z}) with $\rho_x\, \rho_y$ and integrating by parts
over $\rho$ gives
\begin{eqnarray}
{d\over d\, y}\,<\rho_u\rho_v>=
\alpha_s \, <\chi_{u v}+\rho_u\sigma_v+\rho_v\sigma_u>.
\label{eq:rg_2pt}
\end{eqnarray}
This equation can be shown to reduce to the BFKL equation in the low density limit 
which will be considered later.

\begin{figure}
\begin{center}
\epsfxsize=11cm
\leavevmode
\hbox{\epsffile{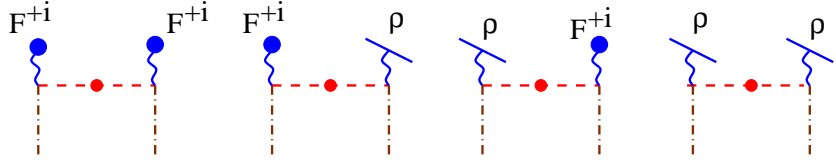}}
\end{center}
\caption{Diagrams contributing to the color charge fluctuation $\chi$.}  
\label{fig:chi}
\end{figure}

To complete the derivation of the JIMWLK equation, we need to calculate the one and two
point fluctuations $\chi$ and $\sigma$ in terms of the color charge density $\rho$. The
details are given in \cite{Jalilian-Marian:1997dw}-\cite{Ferreiro:2001qy}.  Here we
follow the derivation in  \cite{Ferreiro:2001qy} which is done in coordinate space and 
is closer in spirit to the derivation of BK equation in the next section. The starting point
is $\chi$ as defined in (\ref{eq:chisig}) to which only the color charge 
$\rho_1$ given in (\ref{eq:rho1}) contributes. Suppressing the overall coupling constant
and rapidity for the moment, $\chi$ is 
\begin{eqnarray}\label{chiab}
\chi_{ab}( {\un x},{\un y})\,=
\left\langle \left[-2i{\cal F}^{+i}\, 
\delta A^i + \,\rho {1 \over i\partial^-} \, \delta A^- \right]_x^a
\left[
2i \, \delta A^i {\cal F}^{+i} + \delta A^- {1 \over 
i\partial^-} \rho \right]_y^b \right\rangle
\end{eqnarray}
where ${\cal F}^{+i}=\delta (x^-) \alpha_i({\un x})$ (integration over
$x^-$ and $y^-$ is implied above). \eq{chiab} can be rewritten in
terms of the propagator of the hard fluctuations $\delta A$ as
\begin{eqnarray}
\chi_{ab}( {\un x},{\un y})\,&=&
2{\cal F}^{+i}_x\, \langle x|G^{ij}|y\rangle \,2{\cal F}^{+j}_y \, +\, 
2{\cal F}^{+i}_x\,\langle x|G^{i-}\,{1 \over i\partial^-}|y\rangle \, \rho_y 
\nonumber\\
&-&\,
\rho_x \, \langle x|{1 \over i\partial^-}\, G^{-i}|y\rangle \, 
2{\cal F}^{+i}_y\,+\,i
\rho_x \langle x|{1 \over i\partial^-} G^{--} {1 \over i\partial^-}
|y\rangle\,\rho_y.
\label{eq:chi_4}
\end{eqnarray}
The four terms in (\ref{eq:chi_4}) are depicted in
Fig. (\ref{fig:chi}) and correspond to real corrections to the
evolution equations. The dashed lines denote the propagator of the
hard modes in the background field while the dash-dotted lines denote
the soft modes for which the effective theory is written. The thick
wavy lines attached to a circle denote the classical field while
the thick wavy lines attached to the solid lines represent the color 
charge density $\rho$.

%%%%%%%%%%%%%%%%%%%%%%%%%%%%%
%%%%%%%%%%%%%%%%%%%%%%%%%%%%%%

The hard fluctuation propagators $G^{\mu\nu}$
were derived in \cite{Iancu:2000hn,ajmv1,hw}. Here we give the final result
for $\chi$
\begin{eqnarray}
&&\chi({\un x}, {\un y})=
4 \Biggl\{
{\cal F}^{+i}_x\, \delta^{ij}_{\perp}({\un x}-{\un y})\, {\cal F}^{+j}_y\, -
\nonumber \\ &&
\biggl[2 {\cal F}^{+i} \biggl({\cal D}^i - {\partial^i \over 2}\biggr)
+\rho\biggr]_x\,\langle
{\un x} | {1 \over \partial_{\perp}^2} | {\un y}
\rangle\,
\biggl[2\biggl({\cal D}^{\dagger j} - {\partial^{\dagger j} \over 2}\biggr)
{\cal F}^{+j}+ \rho\biggr]_y\Biggr\}
\label{eqn:chi_final}
\end{eqnarray}
with ${\cal D}^i\equiv\partial^i -ig{\cal A}^i$ and ${\cal D}^{\dagger
j}=\partial^{\dagger j} +ig{\cal A}^j$ the covariant derivatives
constructed with the background field ${\cal A}^i$, where the
derivative $\partial^{\dagger}$ is acting on the function to its
left. Furthermore, we have used the following short hand notation
\begin{eqnarray}
\delta^{ij}_{\perp}({\un x}-{\un y})&=&
\int {d^2p_\perp \over (2 \pi)^2}\,
\left[\delta^{ij}- {p^i p^j \over p_{\perp}^2}\right]\
{\rm e}^{ip_{\perp}\cdot({\un x}-{\un y})}\,\\
\langle
{\un x} | {1 \over \partial_{\perp}^2} | {\un y}
\rangle &=&-
\int {d^2p_\perp \over (2 \pi)^2}\, {1 \over p_{\perp}^2}\
{\rm e}^{ip_{\perp}\cdot({\un x}-{\un y})} .
\label{eq:inchi_def}
\end{eqnarray}

Finally, we note that the above expressions look much simpler when one 
expresses the background field $A$ in terms of the covariant gauge color
charge density $\tilde{\rho}$. This is allowed since both the functional measure
and the weight function are gauge invariant. The classical solution is now
simple and reads $\tilde{A^{\mu}_a} = \delta^{\mu +} \alpha_a$ where 
$\alpha^a = - {1\over \partial_{\perp}^2} \tilde{\rho}^a$. The non-linear evolution equation
(\ref{eq:rg_Z}), written in terms of the covariant fields $\alpha$ looks like
\begin{eqnarray}
{\partial Z[\alpha] \over \partial y}\,=\,\alpha_s
\left\{ {1 \over 2} {\delta^2 \over {\delta
\alpha^a({\un x}) \delta \alpha^b({\un y})}} 
[Z \, \eta_{xy}^{ab}] - 
{\delta \over {\delta \alpha^a({\un x})}}
[Z \, \nu_{x}^a] \right\}
\label{eq:rg_alpha}
\end{eqnarray}
where we have defined 
\begin{eqnarray}
\!\!\!\!\!\!\!\!\!\!\eta^{ab}({\un x},{\un y})&\equiv&\int d^2z_\perp \int d^2u_\perp\,
\langle {\un x}|\,\frac{1}{\partial^2_\perp}\,|{\un z}\rangle\,
\tilde\chi^{ab}({\un z},u_\perp)\,
\langle u_\perp|\,\frac{1}{\partial^2_\perp}\,|{\un y}\rangle \\
\nu^a ({\un x})&\equiv&- \int d^2z_\perp\,
\langle {\un x}|\,\frac{1}{\partial^2_\perp}\,|{\un z}\rangle\,
\tilde\sigma^a ({\un z})
\label{eq:nueta}
\end{eqnarray}
and $\tilde{\chi}$, $\tilde{\sigma}$ are related to $\chi$, $\sigma$ via
\begin{eqnarray}
\!\!\!\!\!\!\!\!\!\!\!\!\!\!\!\!\!\!\!\!
\tilde\sigma_a ({\un x})&\equiv&
 V^{\dagger}_{ab}({\un x})\,\sigma_b({\un x}) +
{1\over 2}\,f^{abc}\int d^2 y_\perp \,\,
\tilde\chi_{cb}({\un x},{\un y})\,
\langle {\un y}|\,\frac{1}{\partial^2_\perp}\,|
{\un x}\rangle \nonumber \\
\tilde \chi_{ab}({\un x},{\un y})&\equiv&
V^{\dagger}_{ac}({\un x})\,
\chi_{cd}({\un x},{\un y})\,V_{d b}({\un y}) 
\label{eq:chisig_tilde}
\end{eqnarray}
with  
\begin{eqnarray}\label{eq:v_fun}
V^{\dagger}_x \equiv V^\dagger({\un x})\,=\,{\rm P} \exp \left \{ ig
\int_{-\infty}^{\infty} dz^-\,\alpha (z^-,{\un x})  \right \}.
\end{eqnarray}
Combining all the expressions above, the final expression for $\eta_{xy}^{ab}$ is
given by
\begin{eqnarray}
\eta^{ab}_{xy}= {1\over \pi}
\int\! \!{d^2z_\perp\over (2\pi)^2}
\frac{(x^i-z^i)(y^i-z^i)}{({\un x}-{\un z})^2({\un y}-{\un z})^2 }
\Bigl[1 + V^\dagger_x V_y-V^\dagger_x V_z - V^\dagger_z
V_y\Bigr]^{ab}.
\label{eq:eta_final}
\end{eqnarray}
For completeness, we also give the final expression for the virtual
corrections denoted $\nu$
\begin{eqnarray}
\nu^a_x&=&{ig\over 2\pi}
\int {d^2z_\perp\over (2\pi)^2}\,\frac{1}{({\un x}-{\un z})^2}
\,{\rm Tr}\Bigl[T^a V^\dagger_x V_z\Bigr].
\label{eq:nu_final}
\end{eqnarray}
Eq. (\ref{eq:rg_alpha}) with the coefficient functionals given by
(\ref{eq:eta_final},\ref{eq:nu_final}) is known as the JIMWLK
equation.  To conclude this section, we note that it is possible to
rewrite the non-linear evolution equation in a more compact form,
eliminating $\sigma$ in favor of $\chi$, as was done in
\cite{Weigert:2000gi}. Furthermore, it is possible to recast the
JIMWLK equation as a Langevin equation which lends itself easily to
numerical investigation, for example, on a lattice \cite{RW}. For a
review of the most recent progress in understanding the JIMWLK
equation and its various forms, we refer the reader to
\cite{Heribert}.

\subsubsection{The Balitsky-Kovchegov Equation}
\label{BKsect}

Let us approach the problem of resumming quantum evolution corrections
from a different side: instead of including small-$x$ evolution
corrections in the gluon distribution function, as was done in
Sect. \ref{JIMWLKsect}, we will consider quantum corrections to the
dipole-nucleus cross section from Sect. \ref{GM}. In the
quasi-classical limit, the forward amplitude of the dipole-nucleus
scattering is given by \eq{glaN2} \cite{GlaMue}, obtained by resumming
multiple rescatterings of \fig{qqbar}. Now let us study how the
quantum corrections come into this multiple rescatterings picture.
%%%%%%%%%%%%%%%%%%%%%%%%%%%%%
\begin{figure}[ht]
\begin{center}
\epsfxsize=10cm
\leavevmode
\hbox{\epsffile{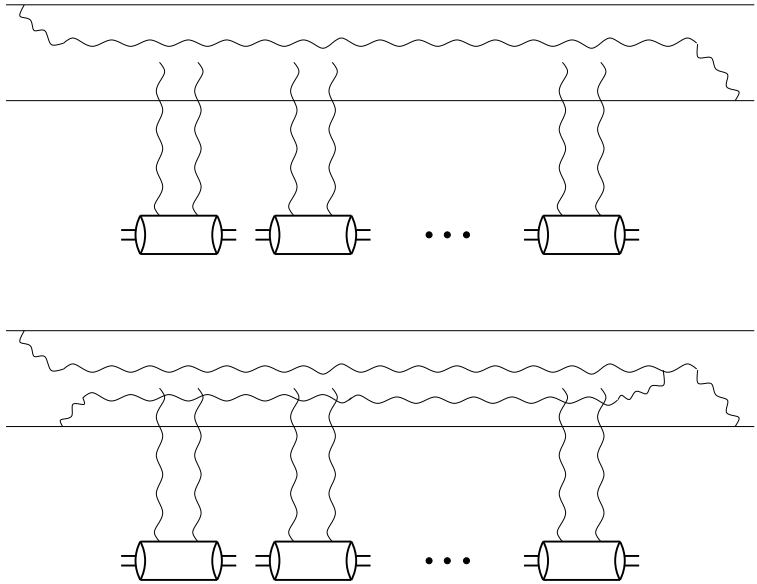}}
\end{center}
\caption{Quantum corrections to dipole-nucleus scattering. }
\label{fock}
\end{figure}
%%%%%%%%%%%%%%%%%%%%%%%%%%%%%%

Similar to the BFKL evolution equation \cite{BFKL} and to JIMWLK
equation, we are interested in quantum evolution in the leading
longitudinal logarithmic approximation resumming the powers of
\be
\as \, \ln \frac{1}{x_{Bj}} \, \sim \, \as \, Y
\ee
with $Y$ the rapidity variable. Again we will be working in the rest
frame of the nucleus, but this time we choose to work in the light
cone gauge of the projectile $A^+ = 0$ if the dipole is moving in the
light cone $+$ direction. This gauge is equivalent to covariant gauge
for the multiple rescatterings of \fig{qqbar}: therefore, all our
discussion in Sect. \ref{GM} remains valid in this new gauge.

We need to identify radiative corrections that bring in powers of $\as
\, Y$.  As we have seen in Sect. \ref{GM}, instantaneous multiple
rescatterings bring in only powers of $\as$ not enhanced by factors of
$Y$. Therefore, additional Coulomb gluon exchanges would not generate
any logarithms of $x_{Bj}$ bringing in only extra powers of $\as$. We
are not interested in such corrections. Other possible corrections in
the light cone gauge of the projectile dipole are modifications of the
dipole wave function. The incoming dipole may have some gluons (and
other quarks) present in its wave function. For instance, the
dipole may emit a gluon before interacting with the target, and then
the whole system of quark, anti-quark and the gluon would rescatter in
the nucleus, as shown in the top diagram of \fig{fock}. The dipole may
emit two gluons which would then interact with the nucleus, along with
the original $q\bar q$ pair, as shown in the bottom diagram of
\fig{fock}. In principle we can have as may extra gluon emissions, as 
well as generation of extra $q\bar q$ pairs in the dipole's wave
function. As we will shortly see, these gluonic fluctuations from
\fig{fock} actually do bring in the factors of $\as$ enhanced by powers 
of rapidity $Y$, i.e., they do generate leading logarithmic
corrections. Fluctuations leading to formation of $q\bar q$ pairs
actually enter at the subleading logarithmic level bringing in powers
of $\as^2 \, Y$ \cite{FL,CC,BB} and are not important for the leading
logarithmic approximation used in this Section.
%%%%%%%%%%%%%%%%%%%%%%%%%%%%%
\begin{figure}[ht]
\begin{center}
\epsfxsize=10cm
\leavevmode
\hbox{\epsffile{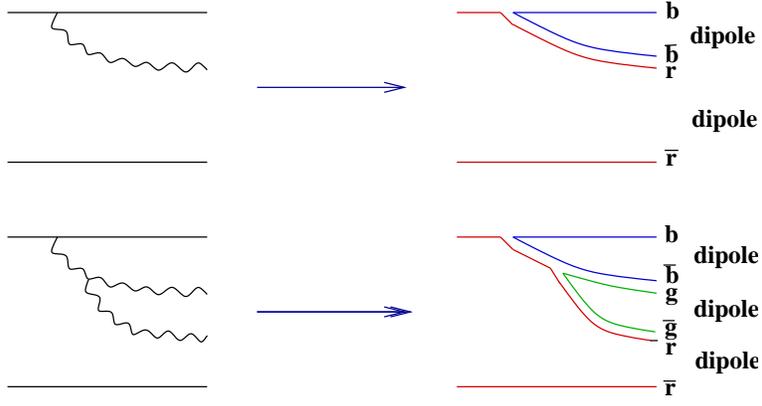}}
\end{center}
\caption{In the large-$N_c$ limit of the  Mueller's dipole model 
\cite{dip1,dip2,dip3} the gluon cascade is represented as a color dipole 
cascade. Quark colors are labeled r, b, g for red, blue and green
correspondingly.}
\label{dip}
\end{figure}
%%%%%%%%%%%%%%%%%%%%%%%%%%%%%%

The gluons in the dipole wave function have coherence length of the
order of (see
\eq{lctime})
\be\label{lcoh}
l_{coh} \, \approx \, \frac{2 \, k^+}{{\un k}^2}
\ee
if the incoming $q\bar q$ pair is moving in the light cone $+$
direction. If $k^+$ is large enough, the coherence lengths of these
gluons would be much larger than the nuclear radius, $l_{coh} \gg R$,
so that each gluon would coherently rescatter on the nucleons in the
nucleus, just like the original dipole in \fig{qqbar}. This is indeed
what is also shown in \fig{fock}. Note that gluons are emitted by the
incoming dipole only before the multiple rescattering interaction (or
after the interaction in the forward amplitude). Emissions during the
interaction are suppressed by the powers of $x_{Bj}$ (or,
equivalently, by inverse powers of center of mass energy of the
scattering system) \cite{KM}. This can be checked via an explicit
calculation in the covariant Feynman perturbation theory or by
performing the calculation in the light cone perturbation theory
\cite{BL,BPP}. In the latter case, the emission of a gluon is 
allowed and is equally probable at any point throughout the coherence
length of the parent dipole $l_{coh}^{q\bar q} = 2 \, p^+ / {\un p}^2$
with $p$ the momentum of the dipole and $p^+$ being very large. The
probability of emitting the gluon inside the nucleus is given by
$R/l_{coh}^{q\bar q} \sim 1/p^+ \sim 1/\sqrt{s}$, i.e., it is
suppressed by powers of the center of mass energy $s$ compared to
emission outside the nucleus and can be neglected in the eikonal
approximation considered here.

Therefore, our goal is to resum the cascade of long-lived gluons,
which the dipole in \fig{fock} develops before interacting with the
nucleus, and then convolute this cascade with the interaction
amplitudes of the gluons with the nucleus. To resum the cascade we
will use the dipole model developed by Mueller in
\cite{dip1,dip2,dip3}. Mueller's dipole model makes use of the 't Hooft's 
large-$N_c$ limit \cite{'tHooft}, taking $N_c$ to be very large while
keeping $\as \, N_c$ constant. In the large-$N_c$ limit only the
planar diagrams contribute, with the gluon line replaced by a double
line, corresponding to replacing the gluon by a quark--anti-quark pair
in the adjoint representation. The diagrams of \fig{fock} translate
into the planar large-$N_c$ diagrams shown in \fig{dip}. The top
diagram there represents emission of a single gluon in the original
incoming dipole. After replacing the gluon by the double quark line,
as shown on the top right of \fig{dip}, the original dipole splits
into two new dipoles, formed by the original quark line combined
with the anti-quark line in the gluon (blue--anti-blue dipole in
\fig{dip}) and by the original anti-quark line combined with the
quark line of the gluon (red--anti-red dipole in
\fig{dip}). Successive gluon emissions would only split the dipoles 
generated in the previous step into more dipoles, as depicted in the
bottom diagram of \fig{dip}.
%%%%%%%%%%%%%%%%%%%%%%%%%%%%%
\begin{figure}[h]
\begin{center}
\epsfxsize=10cm
\leavevmode
\hbox{\epsffile{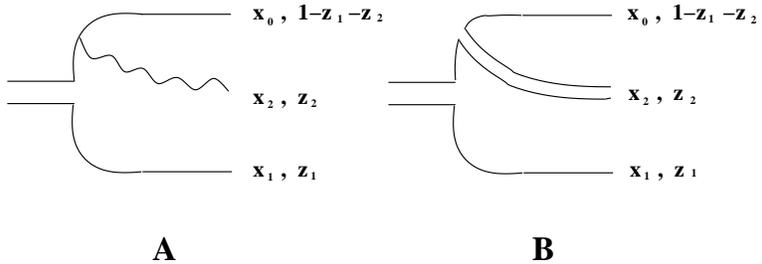}}
\end{center}
\caption{One (real) step of dipole evolution.}
\label{dipole}
\end{figure}
%%%%%%%%%%%%%%%%%%%%%%%%%%%%%%

To understand how the dipole model \cite{dip1,dip2,dip3} works let us
first calculate one step of the evolution, shown in
\fig{dipole}. There the original dipole consists of a quark--anti-quark 
pair, with the quark (top line) having transverse coordinate ${\un
x}_0$ and the anti-quark (bottom line) having transverse coordinate
${\un x}_1$. Suppose the initial dipole as a whole carries large light
cone momentum $p^+$ and the anti-quark carries the momentum
$k^+_1$. It will be convenient to define ratios of these momenta, such
that the anti-quark would carry the fraction $z_1 = k^+_1 /p^+$ of the
dipole's light cone momentum and the quark would carry the fraction
$1-z_1$. Now suppose we emit a gluon with transverse coordinate ${\un
x}_2$ and the light cone momentum fraction $z_2 = k^+_2/p^+$ as shown
in \fig{dipole}. As can be shown by an explicit calculation, such an
emission generates a logarithm of energy only if $z_2 \ll z_1,
1-z_1$. In this approximation the transverse coordinates of the
initial quark and anti-quark do not change: the emission is
recoilless. A straightforward calculation of the diagram in
\fig{dipole} using the rules of the light cone perturbation theory 
\cite{BL} yields \cite{dip1}
\be\label{dip1}
- \frac{i \, g \, t^a}{\pi} \, \left( \frac{{\un x}_{20}}{{\un x}_{20}^2} - 
\frac{{\un x}_{21}}{{\un x}_{21}^2} \right) \cdot {\un \epsilon}^\lambda_2
\ee
where we added the diagram where the gluon is emitted by the
anti-quark to the graph in \fig{dipole}. In \eq{dip1} ${\un
\epsilon}^\lambda_2$ is the polarization of the emitted gluon,
$t^a$ is the color matrix in fundamental representation and ${\un
x}_{ij} = {\un x}_i - {\un x}_j$. Squaring the contribution of
\eq{dip1}, summing over all polarizations and colors of the emitted
gluon, we obtain the following modification of the initial dipole's
wave function
\be\label{dip2}
\Phi^{(1)} ({\un x}_0, {\un x}_1, z_1) \, = \, \frac{\as \, C_F}{\pi^2} \, 
\int d^2 x_2 \, \int_{z_{in}}^{\mbox{min} \{z_1, 1-z_1\} } \frac{d z_2}{z_2} \, 
\frac{x_{01}^2}{x_{20}^2 \, x_{21}^2} \, \Phi^{(0)} ({\un x}_0, {\un x}_1, z_1)
\ee
where $x_{ij} = |{\un x}_{ij}|$ and $\Phi^{(0)}$ and $\Phi^{(1)}$ are
squares of the dipole light cone wave functions before and after the
gluon emission correspondingly. In the spirit of the leading
logarithmic approximation we cut off the $z_2$-integral from above by
min$\{z_1, 1-z_1\}$ to indicate that $z_2 \ll z_1, 1-z_1$: however for
the {\sl gluon} evolution the exact choice of this cutoff does not
matter. (A choice of the cutoff actually does matter for formulating
the evolution of the non-singlet structure functions, the so-called
reggeon evolution, in the language of the dipole model
\cite{IKMT}.) $z_{in}$ is some lower cutoff of the $z_2$ integration
which is not important for our purposes here.
%%%%%%%%%%%%%%%%%%%%%%%%%%%%%
\begin{figure}[h]
\begin{center}
\epsfxsize=10cm
\leavevmode
\hbox{\epsffile{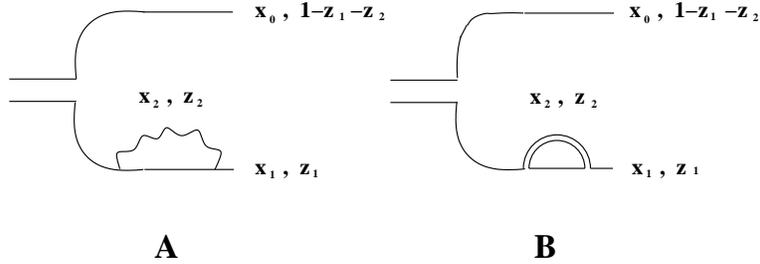}}
\end{center}
\caption{One (virtual) step of dipole evolution.}
\label{dipolev}
\end{figure}
%%%%%%%%%%%%%%%%%%%%%%%%%%%%%%

In addition to the real emission term of \fig{dipole}, we need to add
virtual corrections, one of which is demonstrated in
\fig{dipolev}. Other virtual diagrams involve the gluon coupling to 
both the quark and the anti-quark lines, as well as only to quark
line. The virtual corrections should cancel the real emission graph of
\fig{dipole} in the wave function squared if there is no interaction
with the target (for a more detailed analysis of real-virtual
cancellations see \cite{CM}). Therefore, to calculate the corrections
we just have to perform the ${\un x}_2$-integration in
\eq{dip2} inserting the overall minus sign for the virtual term. This yields 
the following modification of the dipole wave function
\be\label{dip3}
\Phi^{(1)} ({\un x}_0, {\un x}_1, z_1) \, = \, - \frac{4 \, \as \, C_F}{\pi} \, \ln  
\left( \frac{x_{01}}{\rho} \right) \ \int_{z_{in}}^{\mbox{min} \{z_1, 1-z_1\} } 
\frac{d z_2}{z_2} \ \Phi^{(0)} ({\un x}_0, {\un x}_1, z_1),
\ee
where $\rho$ is some ultraviolet cutoff needed to regulate the ${\un
x}_2$-integration in \eq{dip2}: it will cancel out for physical
quantities.

%%%%%%%%%%%%%%%%%%%%%%%%%%%%%
\begin{figure}[ht]
\begin{center}
\epsfxsize=15cm
\leavevmode
\hbox{\epsffile{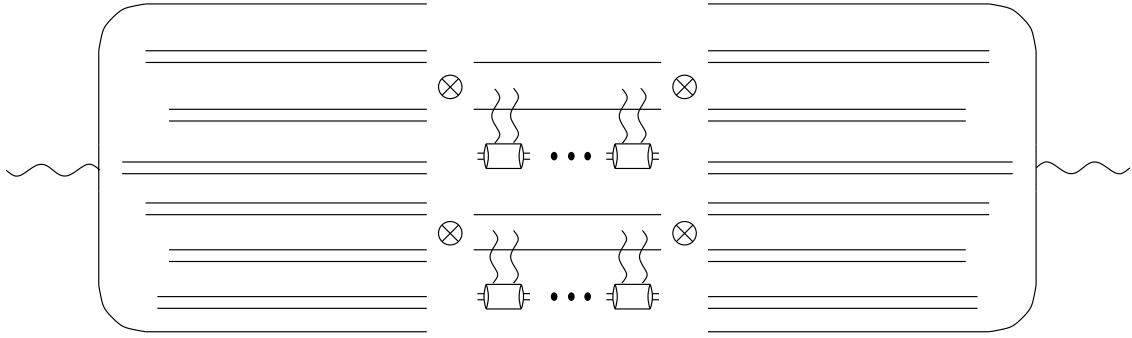}}
\end{center}
\caption{DIS as a dipole cascade interacting with the target (see text).}
\label{dis1}
\end{figure}
%%%%%%%%%%%%%%%%%%%%%%%%%%%%%%

Now resummation of the gluon cascade of \fig{fock} becomes more
tractable \cite{yuri_bk}. First of all, in the large-$N_c$ limit the
gluon cascade translates into a dipole cascade of \fig{dip}. As we
have seen in arriving at Eqs. (\ref{dip2}) and (\ref{dip3}), in the
leading logarithmic approximation the gluon emissions do not change
the transverse coordinates of the quark and anti-quark lines off which
the gluons were emitted. Therefore, the color dipoles have the same
transverse coordinates throughout the whole process: once they are
created the transverse coordinates do not change. Now resummation of
the dipole cascade reduces to the set of diagrams represented in
\fig{dis1}, which is a generalization of \fig{qqbar} for the case of 
quantum evolution corrections. (The cascade in \fig{dis1} is very
similar to the one discussed in Sect. \ref{BFKLeq} shown in
\fig{bfkl_phys}.) In DIS the incoming virtual photon splits into a
$q\bar q$ pair, which we will refer to as the original parent
dipole. The parent dipole may emit a soft gluon, like in \fig{dipole},
splitting itself into a pair of dipoles. (In \fig{dis1} the sum of
gluon emission from the quark and from the anti-quark lines is denoted
by the gluon (double quark) line disconnected from the parent quark
and anti-quark lines.) The dipole may also emit and re-absorb a gluon,
generating virtual corrections of \fig{dipolev}. The produced dipoles
may also evolve, generating more and more dipoles, as shown in
\fig{dis1}.

In the end the evolved system of dipoles interacts with the
nucleus. In the large-$N_c$ limit each dipole does not interact with
other dipoles during the evolution which generates all the
dipoles. For a large nucleus the dipole-nucleus interaction was
calculated in Sect. \ref{GM} yielding the answer in \eq{glaN2}. That
result resums powers of $\as^2 \, A^{1/3}$, as shown in
\eq{mrp}. Analyzing the diagrams for the interaction of several dipoles 
with the nucleus we see that interaction of, say, two dipoles with a
single nucleon is suppressed by a power of $A^{1/3}$ and is therefore
subleading and can be neglected. Interaction of two dipoles with two
nucleons in the large-$N_c$ limit is dominated by the diagram where
each of the dipoles interacts with only one of the nucleons (if we
require both dipoles to interact). In general one can argue that, in
the large-$N_c$ limit and at the leading order in $A$ (or,
equivalently, resumming powers of $\as^2 \, A^{1/3}$), the interaction
of any number of dipoles with the nucleus is dominated by {\sl
independent} interactions of each of the dipoles with a different set
of nucleons in the nucleus through multiple rescatterings shown in
\fig{qqbar}. This is depicted in \fig{dis1}: there each of the dipoles
present in the dipole wave function by the time it hits the nucleus
may interact with different nucleons in the nucleus by exchanging
pairs of gluons. (Only interactions of some of the dipoles are shown.) 
Therefore, the dipoles are completely non-interacting with each other:
they do not exchange gluons in the process of evolution, since those
corrections would be suppressed by powers of $N_c$, and they interact
with {\sl different} nucleons in the nucleus, which is correct at the
leading order in $A$
\cite{yuri_bk}.
%%%%%%%%%%%%%%%%%%%%%%%%%%%%%
\begin{figure}[ht]
\begin{center}
\epsfxsize=15cm
\leavevmode
\hbox{\epsffile{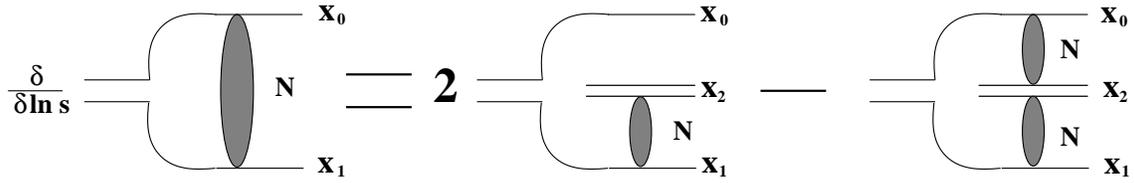}}
\end{center}
\caption{ Diagrammatic representation of the nonlinear evolution equation (\ref{eqN}).}
\label{dipeqn}
\end{figure}
%%%%%%%%%%%%%%%%%%%%%%%%%%%%%%

Summation of the dipole cascade of \fig{dis1} now becomes
straightforward and is illustrated in \fig{dipeqn}
\cite{yuri_bk}. There the dipole cascade and its interaction with the
target are denoted by a shaded oval. In one step of the evolution in
energy (or rapidity) a soft gluon is emitted in the dipole. If the
gluon is real (see \fig{dipole}), than the original dipole would be
split into two dipoles, as shown in \fig{dipeqn}. Either one of these
dipoles can interact with the nucleus with the other one not
interacting, which is shown by the first term on the right hand side
of \fig{dipeqn} with the factor of $2$ accounting for the fact that
there are two dipoles in the wave function now. Alternatively, both
dipoles may interact simultaneously, which is shown by the second term
on the right hand side of \fig{dipeqn}. This term comes in with the
minus sign due to the minus sign in our definition of the amplitude
$N$ in terms of the $S$-matrix, $S = 1-N$. The emitted gluon in one
step of evolution may be a virtual correction of \fig{dipolev}, which
is not shown in \fig{dipeqn}: in that case the original dipole would
not split into two, it would remain the same and would interact with
the target.

Combining the terms from \fig{dipeqn} with the virtual correction term
of \fig{dipolev}, and using Eqs. (\ref{dip2}) and (\ref{dip3}) we
write \cite{yuri_bk,balitsky}
\ben
\frac{\partial N ({\un x}_{01}, {\un b}, Y)}{\partial Y} \, = \, 
\frac{\as \, C_F}{\pi^2} \, 
\int d^2 x_2 \, \frac{x_{01}^2}{x_{20}^2 \, x_{21}^2} \, 
\left[ N ({\un x}_{02}, {\un b} + \frac{1}{2} \, {\un x}_{21}, Y) + 
N ({\un x}_{12}, {\un b} + \frac{1}{2} \, {\un x}_{20}, Y) - N ({\un
x}_{01}, {\un b}, Y) \right. 
\een
\be\label{eqN}
- \left. N ({\un x}_{02}, {\un b} + \frac{1}{2} \, {\un
x}_{21}, Y) \, N ({\un x}_{12}, {\un b} + \frac{1}{2} \, {\un x}_{20}, Y)
\right].
\ee
Note that since the impact parameter of the original dipole is ${\un
b} = ({\un x}_0 + {\un x}_1)/2$, the impact parameters of the produced
dipoles $x_{02}$ and $x_{12}$ are different from ${\un b}$ on the
right hand side of \eq{eqN}.

Equation (\ref{eqN}) is a nonlinear evolution equation, a solution of
which gives us the forward amplitude of the dipole-nucleus scattering
at high energy. The initial condition for \eq{eqN} at some initial
rapidity which we denote $Y=0$ is given by \eq{glaN2}
\cite{yuri_bk}. Therefore, \eq{eqN} resums all powers of the multiple 
rescattering parameter $\as^2 \, A^{1/3}$ along with the leading
logarithms of energy in the large-$N_c$ limit given by powers of $\as
\, N_c \, Y$. \eq{eqN} was originally derived by Balitsky in the framework 
of the effective theory of high energy interactions in \cite{balitsky}
and, independently, by one of the authors in \cite{yuri_bk} using the
formalism of the dipole model \cite{dip1,dip2,dip3}. It is commonly
referred to as the BK equation. It was also re-derived by Braun in
\cite{Braun1} using the expression for the triple pomeron vertex from
\cite{BW} in resummation of fan diagrams in \fig{fan}.

\eq{eqN} has a structure similar to the GLR equation (\ref{glreq}) \cite{GLR,GLR2} 
and appears to correspond to summation of the fan diagrams of
\fig{fan}.  If one neglects the impact parameter dependence
of the dipole amplitude $N$ and neglects the dependence of $n$ on the
angle of ${\un x}_{01}$, such that $N ({\un x}_{01}, {\un b}, Y)
\approx N (x_{01}, Y)$ one can perform a Fourier transform
\begin{eqnarray}\label{trfm}
  N(x_\perp, Y) = x_\perp^2 \int \frac{d^2 k}{2 \pi} \, e^{i {\un k}
    \cdot {\un x}} \, {\tilde N} (k, Y) 
\end{eqnarray}
to obtain \cite{yuri_bk}
\begin{eqnarray}\label{bsp}
  \frac{\partial {\tilde N} (k, Y)}{ \partial Y} = \frac{2 \, \as \,
  N_c}{\pi} \, \chi \left[ 0, \frac{i}{2} \left( 1 +
  \frac{\partial}{\partial \ln k} \right) \right] \, {\tilde N} (k, Y)
  - \frac{\as \, N_c}{\pi} \, {\tilde N}^2 (k, Y),
\end{eqnarray}
where $\chi (n, \nu)$ was defined in \eq{chi}. One can show that
\eq{bsp} is equivalent to \eq{glreq} if one identifies
\be
\phi (x, {\un k}^2) \, = \, \frac{N_c \, S_\perp}{\as \, \pi^2} \, 
{\tilde N} (k, Y = \ln 1/x_{Bj}),
\ee
which, combined with \eq{trfm}, leads to
\be\label{fakephi}
\phi (x, {\un k}^2) \, = \, \frac{N_c \, S_\perp}{\as \, 2 \, \pi^3} \, 
\int d^2 x \, e^{- i \, {\un k} \cdot {\un x}} \, \frac{1}{x_\perp^2} \, 
N(x_\perp, Y= \ln 1/x_{Bj}).
\ee
For $N$ from \eq{glaN2} the unintegrated gluon distribution function
from \eq{fakephi} looks similar to the gluon distribution from
\eq{ww2}. However, the important difference is that the
Glauber-Mueller dipole amplitude $N_G$ which enters \eq{ww2} is the
amplitude for a {\sl gluon} (adjoint) dipole, while $N$ in
\eq{fakephi} is the amplitude for a {\sl quark} (fundamental) dipole. Even 
this formal difference notwithstanding, it is not at all clear whether
\eq{fakephi}, or even \eq{ww2}, would still be valid for the dipole amplitude $N$
from \eq{eqN} with all the evolution effects included: after all,
\eq{ww2} was derived only in the quasi-classical limit. We will return
to this question in Sect. \ref{clpa} where we will discuss particle
production.

%%%%%%%%%%%%%%%%%%%%%%%%%%%%%%%%%%%%%%%%%%%%%%%%%%%%%%%%%%%%%%%%%%%%%%%%%%%%%%%%%%%%%%%%%%%%

\subsection{Solving the Evolution Equations}
\label{see}

After the JIMWLK and the BK evolution equation were written down there
has been a number of analytical \cite{LT,yuri_bk,IIM1,MT,Mueller3} and
numerical \cite{Braun1,GBMS,GLLM,Misha,AAMSW,RW} attempts to solve
them. While an exact analytical solution valid both inside and outside
of the saturation region still does not exist, there is a number of
good approximations for various kinematic regions. Here we are {\sl
not} going to give an extensive review of the existing approaches
referring the interested reader to the existing review articles
\cite{ILM,IV,Heribert}. Instead we will outline features of the solution
of the evolution equations which will be most important for our
discussion of particle production below.

\subsubsection{Linear Evolution}
\label{LE}

When the dipole-nucleus interactions are weak and the forward
scattering amplitude is small $N \ll 1$, we can neglect the quadratic
term in \eq{eqN} and write \cite{dip1,dip2}
\be\label{bfkl_dip}
\frac{\partial N ({\un x}_{01}, {\un b}, Y)}{\partial Y} \, = \, \frac{\as \, C_F}{\pi^2} \, 
\int d^2 x_2 \, \frac{x_{01}^2}{x_{20}^2 \, x_{21}^2} \, 
\left[ N ({\un x}_{02}, {\un b} + \frac{1}{2} \, {\un x}_{21}, Y) + 
N ({\un x}_{12}, {\un b} + \frac{1}{2} \, {\un x}_{20}, Y) - N ({\un
x}_{01}, {\un b}, Y) \right].
\ee
This equation is equivalent to the BFKL equation \cite{BFKL}, as has
been shown in the framework of the dipole model in \cite{NW}.
Defining the the eigenfunctions of the Casimir operators of conformal
algebra \cite{Lipatov2,Lipatov3}
\be\label{enn}
E^{n, \nu} (\rho_0, \rho_1) \, = \, \left( \frac{\rho_{01}}{\rho_0 \,
\rho_1} \right)^{\frac{1+n}{2} + i \nu} \ \left(
\frac{\rho_{01}^*}{\rho_0^* \, \rho_1^*} \right)^{\frac{1-n}{2} + i
\nu}
\ee
with (in general) complex coordinates $\rho_i$, one can show that they
are also eigenfunctions of the dipole kernel of
\eq{bfkl_dip} \cite{NW,KSW}
\be\label{dipeig}
\int d^2 x_2 \, \frac{x_{01}^2}{x_{02}^2 \,
x_{12}^2} \, \left[ E^{n, \nu} (x_0, x_2) + E^{n, \nu} (x_2, x_1) -
E^{n, \nu} (x_0, x_1) \right] \, = \, 4 \, \pi \, \chi(n, \nu) \,  
E^{n, \nu} (x_0, x_1)
\ee
with $\chi(n, \nu)$ given by \eq{chi}. \eq{dipeig} demonstrates that
the BFKL equation \cite{BFKL} and \eq{bfkl_dip} have the same
eigenfunctions and eigenvalues, and are, therefore, equivalent.

Let us write down explicitly the solution of \eq{bfkl_dip} for the
case of a large nucleus, where we can neglect the $\un b$-dependence
in $N ({\un x}_{01}, {\un b}, Y)$ along with its dependence on the
azimuthal angle of ${\un x}_{01}$. Using again the Fourier transform
of \eq{trfm} we rewrite \eq{bfkl_dip} as (see \eq{bsp})
\begin{eqnarray}\label{ksp}
  \frac{\partial {\tilde N} (k, Y)}{ \partial Y} = \frac{2 \, \as \,
  N_c}{\pi} \, \chi \left[ 0, \frac{i}{2} \left( 1 +
  \frac{\partial}{\partial \ln k} \right) \right] \, {\tilde N} (k, Y)
\end{eqnarray}
the solution of which is given by
\be\label{dsol1}
{\tilde N} (k, Y) \, = \, \exp \left\{ \frac{2 \, \as \,
  N_c}{\pi} \, Y \, \chi \left[ 0, \frac{i}{2} \left( 1 +
  \frac{\partial}{\partial \ln k} \right) \right] \right\} \, {\tilde C} (k),
\ee
where ${\tilde C}(k)$ is some unknown function, which could be
determined by using the initial conditions (\ref{glaN2}) in
\eq{trfm}. To determine the high energy asymptotics of the solution
(\ref{dsol1}) we write ${\tilde C}(k)$ as a Mellin transform
\be\label{CM}
C (k) \, = \, \int_{-\infty}^\infty d \nu \, \left( \frac{Q_{s0}}{k} 
\right)^{1 + 2 \, i \, \nu} \, {\tilde C}_\nu.
\ee
Here $Q_{s0}$ is the saturation scale in the quasi-classical
approximation given by \eq{qsmv}. From here on we will use the
subscript $0$ to distinguish it from the full energy-dependent
saturation scale which would result from our analysis of the
JIMWLK and BK evolution equations. 

Substituting \eq{CM} in \eq{dsol1} we obtain
\be\label{dsol2}
{\tilde N} (k, Y) \, = \, \int_{-\infty}^\infty d \nu \
e^{2 \, \bas \, \chi (0, \nu) \, Y } \, \left( \frac{Q_{s0}}{k} 
\right)^{1 + 2 \, i \, \nu} \, {\tilde C}_\nu,
\ee
where we have defined
\be\label{bas}
\bas \, \equiv \, \frac{\as \, N_c}{\pi}.
\ee
Substituting the result of \eq{dsol2} into \eq{trfm} and integrating
over $\un k$ yields
\be\label{dsol3}
N (x_T, Y) \, = \, \int_{-\infty}^\infty d \nu \ e^{2 \, \bas \,
\chi (0, \nu) \, Y } \, (x_T Q_{s0})^{1 + 2 \, i \, \nu} \, C_\nu
\ee
where
\be
C_\nu \, \equiv \, {\tilde C}_\nu \, 2^{- 2 \, i \, \nu} \, \frac{\Gamma \left( 
\frac{1 - 2 \, i \, \nu}{2}\right)}{\Gamma \left( 
\frac{1 + 2 \, i \, \nu}{2}\right)}.
\ee

%%%%%%%%%%%%%%%%%%%%%%%%%%%%%
\begin{figure}
\begin{center}
\epsfxsize=10cm
\leavevmode
\hbox{\epsffile{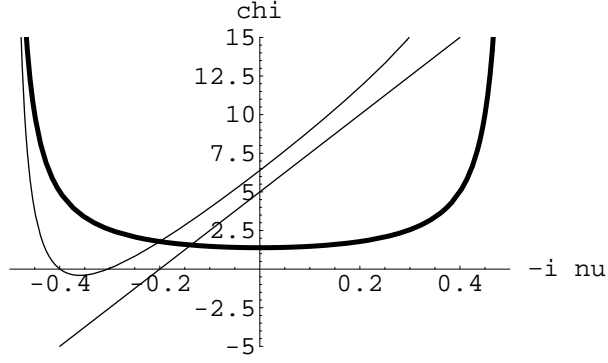}}
\end{center}
\caption{The eigenvalue of the BFKL kernel $\chi (0, \nu)$ plotted as a function of 
$-i \, \nu$ (thick line). The thin straight line is due to the linear
term $2 \, i \, \nu \, \ln (x_T Q_{s0})$ in the exponent of
\eq{dsol4}. The thin curve is a sum of the thick line and the thin
straight line and it represents the complete expression in the
exponent of \eq{dsol4}. }
\label{chi_fig}
\end{figure}
%%%%%%%%%%%%%%%%%%%%%%%%%%%%%%
\eq{dsol3} gives us the solution of the dipole version of the BFKL equation 
(\ref{bfkl_dip}) with the coefficients $C_\nu$ to be fixed from the
initial conditions of \eq{glaN2}. Now let us rewrite \eq{dsol3} as
\be\label{dsol4}
N (x_T, Y) \, = \, x_T Q_{s0} \, \int_{-\infty}^\infty d \nu \
\exp \left[ 2 \, \bas \, \chi (0, \nu) \, Y  + 2 \, i \, \nu \, 
\ln (x_T Q_{s0}) \right] \, C_\nu
\ee
The integral in \eq{dsol4} can be approximated by the saddle point
method in the following two important cases.

\begin{itemize}

\item $x_T Q_{s0} \lsim 1$ In this case the term 
$2 \, i \, \nu \, \ln (x_T Q_{s0})$ in the exponent of
\eq{dsol4} is small compared to $\bas \, Y$ and does not significantly 
influence the position of the saddle point. The saddle point is
determined mostly by the function $\chi (0, \nu)$, which is shown in
\fig{chi_fig} by thick line as a function of $-i\, \nu$: there one can
clearly see a saddle point near $\nu = 0$.  We use the expansion of
$\chi (0, \nu)$ around $\nu =0$ (see \eq{sp}) to determine the saddle
point
\be\label{splla}
\nu^*_{LLA} \, \approx \,  \frac{i \, \ln (x_T Q_{s0})}{14 \,
\zeta (3) \, \bas \, Y},
\ee
where the subscript LLA stands for leading logarithmic
approximation. Performing the saddle point integration we write
(cf. \eq{spa})
\be\label{dsol5}
N (x_T, Y) \, = \, x_T Q_{s0} \, C_{\nu^*_{LLA}} \,
\sqrt{\frac{\pi}{14 \, \zeta (3) \, \bas \, Y}} \ \exp \left[ (\alpha_P - 1) \, Y -
\frac{\ln^2 (x_T Q_{s0})}{14 \, \zeta (3) \, \bas \, Y}
\right]
\ee
with the BFKL pomeron intercept $\alpha_P - 1$ given by \eq{bfkl_int}.

\item $x_T Q_{s0} \ll 1$ This case is also known as the double 
logarithmic approximation (DLA), since here transverse logarithms like
$\ln (x_T Q_{s0})$ become important, leading to a new resummation
parameter $\as \, Y \, \ln (x_T Q_{s0})$. Transverse logarithms are of
course resummed by the Dokshitzer, Gribov, Lipatov, Altarelli, Parisi
(DGLAP) evolution equation \cite{DGLAP}. The DLA region is where the
results of DGLAP and BFKL are identical \cite{MuellerCarg}, since
there both equations resum powers of the same parameter $\as \, Y \,
\ln (x_T Q_{s0})$.
%, one by resumming terms with $\as \, \ln (x_T
%Q_{s0})$, the other one by resumming powers of $\as \, Y$. 

As shown in \fig{chi_fig} by the thin curve, when $\ln (1/x_T
Q_{s0})$ becomes large, it shifts the position of saddle point towards
$\nu = - i/2$ \cite{Ryskin}. In that region we approximate $\chi (0,
\nu)$ by
\be\label{spdla}
\chi (0, \nu) \, \approx \, \frac{1}{1-2 \, i \, \nu}
\ee
which leads to the saddle point at 
\be\label{spdla1}
\nu^*_{DLA} \, \approx \, - \frac{i}{2} \, 
\left( 1 - \sqrt{\frac{2 \, \bas \, Y}{\ln 1/(x_T
Q_{s0})}} \right).
\ee
Using \eq{spdla} in \eq{dsol4} and performing the integration over
$\nu$ in the saddle point approximation yields
\be\label{dsol6}
N (x_T, Y) \, = \, (x_T Q_{s0})^2 \, C_{\nu^*_{DLA}} \,
\frac{\sqrt{\pi}}{2} \, (2 \, \bas \, Y)^{1/4} \, \ln^{-3/4} \left( \frac{1}{x_T
Q_{s0}} \right) \ \exp \left[ 2 \, \sqrt{2 \, \bas \, Y \, \ln 1/(x_T
Q_{s0})} \right].
\ee
The coefficient $C_{\nu^*_{DLA}}$ in \eq{dsol6} is in fact important
\cite{KKT}, since it may modify the prefactor: if we require, for instance, 
that at $Y=0$ \eq{dsol4} reduces to the amplitude given by a two-gluon
exchange from \eq{1sc3}, or, equivalently, by the first term in
expansion of \eq{glaN2}, than we would get
\be\label{c2g}
C_{\nu } \, = \, \frac{1}{4 \, \pi \, (1 - 2 \, i \, \nu)^2}.
\ee
Using \eq{c2g} in \eq{dsol6} yields
\be\label{dsol7}
N (x_T, Y) \, = \, (x_T Q_{s0})^2 \,
\frac{1}{8 \, \sqrt{\pi}} \, (2 \, \bas \, Y)^{-3/4} \, \ln^{1/4} \left( \frac{1}{x_T
Q_{s0}} \right) \ \exp \left[ 2 \, \sqrt{2 \, \bas \, Y \, \ln
1/(x_T Q_{s0})} \right].
\ee
\end{itemize}

The general analytic linear solution found above in \eq{dsol4} allows
one to estimate the boundary of the saturation region by determining
the limit of applicability of the linear equation (\ref{bfkl_dip})
\cite{LT,IIM1,MT}. Linear evolution is applicable only if $N \ll 1$, 
and, therefore, breaks down when $N \sim 1$. To estimate when that
happens, let us rewrite \eq{dsol4} as
\be\label{dsol8}
N (x_T, Y) \, = \, \int_{-\infty}^\infty d \nu \
\exp \left[ 2 \, \bas \, \chi (0, \nu) \, Y  + (2 \, i \, \nu + 1 ) \, 
\ln (x_T Q_{s0}) \right] \, C_\nu.
\ee
Following \cite{MT} we argue that $N$ becomes of order one when the
value of the exponent in \eq{dsol8} at the saddle point vanishes. This
translates into the following two equations
\be\label{mt1}
2 \, \bas \, \chi' (0, \nu_0) \, Y + 2 \, i \, \ln (Q_{s0}/Q_s) \, =
\, 0
\ee
(saddle point condition) and
\be\label{mt2}
2 \, \bas \, \chi (0, \nu_0) \, Y  + (2 \, i \, \nu_0 + 1 ) \, 
\ln (Q_{s0}/Q_s) \, = \, 0
\ee
(vanishing of the power of the exponent at the saddle point). We have
substituted $x_T \approx 1/Q_s$ in Eqs. (\ref{mt1}) and
(\ref{mt2}), which corresponds to {\sl defining} the saturation scale
$Q_s$ by requiring that $N(1/Q_s, Y) \approx 1$. One can see that this
definition is consistent with the quasi-classical approach of
Sect. \ref{QCA}: if we put $Y=0$ in Eqs. (\ref{mt1}) and (\ref{mt2}),
which corresponds to removing the small-$x$ evolution, we obtain $Q_s
= Q_{s0}$ as the solution, meaning that the full saturation scale
$Q_s$ maps onto the quasi-classical saturation scale $Q_{s0}$ in the
small rapidity limit.

Solving Eqs. (\ref{mt1}) and (\ref{mt2}) yields \cite{MT}
\be\label{qsy0}
Q_s (Y) \, = \, Q_{s0} \, \exp \left\{ 2 \, \bas \, \frac{\chi (0,
\nu_0)}{2 \, i \, \nu_0 + 1} \, Y \right\}
\ee
with 
\be\label{nu0}
\frac{\chi' (0, \nu_0)}{\chi (0, \nu_0)} \, = \, \frac{2 \, i}{2 \, i \, \nu_0 + 1}
\ee
giving 
\be\label{nu0val}
\nu_0 \approx - i \, 0.1276. 
\ee
Using this value of $\nu_0$ in
\eq{qsy0} we get
\be\label{qsy1}
Q_s (Y) \, \approx \, Q_{s0} \, e^{ 2.44 \, \bas \, Y }.
\ee
A more detailed determination of the saturation scale specifying the
prefactor in \eq{qsy0} can be found in \cite{MT,IIM1}.
%%%%%%%%%%%%%%%%%%%%%%%%%%%%%
\begin{figure}
\begin{center}
\epsfxsize=11.5cm
\leavevmode
\hbox{\epsffile{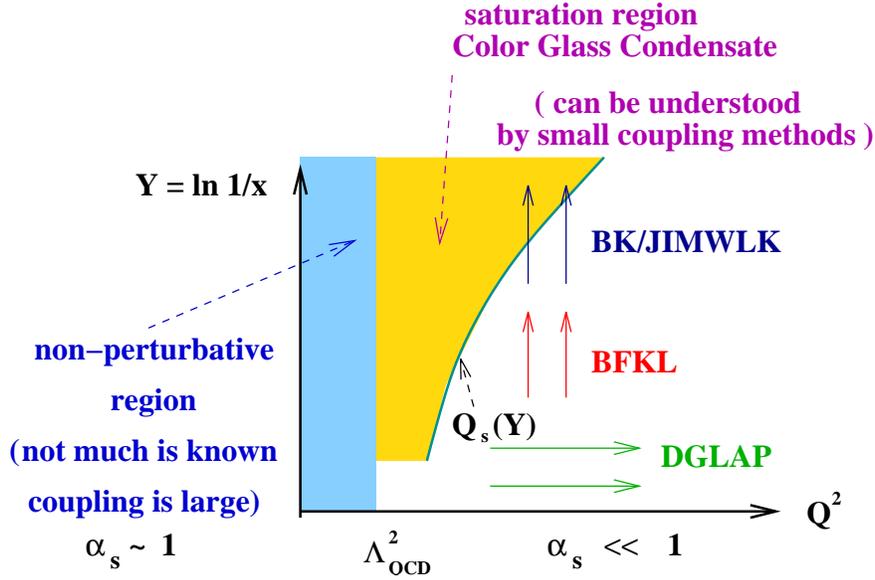}}
\end{center}
\caption{Our understanding of high energy QCD interactions plotted in the plane 
of rapidity $Y = \ln 1/x_{Bj}$ and $\ln Q^2$ (see text).  }
\label{satbk}
\end{figure}
%%%%%%%%%%%%%%%%%%%%%%%%%%%%%%

\eq{qsy0}, or, equivalently, \eq{qsy1}, give us the dependence of the saturation 
scale $Q_s$ on rapidity or energy. We see that as rapidity $Y$
increases, so does the saturation scale. \eq{qsy0} generalizes
\eq{qsmv} for $Q_{s0}$ by including the effects of small-$x$ evolution, which 
bring in rapidity dependence. Combining the two equations one writes
\be\label{qsy2}
Q_s^2 (Y) \, = \, \frac{4 \, \pi \, \as^2 \, C_F}{N_c} \, \rho \,
T({\un b}) \, \exp \left\{ 4 \, \bas \, \frac{\chi (0,
\nu_0)}{2 \, i \, \nu_0 + 1} \, Y \right\}.
\ee
\eq{qsy2} demonstrates that for a large nucleus and at high energies the saturation 
scale is proportional to
\be\label{qsy3}
Q_s^2 \, \sim \, A^{1/3} \, e^{c \, \bas \, Y} \, \sim \,
A^{1/3} \, s^{c \, \bas}
\ee
with $c$ some constant, $c=4.88$ in \eq{qsy1}.  Therefore, $Q_s$ is an
increasing function of both the atomic number $A$ and of center of
mass energy $s$. It only gets larger as we go to higher energies,
making the coupling constant $\as (Q_s)$ smaller. Thus the saturation
approach is better justified as we go to progressively higher
energies.

Our understanding of high energy scattering at this point in the
article is summarized in \fig{satbk}. There we depict the plane of two
variables essential to the high energy scattering: the typical
transverse momentum scale $Q^2$ and the rapidity $Y=\ln
1/x_{Bj}$. Right away the region of $Q < \Lambda_{QCD}$ should be
excluded from our analysis here, since there the coupling is large and
not much can be understood using the perturbative methods we describe
here. At $Q \gg \Lambda_{QCD}$ we may distinguish two important
regions. One region is given by $Q > Q_s (Y)$, where the amplitude $N$
is small and the linear BFKL evolution equation (\ref{bfkl_dip})
\cite{BFKL} applies, along with the DGLAP evolution \cite{DGLAP}, as
shown in \fig{satbk}. In the region with $Q < Q_s (Y)$ the amplitude
$N$ becomes of order $1$ and the non-linear effects of the BK and
JIMWLK equations \cite{jklw1}-\cite{Weigert:2000gi},
\cite{balitsky,yuri_bk} become important: this region is called the
{\sl saturation region} and is given by the shaded area in
\fig{satbk}. Saturation region is also called Color Glass
Condensate. It is important to point out that all this non-linear
dynamics takes place at $Q_s \gsim Q \gg \Lambda_{QCD}$, i.e., in the
perturbative region where the strong coupling constant is small and
our calculations are justified.

\subsubsection{Geometric Scaling}
\label{GS}

Let us now analyze the behavior of the solution of \eq{eqN} deep
inside the saturation region, where the non-linear effects are very
important. Deep inside the saturation region, when the dipole size
$x_T$ becomes large, $x_T \gg 1/Q_s$ (but still $x_T \ll
1/\Lambda_{QCD}$), the quasi-classical Glauber-Mueller amplitude from
\eq{glaN2} approaches $1$. Analyzing \eq{eqN} we easily see that $N=1$ 
is also a stationary solution of that equation. Therefore, we conclude
that
\be\label{bd}
N ({\un x}, {\un b}, Y) \, = \, 1 , \hspace*{1cm} x_T \gg 1/Q_s (Y),
\ee
corresponding to the black disk limit of \eq{black} \cite{LL}. Now let
us determine the asymptotic approach to the black disk limit of
\eq{bd} following \cite{LT,IIM1}. To do that, let us write 
\be\label{NS}
N ({\un x}, {\un b}, Y) \, = \, 1 - S ({\un x}, {\un b}, Y)
\ee
where $S$ is the $S$-matrix of the dipole-nucleus collision, which is
small when the system approaches the black disk limit ($N$ approaches
1). Substituting \eq{NS} in \eq{eqN} and keeping only terms linear in
$S$ yields \cite{LT,IIM1,IV}
\be\label{eqS1}
\frac{\partial S ({\un x}_{01}, {\un b}, Y)}{\partial Y} \, = \, - \frac{\as \, C_F}{\pi^2} \, 
\int d^2 x_2 \, \frac{x_{01}^2}{x_{20}^2 \, x_{21}^2} \, S ({\un x}_{01}, {\un b}, Y),
\ee
where the integral over dipole sizes goes over $x_{02}, x_{12} > 1/Q_s
(Y)$. Hence we have to replace the ultraviolet (UV) cutoff from
\eq{dip3} with $1/Q_s (Y)$ obtaining
\be\label{eqS2}
\frac{\partial S ({\un x}_{01}, {\un b}, Y)}{\partial Y} \, = \, - 
\frac{4 \, \as \, C_F}{\pi} \, 
\ln \left[ x_{01} Q_s (Y) \right] \, S ({\un x}_{01}, {\un b}, Y).
\ee
Defining the {\sl scaling} variable 
\be\label{xidef}
\xi \, \equiv \, \ln \left[ x_{01}^2 Q_s^2 (Y) \right]
\ee
with (cf. \eq{qsy3})
\be\label{cdef}
c \, \bas \, \equiv \, \frac{\partial \xi}{\partial Y} \, = \,
\frac{\partial \ln \left[ x_{01}^2 Q_s^2 (Y) \right]}{\partial Y}
\ee
we rewrite \eq{eqS2} as
\be\label{eqS3}
\frac{\partial S}{\partial \xi} \, = \, - \frac{1}{c} \, \xi \, S.
\ee
Solution of \eq{eqS3} can be straightforwardly written as
\cite{LT,IIM1}
\be\label{LTs}
S (\xi) \, = \, S_0 \, e^{-\xi^2 /c} \, = \, S_0 \, e^{-\ln^2 \left[
x_{01}^2 Q_s^2 (Y) \right]/c}
\ee
with $S_0 <1$ a constant. Corresponding dipole amplitude $N$ is given
by
\be\label{LTf}
N (\xi)  \, = \, 1 - S_0 \, e^{-\xi^2 /c} \, = \, 1 - S_0 \, e^{-\ln^2 \left[
x_{\perp}^2 Q_s^2 (Y) \right]/c}, \hspace*{1cm} x_T \gg 1/Q_s (Y).
\ee
\eq{LTf} is knows as Levin-Tuchin formula \cite{LT}.

\begin{figure}[t]
\begin{center}
\epsfxsize=12cm
\leavevmode
\hbox{\epsffile{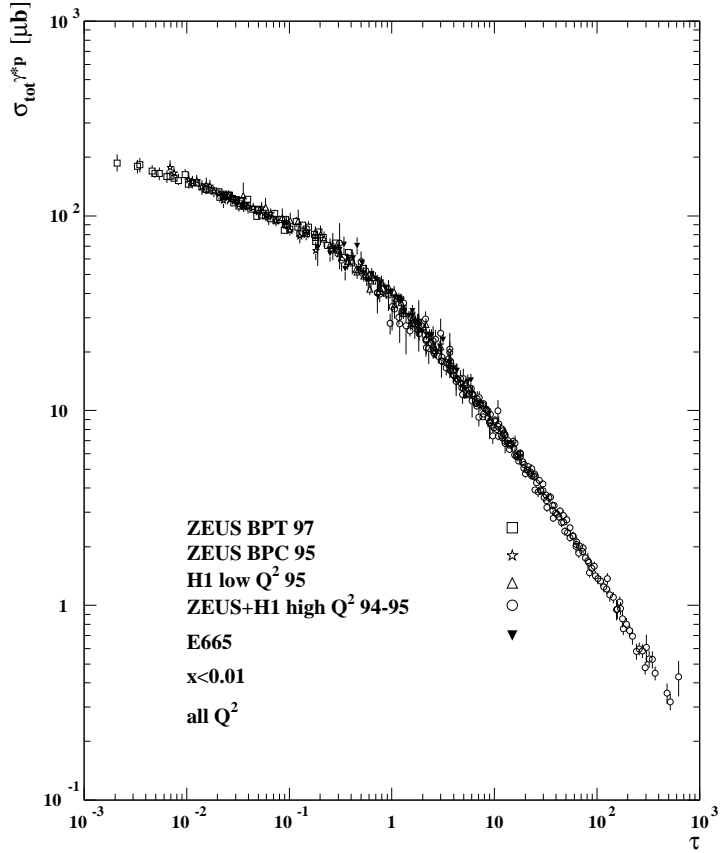}}
\end{center}
\caption{HERA data on the total DIS $\gamma^* p$ cross section plotted in 
\protect\cite{geom} as a function of the scaling variable $\tau = Q^2/Q_s^2 (x_{Bj})$. }
\label{gsdis}
\end{figure}

Note that the $S$-matrix and the amplitude $N$ of the dipole-nucleus
scattering given by Eqs. (\ref{LTs}) and (\ref{LTf}) are functions of
a single variable $\xi$, or, more precisely, of the combination
$x_{\perp} Q_s (Y)$. This phenomenon is known as {\sl geometric
scaling}. The original argument for geometric scaling has been given
by McLerran and Venugopalan in \cite{MV}, where it was suggested that
small-$x$ nuclear or hadronic wave functions are described by a
momentum scale $Q_s$ being the only scale in the problem and,
therefore, all transverse coordinate (or momentum) dependent physical
observables should depend on the combination $x_{\perp} Q_s (Y)$ (or
$k_T/Q_s$). Geometric scaling has been demonstrated in an analysis of
the HERA DIS data by Stasto, Golec-Biernat and Kwiecinski in
\cite{geom}, presenting one of the strongest arguments for observation 
of saturation phenomena at HERA. These results are shown here in
\fig{gsdis} from \cite{geom}, where the authors of \cite{geom} plot combined HERA 
data on the total DIS $\gamma^* p$ cross section $\sigma^{\gamma^*
p}_{tot}$ for $x_{Bj} < 0.01$ as a function of the scaling variable
$\tau = Q^2/Q_s^2 (x_{Bj})$. One can see that, amazingly enough, all
the data falls on the same curve, indicating that $\sigma^{\gamma^*
p}_{tot}$ is a function of a single variable $Q^2/Q_s^2 (x_{Bj})$! 
This gives us the best to date experimental proof of geometric
scaling. (For a similar analysis of DIS data on nuclear targets see
\cite{FRWS} and the first reference of \cite{AAMSW}.)
%%%%%%%%%%%%%%%%%%%%%%%%%%%%%
%%%%%%%%%%%%%%%%%%%%%%%%%%%%%%

The fact that geometric scaling is a property of the solution of the
BK equation has been later demonstrated in \cite{LT,IIM1}. In the next
Section we will also show that such scaling is also valid outside of
the $Q < Q_s (Y)$ saturation region.

\subsubsection{Extended Geometric Scaling}

In \cite{IIM1} Iancu, Itakura and McLerran noticed that the geometric
scaling of \cite{LT}, which was originally derived as a solution of
the BK equation inside the saturation region, also applies in an area
outside of that region. To see this let us first remember that outside
of saturation region ($x_T < 1/Q_s (Y)$) the nonlinear BK equation
(\ref{eqN}) reduces to the linear BFKL evolution
(\ref{bfkl_dip}). Therefore, the dipole-nucleus amplitude $N$ is still
given by \eq{dsol8}, which we rewrite here as
\be\label{dsol9}
N (x_T, Y) \, = \, \int_{-\infty}^\infty d \nu \, e^{P (\nu)} \,
C_\nu
\ee
with
\be
P (\nu) \, = \, 2 \, \bas \, \chi (0, \nu) \, Y + (2 \, i \, \nu + 1 )
\, \ln (x_T Q_{s0}).
\ee
In order to perform the integration in \eq{dsol9} around the saddle
point from \eq{nu0} we expand $P (\nu)$ around $\nu_0$
\be\label{ptay}
P(\nu) \, = \, P (\nu_0) + P'(\nu_0) \, (\nu - \nu_0) + \frac{1}{2} \,
P'' (\nu_0) \, (\nu - \nu_0)^2 + \ldots .
\ee
Following \cite{IIM1} we first assume that the integral in \eq{dsol9}
is dominated by its value at the saddle point $\nu_0$. Then noticing
that
\be\label{p0}
P (\nu_0) \, = \, (2 \, i \, \nu_0 + 1 ) \, \ln \left[ x_T Q_{s}
(Y) \right] \, = \, \left( \frac{1}{2} + i \, \nu_0 \right) \, \xi
\ee
we obtain \cite{IIM1}
\be\label{dsol10}
N (x_T, Y) \, \approx \, C_{\nu_0} \, e^{P (\nu_0)} \, = \, \,
C_{\nu_0} \, \left[ x_T Q_{s} (Y) \right]^{1 + 2 \, i \, \nu_0} \,
= \, C_{\nu_0} \, e^{\left( \frac{1}{2} + i \, \nu_0 \right) \, \xi}.
\ee
\eq{dsol10} obviously exhibits geometric scaling, since the amplitude 
$N$ in it is a function of $\xi$ only. Therefore geometric scaling
also works outside of the saturation region for $x_T < 1/Q_s (Y)$
($Q > Q_s (Y)$) \cite{IIM1}. This effect is called {\sl extended
geometric scaling}.

To determine the applicability region of \eq{dsol11} we use the
formula (\ref{ptay}), truncate the series in it at the quadratic
order, substitute the result in \eq{dsol9} and integrate over $\nu$
obtaining
\be\label{dsol11}
N (x_T, Y) \, = \, C_{\nu_0} \, \sqrt{\frac{2 \, \pi}{- P''
(\nu_0)}} \, \exp \left\{P (\nu_0) - \frac{[P' (\nu_0)]^2}{2 \, P''
(\nu_0)} \right\} .
\ee
Since
\be\label{p1}
P' (\nu_0) \, = \, 2 \, i \, \ln \left[ x_T Q_{s} (Y) \right] \, =
\, i \, \xi
\ee
and
\be\label{p2}
P'' (\nu_0) \, = \, 2 \, \bas \, \chi'' (0, \nu_0) \, Y
\ee
the second (diffusion) term in the exponent of \eq{dsol11} along with
the prefactor in \eq{dsol11} violate geometric scaling. These scaling
violation corrections are small and can be neglected as long as
\be\label{pcond}
\frac{[P' (\nu_0)]^2}{2 \, P'' (\nu_0)} \, \ll \, P (\nu_0).
\ee
%(The prefactor in the exponent will be absorbed in the definition of
%$Q_s (Y)$ in a more complete treatment than the one outlined here in
%Sect. \ref{LE} \cite{MT}.) 
Using Eqs. (\ref{p0}), (\ref{p1}) and (\ref{p2}) in \eq{pcond} yields
\be
x_T \, > \, \frac{1}{Q_s (Y)} \, e^{\bas \, \chi'' (0, \nu_0) \,
(2 \, i \, \nu + 1 ) \, Y}.
\ee
Replacing $x_T$ by $1/Q$ this translates into
\be\label{gs1}
Q \, < \, Q_s (Y) \, e^{- \bas \, \chi'' (0, \nu_0) \, (2 \, i \,
\nu_0 + 1 ) \, Y} \, = \, Q_s (Y) \, e^{30.4 \, \bas \, Y}.
\ee
However, the condition in \eq{gs1}, while being necessary for extended
geometric scaling to apply, is not sufficient. What we learn from
\eq{dsol11} is that geometric scaling is violated when the regions of 
$\nu$ further away from $\nu_0$ start to contribute in the integral of
\eq{dsol9}. This is also demonstrated in the DLA approximation: clearly 
\eq{dsol7} can not be written as a function of a single variable $\xi$ and 
thus violates geometric scaling. We may, therefore, define the border
of the extended geometric scaling region by the transition point
between the LLA saddle point of \eq{splla} and the DLA saddle point of
\eq{spdla1} \cite{IIM1}. The point of closest approach of the two saddle 
points in Eqs. (\ref{splla}) and (\ref{spdla1}) is at 
\be
\ln \frac{1}{x_T \, Q_{s0}} \, \approx \, 3.28 \ \bas \, Y 
\ee
such that the border of extended geometric scaling region is defined
by the scale
\be
k_{\rm geom} \, \equiv \, Q_{s0} \, e^{3.28 \ \bas \, Y}  
\ee
which, using \eq{qsy1} can be rewritten as \cite{IIM1,KKT}
\be\label{kgeom}
k_{\rm geom} \, = \, Q_{s} (Y) \, \left( \frac{Q_{s} (Y)}{Q_{s0}}
\right)^{0.34}.
\ee
Therefore, extended geometric scaling is valid up to 
\be
Q \, \leq \, k_{\rm geom},
\ee
which is a more restrictive condition than \eq{gs1}. The exact value
of $k_{\rm geom}$ may still be slightly different from the one given by
\eq{kgeom} \cite{IIM1,MT,Mueller3}: what is important is that at large $Y$ this 
scale is much larger than the saturation scale, $Q_s (Y) \ll k_{\rm
geom}$, allowing for a parametrically wide region of extended
geometric scaling. 

%%%%%%%%%%%%%%%%%%%%%%%%%%%%%
\begin{figure}[t]
\begin{center}
\epsfxsize=11.5cm
\leavevmode
\hbox{\epsffile{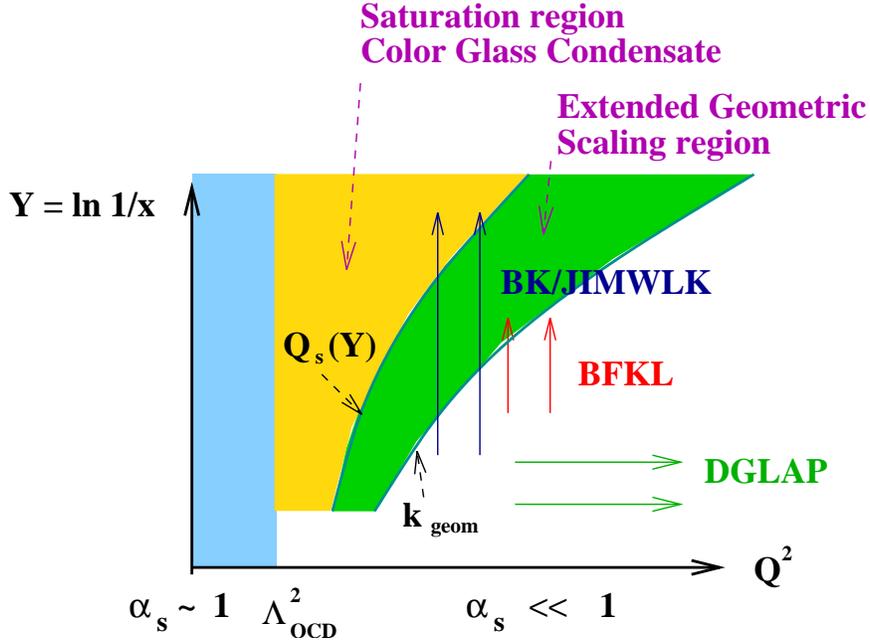}}
\end{center}
\caption{The summary of our knowledge of high energy QCD interactions 
plotted again in the plane of rapidity $Y = \ln 1/x_{Bj}$ and $\ln Q^2$.}
\label{satgs}
\end{figure}
%%%%%%%%%%%%%%%%%%%%%%%%%%%%%%

The summary of our current knowledge of high energy interactions is
shown in \fig{satgs}. One can see there that, on top of the saturation
region $Q \leq Q_s (Y)$ of \fig{satbk}, one now has the extended
geometric scaling region $Q_s (Y) \, < \, Q \, \leq \, k_{\rm geom}$,
where the linear BFKL evolution still applies with its solution having
the property of geometric scaling due to the presence of saturation
region.

Indeed the property of extended geometric scaling was derived here for
the case of DIS where a small projectile ($q\bar q$ dipole) scatters
on a nucleus. Saturation effects were only present in the nuclear wave
function. At extremely high energies gluon saturation may take place
even in the projectile's wave function: to take this into account
pomeron loop resummation needs to be performed
\cite{loop1}-\cite{loop5}. While such resummation is still an open problem, 
it has been argued in \cite{IMM,DDD} that such pomeron loop effects
along with energy conservation constraints may potentially lead to a
violation of geometric scaling at very high energies.

The interested reader, who would like to learn more about how to
obtain the above analytical solutions in a more complete way keeping
all the prefactors and on how to include the running coupling effects
into the problem is referred to \cite{IIM1,MT,Mueller3} for further
reading. An elegant solution of the BK equation, reducing it in the
geometric scaling region to Kolmogorov-Petrovsky-Piskunov (KPP)
equation, which has traveling wave solutions, was found in \cite{MP}.

\subsubsection{Numerical Solutions}
%%%%%%%%%%%%%%%%%%%%%%%%%%%%%
\begin{figure}[b]
\begin{center}
\epsfxsize=11.5cm
\leavevmode
\hbox{\epsffile{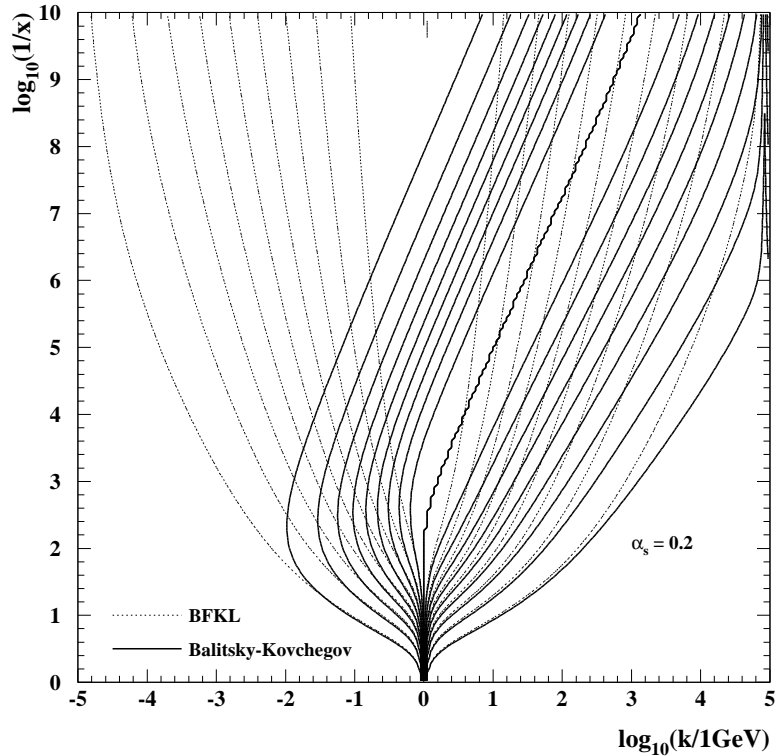}}
\end{center}
\caption{The contour plot of the numerical solutions of the BFKL and BK evolution equations 
in momentum space ($k_T \, {\tilde N} ({\un k}, Y)$) as a function of
transverse momentum $k$ and rapidity $Y = \ln 1/x_{Bj}$ from
\protect\cite{GBMS}.}
\label{bkcont}
\end{figure}
%%%%%%%%%%%%%%%%%%%%%%%%%%%%%%

As we have mentioned before, both the BK and the JIMWLK evolution
equations have been studied numerically in
\cite{Braun1,GBMS,GLLM,Misha,AAMSW,RW}. Here we are not going to give 
a comprehensive review of these solutions, but will just show the
results of one of them which summarizes all the important qualitative
features of the nonlinear evolution and shows how it resolves the
problems of the BFKL equation stated in the end of Sect. \ref{BFKLeq}.

As we have already seen in Sect. \ref{GS}, the solution of BK
evolution equation approaches $N = 1$ at very high energy. Using
\eq{sigtot} we see that this behavior corresponds to the black disk 
limit for total cross sections given by \eq{black}. The total cross
section in \eq{black} is constant and, therefore, does not violate the
Froissart bound of \eq{fro}. Therefore the nonlinear evolution appears
to resolve the unitarity problem of the BFKL equation, posed as
question {\bf (i)} in Sect. \ref{BFKLeq}
\cite{yuri_bk,Braun1,LT,IIM1,MT,Mueller3,GBMS,GLLM,Misha,AAMSW,RW}. 
(More precisely, unitarity is restored at a given impact parameter
${\un b}$: integration over ${\un b}$ in \eq{sigtot} may still give a
cross section growing with energy due to the diffusion of the boundary
of the black disk which increases its radius $R$ and the area $\pi
R^2$ and, hence, the total cross section $\sigma_{tot} = 2 \, \pi \,
R^2$ (see \cite{KW1,FIIM} and the second reference in \cite{GBMS}).)

To answer the question {\bf (ii)} in Sect. \ref{BFKLeq} regarding the
diffusion into the infrared shown by the Bartels cone of \fig{cone} we
will present one result from the numerical solution of the BK equation
done in \cite{GBMS}. In \fig{bkcont} the authors of \cite{GBMS} plot
the lines of constant value for the numerical solution of the BFKL and
BK equations in momentum space. Namely they make a contour plot of
$k_T \, {\tilde N} ({\un k}, Y)$ from \eq{trfm} scaled down by the
maximum value reached by that function in the phase space region
considered \cite{GBMS} as a function of transverse momentum $k$ and
rapidity $Y = \ln 1/x_{Bj}$. One can see that the solution of the BFKL
equation (dashed lines in \fig{bkcont}) spreads out as the rapidity
increases ($x_{Bj}$ decreases). This is the diffusion discussed in
Sect. \ref{BFKLeq} \cite{Bartels}, which is dangerous because it
generated non-perturbative low-$k_T$ gluons, for which our small
coupling treatment would not apply. \fig{bkcont} shows that the
nonlinear BK evolution (shown by solid lines in \fig{bkcont}) avoids
this problem. The effect of non-linear term in \eq{eqN} is to drive
the constant value lines of the solution towards higher momenta, which
is consistent with increase of the saturation scale in \eq{qsy0}, and
to eliminate the diffusion spread of the solution: as one can see from
\fig{bkcont} the width of the $k_T$-distribution of the BK solution is 
roughly independent of rapidity. This solves the IR diffusion problem
posed as question {\bf (ii)} in Sect. \ref{BFKLeq}. Similar results
were obtained by other numerical simulations of the BK
\cite{Braun1,GLLM,Misha,AAMSW} and the JIMWLK \cite{RW} evolution equations.

%%%%%%%%%%%%%%%%%%%%%%%%%%%%%%%%%%%%%%%%%%%%%%%%%%%%%%%%%%%%%%%%%%%%%%%%%%%%%%%%%%%%%%%%%%%%%%%%%

\section{Particle Production in pA Collisions}
\label{pppa}

In this Section we apply the formalism of saturation/Color Glass
Condensate physics to the problem of particle production in
proton-nucleus ($pA$) collisions, which is directly relevant to
deuteron-gold collision experiments at RHIC advertised in our paper's
title. The Section is structured in the following way: first, in
Sect. \ref{clpa}, we calculate gluon production in the quasi-classical
approximation (McLerran-Venugopalan model) of Sect. \ref{QCA}. We will
then continue in Sect. \ref{qepa} by including the effects of BK and
JIMWLK quantum evolution from Sect. \ref{QE} into the obtained
expressions for gluon production cross section. We then calculate
valence quark production in Sect. \ref{clquark} in the approach where
inclusion of quantum evolution is straightforward. We move on to
electromagnetic probes in Sect. \ref{ep} by deriving prompt photon and
dilepton production cross sections. In Sect. \ref{2pi} we study
two-particle correlations by calculating two-gluon, gluon-valence
quark and $q\bar q$ pair production cross sections.

\subsection{Gluon Production in the Classical Approximation}
\label{clpa}

\subsubsection{Gluon Production Cross Section}
\label{clglue}

Here we are interested in calculating the inclusive single gluon
production cross section in $pA$ collisions. We assume that, in the
center of mass frame, the gluons light cone momentum components are
much smaller than that of the proton and the nucleus. If in the center
of mass frame the proton has a large $p^-$ component of its light cone
momentum, we will be interested in produced gluons with $k^- \ll
p^-$. Conversely, if a nucleon in the nucleus has a large $p'^+$
component of its momentum, then $k^+ \ll p'^+$. 
%%%%%%%%%%%%%%%%%%%%%%%%%%%%%
\begin{figure}[ht]
\begin{center}
\epsfxsize=17cm
\leavevmode
\hbox{\epsffile{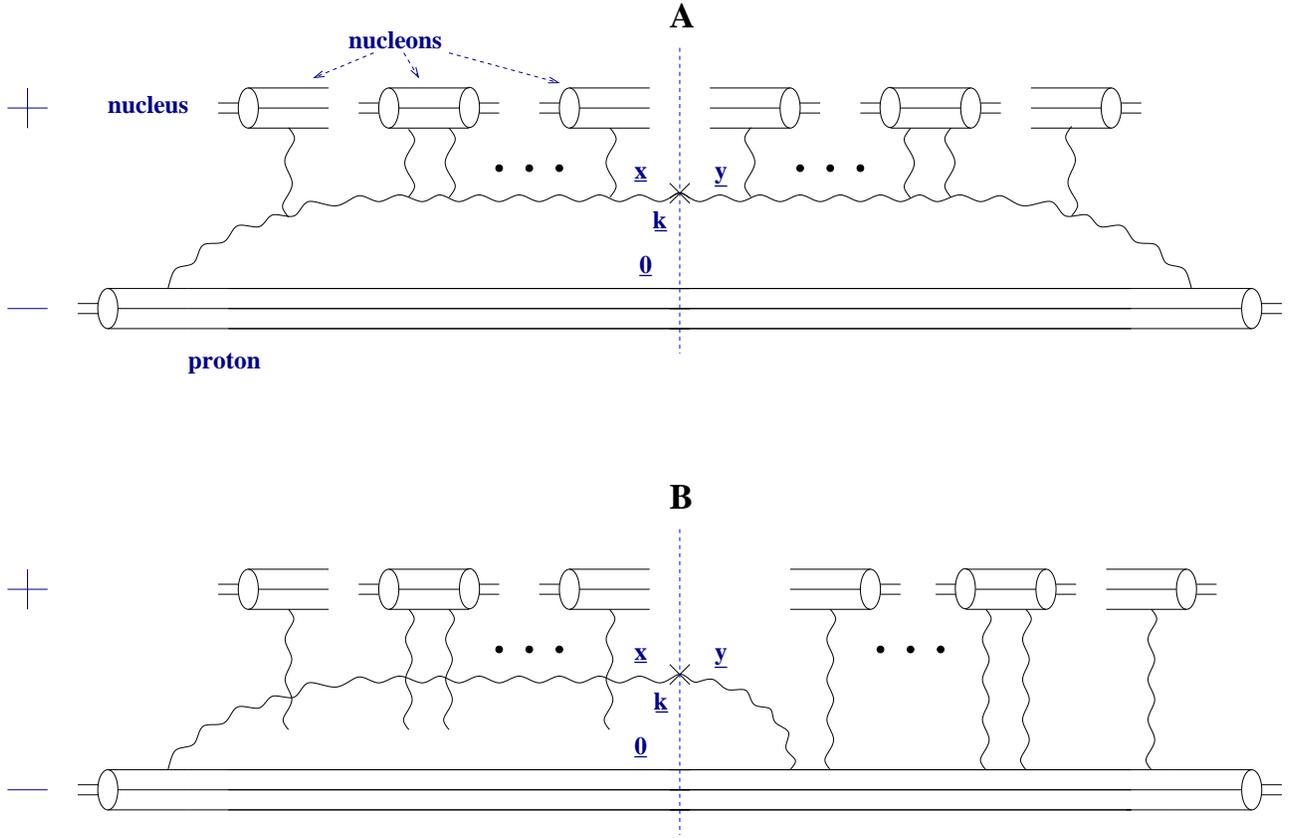}}
\end{center}
\caption{Diagrams contributing to gluon production in $pA$ collision in 
$A^+ =0$ (proton) light cone gauge in the quasi-classical
approximation. }
\label{pa}
\end{figure}
%%%%%%%%%%%%%%%%%%%%%%%%%%%%%%

Just like for the gluon field of a single nucleus in Sect. \ref{MV},
the problem of gluon production in $pA$ collisions in the
quasi-classical approximation can be formulated as the problem of
finding the classical gluon field from the Yang-Mills equations
[\eq{YMeq}] with the source current now given by both the nucleus and
the proton \cite{KMW1,KMW2,KR,GM,DM}. The outgoing gluon line of the
resulting gluon field can be truncated and the gluon can be put on the
mass shell, giving the production amplitude needed for constructing
production cross section. The classical gluon field of the proton and
the nucleus, or, equivalently, the resulting production cross section,
would resum powers of the parameter $\as^2 \, A^{1/3}$ (see \eq{mrp},
just like the classical gluon distribution from \eq{ww2} or the
dipole-nucleus forward scattering amplitude in \eq{glaN2}. This
program has been carried out in \cite{DM} and the classical gluon
field produced in $pA$ has been found there. However, here we use a
different strategy: instead of calculating the classical gluon field,
we calculate the production cross section directly by summing
diagrams, following \cite{KM} (see also \cite{KTS,yuriaa,yuridiff}).

We begin by choosing the gauge: it turns out to be much simpler to
work in the light cone gauge of the projectile (proton). Let us agree
that, in the center of mass frame, the proton is moving in the light
cone $-$ direction: that means we will be working in $A^- = 0$ light
cone gauge. Again we will use the light cone perturbation theory
(LCPT) \cite{BL,BPP}. The diagrams contributing to gluon production in
$A^- = 0$ light cone gauge in the LCPT framework are shown in
\fig{pa}, where one should also add (but we do not show) the complex 
conjugate of the graph in \fig{pa}B, which is obtained from the graph
in \fig{pa}B by mirror-reflecting it with respect to a vertical axis.

The physical picture of the gluon production is the following: the
incoming proton may already have a gluon in its light cone wave
function before the collision with the nucleus and the system of the
proton and gluon would multiply rescatter on the nucleons in the
nucleus. Alternatively the proton can emit the gluon after the
multiple rescatterings in the nucleus. Just like in derivation of the
BK equation in Sect. \ref{BKsect}, the diagrams where the gluon is
emitted during the proton's passing through the nucleus are suppressed
by powers of its large light cone momentum $p^-$, i.e., by powers of
the center of mass energy of the system (see the discussion following
\eq{lcoh}), and can be neglected in the spirit of the eikonal approximation. 
The graph in \fig{pa}A corresponds to the square of the amplitude
given by the case when the gluon is present in proton's wave function
before the collision. As one can see in \fig{pa}A, in the gluon-proton
system only the gluon interacts with the nucleons in the nucleus:
interactions of the proton cancel by real-virtual cancellations
\cite{KM}. Moving a gluon exchanged between the nucleus and the proton 
across the cut does not change the momentum of the produced gluon in
\fig{pa}A, but does change the sign of the whole term, causing the
cancellation. That is why we have to consider only the interactions
with the {\sl gluon} in \fig{pa}A.  The diagram in \fig{pa}B gives the
interference term between the amplitude from \fig{pa}A with the gluon
present in the proton's wave function before the interaction and the
amplitude in which the gluon is emitted by the proton after the
collision. Of course a diagram complex conjugate to
\fig{pa}B should also be included. Real-virtual cancellations of \fig{pa}A 
do {\it not} happen in \fig{pa}B. Moving an exchanged (Coulomb) gluon
across the cut would force us to move it across the gluon emission
vertex for the produced gluon on the right hand side, thus changing
the momentum of the produced gluon. Thus the interactions of the
nucleons with both the proton and a gluon have to be included in
\fig{pa}B. On the right hand side of the diagram in \fig{pa}B only
the interactions with the proton are possible. It can also be shown
that the square of the diagram with late gluon emission (after the
proton passes through the system, as shown on the right hand side of
\fig{pa}B) does not have any interactions in it and can be
neglected. (The gluon exchanges between the proton and the nucleus
cancel by real-virtual cancellations \cite{KM}.)

Note that in the quasi--classical approximation depicted in
\fig{pa} the interaction is modeled by single and double gluon
exchanges. Similar to \fig{qqbar} and Sect. \ref{QCA} we have to
impose the limit of no more than two gluons per nucleon
\cite{yuri2}. If a particular nucleon exchanges a gluon with the
incoming proton and/or gluon in the amplitude then it has to exchange
a gluon in the complex conjugate amplitude to ``remain'' color
neutral. Alternatively the nucleon can exchange two gluons in the
amplitude (complex conjugate amplitude), but then it would not be able
to interact in the complex conjugate amplitude (amplitude). This is
done in the spirit of the quasi--classical approximation resumming all
powers of $\as^2 \ A^{1/3}$, as was discussed in Sect. \ref{QCA}.

The calculation is easier to perform in transverse coordinate
space. We assume that the outgoing gluon has a transverse coordinate
${\un x}$ to the left of the cut in \fig{pa} and a transverse
coordinate ${\un y}$ to the right of the cut. To calculate the
production cross section we will first calculate the amplitude $M
({\un x})$ and its conjugate $M^* ({\un y})$ in the transverse
coordinate space, Fourier-transform them into transverse momentum
space and take their product, which would be the square of the
amplitude in the momentum space ${\tilde M} ({\un k})$, giving the
production cross section
\be
\frac{d \sigma^{pA}}{d^2 k \ d^2 b \ dy} \ = \ \frac{1}{2 \, (2 \pi)^3} 
\, | {\tilde M} ({\un k})|^2 
\, =  \, \frac{1}{2 \, (2 \pi)^3} \, \int \, d^2 x \, d^2 y \,  e^{- i {\underline k} 
\cdot ({\underline x} - {\underline y})} \, M ({\un x}) \, M^* ({\un y}).
\ee
To obtain the answer for the gluon production cross section in pA in
the quasi--classical approximation we have to convolute the wave
function of the proton having a soft gluon in it with the interaction
amplitude of a gluon scattering in the nucleus. Below we will model
the proton by a single valence quark: generalization back to the real
proton case will be manifest. The soft gluon wave function of a single
quark is given by the first term in the parenthesis of \eq{dip1}
corresponding to the diagram in \fig{dipole}, where we should
disregards the anti-quark (lower) line. The interaction amplitude
between the gluon and the nucleus squared can be calculated directly
following \cite{KM,BDMPS1}. However, it is easier to use crossing
symmetry \cite{Mueller_cross} and ``reflect'' the gluon having
transverse coordinate ${\un y}$ from the complex conjugate amplitude
with respect to the cut placing it into the amplitude. One would then
see that the interaction amplitude squared would reduce to the forward
amplitude of a gluon dipole formed by gluons at ${\un x}$ and ${\un
y}$ scattering on the nucleus. In the quasi-classical approximation
such amplitude is given by \eq{NGgm}. Combining the first term in
\eq{dip1} squared with the amplitude from \eq{NGgm} yields the
following contribution to production cross section coming from the
diagram in \fig{pa}A \cite{KM}
\be\label{pasolA}
\frac{d \sigma^{pA}_A}{d^2 k \ dy} \ = \ \frac{1}{\pi} \, \int \ d^2 b \, d^2 x 
\, d^2 y \, \frac{1}{(2 \pi)^2} \, \frac{\as C_F}{\pi} \frac{{\underline x} 
\cdot {\underline y}}{{\underline x}^2 {\underline y}^2} \, e^{- i {\underline k} 
\cdot ({\un x} - {\un y})} \, \left( e^{- ({\un x} - 
{\un y})^2 \ Q_{s0}^2 \, \ln (1/|{\un x} - {\un y}| \Lambda)/4 } - 1
\right)
\ee
with the gluon saturation scale $Q_{s0}$ given by \eq{qsmvg}, where
now the subscript $0$ denotes the saturation scale of the
quasi-classical McLerran-Venugopalan model without quantum
evolution/rapidity dependence. In \eq{pasolA} we have also assumed
that the proton's transverse coordinate is ${\un 0}$.

The diagram in \fig{pa}B is calculated in a similar way. The incoming
proton wave function there is the same as in \fig{pa}A. To resum the
interactions with the nucleus we can either follow direct calculations
of \cite{KM,BDMPS1} or reflect the quark line from the complex
conjugate amplitude with respect to the cut into the amplitude, making
it an anti-quark in the amplitude. The reflected quark line would then
be on top of the quark line in the amplitude, since they both have the
same transverse coordinate ${\un 0}$. However the quark and the
anti-quark do not cancel each other: in fact, since they emit a gluon,
the pair of quarks should be in the color octet state. We again have
an interaction of a color-octet dipole, formed by the gluon at $\un x$
and the quark--anti-quark pair at ${\un 0}$. Adding the diagram
complex conjugate to \fig{pa}B we obtain the following contribution to
the cross section
\be\label{pasolB}
\frac{d \sigma^{pA}_{B+C}}{d^2 k \ dy} \ = \ \frac{1}{\pi} \, \int \ d^2 b \, d^2 x 
\, d^2 y \, \frac{1}{(2 \pi)^2} \, \frac{\as C_F}{\pi} \frac{{\underline x} 
\cdot {\underline y}}{{\underline x}^2 {\underline y}^2} \, e^{- i {\underline k} 
\cdot ({\underline x} - {\underline y})} \, \left( 1 -  e^{- {\underline x}^2 \ 
Q_{s0}^2 \ln (1/|{\un x}| \Lambda)/4 } + 1 - e^{- {\underline y}^2 \
Q_{s0}^2 \ln (1/|{\un y}| \Lambda)/4 } \right).
\ee
Combining the terms in Eqs. (\ref{pasolA}) and (\ref{pasolB}) yields \cite{KM}
\ben
\frac{d \sigma^{pA}}{d^2 k \ dy} \ = \ \frac{1}{\pi} \, \int \ d^2 b \, d^2 x 
\, d^2 y \, \frac{1}{(2 \pi)^2} \, \frac{\as C_F}{\pi} \frac{{\underline x} 
\cdot {\underline y}}{{\underline x}^2 {\underline y}^2} \, e^{- i {\underline k} 
\cdot ({\underline x} - {\underline y})} 
\een
\be\label{paclg}
\times \, \left( 1 -  e^{- {\underline x}^2 \ 
Q_{s0}^2 \ln (1/|{\un x}| \Lambda)/4 } - e^{- {\underline y}^2 \ Q_{s0}^2
\ln (1/|{\un y}| \Lambda)/4 }
+ e^{- ({\un x} - {\un y})^2 \ Q_{s0}^2 \, \ln (1/|{\un x} - {\un y}|
\Lambda)/4 } \right).
\ee
\eq{paclg} gives us the single gluon inclusive production cross section for a 
scattering of a quark on a nucleus in the quasi-classical
approximation, resumming multiple rescatterings of \fig{pa}. Before we
proceed to derive the properties of \eq{paclg} let us first show which
diagrams it corresponds to in a different gauge, the light cone gauge
of the nucleus, and demonstrate an interesting duality between
multiple rescatterings of \fig{pa} and the non-Abelian
Weizs\"{a}cker-Williams wave function of the nucleus on the light cone
from \fig{ww} first observed in \cite{KM}. \\

{\sl Initial Versus Final State Interactions} \\

Let us now consider $pA$ scattering in $A^+ = 0$ gauge, which, in our
convention, is the light cone gauge of the nucleus. We also boost the
system into a frame where the proton is at rest and the incoming
nucleus is ultrarelativistic. The analysis of $pA$ process in the
light cone gauge of the nucleus is sometimes referred to as $Ap$
scattering \cite{KW2}.

\begin{figure}[ht]
\begin{center}
\epsfxsize=17cm
\leavevmode
\hbox{\epsffile{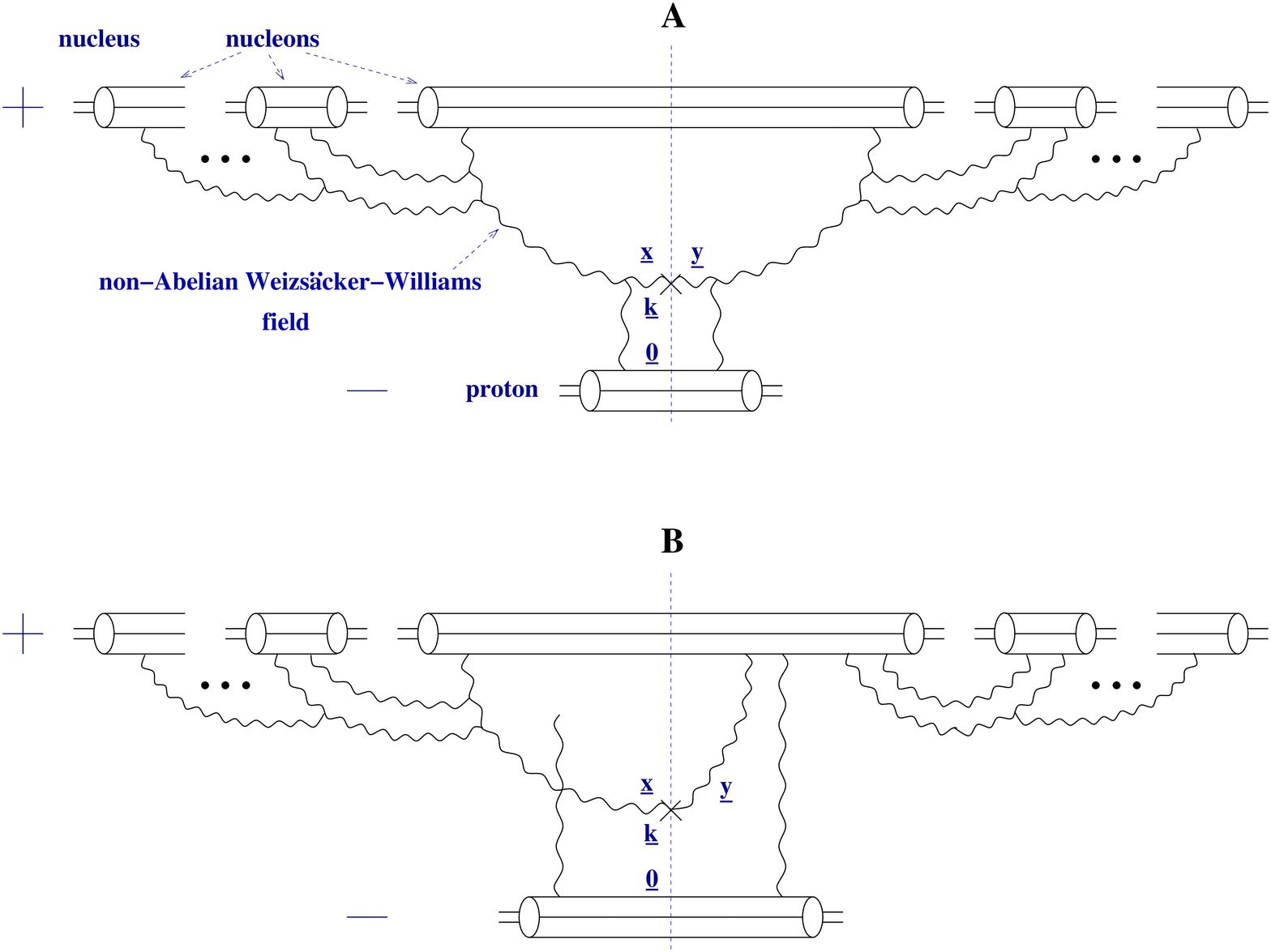}}
\end{center}
\caption{Diagrams contributing to the same process of gluon production 
in $pA$ collision as in \fig{pa}, but now working in $A^+ =0$
(nucleus) light cone gauge in the quasi-classical approximation. }
\label{palc}
\end{figure}

The light cone gauge diagrams contributing to the gluon production
cross section in proton--nucleus collisions are depicted in
\fig{palc}. Again we are going to perform the calculation in the 
framework of the light cone perturbation theory \cite{BL,BPP}. Similar
to $A^- =0$ gauge case considered above the incoming nucleus can emit
a gluon in its wave function either before or after the collision with
the proton. The one gluon light cone wave function of an
ultrarelativistic nucleus is given by ${\underline A}^{WW}
({\underline x}) \cdot {\underline\epsilon}$, with ${\underline
A}^{WW}$ the non-Abelian Weizs\"{a}cker-Williams field of the nucleus
given by \eq{clsol} of Sect. \ref{MV} with suppressed $x^-$ dependence
(taking $x^- \rightarrow \infty$) and with ${\underline\epsilon}$ the
gluon polarization vector in $A^+ = 0$ light cone gauge. The
correspondence between the light-cone wave function and the classical
gluon field was shown above to be true at the lowest order, which
could be seen by comparing $x^- \rightarrow \infty$ limit of the
lowest order term in \eq{clsol} with the light cone wave function in
\eq{dip1}. Diagrammatically the light cone wave function corresponds
to the same set of diagrams as was depicted in \fig{ww}. The
fields of the nucleons in the nucleus ``gauge rotate'' the
Weizs\"{a}cker-Williams field of one of the nucleons
\cite{yuri2}. Therefore, here we have strong reasons to assume that 
this light cone wave function can be obtained from the non-Abelian
Weizs\"{a}cker-Williams field of \eq{clsol} by taking $x^- \rightarrow
\infty$ in it and multiplying the resulting field by
${\underline\epsilon}$. The interaction of this non-Abelian
Weizs\"{a}cker-Williams (WW) wave function with the proton can only be
by the means of single or double gluon exchanges, since having more
gluons in interaction of the WW gluon field and nucleons from the wave
function of \fig{ww} with the same proton would bring in extra powers
of $\as$ not enhanced by $A^{1/3}$, which would be subleading in terms
of resummation of the parameter $\as^2 A^{1/3}$.

%%%%%%%%%%%%%%%%%%%%%%%%%%%%%
%%%%%%%%%%%%%%%%%%%%%%%%%%%%%%
Before colliding with the proton the nucleus can develop the
Weizs\"{a}cker-Williams one gluon light cone wave function which then
interacts with the proton by means of one or two gluon exchanges,
according to the rules of the quasi--classical approximation.  The
square of the graph corresponding to this scenario is shown in
\fig{palc}A. As in \fig{pa} the direct interactions of the proton with the
nucleons in the nucleus cancel through the real--virtual cancellation leaving only
the interactions with the gluon line. One may notice that the final
state interactions, where another WW gluon merges with the outgoing
gluon in \fig{palc} are left out in the diagram of
\fig{palc}A, but as we will show below we do not need them to
reproduce the contribution of the graph in \fig{pa}A, which implies
that they cancel.

The second possible scenario corresponds to the case when there is no
gluon in the nuclear wave function by the time the collision happens
and the gluon is emitted by the nucleus after the interaction with the
proton. Then the nuclear wave function without an emitted gluon
corresponds to the fields of the nucleons ``rotating'' the current
(quark line) of one of the nucleons in the nucleus. This is shown on
the right hand side of \fig{palc}B. The nucleon then interacts with
the proton by exchanging one or two gluons with it. After that the
nucleus can emit a gluon to be produced in the final state. Another
possibility which is not shown in \fig{palc}B but which contributes to
the gluon production corresponds to the case when the
Weizs\"{a}cker-Williams gluon is present in the nuclear wave function
by the time of the collision, similar to \fig{palc}A, but after the
interaction with the proton the gluon merges into the quark line of
one of the nucleons, which later re-emits the gluon. We could not find
an {\it a priori} argument prohibiting an emission of the whole
Weizs\"{a}cker-Williams field after the interaction. However, as we
will see below one needs to emit only one gluon to be able to
reproduce \eq{paclg}. The square of the diagram on the right hand side
of \fig{palc}B is zero since the interactions cancel due to
real--virtual cancellation \cite{KM}. The only contribution we get
from it is the interference term depicted in
\fig{palc}B. There on the left hand side we have the same diagram as
in \fig{palc}A except that now interactions of the proton with the
``last'' nucleon in the nucleus do not cancel. One can show that the
diagram of \fig{palc}B provides us with the contribution equal to that
of the graph in \fig{pa}B.

Let us now calculate the diagrams in \fig{palc}. The contribution of
\fig{palc}A can be obtained by convoluting the correlation function of
the fields on both sides of the cut with the gluon--proton
interactions amplitude, which can be obtained from \eq{1sc3} by
replacing $C_F \rightarrow N_c$ in the gluon distribution function
from \eq{xglo}. The result yields
\be\label{lc11}
\frac{d \sigma^{pA}_{LC \, A}}{d^2 k \ dy} \ = \ \int \frac{d^2 x \ 
d^2 y}{(2 \pi)^2} \, e^{i {\underline k} \cdot ({\underline x} -
{\underline y})}\, \frac{2}{\pi} \, \mbox{Tr} \, \left< {\underline A}^{WW}
({\underline x}) \cdot {\underline A}^{WW} ({\underline y}) \right> \,
\frac{- \as \pi^2 N_c}{N_c^2 - 1} \, ({\underline x} - {\underline y})^2
\, xG (x, 1/({\underline x} - {\underline y})^2).
\ee
Employing the correlation function of two WW fields from \eq{gla2} in
\eq{lc11} and defining new variables ${\underline z} = {\underline x}
- {\underline y}$ and ${\underline b} = {\underline y}$ we obtain
\be\label{lc12}
\frac{d \sigma^{pA}_{LC \, A}}{d^2 k \ dy} \ = \ \int \, d^2 b \, d^2 z \, 
e^{i {\underline k} \cdot {\underline z}} \, \frac{1}{(2 \pi)^2} \,
\frac{\as C_F}{\pi} \, \ln \frac{1}{{\underline z}^2 \Lambda^2} \, \left( 
e^{ - {\underline z}^2 \, Q_s^2 \, \ln (1/|{\un z}| \, \Lambda)/ 4 } -
1 \right).
\ee
We will need the following mathematical formula \cite{yuriaa}
\be\label{for}
\ln \frac{1}{{\underline z}^2 \Lambda^2} \, = \, \frac{1}{\pi} \, \int 
d^2 y \, \frac{{\underline y} \cdot ({\underline z} + {\underline
y})}{{\underline y}^2 ({\underline z} + {\underline y})^2}
\ee
where the $y$ integration is cut off by $1/\Lambda$ in the infrared
near ${\un y} = {\un 0}$ and ${\un y} = - {\un z}$. Inserting \eq{for}
into \eq{lc12} and comparing the result to Eq. (\ref{pasolA}) one can
see that
\be
\frac{d \sigma^{pA}_{LC \, A}}{d^2 k \ dy} \ = \ \frac{d \sigma^{pA}_{A}}{d^2 k \ dy}.
\ee
Thus we have shown that the contribution of the diagrams in
\fig{palc}A is equal to the contribution of the diagrams in
\fig{pa}A.

The calculation of the graphs depicted in \fig{palc}B is a little more
complicated. We refer the interested reader to \cite{yuriaa}, where
the estimate is done in detail. After performing the calculation one
demonstrates that the contribution to the inclusive gluon production
cross section from the diagram in \fig{palc}B is equal to the
contribution of the diagrams in \fig{pa}B. We can now conclude that
the diagrams in \fig{palc} in the $A^+ =0$ gauge reproduce the
production cross section from \eq{paclg}.

Comparing \fig{pa} with \fig{palc} we see that, indeed, depending on
the choice of gauge, different sets of diagrams become dominant in the
production cross section. Diagrammatic representation of the process
is gauge dependent. In the light cone gauge of the projectile, $A^-
=0$, the gluon production in proton-nucleus interactions is dominated
by multiple rescatterings, as we see from \fig{pa}. These multiple
rescattering are (almost) instantaneous, happening only during the
passage of the proton through nuclear matter. One can refer to them as
{\sl final} state interactions.  On the other hand, the same process
of gluon production looks different in the light cone gauge of the
nucleus, $A^+ =0$, as shown in \fig{palc}. The multiple rescatterings
of \fig{pa} become incorporated into the WW wave function of the
nucleus in \fig{palc}. Therefore, {\sl final} state interactions from
one gauge become {\sl initial} state interactions in another
gauge. This {\sl duality} (or dichotomy) was first observed in
\cite{KM}.\\

{\sl $k_T$-factorization} \\

\begin{figure}[ht]
\begin{center}
\epsfxsize=5cm
\leavevmode
\hbox{\epsffile{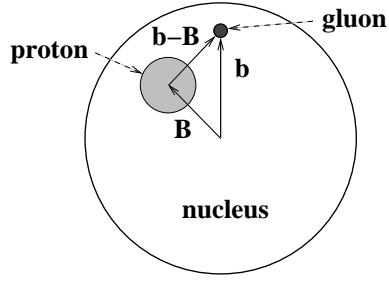}}
\end{center}
\caption{Gluon production in pA collisions as seen in the transverse 
plane. To make the picture easier to read the gluon is placed far away
from the proton which is a highly improbable configuration.}
\label{paimp}
\end{figure}

It is interesting to show that one can recast \eq{paclg} in a
$k_T$-factorized form. Following \cite{KT,KKT} we first use the
amplitude from
\eq{NGgm} to rewrite \eq{paclg} as
\be\label{paclg1}
\frac{d \sigma^{pA}}{d^2 k \ dy} \ = \ \frac{1}{\pi} \, \int \ d^2 b \, d^2 x 
\, d^2 y \, \frac{1}{(2 \pi)^2} \, \frac{\as C_F}{\pi} \frac{{\underline x} 
\cdot {\underline y}}{{\underline x}^2 {\underline y}^2} \, e^{- i {\underline k} 
\cdot ({\underline x} - {\underline y})} 
\, \left[ N_G ({\underline x}, {\un b}, 0) + N_G ({\underline y}, {\un b}, 0) 
- N_G ({\un x} - {\un y}, {\un b} , 0) \right],
\ee
where we have for simplicity assumed that $N_G$ is a slowly varying
function of the impact parameter ${\un b}$. (In principle gluon
dipoles $\un x$, $\un y$ and ${\un x} - {\un y}$ have slightly
different impact parameters, the spread in which we neglected in
\eq{paclg1} by putting them all equal to $\un b$.) Integrating over 
$\un y$ in the first term in the brackets of \eq{paclg1}, over $\un x$
in the second term in the brackets of \eq{paclg1}, and over, say, $\un
x$ in the third term in the brackets of \eq{paclg1} (keeping ${\un x}
- {\un y}$ fixed) yields \cite{KT,KKT}
\be\label{paclg2}
\frac{d \sigma^{pA}}{d^2 k \ dy} \, = \, \frac{1}{2 \pi^2} \, 
\frac{\as C_F}{\pi} \, \int d^2 b \, d^2 z \, 
e^{- i {\un k} \cdot {\un z}} \, \left[ 2 \, i \, 
\frac{{\un z} \cdot {\un k}}{{\un z}^2 {\un k}^2} -  
\ln \frac{1}{z_T \Lambda} \right] \, N_G ({\un z}, {\un b}, 0),
\ee
where $\un z$ was chosen to replace $\un x$, $\un y$ and ${\un x} -
{\un y}$ in each of the terms in the brackets of \eq{paclg1}
correspondingly. We also use $z_T \equiv |{\un z}|$. In arriving at
\eq{paclg2} we have used \eq{for} along with
\be
\int \frac{d^2 z}{ (2 \pi)^2} \, e^{- i \, {\un k} \cdot {\un z}} \, 
\frac{\un z}{{\un z}^2} \, = \, - \frac{i}{2 \pi} \, \frac{\un k}{{\un k}^2}.
\ee
Using the fact that $N_G ({\un z}=0, {\un b}, 0) = 0$ we write \eq{paclg2} as
\be\label{paclg3}
\frac{d \sigma^{pA}}{d^2 k \ dy} \, = \, \frac{1}{2 \pi^2} \, 
\frac{\as C_F}{\pi} \, \frac{1}{{\un k}^2} \, \int d^2 b \, d^2 z \, 
N_G ({\un z}, {\un b}, 0) \, \nabla^2_z \, \left( e^{- i {\un k}
\cdot {\un z}} \, \ln \frac{1}{z_T \Lambda} \right).
\ee
Eqs. (\ref{1sc3}) and (\ref{xglo}) allow us to derive the forward
scattering amplitude of a gluon dipole on a single quark (or proton),
which we will denote $n_G$, such that
\be\label{ng}
\int d^2 b' \, n_G ({\un z}, {\un b}', y=0) \, = \, \pi \, \as^2 \,
{\un z}^2 \, \ln \frac{1}{ z_T \, \Lambda}.
\ee
\eq{ng} corresponds to the two gluon exchange interaction between
the dipole and the proton. With the help of \eq{ng} we rewrite
\eq{paclg3} as \cite{KT}
\be\label{paclkt}
\frac{d \sigma^{pA}}{d^2 k \ dy} \, = \, \frac{C_F}{\as \, 
\pi \, (2 \pi)^3} \, 
\frac{1}{{\un k}^2} \, \int d^2 B \, d^2 b \, d^2 z \, 
\nabla^2_z \, n_G ({\un z}, {\un b} - {\un B}, 0) \, e^{- i {\un k}
\cdot {\un z}} \, \nabla^2_z \, N_G ({\un z}, {\un b}, 0).
\ee
Now ${\un B}$ is the impact parameter of the proton with respect to
the center of the nucleus and ${\un b}$ is the impact parameter of the
gluon with respect to the center of the nucleus as shown in \fig{paimp}.
%%%%%%%%%%%%%%%%%%%%%%%%%%%%%
%%%%%%%%%%%%%%%%%%%%%%%%%%%%%%
%%%%%%%%%%%%%%%%%%%%%%%%%%%%%
\begin{figure}[hb]
\begin{center}
\epsfxsize=7cm
\leavevmode
\hbox{\epsffile{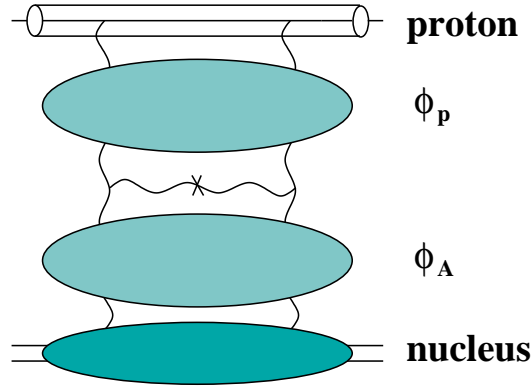}}
\end{center}
\caption{$k_T$-factorization in $pA$ collisions, as implied by \eq{ktcl}. Unfortunately 
in no known gauge do the contributing diagrams look as shown.}
\label{ktfact}
\end{figure}
%%%%%%%%%%%%%%%%%%%%%%%%%%%%%%

\begin{figure}[t]
\begin{center}
\begin{tabular}{cc}
\epsfig{file=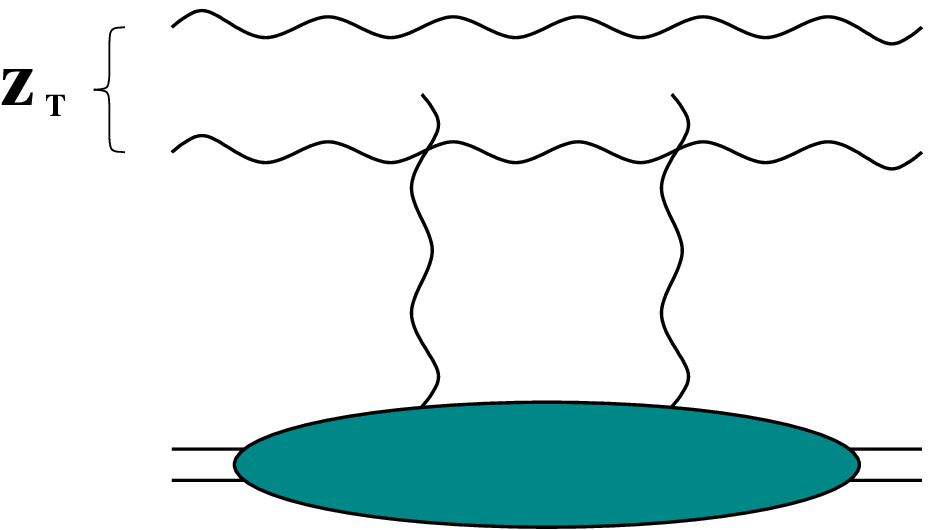, width=7cm}
& \hspace*{1cm}
\epsfig{file=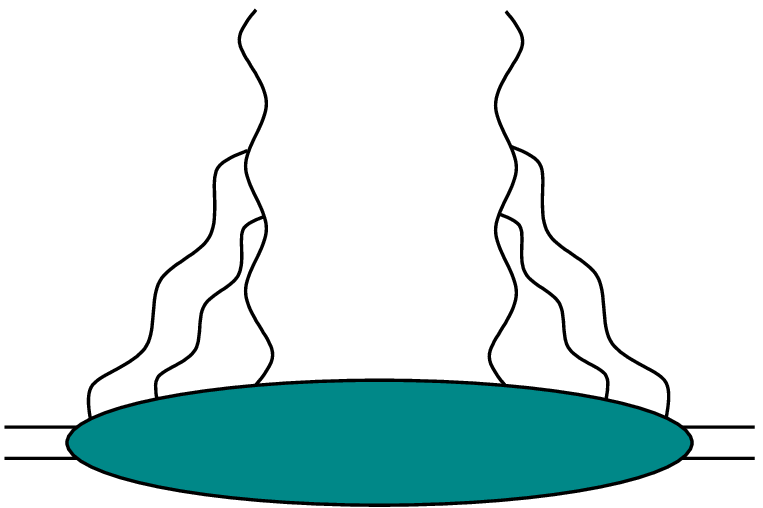, width=6cm}\\
\hspace*{1cm} {\bf A} & \hspace*{1cm} {\bf B}
\end{tabular}
\end{center}
\caption{{\bf A.} The definition of unintegrated gluon distribution 
relating it to the gluon dipole cross section from \eq{ktglue}. The
exchanged gluon lines can connect to either gluon in the dipole. {\bf
B.} The non-Abelian Weizs\"{a}cker-Williams gluon distribution
obtained by calculating the correlator of two WW fields, given by
\eq{ww2}.}
\label{glue}
\end{figure}

Defining the unintegrated gluon distributions for the nucleus 
\be\label{ktglue}
\phi_A (x, \un k^2) \, = \, \frac{C_F}{\as \, (2 \pi)^3} \, \int d^2 b \, 
d^2 z \, e^{- i {\un k} \cdot {\un z}} \ \nabla^2_z \, N_G ({\un z},
{\un b}, y = \ln 1/x)
\ee
and for the proton
\be\label{ktgluep}
\phi_p (x, \un k^2) \, = \, \frac{C_F}{\as \, (2 \pi)^3} \, \int d^2 b \, 
d^2 z \, e^{- i {\un k} \cdot {\un z}} \ \nabla^2_z \, n_G ({\un z},
{\un b}, y = \ln 1/x),
\ee
we transform \eq{paclkt} into
\be\label{ktcl}
\frac{d \sigma^{pA}}{d^2 k \ dy} \, = \, \frac{2 \, \as}{C_F} \, 
\frac{1}{{\un k}^2} \, \int d^2 q \, \phi_p ( {\un q},  y=0) \, \phi_A ({\un k}
- {\un q}, y=0  ). 
\ee

\eq{ktcl} is the well-known $k_T$-factorization expression for gluon 
production cross section \cite{CCH,GLR,GLR2,Ryskin,Braun2}. It has
been proven in \cite{CCH} for a gluon production from a single BFKL
ladder of \fig{ladder}. A graphical representation of
$k_T$-factorization implied by \eq{ktcl} is shown in
\fig{ktfact}. $k_T$-factorization, as shown in \eq{ktcl}, is the
statement of separation of the inclusive cross section into a
convolution of unintegrated gluon distribution functions of the target
and of the projectile with the square of Lipatov vertex for gluon
production from \eq{lipa}. This is what is shown in \fig{ktfact} for
$pA$ collisions at hand. It is very surprising to obtain this
factorization formula after resumming multiple rescatterings of
\fig{pa}, which seem to explicitly violate factorization of \fig{ktfact}. It may 
appear that the representation of gluon production in the light cone
gauge of the nucleus, shown in \fig{palc}, may help cast the process
in a factorized form of \fig{ktfact}. Unfortunately that is not the
case: the reason is that the unintegrated distribution function used
in \eq{ktcl}, given by \eq{ktglue}, is different from the unintegrated
gluon distribution in \eq{ww2}, which we obtained by calculating the
correlator of two WW fields. The ``factorization'' of \fig{palc}
appears to require the nucleus to be described by the gluon
distribution of \eq{ww2} (see \fig{glue}B), which is the correct
correlator of two WW fields which provides us with the number of gluon
quanta in the wave function. Unfortunately \eq{paclkt} demands that
one uses the gluon distribution from \eq{ktglue} in the factorization
of \eq{ktcl} instead of \eq{ww2}. Therefore, the $A^+ =0$ light cone
gauge representation of the interaction in \fig{palc} does not appear
to lead to \eq{ktcl} with the gluon distribution from \eq{ktglue}. In
general, no gauge is known to the authors in which the dominant
diagrams responsible for gluon production could be cast in the form
shown in \fig{ktfact}.
%%%%%%%%%%%%%%%%%%%%%%%%%%%%%
%%%%%%%%%%%%%%%%%%%%%%%%%%%%%%

Finally, we would like to point out the difference between the two
unintegrated gluon distribution functions in Eqs. (\ref{ktglue}) and
(\ref{ww2}). We show the definitions of both distributions in
\fig{glue}. As was mentioned above, the WW gluon distribution from 
\eq{ww2} is obtained by constructing a correlator of two WW gluon 
fields of \eq{clsol}. It is shown in \fig{glue}B. It is this
distribution which counts the number of gluon quanta and, in that
sense, is the true unintegrated gluon distribution function.  (The
distribution from
\eq{ww2} should not be confused with the one we defined in
\eq{fakephi} to connect the BK and GLR equations: the former is 
defined in terms of a gluon dipole, while the latter employs the quark
dipole.)  The gluon distribution from \eq{ktglue} was required to
recast the gluon production cross section (\ref{paclg}) in the
$k_T$-factorization form of \eq{ktcl}.  However, its diagrammatic
representation exists and is shown in \fig{glue}A. There we consider
the scattering amplitude of a gluon dipole on a nucleus. To define the
gluon distribution from \eq{ktglue} one has to require that the
interaction happens only via a two-gluon exchange, with some
complicated gluon distribution in the nuclear wave function. It is not
clear whether diagrammatic representation of the dipole-nucleus
interaction from \fig{glue}A is achievable in any known gauge. The
bottom line is that it is rather puzzling that the $k_T$-factorization
of \eq{ktcl} arises for the single gluon production cross section in
$pA$, and it is also not clear why the unintegrated gluon distribution
used in that formula should be given by \eq{ktglue}, which the
calculations lead to, instead of the physical gluon distribution of
\eq{ww2}.

\eq{ktglue}, along with Eqs. (\ref{ktgluep}), allows one to generalize 
the gluon production cross section (\ref{paclg}), which was originally
derived for quark-nucleus scattering, to the case of proton-nucleus
scattering. This could be done by using the gluon dipole--proton
amplitude $n_G$ instead of the dipole-quark one from \eq{ng}. However,
the resulting two gluon distribution functions are different only by
an overall prefactor, which is not important for our discussion below.

%%%%%%%%%%%%%%%%%%%%%%%%%%%%%%%%%%%%%%%%%%%%%%%%%%%%%%%%%%%%%%%%%%%%%%%%%

\subsubsection{Nuclear Modification Factor: Low-$p_T$ Suppression}

To analyze the properties of the particle production it is convenient
to construct {\sl nuclear modification factor}, which is defined
experimentally as
\be\label{rpaex}
R^{pA} (k_T, y) \, \equiv \, \frac{\frac{d N^{pA}}{d^2 k \ dy}}{N_{coll}
\, \frac{d N^{pp}}{d^2 k \ dy}},
\ee
where $N_{coll}$ is the number of proton-nucleon collisions in the
$pA$ scattering process, $d N^{pA} / d^2 k dy$ is the particle
multiplicity in $pA$ scattering and $d N^{pp}/ d^2 k dy$ is the
particle multiplicity in $pp$. $R^{pA}$ can be expressed in terms of
particle production cross section as (for subtleties of this
redefinition see \cite{Boris03})
\be\label{rpa}
R^{pA} (\un k, y) \, = \, \frac{\frac{d \sigma^{pA}}{d^2 k \ dy}}{A \, \frac{d
\sigma^{pp}}{d^2 k \ dy}}.
\ee

In this and the following Sections we will study $R^{pA}$ given by the
quasi-classical gluon production cross section derived above in
Sect. \ref{clglue}. $R^{pA}$ for quark production, that will be
derived later, can be constructed by analogy. To construct
gluon $R^{pA}$ using \eq{rpa} we will need the gluon production cross
section (\ref{paclg}), which we will use in the form given by
\eq{paclg2}, since it is more convenient for deriving various
asymptotic regimes.  \eq{rpa} also requires us to know the gluon
production in $pp$ collisions, which could be obtained by expanding
$N_G$ from \eq{NGgm} to the lowest non-trivial order, corresponding to
rescattering on a single nucleon in the nucleus, putting $A=1$ in it
and using the resulting expression in
\eq{paclg2}. The result yields
\be\label{pp}
A \, \frac{d \sigma^{pp}}{d^2 k \ dy} \, = \, \frac{\as \, C_F}{\pi^2} \,
\int d^2 b \, \frac{Q_{s0}^2}{{\un k}^4} \, 
\left( \ln \frac{{\un k}^2}{4 \, \Lambda^2} + 2 \, \gamma - 1 \right) \,
\approx \, \frac{\as \, C_F}{\pi^2} \,
\int d^2 b \, \frac{Q_{s0}^2}{{\un k}^4} \, 
\ln \frac{{\un k}^2}{\Lambda^2}
\ee
with $\Lambda$ some infrared cutoff. In \eq{pp} we have included the
factor of $A$ on the left hand side to make it ready to use in
\eq{rpa}.

In this Section we are interested in the behavior of $R^{pA} ({\un k},
y)$ at low values of the gluon transverse momentum $k_T$, namely for
$k_T \lsim Q_{s0}$. Analyzing \eq{paclg2} we note that low $k_T \lsim
Q_{s0}$ corresponds to dominance of large dipole sizes $z_T \gsim
1/Q_{s0}$ in the integral. For $z_T \gsim 1/Q_{s0}$ the cross section
is likely to be black, and we can approximate $N_G$ in \eq{paclg2} by
$N_G (\un z, \un b, 0) \approx 1$. Integrating the resulting
expression over $\un z$ we obtain
\be\label{paclglkt}
\frac{d \sigma^{pA}}{d^2 k \ dy} \, \approx \, \frac{\as \, C_F}{\pi^2} \,
\int d^2 b \, \frac{1}{{\un k}^2}, \hspace*{1cm} k_T < Q_{s0}.
\ee

Before we proceed, let us point a very important property of the
classical gluon production cross section (\ref{paclg}). \eq{pp} in
fact reflects a high-$k_T$ asymptotics of the gluon production cross
section: at high-$k_T$ it scales as $\sim 1/k_T^4$, in agreement with
the lowest order perturbative calculations
\cite{GB,KMW1,KMW2,KR,GM,Lipatov1}. This result is badly infrared 
divergent, and demonstrates a problem of lowest order perturbative
calculation of gluon production, shown in \fig{lvert}: if one wants to
obtain the integral production cross section $d \sigma /dy$ from it,
one would have to introduce an infrared cutoff $\Lambda$ for
$k_T$-integration. The result would depend on the cutoff as $\sim
1/\Lambda^2$, making the resulting $d \sigma /dy$ very sensitive to
the value of the cutoff and demonstrating that the lowest order cross
section is dominated by non-perturbative infrared effects, which we do
not know how to account for theoretically. On the other hand the full
classical cross section of \eq{paclg} is (almost) free of such
problems. \eq{paclglkt} describes the low-$k_T$ asymptotics of the
cross section in \eq{paclg}. As one can see it leads to the cross
section scaling as $\sim 1/k_T^2$. Thus, similar to the classical
gluon distribution of Sect. \ref{MV} (see \fig{mv1}), the multiple
rescattering effects regulate the infrared singularity, changing it
from $\sim 1/k_T^4$ to $\sim 1/k_T^2$, and making the resulting cross
section less infrared divergent. The residual infrared divergence
$\sim 1/k_T^2$ is due to the fact that we have assumed that saturation
behavior with multiple rescatterings takes place only in the nuclear
wave function: there is no saturation in the proton wave function in
the approach used here. The case of multiple rescatterings in both the
target and the projectile is usually studied in the context of
nucleus-nucleus ($AA$) collisions. There the residual $\sim 1/k_T^2$
divergence is regulated completely \cite{KV,Lappi,yuriaa}.

Using Eqs. (\ref{paclglkt}) and (\ref{pp}) in \eq{rpa} yields
\be\label{rpalkt}
R^{pA} (\un k, y) \, \approx \, \frac{k_T^2}{Q_{s0}^2 \, \ln (k_T^2
/\Lambda^2)}, \hspace*{1cm} k_T < Q_{s0},
\ee
where, for simplicity, we assumed again that the nuclei are
cylindrical in the beam ($z$) direction, such that the impact
parameter integration over $\un b$ in Eqs. (\ref{paclglkt}) and
(\ref{pp}) would give a trivial overall factor of the nuclear
transverse area $S_\perp$, which cancels in $R^{pA}$. From \eq{rpalkt}
we see that
\cite{KTS,BKW,KNST,KKT,KM,ktbroadening1,ktbroadening2,Vitev03,ktbroadening3}
\be\label{rpaglkt}
R^{pA} (\un k, y) \, < \, 1 ,  \hspace*{1cm} k_T < Q_{s0}.
\ee
Therefore we conclude that classical gluon production in $pA$ leads to
{\sl suppression} of low-$k_T$ gluons.

%%%%%%%%%%%%%%%%%%%%%%%%%%%%%%%%%%%%%%%%%%%%%%%%%%%%%%%%%%%%%%%%%%%%%%%%%%%%%%%%%%

\subsubsection{Nuclear Modification Factor: Cronin Effect}

Our goal now is to study the nuclear modification factor $R^{pA}$ for
high-$k_T$ particles with $k_T > Q_{s0}$. Starting with gluon
production again we note that, in the quasi-classical case one can
prove the following sum rule \cite{KKT}
\be\label{sumr}
\int d^2 k \, {\un k}^2 \, \frac{d \sigma^{pA}}{d^2 k \ dy} \, = \, 
A \, \int d^2 k \, {\un k}^2 \, \frac{d \sigma^{pp}}{d^2 k \ dy}, 
\ee
which could be obtained from \eq{paclg2} using the fact that, for the
quasi-classical amplitude (\ref{NGgm}) and the two-gluon exchange
amplitude from \eq{ng} the following relation is true
\be\label{subtr}
\lim_{z_T \rightarrow 0} \left\{ 
\left[ \nabla^2_z \, N_G ({\un z}, {\un b}, 0) \right] - A^{1/3} \,
\left[ \nabla^2_z \, n_G ({\un z}, {\un b}, 0) \right] \right\} \, = \, 0.
\ee
\eq{subtr} and the sum rule of \eq{sumr} are valid in the quasi-classical 
case of McLerran-Venugopalan model {\sl only}. (When quantum small-$x$
evolution is included, \eq{subtr} ceases to be valid.)

The sum rule (\ref{sumr}) insures that if the quasi-classical gluon
production cross section in $pA$ collisions is, in some region of
$k_T$, smaller than $A$ times the gluon production cross section in
$pp$ than there should be some other region of $k_T$ in which their
roles are reversed. For $R^{pA}$ defined in \eq{rpa} that means that
if, in some region of $k_T$, it is less than $1$ there must be some
other region of $k_T$ in which it is greater than $1$. Since in the
previous section we have proven that $R^{pA} < 1$ for $k_T < Q_{s0}$,
the sum rule (\ref{sumr}) demands that $R^{pA} > 1$ for some region of
$k_T$ in the $k_T > Q_{s0}$ interval. Enhancement of particle
production in $pA$ collisions leading to $R^{pA} > 1$ was originally
observed by Cronin et al in \cite{Cronin} and is usually referred to
as Cronin effect. Eqs. (\ref{sumr}) and (\ref{rpaglkt}) prove that the
cross section (\ref{paclg}) leads to Cronin enhancement of produced
gluons in $pA$ collisions.
%%%%%%%%%%%%%%%%%%%%%%%%%%%%%
\begin{figure}[ht]
\begin{center}
\epsfxsize=10cm
\leavevmode
\hbox{ \epsffile{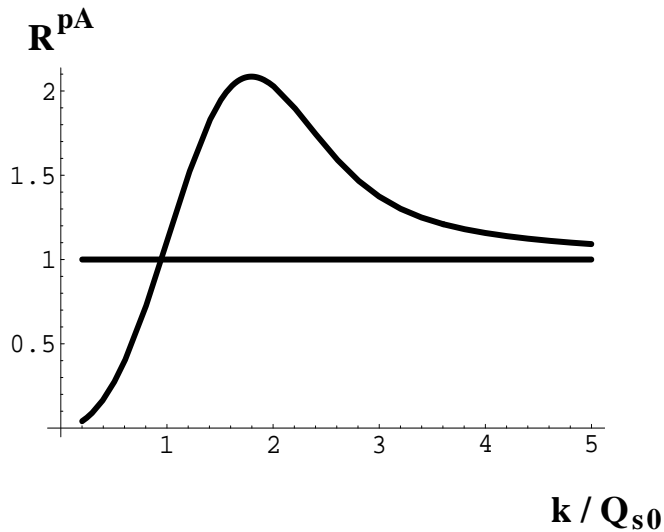}}
\end{center}
\caption{ The ratio $R^{pA}$ for gluons plotted as a function of 
$k_T/Q_{s0}$ in the quasi-classical McLerran-Venugopalan model as
found in \protect\cite{KM,KTS,BKW,KNST,KKT}. The cutoff is $\Lambda = 0.2 \, Q_s$.}
\label{cron}
\end{figure}
%%%%%%%%%%%%%%%%%%%%%%%%%%%%%%

To analyze the behavior of $R^{pA}$ for $k_T > Q_{s0}$ we follow the
approach originally introduced in \cite{GJD,Gelis:2001da} and find the
first correction to the high-$k_T$ asymptotics of \eq{paclg} given by
\eq{pp}. Expanding $N_G$ from \eq{NGgm} to the second non-trivial order, 
substituting the result in \eq{paclg2} and performing the integration
over $\un z$ we obtain \cite{KKT}
\ben
\frac{d \sigma^{pA}}{d^2 k \ dy} \, = \, \frac{\as \, C_F}{\pi^2} \,
\int d^2 b \, \frac{Q_{s0}^2}{{\un k}^4} \, \left[ 
\left( \ln \frac{{\un k}^2}{4 \, \Lambda^2} + 2 \, \gamma - 1 \right) + \right.
\een
\be\label{paas1} 
+ \left.  \frac{Q_{s0}^2}{4 \, {\un k}^2} \, \left( 6 \, \ln^2 
\frac{{\un k}^2}{4 \, \Lambda^2} - 8 \, (4 - 3 \gamma) \,
\ln \frac{{\un k}^2}{4 \, \Lambda^2} + 29 + 24 \, \gamma^2 - 64 \, \gamma \right) 
+ \ldots \right], \hspace*{1cm} k_T \rightarrow \infty.
\ee
Substituting \eq{paas1} along with \eq{pp} into \eq{rpa} and keeping
only the leading logarithmic ($\ln (k_T^2 /\Lambda^2)$) terms in the
parentheses of \eq{paas1} yields for a cylindrical nucleus \cite{KKT}
\be\label{rpaghkt}
R^{pA} (k_T) \, = \, 1 + \frac{3}{2} \, \frac{Q_{s0}^2}{{\un k}^2} \,
\ln \frac{k_T^2}{\Lambda^2} + \ldots, \hspace*{1cm} k_T \rightarrow \infty.
\ee
\eq{rpaghkt} indicates that $R^{pA}$ approaches $1$ from above at high $k_T$, which
is typical of Cronin enhancement \cite{Cronin}. We therefore conclude
that in the framework of the quasi-classical approximation employed in
arriving at \eq{paclg} the ratio $R^{pA}$ is less than $1$ at small
$k_T \lsim Q_{s0}$ and displays Cronin {\sl enhancement} at high $k_T
\gsim Q_{s0}$. This conclusion has been reached in
\cite{KTS,BKW,KNST,KKT,KM,ktbroadening1,ktbroadening2,Vitev03,ktbroadening3}.

To summarize let us note that, if one neglects the logarithms in the
exponents of \eq{paclg} as a slowly varying functions of transverse
separations, writing for instance $\un x^2\ln(1/x_T\Lambda)\approx \un
x^2$, it would become possible to perform the $\un x$ and $\un y$
integrations in \eq{paclg} exactly, obtaining \cite{KM,KKT}
\be\label{paqs}
\frac{d \sigma^{pA}}{d^2 k \ dy} \, = \, \frac{\as \, C_F}{\pi^2} \,\int d^2 b 
\, \left\{ - \frac{1}{{\un k}^2} +  \frac{2}{{\un k}^2} \, e^{- {\un k}^2 / 
Q_{s0}^2} + \frac{1}{Q_{s0}^2} \, e^{- {\un k}^2 / Q_{s0}^2} \, \left[
\ln \frac{Q_{s0}^4}{4 \, \Lambda^2 {\un k}^2} + \mbox{Ei} \left( 
\frac{{\un k}^2}{Q_{s0}^2} \right) \right] \right\}, 
\ee
where $\mathrm{Ei}(x)$ is the exponential integral. Expanding
\eq{paqs} would give the gluon production cross section in $pp$, which 
can be obtained from \eq{pp} by dropping $\ln (k_T^2 /\Lambda^2)$ in
it. The nuclear modification factor resulting from \eq{paqs} is
\cite{KKT}
\be\label{rpa1}
R^{pA} (k_T) \, = \, \frac{{\un k}^4}{Q_{s0}^2} \, 
\left\{ - \frac{1}{{\un k}^2} +  \frac{2}{{\un k}^2} \, e^{- {\un k}^2 / 
Q_{s0}^2} + \frac{1}{Q_{s0}^2} \, e^{- {\un k}^2 / Q_{s0}^2} \, \left[
\ln \frac{Q_{s0}^4}{4 \, \Lambda^2 {\un k}^2} + \mbox{Ei} \left( 
\frac{{\un k}^2}{Q_{s0}^2} \right) \right] \right\}.
\ee
The ratio $R^{pA} (k_T)$ is plotted in \fig{cron} for $\Lambda = 0.2 \
Q_{s0}$. It clearly exhibits an enhancement at high-$k_T$
\cite{KTS,BKW,KNST,KKT,KM,ktbroadening1,ktbroadening2,Vitev03,ktbroadening3}
typical of Cronin effect \cite{Cronin}. The height and position of
Cronin maximum are increasing functions of $Q_s$, and, therefore, of
collision centrality \cite{BKW,KKT}.

%%%%%%%%%%%%%%%%%%%%%%%%%%%%%%%%%%%%%%%%%%%%%%%%%%%%%%%%%%%%%%%%%%%%%%%%%%%%%%%%%%%%%%%%%%%%%%%

\subsection{Gluon Production Including Quantum Evolution}
\label{qepa}

Here we are going to show how to include the effects of quantum BK
evolution equation from Sect. \ref{QE} into the gluon production cross
section derived in Sect. \ref{clpa}. We will then analyze the impact
of small-$x$ evolution on the nuclear modification factor $R^{pA}$,
demonstrating that it eliminates Cronin enhancement leading to {\sl
suppression} of gluon production at all transverse momenta $k_T$.

%%%%%%%%%%%%%%%%%%%%%%%%%%%%%%%%%%%%%%%%%%%%%%%%%%%%%%%%%%%%%%%%%

\subsubsection{Gluon Production Cross Section}

Starting with the gluon production cross section, let us first solve
the problem of gluon production in deep inelastic scattering (DIS)
instead of $pA$ collisions. The advantage of DIS is that we know
explicitly how to include the non-linear small-$x$ evolution into the
dipole wave function, as shown in \fig{dis1} and given by
\eq{eqN}. The answer we will obtain this way would be easy to
generalize to the case of $pA$ collisions.
%%%%%%%%%%%%%%%%%%%%%%%%%%%%%%%%%%%%%%%%%%%%%%%%
\begin{figure}[ht]
\begin{center}
\epsfxsize=13cm
\leavevmode
\hbox{ \epsffile{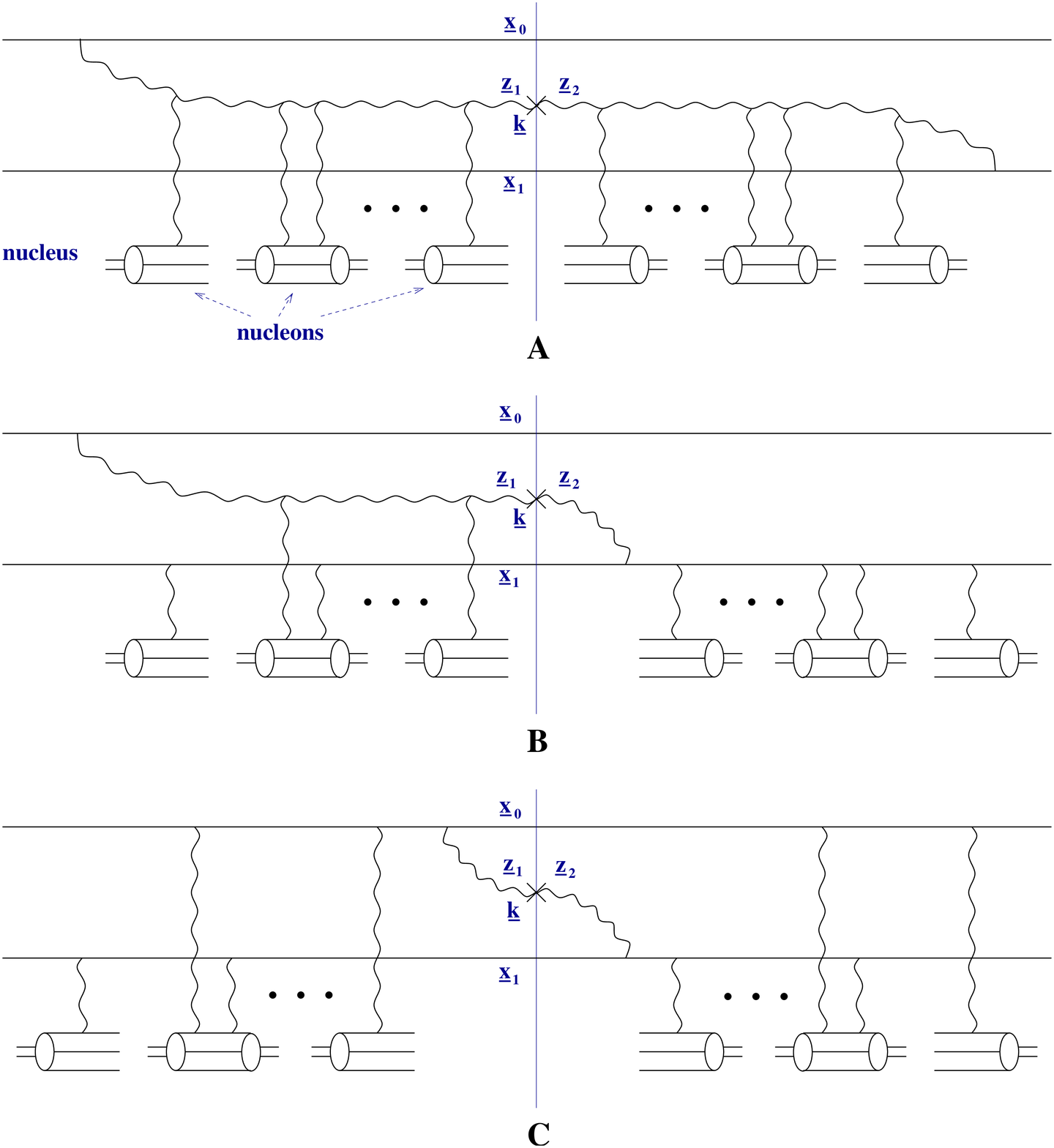}}
\end{center}
\caption{Gluon production in DIS in the quasi-classical approximation. 
The produced gluon may be emitted either off the quark or off the
antiquark lines both in the amplitude and in the complex conjugate
amplitude. Only one connection is shown. }
\label{qcdis}
\end{figure}
%%%%%%%%%%%%%%%%%%%%%%%%%%%%%%%%%%%%%%%%%%%%%%%%

The problem of including the effects of BK evolution into the
quasi-classical gluon production cross section from \eq{paclg} could
be thought of as producing one of the gluons in the non-linear cascade
of \fig{dis1}. However, the cascade we needed for calculation of the
total DIS cross section is somewhat different from what we need now
for the inclusive cross section. The difference is in the fact that
the cascade of \fig{dis1} develops from early times, with light cone
time $\tau \equiv x^- = -\infty$, till the interaction with the
nucleus at $\tau =0$.\footnote{Remember, all the multiple
rescatterings with the nucleus in \fig{qqbar} happen very fast, hence
the whole multiple rescattering process can be viewed as instantaneous
when compared to the time it takes to develop the cascade of
gluons/dipoles in \fig{dis1}: therefore we denote the time at which
all the multiple rescatterings take place as $\tau=0$.} Since
\fig{dis1} represents the forward scattering amplitude, the evolution
of the cascade after the interaction, during the times between $\tau
=0$ and $\tau = +\infty$, is restricted to one simple case when the
system returns back to the initial state and nothing is produced. This
is the definition of the forward amplitude and this is all we need to
calculate the total cross section. To calculate the inclusive cross
section we need to square the production amplitude, as was shown in,
for instance, \fig{pa}, for the quasi-classical case. There, already
in the amplitude, we had to include the diagrams where the incoming
proton would ``cascade'', generating a gluon both {\sl before} the
interaction in the time interval $\tau = -\infty \ldots 0$ and {\sl
after} the interaction, at $\tau = 0 \ldots +\infty$. Therefore, to
construct a gluon cascade making sure that we always have the gluon we
are trying to produce in the final state, we have to include gluon
emissions bringing in logarithms of energy both at $\tau = -\infty
\ldots 0$ and $\tau = 0 \ldots +\infty$ in the amplitude and in its 
complex conjugate.

To do that let us first generalize the quasi-classical expression from
\eq{paclg} to the case of DIS. The relevant diagrams are shown in \fig{qcdis}. 
They are similar to the case of quark-nucleus scattering pictured in
\fig{pa}. The major difference is that now we have an incoming $q\bar q$ 
dipole instead of just a single quark. The produced gluon can,
therefore, be emitted off of either quark and anti-quark lines on both
sides of the cut. In \fig{qcdis} we show only one particular way of
emitting the gluon. Diagrams A and B in \fig{qcdis} correspond to
diagrams A and B in \fig{pa}. Another difference now is the diagram in
\fig{qcdis}C, which used to cancel through real-virtual cancellations 
in quark--nucleus scattering case and was not even shown in \fig{pa},
is now non-zero and has to be included. This is due to the fact that
moving a $t$-channel exchanged gluon across the cut in \fig{qcdis}C
would change the color factor of the diagram, thus preventing the
cancellation. Adding all the diagrams in \fig{qcdis}, and summing over
all possible emissions of the gluon off the quark and anti-quark
lines, we write for the gluon production cross section in the
dipole-nucleus scattering \cite{yuridiff,KT}
\begin{eqnarray}\label{qcincl}
\frac{d {\hat \sigma}^{q{\bar q}A}}{d^2 k \, dy \, d^2 B}({\underline x}_{01}) 
\, = \,  \frac{\as C_F}{\pi^2} \, \frac{1}{(2 \pi)^2} \, \int \, 
d^2 z_1 \, d^2 z_2 \, e^{- i {\underline k} \cdot ({\underline z}_1 -
{\underline z}_2)} \, \sum_{i,j=0}^1 (-1)^{i+j}
\frac{{\underline z}_1- {\underline x}_i}{|{\underline z}_1-
{\underline x}_i|^2} \cdot
\frac{{\underline z}_2- {\underline x}_j}{|{\underline z}_2- {\underline
x}_j|^2} \nonumber \\
\times \bigg[ N_G \left({\underline z}_1 - {\underline x}_j,
\frac{1}{2} ({\underline z}_1 + {\underline x}_j) , 0\right) + N_G
\left({\underline z}_2 - {\underline x}_i, \frac{1}{2} ({\underline
z}_2 + {\underline x}_i), 0\right) - N_G
\left({\underline z}_1 - {\underline z}_2, \frac{1}{2} ({\underline
z}_1 + {\underline z}_2) , 0\right) \nonumber \\
- N_G \left({\underline x}_i -
{\underline x}_j, \frac{1}{2} ({\underline x}_i + {\underline x}_j),
0\right)\bigg],
\end{eqnarray}
where, just like in Sect. \ref{BKsect}, the quark and the anti-quark
have transverse coordinates ${\un x}_0$ and ${\un x}_1$
correspondingly. The rest of the notation of \eq{qcincl} is explained
in \fig{qcdis}. $N_G$ in \eq{qcincl} is taken from \eq{NGgm}, and now
we put the correct impact parameters for all the dipoles.

To include the effects of small-$x$ evolution we should add to the
graphs in \fig{qcdis} the diagrams with more gluon emissions before
and after the interaction. Let us denote the rapidity of the target
nucleus as $0$ and the rapidity of the incoming $q\bar q$ dipole as
$Y$. The produced gluon would have rapidity $y$. Following \cite{KT}
we will divide all possible extra gluon emission into two categories:
the gluons can have rapidities larger (harder) or smaller (softer)
than $y$.
%%%%%%%%%%%%%%%%%%%%%%%%%%%%%%%%%%%%%%%%%%%%%%%%%%%%%%%
\begin{figure}[ht]
\begin{center}
\epsfxsize=15cm
\leavevmode
\hbox{ \epsffile{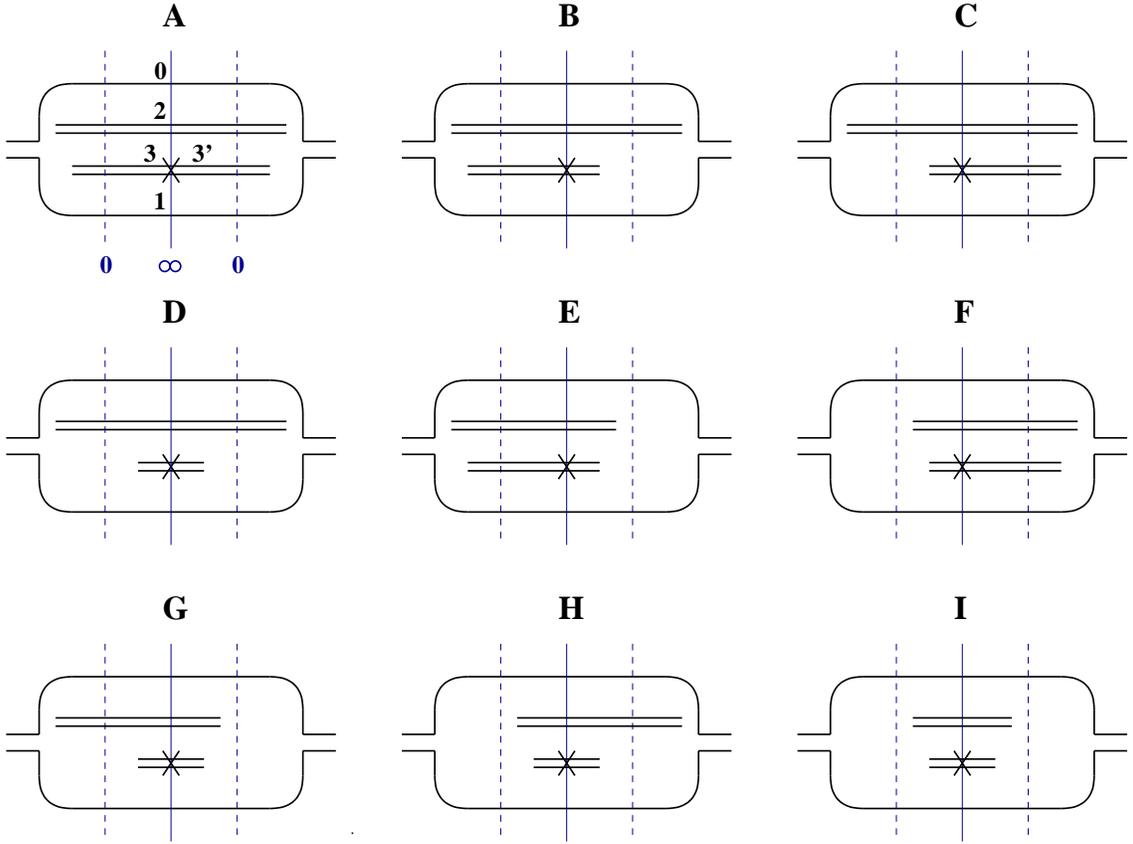}}
\end{center}
\caption{Emission of a harder gluon in the dipole evolution, as considered in \cite{KT}. 
Gluons are denoted by double lines in the large $N_c$ limit. The
produced gluon is marked by a cross. Dashed lines comprise multiple
rescatterings with the target from \fig{qqbar}, which are
instantaneous on the time scale of the dipole evolution. }
\label{eem}
\end{figure}
%%%%%%%%%%%%%%%%%%%%%%%%%%%%%%%%%%%%%%%%%%%%%%%%%%%%%%%%%

First we analyze the emissions of {\sl harder} gluons. Some of the
relevant diagrams are shown in \fig{eem}. Since we are trying to apply
quantum evolution from Mueller's dipole model \cite{dip1,dip2,dip3} we
are again working in the large-$N_c$ limit. Similar to \fig{dipeqn}
the gluons are denoted by double lines which do not connect to any
particular quark or anti-quark lines in the dipole in which a
particular gluon is emitted: this denotes summation over both
connections on both sides of the cut. In \fig{eem} the gluon $3$ is
the one that we measure in the final state, the gluon $2$ is a harder
gluon emitted before gluon $3$. Solid vertical lines in \fig{eem}
denote the $\tau = +\infty$ final state, while the dashed vertical
lines denote the multiple rescattering interactions with the target
from \fig{qqbar} taking place at $\tau =0$ on both sides of the cut.

In Sect. \ref{BKsect} we have demonstrated that successive emissions
of progressively softer gluons before the interaction (for $\tau =
-\infty \ldots 0$) give the leading logarithmic contribution, in the
sense of giving a factor of $Y$ for each factor of $\as$ resumming the
parameter $\as Y$, as is accomplished by \eq{eqN}. Interestingly
enough, this ordering has to be reversed for emissions after the
interaction (for $\tau = 0 \ldots +\infty$): to obtain the leading
logarithmic contribution ($\as Y$), softer gluons have to be emitted
before harder ones at $\tau = 0 \ldots +\infty$ \cite{KT}. More
details on how this happens could be found in \cite{KT}. Indeed, if
emission of a harder gluon happened after the emission of the softer
one, the harder gluon can not be emitted off the softer one: it is
emitted by hard quark lines of the original dipole \cite{KT}. Since
the diagrams E--I in \fig{eem} violate such inverse ordering, they can
be discarded. Only the diagrams A--D in \fig{eem} contribute. Those
are {\sl real} contributions to the dipole evolution leading to
production of the dipole in which gluon $3$ was emitted. Note that all
of these surviving diagrams have early time ($\tau = -\infty \ldots
0$) emissions only: they correspond to standard dipole evolution of
\fig{dis1}. A similar analysis shows that out of the rest of the
diagrams with possible emissions of gluon $2$, the ones that bring in
leading logarithmic contribution are the ones which give {\sl virtual}
corrections to the dipole evolution at light cone times $\tau =
-\infty \ldots 0$ \cite{KT}. Emissions of other harder gluons do not
change the conclusions: the surviving diagrams are the ones which
contribute to dipole evolution leading to production of the dipole in
which gluon $3$ was emitted. Such evolution is described by the
quantity $n_1 ({\un x}_{01}, {\un x}_{0'1'}, {\un B} - {\un b}, Y-y)$,
originally introduced in \cite{dip1,dip2}, which has the meaning of
the number of dipoles of size ${\un x}_{0'1'}$ at rapidity $y$ and
impact parameter $\un b$ generated by evolution from the original
dipole ${\un x}_{01}$ having rapidity $Y$ and impact parameter $\un
B$. This quantity obeys the dipole model analogue of the BFKL equation
\cite{BFKL}, which is just the linear part of \eq{eqN}
\cite{dip1,dip2}
\ben
\frac{\partial n_1 ({\un
x}_{01}, {\un x}_{0'1'}, {\un b}, y)}{\partial y} \, = \, 
\frac{\as \, N_c}{2 \, \pi^2} \, 
\int d^2 x_2 \, \frac{x_{01}^2}{x_{20}^2 \, x_{21}^2} \, 
\bigg[ n_1 ({\un x}_{02}, {\un x}_{0'1'}, {\un b} + \frac{1}{2} \, {\un x}_{21}, y) + 
n_1 ({\un x}_{12}, {\un x}_{0'1'}, {\un b} + \frac{1}{2} \, {\un x}_{20}, y) 
\een
\be\label{eqn}
- n_1 ({\un x}_{01}, {\un x}_{0'1'}, {\un b}, y) \bigg] 
\ee
with the initial condition 
\be
n_1 ({\un x}_{01}, {\un x}_{0'1'}, {\un b}, y=0) \, = \, \delta ({\un
x}_{01} - {\un x}_{0'1'}) \, \delta ({\un b}).
\ee

The inclusion of the effects of harder gluons in \eq{qcincl} is then
accomplished by replacing the cross section in \eq{qcincl} by
\cite{KT}
\be\label{hard}
\frac{d {\hat \sigma}^{q{\bar q}A}}{d^2 k \, dy \, d^2 B}({\underline x}_{01})  
\rightarrow \int d^2 b \, d^2 x_{0'1'} \ 
n_1 ({\un x}_{01}, {\un x}_{0'1'}, {\un B} - {\un b}, Y-y) \ \frac{d
{\hat \sigma}^{q{\bar q}A}}{d^2 k \, dy \, d^2 b}({\underline
x}_{0'1'}).
\ee

Now we have to resum emissions of softer gluons having rapidity less
than the rapidity of the produced gluon $y$.
%%%%%%%%%%%%%%%%%%%%%%%%%%%%%%%%%%%%%%%%%%%%%%%%%%%%%%%%%%%%%%%
\begin{figure}[ht]
\begin{center}
\epsfxsize=16cm
\leavevmode
\hbox{ \epsffile{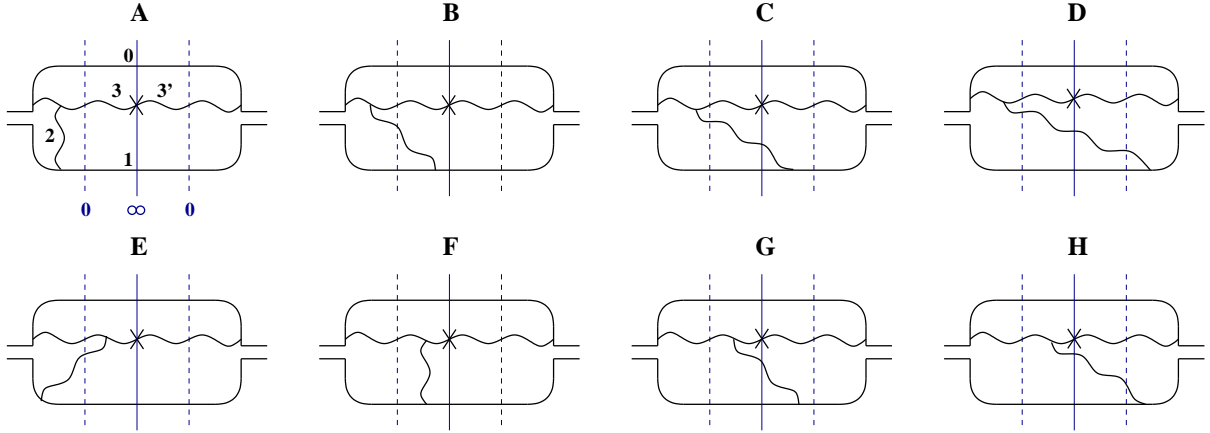}}
\end{center}
\caption{Diagrams including one softer gluon produced after the emission 
of the measured gluon \cite{KT}.}
\label{lem}
\end{figure}
%%%%%%%%%%%%%%%%%%%%%%%%%%%%%%%%%%%%%%%%%%%%%%%%%%%%%%%%%%%%%%%%%
Such emissions are demonstrated in \fig{lem}, where now the gluon $2$
is softer than the gluon $3$. \fig{lem} only has emissions where gluon
$2$ interacts with gluon $3$ and with either the quark or the
anti-quark lines in the dipole. As was shown in \cite{KT}, such
diagrams cancel pairwise via real-virtual cancellations:
\be
A + D =0, \hspace*{5mm} B + C =0, \hspace*{5mm} F+G=0 , \hspace*{5mm}
E+H=0.
\ee
The diagrams where gluon $2$ connects only to the quark and the
anti-quark lines cancel analogously \cite{KT}. The only remaining
diagrams are where the gluon $2$ is emitted and reabsorbed by gluon
$3$ only. Using the crossing symmetry \cite{Mueller_cross} we reflect
the line of gluon $3$ into the complex conjugate amplitude (denoted by
$3'$) with respect to the cut, obtaining a gluon dipole in the
amplitude, just like we did in Sect. \ref{clglue} to obtain
\eq{paclg}. Then it becomes manifest that emission of gluon $2$ can be 
thought of as one iteration of dipole evolution in the gluon dipole
$33'$. Successive emissions of even softer gluon would not modify this
conclusion. Therefore, to include the effects of softer gluon
emissions in \eq{qcincl} we have to replace \cite{KT}
\be\label{soft}
N_G ({\un x}, {\un b}, 0) \rightarrow N_G ({\un x}, {\un b}, y)
\ee
on its right hand side, where now $N_G ({\un x}, {\un b}, y)$ is the
amplitude of a gluon dipole interacting with the nucleus including the
small-$x$ evolution of \eq{eqN}. In the large-$N_c$ limit $N_G$ can be
expressed in terms of the quark dipole amplitude $N$ from \eq{eqN} as
\be\label{ggqq}
N_G ({\underline x}, {\underline b}, y) \, = \, 2 \, N ({\underline
x}, {\underline b}, y) - N^2 ({\underline x}, {\underline b}, y),
\ee
since a gluon dipole in the large-$N_c$ limit can be thought of as a
pair of quark dipoles, with either one of them or both quark dipoles
interacting.

Combining the prescriptions for including quantum evolution for hard
(\ref{hard}) and soft (\ref{soft}) emissions we obtain the following
expression for inclusive gluon production cross section in DIS
\cite{KT}
\ben
\frac{d \sigma^{q{\bar q}A}}{d^2 k \ dy \, d^2 B} ({\un x}_{01}) \ = \ 
 \int  \, d^2 x_{0'1'} \,
n_1 ({\underline x}_{01}, {\underline x}_{0'1'}, {\underline B} - {\underline
b}, Y - y) \,  \frac{\as
C_F}{\pi^2} \, \frac{1}{(2 \pi)^2} \, d^2 b
\een
\ben
\times \, d^2 z_1 \, d^2 z_2 \, e^{- i {\underline k} \cdot ({\underline z}_1 -
{\underline z}_2 )} \, \sum_{i,j = 0'}^{1'} (-1)^{i+j} \,
\frac{{\underline z}_1 - {\underline x}_i}{|{\underline z}_1 - 
{\underline x}_i|^2} \cdot \frac{{\underline z}_2 - {\underline
x}_j}{|{\underline z}_2 - {\underline x}_j|^2} \left[ N_G \left({\underline
z}_1 - {\underline x}_j, \frac{1}{2} ({\underline z}_1 + {\underline
x}_j) , y\right) + \right.
\een
\be\label{incl}
+ \left. N_G \left({\underline z}_2 - {\underline x}_i, \frac{1}{2}
({\underline z}_2 + {\underline x}_i), y\right) - N_G
\left({\underline z}_1 - {\underline z}_2, \frac{1}{2} ({\underline
z}_1 + {\underline z}_2) , y\right) - N_G \left({\underline x}_i -
{\underline x}_j, \frac{1}{2} ({\underline x}_i + {\underline x}_j),
y\right)\right].
\ee
\eq{incl} provides us with the single inclusive gluon production cross section 
for scattering of the quark--anti-quark dipole $01$ on the target
nucleus, which includes all multiple rescatterings (powers of $\as^2
A^{1/3}$) and small-$x$ evolution corrections (powers of $\as Y$ and
$\as y$) \cite{KT}.

%%%%%%%%%%%%%%%%%%%%%%%%%%%%%%%%%%%%%%%%%%%%%%%%%%%%%%%%%%%%%%%
\begin{figure}[ht]
\begin{center}
\epsfxsize=10cm
\leavevmode
\hbox{ \epsffile{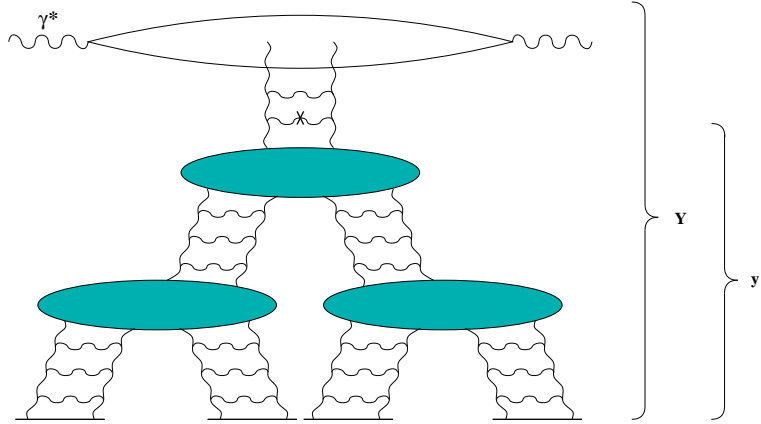}}
\end{center}
\caption{Interpretation of the inclusive gluon production in $pA$ collisions from 
\eq{incl} in terms of fan diagrams.}
\label{fan1}
\end{figure}
%%%%%%%%%%%%%%%%%%%%%%%%%%%%%%%%%%%%%%%%%%%%%%%%%%%%%%%%%%%%%%%%%

If one associates \eq{eqN} with summation of fan diagrams from
\fig{fan}, than one can represent \eq{incl} in the way shown in 
\fig{fan1}. There the interaction between the quark--anti-quark 
dipole and the target is mediated by exchange of fan diagrams, where
we trigger on one of the produced gluons, which is denoted by the
cross. The part of the diagram between the produced gluon and the
projectile dipole is described in \eq{incl} by the quantity $n_1
({\underline x}_{01}, {\underline x}_{0'1'}, {\underline B} -
{\underline b}, Y - y)$, which satisfies linear BFKL evolution
equation (\ref{eqn}). In terms of ladder and fan diagrams, that
corresponds to having just a single ladder exchange between the dipole
and the produced gluon, as shown in \fig{fan1}. In principle one could
draw gluon production diagrams where the ladder between the dipole and
produced gluon would split into several ladders interacting with the
target. However, \eq{incl} implies that such graphs cancel. The same
cancellation is expected if one applies AGK cutting rules to QCD
diagrams \cite{AGK,BR,Braun2}. It is intriguing to see AGK rules work
in perturbative QCD in the framework of the dipole model. They have
also been found to work previously for single diffractive cross
sections in DIS calculated in the dipole model approach
\cite{KL}. However, the AGK rules appear to break down in QCD for inclusive 
two-particle production cross sections \cite{JMK,BGV,NSZZ}.

In the limit of a very large target nucleus the momentum transfer to
the nucleus is cut off by inverse nuclear radius and is very small. We
can therefore take the scattering amplitudes $N_G$ in
\eq{incl} at $t=0$, which in coordinate space is equivalent to
neglecting (factoring out) the impact parameter dependence since the
shifts in impact parameter in \eq{incl} are small compared to the
nuclear radius \cite{FIIM,BKL}. We, therefore, put all the impact
parameters in $N_G$'s of \eq{incl} to be equal to $\un b$. Repeating
the steps outlined at the end of Sect. \ref{clglue} we arrive at \cite{KT}
\be\label{ktdip}
\frac{d \sigma^{q{\bar q}A}}{d^2 k \ dy} \, = \, \frac{2 \, \as}{C_F} \, 
\frac{1}{{\un k}^2} \, \int d^2 q \, \phi_{q\bar q} ({\un q}, Y-y) \, \phi_A ({\un k} 
- {\un q}, y)
\ee
with $\phi_A ({\un k} - {\un q}, y)$ given by \eq{ktglue} and the
unintegrated gluon distribution in the dipole given by \cite{KT,JMK}
\be\label{dipglue}
\phi_{q\bar q} ({\un q}, y) \, = \, \frac{2 \, \as \, C_F}{\pi} \, 
\int d^2 b \, d^2 x_{0'1'} 
\, e^{- i {\un q} \cdot {\un x}_{0'1'}} \, \frac{1}{\nabla^2_{x_{0'1'}}} \, 
n_1 ({\underline x}_{01}, {\underline x}_{0'1'}, {\underline b}, y).
\ee
Interestingly enough, as we see from \eq{ktdip}, the small-$x$
evolution does not lead to break-down of the $k_T$-factorization
formula (\ref{ktcl})! In fact, \eq{ktdip}, with the gluon distribution
functions given by \eq{dipglue} and by \eq{ktglue}, only with twice
the quark dipole amplitude $2 \, N$ instead of $N_G$, was postulated
as the answer for inclusive gluon production cross section by Braun in
\cite{Braun2} using $k_T$-factorization approach. Still, \eq{ktdip},
obtained in \cite{KT}, differs from the result of \cite{Braun2} by the
difference in $N_G$ and $2 \, N$ (see \eq{ggqq}), which in the
standard pomeron language appears to corresponds to emission of the
produced gluon from the triple pomeron vertex \cite{Braun3}. This
difference may become numerically significant for low transverse
momenta of the produced gluon, as was shown in \cite{Braun3}.

Now generalization to the case of $pA$ becomes manifest: we have to
replace the unintegrated gluon distribution function of a dipole,
$\phi_{q\bar q} ({\un q}, y)$, by the one for the proton, $\phi_p
({\un q}, y)$, given by \eq{ktgluep} with $n_G$ satisfying the linear
part of \eq{eqN}
\be\label{eqnG}
\frac{\partial n_G ({\un x}_{01}, {\un b}, y)}{\partial y} \, = \, 
\frac{\as \, N_c}{2 \, \pi^2} \, 
\int d^2 x_2 \, \frac{x_{01}^2}{x_{20}^2 \, x_{21}^2} \, 
\bigg[ n_G ({\un x}_{02}, {\un b} + \frac{1}{2} \, {\un x}_{21}, y) + 
n_G ({\un x}_{12}, {\un b} + \frac{1}{2} \, {\un x}_{20}, y) - n_G
({\un x}_{01}, {\un b}, y) \bigg]
\ee
with \eq{ng} as the initial condition. To summarize, the inclusive
cross section from Eq. (\ref{ktdip}) becomes in the case of $pA$
\be\label{ktpa}
\frac{d \sigma^{pA}}{d^2 k \ dy} \, = \, \frac{2 \, \as}{C_F} \, 
\frac{1}{{\un k}^2} \, \int d^2 q \, \phi_p ({\un q}, Y-y) \, \phi_A ({\un k} 
- {\un q}, y)
\ee
with the unintegrated gluon distributions given by Eqs. (\ref{ktglue})
and (\ref{ktgluep}) with the dipole cross sections evolved with
Eqs. (\ref{eqN}) and (\ref{eqnG}) correspondingly.

Indeed \eq{ktdip} was derived in the large-$N_c$ limit. It appear hard
(though maybe not impossible) to generalize the large-$N_c$ dipole
model to the case of projectile proton: therefore, we can not prove
the suggested generalization to the proton case. However, this
generalization is probably a good approximation of the exact answer
for the following reasons.  We showed that $k_T$-factorization works
in the quasi-classical case to all orders in $N_c$ (see \eq{ktcl}). We
have also shown that it survives evolution corrections
(\eq{ktdip}). Therefore, factorization formula (\ref{ktpa}), which
expresses the production cross section in terms of gluon dipole
amplitudes scattering on the projectile and on the target, is probably
valid in the general case of including small-$x$ evolution beyond the
large-$N_c$ limit. The results of numerical solution of the full
JIMWLK evolution equation for the dipole amplitude were shown to be
amazingly close to the solution of BK equation \cite{RW}. The linear
BFKL evolution analogue in the dipole model \cite{dip1} gives the same
value for the pomeron intercept as the exact BFKL equation
\cite{BFKL}. Therefore, it is probably safe to use \eq{ktpa} for the 
gluon production in $pA$, even though strictly speaking it should
work only for the case when the proton is approximated by a
di-quark. Both pA and DIS processes are considered here as scatterings
of an unsaturated projectile (proton or $q\bar q$ pair) on a saturated
target (nucleus), and, therefore, gluon production cross sections for
them should be similar. Therefore, we will use \eq{ktpa} for the
analysis outlined below of the gluon production in $pA$.

%%%%%%%%%%%%%%%%%%%%%%%%%%%%%%%%%%%%%%%%%%%%%%%%%%%%%%%%%%%%%%%%%

\subsubsection{Nuclear Modification Factor: Double Logarithmic Region}

Let us now analyze what happens to nuclear modification factor
(\ref{rpa}) when quantum evolution is included. To do that, we use
Eqs. (\ref{ktglue}) and (\ref{ktgluep}) to rewrite \eq{ktpa} as
\be\label{paevc1}
\frac{d \sigma^{pA}}{d^2 k \ dy} \, = \, \frac{C_F}{\as \, 
\pi \, (2 \pi)^3} \, 
\frac{1}{{\un k}^2} \, \int d^2 B \, d^2 b \, d^2 z \, 
\nabla^2_z \, n_G ({\un z}, {\un B} - {\un b}, Y-y) \, e^{- i {\un k}
\cdot {\un z}} \, \nabla^2_z \, N_G ({\un z}, {\un b}, y).
\ee
For simplicity, let us neglect the impact parameter dependence of
$N_G$ throughout the nucleus and of $n_G$ over the transverse area of
the proton. Suppressing the impact parameter dependence in the
arguments of $N_G$ and $n_G$ in \eq{paevc1} we rewrite it as
\be\label{paevc}
\frac{d \sigma^{pA}}{d^2 k \ dy} \, = \, \frac{C_F}{\as \, 
\pi \, (2 \pi)^3} \, 
\frac{S_p \, S_A}{{\un k}^2} \, \int d^2 z \, 
\nabla^2_z \, n_G ({\un z}, Y-y) \, e^{- i {\un k}
\cdot {\un z}} \, \nabla^2_z \, N_G ({\un z}, y),
\ee
with $S_A$ and $S_p$ the cross sectional areas of the nucleus and the
proton correspondingly. In this Section, and in the one below, we will
be interested in gluon production with large $k_T \gsim Q_s (y)$, such
that nonlinear effects could be neglected in $N_G$. Outside of the
saturation region we write, similar to \eq{dsol3},
\be\label{NGlin}
N_G (z_T, y) \, = \, \int_{-\infty}^\infty d \nu \ e^{2 \, \bas \,
\chi (0, \nu) \, y } \, (z_T Q_{s0})^{1 + 2 \, i \, \nu} \, C_\nu^A,
\ee
where $C_\nu^A$ is determined by the initial condition of
\eq{NGgm}. Similarly, for $n_G$ one writes the general $\un b$-independent 
solution of \eq{eqnG} as
\be\label{nGlin}
n_G (z_T, y) \, = \, \int_{-\infty}^\infty d \nu \ e^{2 \, \bas \,
\chi (0, \nu) \, y } \, (z_T \Lambda_p)^{1 + 2 \, i \, \nu} \, C_\nu^p,
\ee
where, to satisfy the initial condition given by \eq{ng}, 
\be\label{Cp}
C_{\nu}^p \, = \, \frac{1}{4 \, \pi \, (1 - 2 \, i \, \nu)^2}
\ee
and
\be
\Lambda^2_p \, = \, 4 \, \pi \, \as^2 \, \frac{1}{S_p}.
\ee
Substituting Eqs. (\ref{NGlin}) and (\ref{nGlin}) into \eq{paevc} and
integrating over $\un z$ yields
\ben
\frac{d \sigma^{pA}}{d^2 k \ dy} \, = \, \frac{C_F \, S_p \, S_A}{\as \, 
(2 \pi)^3} \, \int_{-\infty}^\infty d \nu \, d \nu' \, C_\nu^A \,
C_{\nu'}^p \, (1+2 i \nu)^2 \, (1+2 i \nu')^2 \, 2^{2 i (\nu + \nu')}
\, \frac{\Gamma [i  (\nu + \nu')]}{\Gamma [1- i  (\nu + \nu')]} 
\een
\be\label{dlapa1}
\times \, \left( \frac{Q_{s0}}{k_T} \right)^{1+2 i \nu} \, \left( 
\frac{\Lambda_p}{k_T} \right)^{1+2 i \nu'} \, e^{2 \, \bas \,
\chi (0, \nu) \, y + 2 \, \bas \,
\chi (0, \nu') \, (Y-y)}
\ee
with $\bas$ given by \eq{bas}.

In this Section we are interested in the double logarithmic region
(DLA) where $k_T \gsim k_{\rm geom} \gg Q_s (y)$. Evaluating the $\nu$ and
$\nu'$ integrals in \eq{dlapa1} in the stationary phase approximation
around the saddle point from \eq{spdla1} with $\chi (0,\nu)$ (and
$\chi (0,\nu')$) given by \eq{spdla} we obtain \cite{Ryskin}
\ben
\frac{d \sigma^{pA}}{d^2 k \ dy} \bigg|_{DLA} \, \approx \, 
\frac{C_F \, S_p \, S_A}{\as \, (2 \pi)^4} \, 
\frac{Q_{s0}^{2} \, \Lambda_p^2}{{\un k}^4} \, \frac{1}{2 \, \bas} \, 
\left[ \frac{\ln \frac{k_T}{Q_{s0}} \, \ln \frac{k_T}{\Lambda_p}}{y^3 \, 
(Y-y)^3} 
\right]^{1/4} \, \left( \sqrt{\frac{y}{\ln \frac{k_T}{Q_{s0}}}} + 
\sqrt{\frac{Y-y}{ \ln \frac{k_T}{\Lambda_p}}} \right)
\een
\be\label{dla}
\times \exp\left( 2 \sqrt{2 \, \bas \, y \, \ln \frac{k_T}{Q_{s0}}} +
 2 \sqrt{2 \, \bas \, (Y-y) \, \ln \frac{k_T}{\Lambda_p}} \ \right).
\ee
In arriving at \eq{dla} we made use of \eq{Cp}, and have also used the
fact that near the saddle point of \eq{spdla1} $C_\nu^A$ is
approximately given by \eq{Cp} as well \cite{KKT}.

To calculate $R^{pa}$ we also need the gluon production cross section
in $pp$ collisions in the same kinematic DLA region, which can be
obtained from \eq{dla} by replacing $S_A \rightarrow S_p$ and $Q_{s0}
\rightarrow \Lambda_p$ in it \cite{KKT}
\be\label{dlapp}
\frac{d \sigma^{pp}}{d^2 k \ dy} \bigg|_{DLA} \, \approx \, 
\frac{C_F \, S_p^2}{\as \, (2 \pi)^4} \
\frac{\Lambda_p^4}{{\un k}^4} \, \frac{1}{2 \, \bas} \, 
\frac{\sqrt{y} + \sqrt{Y-y}}{y^{3/4} \, (Y-y)^{3/4}} 
\, \exp\left[ 2 \sqrt{2 \, \bas \, \ln \frac{k_T}{\Lambda_p}} \left( \sqrt{y} 
+ \sqrt{Y-y} \right) \right].
\ee
Substituting Eqs. (\ref{dla}) and (\ref{dlapp}) in \eq{rpa} and
remembering that $S_A = A^{2/3} S_p$ and $Q_{s0}^2 = A^{1/3}
\Lambda_p^2$ yields
\ben
R^{pA} (k_T, y) \bigg|_{k_T \gg Q_s} \, = \, \frac{\left( \ln
\frac{k_T}{Q_{s0}} \, \ln
\frac{k_T}{\Lambda_p} \right)^{1/4}}{\sqrt{y} + \sqrt{Y-y}} \, 
\left( \sqrt{\frac{y}{\ln \frac{k_T}{Q_{s0}}}} + 
\sqrt{\frac{Y-y}{ \ln \frac{k_T}{\Lambda_p}}} \right) 
\een
\be\label{dlar}
\times \, \exp\left[ 2 \sqrt{2 \, \bas \, y} \left( \sqrt{\ln
\frac{k_T}{Q_{s0}}} - \sqrt{\ln \frac{k_T}{\Lambda_p}} \right) \right].
\ee
The behavior of $R^{pA} (k_T, y)$ from \eq{dlar} at high $y$ is
dominated by the exponent in it \cite{KKT}
\be\label{dlaexp}
R^{pA} (k_T, y) \bigg|_{k_T \gg Q_s} \, \approx \, \exp\left[ 2
\sqrt{2 \, \bas \, y} \left( \sqrt{\ln
\frac{k_T}{Q_{s0}}} - \sqrt{\ln \frac{k_T}{\Lambda_p}} \right) \right] \, < \, 1.
\ee
Since $Q_{s0} \gg \Lambda_p$, the exponent in \eq{dlaexp} becomes less
than $1$, driving $R^{pA} (k_T, y)$ below one as well \cite{KKT}. We
therefore conclude that in the DLA region with $k_T \gsim k_{\rm geom}$,
the nuclear modification factor becomes smaller than $1$ due to
quantum evolution. This indicates {\sl suppression} of gluon
production due to quantum evolution. A suppression of $R^{pA}$ by
small-$x$ evolution effects was originally suggested in \cite{KLM},
though the authors of \cite{KLM} analyzed particle (gluon) production
only in the geometric scaling region $k_T \lsim k_{\rm geom}$, which is
what we are going to do now.

%%%%%%%%%%%%%%%%%%%%%%%%%%%%%%%%%%%%%%%%%%%%%%%%%%%%%%%%%%%%%%%%%

\subsubsection{Nuclear Modification Factor: Extended Geometric Scaling Region}

In the extended geometric scaling region, $Q_s (Y) < k_T \lsim k_{\rm
geom}$, multiple rescatterings are still unimportant and
\eq{dlapa1} from the previous Section still applies. To obtain the 
leading behavior of \eq{dlapa1} in the geometric scaling region we
will evaluate the $\nu$-integral in it by simply replacing $\nu$ with
$\nu_0$ from \eq{nu0val} in the integrand, as it was done in obtaining
\eq{dsol10}. We obtain
\ben
\frac{d \sigma^{pA}}{d^2 k \ dy} \, = \, \frac{C_F \, S_p \, S_A}{\as \, 
(2 \pi)^3} \ C_{\nu_0}^A \, (1+2 i \nu_0)^2 \, 2^{2 i \nu_0} \, \left(
\frac{Q_{s0}}{k_T} \right)^{1+2 i \nu_0} \, e^{2 \, \bas \,
\chi (0, \nu_0 ) \, y} \, 
\een
\be\label{llapa}
\times  \, \int_{-\infty}^\infty d \nu' \,  C_{\nu'}^p \, (1+2 i \nu')^2 \, 2^{2 i \nu'}
\, \frac{\Gamma [i  (\nu_0 + \nu')]}{\Gamma [1- i  (\nu_0 + \nu')]} \, \left( 
\frac{\Lambda_p}{k_T} \right)^{1+2 i \nu'} \, e^{2 \, \bas \,
\chi (0, \nu') \, (Y-y)}.
\ee
The gluon production cross section in $pp$ collisions is given by
\ben
\frac{d \sigma^{pp}}{d^2 k \ dy} \, = \, \frac{C_F \, S_p^2}{\as \, 
(2 \pi)^3} \, \int_{-\infty}^\infty d \nu \, d \nu' \, C_\nu^p \,
C_{\nu'}^p \, (1+2 i \nu)^2 \, (1+2 i \nu')^2 \, 2^{2 i (\nu + \nu')}
\, \frac{\Gamma [i  (\nu + \nu')]}{\Gamma [1- i  (\nu + \nu')]} 
\een
\be\label{pp1}
\times \, \left( \frac{\Lambda_p}{k_T} \right)^{1+2 i \nu} \, \left( 
\frac{\Lambda_p}{k_T} \right)^{1+2 i \nu'} \, e^{2 \, \bas \,
\chi (0, \nu) \, y + 2 \, \bas \,
\chi (0, \nu') \, (Y-y)}.
\ee

To evaluate the $\nu'$-integral in \eq{llapa} and both $\nu$ and
$\nu'$ integrals in \eq{pp1} we need to know whether we are inside or
outside of the extended geometric scaling region for the
proton. Recall that, while \eq{pp1} and \eq{ktpa} are valid only
outside the saturation region of the proton, such that the evolution
between the produce gluon and the proton can only be linear, that
evolution may still be dominated either by the double logarithmic
saddle point $\nu^*_{DLA}$ from \eq{spdla1} or by the leading
logarithmic saddle point $\nu_0$ from \eq{nu0val}. The dominant saddle
point is determined by scale of the geometric scaling in the proton,
$k_{\rm geom}^p$, which can be obtained from \eq{kgeom} by using
proton saturation scale in it, which, in turn, is obtained from
\eq{qsy2} by putting $A=1$. In the problem we are considering here,
both the atomic number $A$ and the rapidity intervals $y$ and $Y-y$
are large. Therefore, in principle, $k_{\rm geom}^p$ can be above the
nuclear saturation scale $Q_s (y)$. Since we are interested in the
extended geometric scaling region of the nucleus, $Q_s (Y) < k_T \lsim
k_{\rm geom}$, we have to consider two cases: (a)~$k^p_{\rm geom}
\lsim k_T
\lsim k_{\rm geom}$ and (b)~$Q_s (Y) < k_T \lsim k^p_{\rm geom}$.

\begin{itemize}

\item[(a)] $k^p_{\rm geom} \lsim k_T \lsim k_{\rm geom}$ If $k_T$ of the 
produced gluon is outside of the proton extended geometric scaling
region, then $\nu'$-integral in \eq{llapa} and $\nu$ and $\nu'$
integrals in \eq{pp1} have to be evaluated around the saddle point
$\nu^*_{DLA}$ from \eq{spdla1}, which in momentum space for the case of
the proton is given by
\be
\nu^*_{DLA} \, \approx \, - \frac{i}{2} \, \left( 1 - 
\sqrt{\frac{2 \, \bas \, Y}{\ln (k_T/\Lambda_p)}} \right).
\ee
Evaluating both \eq{llapa} and \eq{pp1} using \eq{Cp}, substituting the
results into \eq{rpa} and dropping the slowly varying prefactors, we
obtain the nuclear modification factor \cite{KKT}
\be\label{llar2}
R^{pA} (k_T, y) \bigg|_{k^p_{\rm geom} \lsim k_T \lsim k_{\rm geom}} \, \approx \,
\frac{k_T}{Q_{s0}} \, \exp \left[{(\alpha_P - 1) \, y - 2 \, 
\sqrt{2 \, \bas \, y \, \ln \frac{k_T}{\Lambda}} - \frac{\ln^2 
\frac{k_T}{Q_{s0}}}{14 \, \zeta (3) \, \bas \, y}}\right].
\ee
To estimate the value of $R^{pA}$ in \eq{llar2} in the region
$k^p_{\rm geom} \lsim k_T \lsim k_{\rm geom}$ we substitute $k_T =
k_{\rm geom}$ into \eq{llar2} with $k_{\rm geom}$ from \eq{kgeom}. The
result yields an asymptotically small value
\be\label{rpae}
R^{pA} (k_T = k_{\rm geom}, y) \, \approx \, e^{- 1.65 \, \bas \, y}
\, \ll \, 1,
\ee
where we used $A=197$ for gold nucleus. For other values of $A$ and
for other values of $k_T$ in the region $k_{\rm geom}^p \lsim k_T
\lsim k_{\rm geom}$ one still gets exponential suppression for $R^{pA}
(k_T, y)$.  Therefore we conclude that $R^{pA} (k_T, y) < 1$ in the
region $k_{\rm geom}^p \lsim k_T \lsim k_{\rm geom}$.

\item[(b)] $Q_s (Y) < k_T \lsim k^p_{\rm geom}$ If $k_T$ of the produced 
gluon is inside the geometric scaling regions of both the nucleus and
the proton, we evaluate the $\nu'$-integral in \eq{llapa} and $\nu$
and $\nu'$ integrals in \eq{pp1} by putting $\nu = \nu' = \nu_0$ in
them with $\nu_0$ given by \eq{nu0val}. Noting that almost all the
prefactors in Eqs. (\ref{llapa}) and (\ref{pp1}) would then be
identical, we write for the nuclear modification factor
\be\label{lla1}
R^{pA} (k_T, y) \bigg|_{Q_s (y) < k_T \lsim k^p_{\rm geom}} \, \approx
\, \frac{C^A_{\nu_0}}{C^p_{\nu_0}} \, A^{-\frac{1}{3}} \, \left(
\frac{Q_{s0}}{\Lambda_p} \right)^{1+2 i \nu_0}.
\ee
As $Q_{s0}^2 = A^{1/3} \Lambda_p^2$ \eq{lla1} leads to 
\be\label{lla2}
R^{pA} (k_T, y) \bigg|_{Q_s (y) < k_T \lsim k^p_{\rm geom}} \, \approx
\, \frac{C^A_{\nu_0}}{C^p_{\nu_0}} \ A^{- \frac{1}{6} + \frac{i \, \nu_0}{3}}.
\ee
Since $\nu_0$ is just a number, $C^A_{\nu_0}/C^p_{\nu_0}$ does not
depend on rapidity and varies weakly with $A$. The nuclear
modification factor in \eq{lla2} is, therefore, driven by the power of
$A$ in it. Using the numerical value of $\nu_0$ from \eq{nu0val} we
obtain \cite{KKT}
\be\label{lla3}
R^{pA} (k_T, y) \bigg|_{Q_s (y) < k_T \lsim k^p_{\rm geom}} \, \approx
\, A^{- \frac{1}{6} + \frac{i \, \nu_0}{3}} \, = \, A^{-0.124} \, < \, 1.
\ee
This is the suppression of particle production which was originally
predicted in \cite{KLM}. 

\end{itemize}

We conclude that small-$x$ evolution leads to suppression of $R^{pA}$
in the whole extended geometric scaling region $Q_s (y) < k_T \lsim
k_{\rm geom}$ \cite{KLM,KKT,BKW,AAKSW}.

%%%%%%%%%%%%%%%%%%%%%%%%%%%%%%%%%%%%%%%%%%%%%%%%%%%%%%%%%%%%%%%%%

\subsubsection{Nuclear Modification Factor: Saturation Region}

Above we have shown that for $k_T > Q_s (y)$ quantum evolution
introduces suppression of $R^{pA}$. Here we will study the saturation
region, $k_T \lsim Q_s (y)$. The important question now is to
understand what happens to Cronin maximum from \fig{cron} as the
effects of small-$x$ evolution become important. Since the Cronin peak
is located at $k_T = Q_s (y)$, we need to find out what happens with
$R^{pA} (k_T = Q_s (y), y)$ as $y$ increases. 

Inside the geometric scaling region ($z_T > 1/k_{\rm geom}$) the
dipole amplitude $N_G$ depends only on the parameter $\xi$ from
\eq{xidef}. Therefore one can write
\be\label{NGgeom}
N_G (z_T, y) \, = \, \int_{-\infty}^\infty d \nu \ [z_T Q_{s}
(y) ]^{1 + 2 \, i \, \nu} \, {\tilde C}_\nu^A
\ee
with ${\tilde C}_\nu^A$ some unknown coefficient. Substituting
\eq{NGgeom} along with \eq{nGlin} into \eq{paevc} and integrating over 
${\un z}$ yields
\ben
\frac{d\sigma^{pA}}{d^2k \, dy}\bigg|_{k_T=Q_s (y)} \, = \,
 \frac{C_F\, S_p\, S_A }{\as (2\pi)^3} \int \, d \nu \, d \nu' \,
 C_{\nu'}^p \, \tilde C_\nu^A \, (1+2 i \nu)^2 \, (1+2 i \nu')^2 \,
 2^{2 i (\nu + \nu')} \, \frac{\Gamma [i (\nu + \nu')]}{\Gamma [1- i
 (\nu + \nu')]} 
\een
\be\label{fm} 
\times \, \left( \frac{\Lambda_p}{Q_s (y)} \right)^{1 + 2 \, i \, \nu'} \, 
e^{2 \, \bas \, \chi (0, \nu') \, (Y-y)}.  
\ee
(In arriving at \eq{fm} we have assumed that extended geometric
scaling region $z_T > 1/k_{\rm geom}$ dominates in the integral of
\eq{paevc}: this is explicitly shown in the Appendix of \cite{KKT}.) 
To evaluate the $\nu'$-integral in \eq{fm} we again need to know
whether $k_T$ of the produced gluon is above or below the scale of
extended geometric scaling in the proton $k^p_{\rm geom}$. This may be
an issue when $k^p_{\rm geom} < Q_s (y)$.  Assuming for simplicity
that $k^p_{\rm geom} > Q_s (y)$ we evaluate the $\nu'$-integration in
\eq{fm} by putting $\nu' = \nu_0$ with $\nu_0$ given by \eq{nu0val}. 
(The case when the evolution between the proton and the produced gluon
is dominated by DLA saddle point is examined in \cite{KKT} with the
same conclusion as we will reach below.) \eq{fm} becomes
\be\label{fm1} 
\frac{d\sigma^{pA}}{d^2k \, dy}\bigg|_{k_T=Q_s (y)} \, = \,
 \frac{C_F\, S_p\, S_A }{\as (2\pi)^3} \ C_{\nu_0}^p \, {\cal C}_A \,
 (1+2 i \nu_0)^2 \, 2^{2 i \nu_0} \, \left( \frac{\Lambda_p}{Q_s (y)}
 \right)^{1 + 2 \, i \, \nu_0} \, e^{2 \, \bas \, \chi (0, \nu_0) \,
 (Y-y)},
\ee
where we have defined
\be\label{calc}
{\cal C}_A \, \equiv \, \int \, d \nu \,  \tilde C_\nu^A \, (1+2 i \nu)^2 \,  
2^{2 i \nu} \, \frac{\Gamma [i (\nu + \nu_0)]}{\Gamma [1- i
 (\nu + \nu_0)]}.
\ee
It is important to note that all the information on the solution of
the non-linear evolution equation given by $N_G$ is now absorbed into
a {\sl constant} from \eq{calc}, which does not depend on rapidity and
on atomic number of the nucleus, and is thus not important for
behavior of $R^{pA}$ at high energy/rapidity.

Similarly, evaluation the $\nu$ and $\nu'$ integrals in \eq{pp1}
around $\nu_0$ yields
\be\label{pp2}
\frac{d \sigma^{pp}}{d^2 k \ dy} \, = \, \frac{C_F \, S_p^2}{\as \, 
(2 \pi)^3} \, [C_{\nu_0}^p]^2 \, (1+2 i \nu_0)^4 \, 2^{4 i \nu_0}
\, \frac{\Gamma (2 i \nu_0)}{\Gamma (1- 2 i \nu_0)} 
\, \left( \frac{\Lambda_p}{k_T} \right)^{2+4 i \nu_0} \,  e^{2 \, \bas \,
\chi (0, \nu_0) \, Y}.
\ee
Putting $k_T = Q_s (y)$ in \eq{pp2}, and using it together with
\eq{fm1} in \eq{rpa} yield for the nuclear modification factor
\be\label{rpacron1}
R^{pA}(k_T = Q_s (y),y) \, = \, \frac{{\cal C}_A}{C_{\nu_0}^p} \, (1+2
i \nu_0)^{-2} \, 2^{-2 i \nu_0} \, \frac{\Gamma (1- 2 i \nu_0)}{\Gamma
(2 i \nu_0)} \, A^{- \frac{1}{3}} \, \left( \frac{Q_s (y)}{\Lambda_p}
\right)^{1 + 2 i \nu_0} \, e^{- 2 \, \bas \, \chi (0, \nu_0) \, y}.
\ee
Remembering that $Q_s (y)$ is given by \eq{qsy0} with $Q_{s0} =
A^{1/6} \Lambda_p$ we can recast \eq{rpacron1} into the following form
\be\label{rpacron2}
R^{pA}(k_T = Q_s (y),y) \, = \, \frac{{\cal C}_A}{C_{\nu_0}^p} \, (1+2
i \nu_0)^{-2} \, 2^{-2 i \nu_0} \, \frac{\Gamma (1- 2 i \nu_0)}{\Gamma
(2 i \nu_0)} \ A^{- \frac{1}{6} + \frac{i \, \nu_0}{3}}.
\ee
The prefactor of \eq{rpacron2} is just a number. The dynamical
information is carried only by the power of $A$ in it, which, for the
value of $\nu_0$ from \eq{nu0val} gives
\be\label{rpacron3}
R^{pA}(k_T = Q_s (y),y) \, \approx \, A^{- \frac{1}{6} + \frac{i \,
\nu_0}{3}} \, = \, A^{- 0.124} \, \ll 1 \hspace*{1cm} \mbox{for} \hspace*{1cm} A \gg 1.
\ee
\eq{rpacron3} allows us to conclude that small-$x$ evolution tends to 
reduce the Cronin peak turning enhancement into suppression. Since the
suppression in \eq{rpacron3} is (parametrically) of the same order as
given by \eq{lla3} for the extended geometric scaling region, we
conclude that $R^{pA}$ both at $k_T = Q_s (y)$ and in the extended
geometric scaling region is smaller than $1$ at all values of $k_T$ in
that region, approaching the same (energy-independent) limit at very
high energies/rapidities.

Deep inside the saturation region, for $k_T \ll Q_s (y)$, one can use
the above techniques to show that $R^{pA}$ there is also suppressed at
least at the level of \eq{rpacron3} and likely even more than that. We
refer the interested reader to \cite{KKT} for details on this
derivation. Here we can just argue that saturation effects with or
without small-$x$ evolution, reduce the $\sim 1/k_T^4$
infrared-singular scaling of the inclusive production cross section of
\eq{pp} to the more infrared-regular $\sim 1/k_T^2$ scaling of \eq{paclglkt}. 
Hence the suppression of $R^{pA}$ at $k_T \ll Q_s (y)$, shown in the
quasi-classical limit by \eq{rpalkt}, will still be valid in the case
of quantum evolution, though it will be slightly modified by the
anomalous dimension $\nu_0$.

With the help of \eq{rpacron3} we conclude that at high
rapidities/energies the Cronin maximum decreases with energy and
centrality, with $R^{pA} (Q_s (y),y)$ becoming less than $1$.
Eventually, at very high energy, the Cronin peak flattens out and
saturates to an energy independent lower limit given by \eq{rpacron3},
which is parametrically suppressed by powers of $A$. The height of
$R^{pA} (Q_s (y),y)$ is also a decreasing function of collision
centrality/atomic number $A$.

%%%%%%%%%%%%%%%%%%%%%%%%%%%%%%%%%%%%%%%%%%%%%%%%%%%%%%%%%%%%%%%%%

%%%%%%%%%%%%%%%%%%%%%%%%%%%%%
\begin{figure}[ht]
\begin{center}
\epsfxsize=10cm
\leavevmode
\hbox{\epsffile{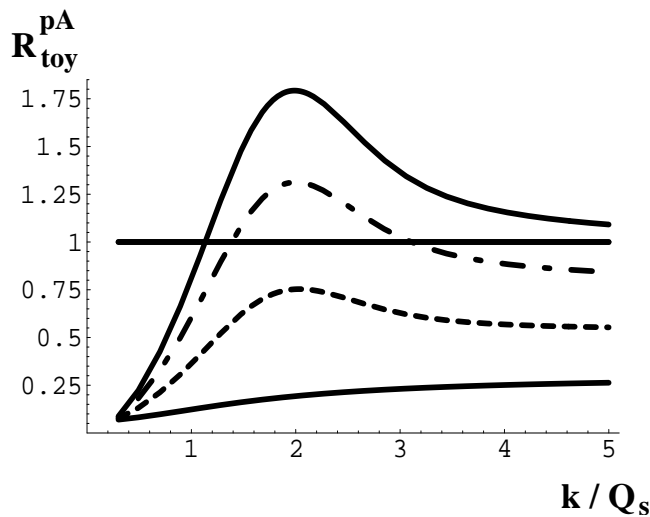}}
\end{center}
\caption{The ratio $R^{pA}$ plotted as a function of $k_T/Q_s (y)$ for 
(i) the quasi-classical approximation of \fig{cron}, which is valid
for moderate energies/rapidities (upper solid line); (ii) the toy
model for very high energies/rapidities from \protect\cite{KKT} (lower
solid line); (iii) an interpolation between the two to intermediate
energies (dash-dotted and dashed lines). The cutoff is $\Lambda = 0.3
\, Q_s$. The figure is from \protect\cite{KKT}.}
\label{toy}
\end{figure}
%%%%%%%%%%%%%%%%%%%%%%%%%%%%%%

\subsubsection{Nuclear Modification Factor: Overall Picture}

In the three above Sections we have demonstrated that small-$x$
evolution leads to suppression of the nuclear modification factor
$R^{pA}$ for the transverse momenta $k_T$ of the produced gluon in the
double logarithmic region $k_T \gsim k_{\rm geom}$ (see
\eq{dlaexp}), in the extended geometric scaling region 
$Q_s (y) < k_T \lsim k_{\rm geom}$ (see \eq{lla3} and \eq{rpae}) and
inside the saturation region $k_T \lsim Q_s (y)$ (see
\eq{rpacron3}). To summarize the effects of quantum evolution, we 
illustrate how $R^{pA} (k_T, y)$ varies with rapidity/energy in
\fig{toy}, which is constructed using a toy model of \cite{KKT}, that 
incorporates the correct {\sl qualitative} behavior of $R^{pA} (k_T, y)$.

The top line in \fig{toy} is borrowed from \fig{cron} and represents
the Cronin enhancement of particle production in the quasi-classical
approximation. As energy (or rapidity) increases the suppression
begins to set in as shown by dash-dotted and dashed lines in
\fig{toy}. The asymptotic energy-independent limit of maximum 
suppression to be reached at very high energies/rapidities given by
Eqs. (\ref{lla3}) and (\ref{rpacron3}) is depicted by the lower solid
line in \fig{toy}. We conclude that as the energy/rapidity increases
the upper solid line in \fig{toy} would decrease eventually turning
into the lower solid line. This result was obtained by numerical
simulations in \cite{AAKSW} and by an analytical analysis similar to
the one outline above in \cite{KKT} (see also \cite{BKW,IIT}).

Indeed \fig{toy} is not yet ready to be compared to data. In $d+Au$
collisions at RHIC the forward rapidity high-$p_T$ particle production
receives a substantial contribution from valence quark production,
along with the gluon production which led to \fig{toy}. On general
grounds, since the gluon suppression in \fig{toy} is driven by
small-$x$ evolution, we expect valence quark production cross section
to also be suppressed at high-$p_T$. To investigate this further and
to be able to make quantitative predictions we will calculate forward
production of valence quarks in the next Section.

%%%%%%%%%%%%%%%%%%%%%%%%%%%%%%%%%%%%%%%%%%%%%%%%%%%%%%%%%%%%%%%%%%%%%%%%%

\subsection{Valence Quark Production}
\label{clquark}

We now consider production of valence (high $x$) quarks in
proton-nucleus collisions, due to scattering of a valence quark on the
nucleus \cite{Dumitru:2002qt}.  In this section, we treat the target
nucleus as a Color Glass Condensate while the projectile proton is
taken to be a dilute system of quarks and gluons in the spirit of
standard pQCD.  This is the appropriate approach if one is probing the
large $x$ components of the projectile proton wave function which is
the case in the very forward proton-nucleus collisions.

\begin{figure}[htb]
\begin{center}
\epsfxsize=7cm
\leavevmode
\hbox{\epsffile{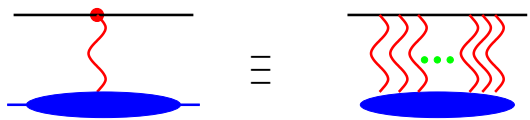}}
\end{center}
\caption{Scattering of a high $x$ quark from a dense target.}
\label{fig:v_quark}
\end{figure}

The leading order diagram is shown in Fig. (\ref{fig:v_quark}).
%%%%%%%%%%%%%%%%%%%%%%%%%%%%%
%%%%%%%%%%%%%%%%%%%%%%%%%%%%%%
Using the LSZ formalism, the scattering amplitude is   
\begin{eqnarray}
\langle q(q)_{out}|q(p)_{in}\rangle =
\langle out|b_{out}(q)b^{\dagger}_{in}(p)|in\rangle 
\label{eq:amp}
\end{eqnarray}
which can be written as
\begin{eqnarray}
\langle out|b_{out}(q)b^{\dagger}_{in}(p)|in\rangle \!\!=\!\!
{ -1 \over Z_2} \int d^4x d^4y  e^{-i(p\cdot x - q\cdot y)}
\bar{u}(q)[i \stackrel{\rightarrow}{\slpartial}_y -m]
\langle out|T \psi (y) \bar{\psi}(x)|in\rangle  
[-i \stackrel{\leftarrow}{\slpartial}_x -m] u(p)
\label{eq:general}
\end{eqnarray}
where $m$ is the quark mass and $Z_2$ is the fermion wave function
renormalization factor while $u(p)$ is the quark spinor with momentum $p$.
The quark propagator $S_F$ in the background of the classical color 
field is defined by 
\begin{eqnarray}
\langle out|T \psi (y) \bar{\psi}(x) |in\rangle  \equiv -i \langle  out |
 in\rangle  S_F (y,x).
\end{eqnarray}
The amplitude then becomes (setting $Z_2 =1$ and $ \langle out |
in\rangle =1$ since we are working to leading order in $\alpha_s$ and
our background field is time independent)
\begin{eqnarray}
\langle q(q)_{out}|q(p)_{in}\rangle \!\!\!\!&=&\!\!\!\!
i \!\!\int d^4x d^4y  e^{-i(p \cdot x - q\cdot  y)} 
\bar{u}(q)[i \stackrel{\rightarrow}{\slpartial}_y -m] 
S_F(y,x) [-i \stackrel{\leftarrow}{\slpartial}_x -m] u(p).
\nonumber
\end{eqnarray}
In momentum space, the fermion propagator $S_F$ can be written
as
\begin{eqnarray}
S_F(q,p)= (2\pi)^4 \delta^4 (q-p) S^0_F (p) -ig S^0_F (q)
\int {d^4k \over (2\pi)^4} \, {\slA}(k) S_F (q+k,p)
\label{eq:Geqmom}   
\end{eqnarray}
where ${\slA}=A^{\mu} \gamma_{\mu}$ is the classical background color
field, and $S^0_F$ is the free fermion propagator. It is useful to
define the interaction part of the fermion propagator from
(\ref{eq:Geqmom}) as
\begin{eqnarray}
S_F(q,p)= (2\pi)^4 \delta^4 (q-p) S^0_F (p) + S^0_F (q)
\tau (q,p) S^0_F (p).
\label{eq:Gint}
\end{eqnarray}
Substituting (\ref{eq:Gint}) into the amplitude leads to
\begin{eqnarray}
\langle q(q)_{out}|q(p)_{in}\rangle  =  \bar{u}(q) \tau (q,p) u(p).
\label{eq:finalamp}
\end{eqnarray}
This simple relation relates the amplitude for scattering of a quark
or anti-quark from the Color Glass Condensate and the quark propagator
in the background color field of the nucleus.

The explicit form of the quark propagator in the background of a
classical color field was calculated in
\cite{hw,Gelis:2001da}. The interaction part, as defined
in (\ref{eq:Gint}) is given by \cite{Gelis:2001da}
\begin{eqnarray}
\tau (q,p)=(2\pi) \delta (p^- - q^-) \gamma^- 
\int d^2 z \bigg [ V ({\un z}) -1 \bigg ]
e^{i({\un q} - {\un p}) \cdot {\un z}}
\label{eq:taures}
\end{eqnarray}
where $V$ is a matrix in the fundamental representation which includes
the multiple scattering of the quark on the nucleus, given by
(\ref{eq:v_fun})
\begin{eqnarray}
V({\un z}) \equiv \hat{P} \exp \bigg [-ig \int^{+\infty}_{-\infty}
d z^- \, A^+_a (z^-,{\un z}) \ t^a \bigg]
\label{eq:udef}
\end{eqnarray}
and $t^a$ are in the fundamental representation. Using (\ref{eq:taures})
in the scattering amplitude (\ref{eq:finalamp}) gives
\begin{eqnarray}
\langle q(q)_{out}|q(p)_{in}\rangle \! =\!
(2\pi) \delta (p^- - q^-)
\bar{u}(q) \gamma^-   u(p) \!\int d^2 z 
[ V({\un z}) -1  ] e^{i({\un q} - {\un p}) \cdot {\un z}}
\label{eq:expamp}
\end{eqnarray}
The presence of the delta function in the amplitude is due to 
the target being (light-cone) time independent which leads
to conservation of the ``minus'' component of momenta and 
can be factored out (for a rigorous treatment of this using 
wave packets, we refer the reader to \cite{Gelis:2002ki}) 
\begin{eqnarray}
\langle q(q)_{out}|q(p)_{in}\rangle  &=& (2\pi) \delta (p^- - q^-) M (p,q)
\label{eq:mamp}
\end{eqnarray}
leading to 
\begin{eqnarray}
d\sigma = \int {d^4 q \over (2\pi)^4} (2\pi) \delta (2 q^+ q^- -q_T^2)
\theta (q^+) {1 \over 2p^-} (2\pi) \delta (p^- - q^-) 
|M (p,q)|^2.
\label{eq:diffcs}
\end{eqnarray}
One can set the transverse momentum of the incoming quark $p_t$ to
zero without any loss of generality. We then get the differential
cross section for production of a quark with transverse momentum $q_t$
per unit area
\begin{eqnarray}
{d\sigma^{qA\to qX} \over d^2 b \, d^2 q} ={2\over (2\pi)^2}
\sigma_{dipole}^F ({\un b}, {\un q})
\label{eq:cs_v_quark}
\end{eqnarray}
where the dipole cross section (which is, via the optical theorem, the
imaginary part of the forward scattering amplitude) is defined as
\begin{eqnarray}
\sigma_{dipole} ^F ({\un b}, {\un q}) \equiv 
\int d^2 r \, e^{i \, {\un q} \cdot {\un r}}
{1 \over N_c} {\rm Tr}_c\left<
V^{\dagger} ({\un b} + {\un r}/2) V({\un b} - {\un r}/2)  - 1
\right>_\rho \, = \, - \int d^2 r \, e^{i \, {\un q} \cdot {\un r}} \, 
N ({\un r}, {\un b},Y)
\label{eq:cs_dipole_F}
\end{eqnarray}
and we have neglected terms which do not contribute at finite quark
transverse momentum. It is trivial to extend this calculation to the
case of incoming on shell gluons \cite{Dumitru:2001jn} scattering on
the dense target. The only difference is the presence of adjoint
matrix $U$ rather than $V$ which have different color factors in the
exponent, appropriate for the given representation. It is given by
\begin{eqnarray}
{d\sigma^{gA\to gX} \over d^2 b \, d^2 q} ={2\over (2\pi)^2}
\sigma_{dipole}^A ({\un b}, {\un q})
\label{eq:cs_v_gluon}
\end{eqnarray}
where the adjoint dipole cross section is defined by 
\begin{eqnarray}
\sigma_{dipole} ^A ({\un b}, {\un q}) \equiv 
\int d^2 r \, e^{i \, {\un q} \cdot {\un r}}
{1 \over N_c^2 - 1} {\rm Tr}_c\left<
U^{\dagger} ({\un b} + {\un r}/2) U({\un b} - {\un r}/2)  - 1
\right>_\rho \, = \, - \int d^2 r \, e^{i \, {\un q} \cdot {\un r}} \, 
N_G ({\un r}, {\un b},Y).
\label{eq:cs_dipole_A}
\end{eqnarray}
Our result in (\ref{eq:cs_v_quark}, \ref{eq:cs_v_gluon}) can be
understood as the generalization of the leading order quark-gluon and
gluon-gluon hard scattering cross section convoluted with the target
gluon distribution function in leading twist pQCD to include multiple
scattering of the incoming quark on the target.  To include the
effects of quantum evolution in the target at high energies, one
should resum large logs of energy or $1/x$. This can be accomplished
by solving the JIMWLK (or BK) equations for the (fundamental or
adjoint) dipole cross section and using the result in
(\ref{eq:cs_v_quark}, \ref{eq:cs_v_gluon}).

The quantum evolution of the quark-nucleus scattering cross section
could have very interesting observable signatures in the forward
rapidity region. At forward rapidities the nuclear saturation scale is
large due to small-$x$ evolution. A typical valence quark in the
incoming proton would have a transverse momentum of order $Q_s$ after
scattering from the nucleus. This momentum is large and the quark will
tend to fragment into hadrons independent of the other valence
quarks. This would reduce the number of baryons produced in the
forward rapidity region of $pA$ collisions as compared to $pp$
collisions, while increasing the number of produced mesons, so that
the meson to baryon ratio could increase in $pA$ compared to $pp$
\cite{markstrikman}.

We note that the qualitative behavior of the nuclear modification
factor $R_{pA}$ for quark production should be very similar to that of
gluons. This is due to the fact that the behavior of $R_{pA}$ is
determined by the properties of the of the underlying (fundamental
vs. adjoint) dipole cross section. There may however be quantitative
differences due to the different color factors for the fundamental and
adjoint dipoles which would also affect the rate of quantum evolution
for the two dipoles.  For example, one can show that in the classical
approximation to quark production, the nuclear modification factor
$R_{pA}$ for quarks \cite{Gelis:2002nn} exhibits an enhancement
similar to that of gluons even though the magnitude of enhancement for
gluon $R_{pA}$ is larger because of the larger value of the gluon
saturation scale due to the color factors.  As one goes to smaller
values of $x$ where quantum evolution in $\ln 1/x$ becomes important,
the nuclear modification factor $R_{pA}$ becomes smaller than unity in
analogy to $R_{pA}$ for gluons even though the rate of decrease of
$R_{pA}$ for quarks is smaller than the rate for the decrease of
$R_{pA}$ for gluons.

To relate this to hadron production in proton-nucleus collisions, one
will need to convolute these hard cross sections with the quark and
gluon distribution functions in a proton and quark-hadron and
gluon-hadron fragmentation functions. At this level, the distribution
and fragmentation functions are delta functions of $x$ and $z$
respectively, given by the parton model. One can include DGLAP
evolution of the projectile parton distributions and hadron
fragmentation by considering radiation of a gluon or a quark
anti-quark pair and then absorbing the resultant collinear divergences
into the bare distribution and fragmentation functions which lead to
DGLAP evolution of the parton distribution functions in the proton and
the hadron fragmentation functions. The detailed results are given in
\cite{aaj}.

%%%%%%%%%%%%%%%%%%%%%%%%%%%%%%%%%%%%%%%%%%%%%%%%%%%%%%%%%%%%%%%%%%%%%%%%%

\subsection{Electromagnetic Probes}
\label{ep}

In this Section we consider the production of photons and dileptons in
proton-nucleus collisions \cite{Gelis:2002ki, Gelis:2002fw,
Baier:2004tj, Jalilian-Marian:2005zw}. Photon and dilepton production
has the advantage that there is no hadronization involved in the final
state, unlike quark and gluon production. Therefore, photons and
dileptons are a cleaner probe of the Color Glass
Condensate. Nevertheless, they have the disadvantage of low production
rate due to the smallness of the electromagnetic coupling vs. the
strong coupling involved in quark and gluon production.

\subsubsection{Photon Production}

We consider the process $q(p) A \rightarrow \gamma(k) \, q(q)\,X$
where $A$ stands for a hadron or nucleus which is treated as a Color
Glass Condensate and the produced quark and photon are real. The
incoming quark has momentum $p$ while $q$ and $k$ denote the momenta
of the outgoing on shell quark and photon respectively. To get the
photon production cross section, we will integrate over the momentum
of the produced quark at the end. The amplitude is given by
\begin{eqnarray}
\left<q( q)\gamma (k)_{\rm out}|q(p)_{\rm in}\right>=
\big<0_{\rm out}\big|a_{\rm out}(k)b_{\rm out}(q)
b^{\dagger}_{\rm in}(p)\big|0_{\rm in}\big>
\label{eq:amp_pho}
\end{eqnarray}
which, using the LSZ formalism and the definition of the quark
propagator, can be written as
\begin{eqnarray}
\big<q(q)\gamma (k)_{\rm out}\big|q(p)_{\rm in}\big> \!\!\!\!&=&\!\!\!\!
e\! \int d^4x\, d^4y\, d^4z\, e^{i(k\cdot x + q\cdot z -p\cdot
y)} \,\overline{u}(q)\nonumber \\
&&(i \stackrel{\rightarrow}{\slpartial}_z -m) 
S_{_{F}}(z,x) {\slvarepsilon} S_{_{F}}(x,y)
(i \stackrel{\leftarrow}{\slpartial}_y -m) u(p)
\label{eq:amplitude_pho}
\end{eqnarray}
where $S_F$ is the quark propagator in the background of the classical
color field given in (\ref{eq:Geqmom}) and $\epsilon^{\mu} (k)$ is the
polarization vector of the produced photon.  Using the decomposition
of the quark propagator as a free part and an interacting part as in
(\ref{eq:Gint}), the amplitude can be written as
\begin{eqnarray}
&&
\big<q(q)\gamma (k)_{\rm out}\big|q(p)_{\rm in}\big>=
-e\, \overline{u}(q) \,\Bigg[ (2\pi)^4 \delta^4 (k+q-p)    
{\slvarepsilon} \nonumber\\
&& + \, 
\tau (q,p-k) \,S_{_{F}}^{0}(p-k)\,
{\slvarepsilon}
 + {\slvarepsilon}\,
S_{_{F}}^{0}(q+k)\,\tau (q+k,p)\nonumber \\
&&
+
\int {{d^4l} \over {(2\pi)^4}} \tau (q,l)\,S_{_{F}}^{0}(l) 
\,{\slvarepsilon}\,S_{_{F}}^{0}(k+l)\,\tau (k+l,p)
\Bigg]\,u( p)
\label{eq:M_pho} 
\end{eqnarray}
where $\tau$ is defined in (\ref{eq:taures}) and $S_F^0$ is the free
quark propagator.
%%%%%%%%%%%%%%%%%%%%%%%%%%%%%
\begin{figure}[htb]
\begin{center}
\epsfxsize=10cm
\leavevmode
\hbox{\epsffile{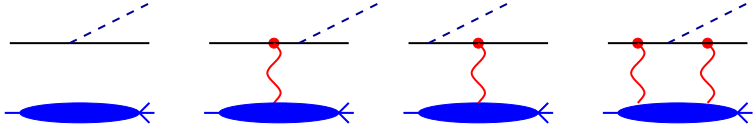}}
\end{center}
\caption{Real photon production in quark-nucleus scattering.}
\label{fig:pho}
\end{figure}
%%%%%%%%%%%%%%%%%%%%%%%%%%%%%%
The various terms in (\ref{eq:M_pho}) are shown in Fig. \ref{fig:pho}
where the thick wavy lines attached to the full circle signify
multiple scattering of the quark from the nucleus and photon is shown
by a dashed line. The first term corresponds to the case when the
incoming quark does not scatter on the target and radiates a
photon. This is kinematically impossible so that this term is
identically zero (as can be seen by noticing that the delta function
has no support). The second term corresponds to the case when the
incoming quark scatters from the nucleus and then radiates a photon
while the third term corresponds to the case when the incoming quark
radiates a photon and then scatters from the target nucleus. The last
term corresponds to the incoming quark scattering from the target,
radiating a photon and then scattering from the target again. In the
high energy (eikonal) limit considered here, the last term is also
zero (or more precisely, suppressed by powers of center of mass
energy) and is therefore dropped (for a proof of this, we refer the
reader to \cite{Gelis:2002ki}). Extracting the overall delta function,
the amplitude is
\begin{eqnarray}
{\cal M}(q,k:p) & = & -i \,e \, \overline{u}(q)\bigg[ 
{\gamma^- ({\slp} - {\slk} + m) \, {\slvarepsilon} \over (p-k)^2 -m^2} +
{{\slvarepsilon}\;({\slq} + {\slk} + m)\gamma^- \over (q+k)^2 -m^2}
\bigg]u(p)\nonumber\\
&& \int d^2 x\, 
e^{i({\un q} + {\un k} - {\un p})\cdot {\un x}} \, 
\Big(V({\un x})- 1\Big).
\label{eq:finalamp_pho}
\end{eqnarray}
To get the cross section, we need to square the amplitude, include the
phase space and flux factors and average (sum) over the initial
(final) state degrees of freedom. We also set the transverse momentum
of the incoming quark and the quark mass to zero for simplicity.  We
get the differential cross section for production of a photon with
transverse momentum $k_T$ and a quark with transverse momentum $q_t$
\begin{eqnarray}
{d\sigma^{q A \to q\,\gamma \, X} \over d^2 b \, d^2 k \, d^2 q \, dz}
 \!\!\!\!&=\!\!\!\!& 
{{e^2} \over {(2\pi)^5} k_T^2}
{[1 + (1-z)^2] \over z}  
{{({\un q} + {\un k})^2 }  \sigma_{dipole}^F ({\un b}, {\un q} + {\un k})    
 \over [ {\un q} + {\un k} - {\un k} /z]^2} \, 
\label{eq:pho_inccs} 
\end{eqnarray}
where $z\equiv k^-/p^-$ and $ {[1 + (1-z)^2]/z}$ is the standard
leading order photon splitting function and the dipole cross section
is defined in (\ref{eq:cs_dipole_F}). To get the cross section for a
hadron $+$ photon cross section, one would need to convolute the above
with the quark-hadron fragmentation function which can then be used to
investigate the correlation between the produced hadron and photon. It
is also remarkable that the cross section is written in terms of the
same degrees of freedom as in the valence quark production cross
section.

To get the single inclusive photon production cross section, we
integrate over the transverse momentum of the produced quark $q_T$. A
quick inspection of Eq. (\ref{eq:pho_inccs}) shows that the
integration over $q_T$ is divergent. This happens when the transverse
momenta of the photon and the final state quark are parallel, ${\un q}
= (1-z){\un k} /z$. This is nothing but the standard collinear
divergence in pQCD calculation of the so called fragmentation photons
(as opposed to direct photons), which occurs when the photon and
produced quark are parallel.  This contribution is formally of order
of $\alpha_s$ compared to direct photon contribution. However, the
large collinear log can be identified as being of order $1/\alpha_s$
and therefore it is customary in pQCD to include fragmentation photons
at the same order of calculation as in direct photons
\cite{Aurenche:1987fs}. This collinear divergence is not affected by
the multiple scattering of the quark on the nucleus.

To do the $q_T$ integration \cite{Jalilian-Marian:2005zw}, we write
the dipole cross section back in the coordinate space and shift the
quark transverse momentum ${\un q} \rightarrow {\un q} + {\un k}
(1-z)/z $ which leads to the following integral
\begin{eqnarray}
\int_0^{\hat{s}} {d^2 q \over q_T^2} \, e^{i {\un q} \cdot {\un r}} \simeq 
\pi \, \ln \, {\hat s}/\Lambda^2_{QCD}
\label{eq:col_div}
\end{eqnarray}
where $\hat{s}$ is the center of mass energy of the quark-nucleus
system. This logarithm is typically written as sum of two pieces; $\ln
\, Q^2/\Lambda^2_{QCD}$ and $\ln \,\hat{s}/Q^2$ where $Q$ is the
factorization scale, usually taken to be the transverse momentum of
the produced photon. With leading log accuracy, only the $\ln
\,Q^2/\Lambda^2_{QCD}$ piece is kept \cite{Aurenche:1987fs} and
absorbed into the quark-photon fragmentation function
$D_{\gamma/q}(z,Q^2)$ which satisfies the DGLAP evolution equation,
evolving with $Q^2$. Identifying the quark-photon splitting function
$P_{\gamma/q} \equiv {e^2 e_q^2 \over 8\pi^2} {1 + (1-z)^2 \over z}$,
we write the photon production cross section as
\begin{eqnarray}
{d\sigma^{q\, A \rightarrow \gamma \, X}
\over d^2b \, d^2k \, dz} &=& {1\over (2\pi)^2} {1\over z^2}\, 
D_{\gamma/q} (z,k_T)\, \sigma_{dipole}^F ({\un b}, {\un k}/z)
\label{eq:frag_pho}
\end{eqnarray}
where the Leading Order (LO) photon fragmentation function is defined as
\be 
D_{\gamma/q} (z,Q^2) \equiv P_{\gamma/q} \, \ln Q^2/\Lambda^2.
\ee 
It is straightforward to include quantum evolution of the target wave
function at high energies in the photon production cross section
here. Again, the dipole cross section is the universal object present
in all single inclusive particle production cross sections in
proton-nucleus collisions. By solving the JIMWLK equations for the
fundamental dipole cross section and using the solution in the above
cross section, one incorporates small $x$ evolution of the target wave
function with Leading Log (of $1/x$) accuracy.

We note that the process considered here is the dominant process at
high energy and forward rapidity. At fixed energy, as one goes towards
mid rapidity, one probes smaller $x$ in the projectile proton where
gluons become as (or more) prominent as valence quarks. One then can
have a gluon split into a quark anti-quark pair, one of which would
then radiate the produced photon. This process is parametrically
suppressed by $\alpha_s$. However, this suppression may be partially
compensated by the dominance of the gluon distribution function in the
proton in mid rapidity and may be numerically important.  Also, as one
goes to very high $p_t$ at fixed center of mass energy, the valence
degrees of freedom in the target wave function will become
important. This is beyond the realm of application of CGC at its
present form and would require extending CGC to high $x$ region.

%%%%%%%%%%%%%%%%%%%%%%%%%%%%%%%%%%%%%%%%%%%%%%%%%%%%%%%%%%%%%%%

\subsubsection{Dilepton Production}

\begin{figure}[htb]
\begin{center}
\epsfxsize=7cm
\leavevmode
\hbox{\epsffile{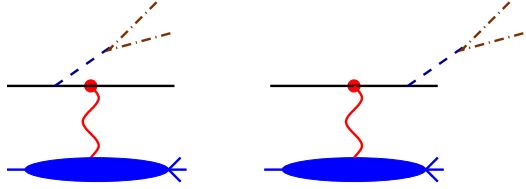}}
\end{center}
\caption{Dilepton production in quark-nucleus scattering.}
\label{fig:dilep}
\end{figure}

We now consider production of dileptons in proton-nucleus collisions
\cite{Gelis:2002fw, Baier:2004tj, Kopeliovich:2001hf, Johnson:2001xf,BGD}. 
This follows simply from our derivation of the photon cross
section. We just need to allow the photon to be off shell and include
the effect of the massive photon splitting into a dilepton pair. We
again start with the production amplitude for this process
\begin{eqnarray}
q(p) + A \to q(q) + l^+ (k_1) +l^- (k_2) + X
\end{eqnarray}
where $k_1,k_2$ are the momenta of the two leptons. The amplitude can
be written as
\begin{eqnarray}
&&\big<0_{\rm out}\big|b_{\rm out} (q)b^{\dagger}_{\rm in}(p)
c_{\rm out}(k_2)d_{\rm out}(k_1)\big|0_{\rm in}\big>=
\int d^4x\, d^4y\, d^4z_1\, d^4z_2\nonumber\\
&& e^{i(q\cdot x-p\cdot y)}\,
e^{i(k_1\cdot z_1 + k_2\cdot z_2)}
\overline{u}(q)\overline{w}(k_2)
(i \stackrel{\rightarrow}{\slpartial}_x -m)
(i \stackrel{\rightarrow}{\slpartial}_{z_2} -m)\nonumber\\
&&
\big<0_{\rm out}\big|{\rm T}\psi(x)\overline{\psi}(y)
\overline{\Psi}(z_1)\Psi(z_2)
\big|0_{\rm in}\big>
(i\stackrel{\leftarrow}{\slpartial}_y +m)
(i\stackrel{\leftarrow}{\slpartial}_{z_1} +m)
v(k_1) u(p)
\end{eqnarray}
where $\psi$ and $\Psi$ are the quark and lepton fields while $u$ and
$w$ ($v$) denote quark and lepton (anti lepton) spinors. The
scattering of the quark on the target is identical to the photon
production case considered earlier. The only difference with the
photon production case is that we need to replace the photon
polarization vectors by the virtual photon propagators which amounts
to the following substitution in the squared amplitude
\begin{eqnarray}
\epsilon_\mu(k)\epsilon_\nu^*(k)\to
{{g_{\mu\rho}}\over{k^2+i\epsilon}}
{{g_{\nu\sigma}}\over{k^2-i\epsilon}}
L^{\rho\sigma}(k_1,k_2)
\end{eqnarray}
with $k\equiv k_1 + k_2$ the virtual photon momentum and the leptonic
tensor $L^{\rho\sigma}(k_1,k_2)$ is the imaginary part of the one loop
leptonic contribution to the photon self energy. Furthermore, we are
interested in the dilepton pair rather than the individual
leptons. Therefore, the leptonic tensor can be written as
\begin{eqnarray}
L^{\rho\sigma}={2\over 3}\alpha_{\rm em}(g^{\rho\sigma}k^2-k^\rho
k^\sigma).
\end{eqnarray}
Extracting the overall delta function again, the amplitude becomes  
\begin{eqnarray}
{\cal M} (p|qk) &=& -ie_q\,\overline{u}(q)\bigg[ 
{\gamma^- ({\slp} - {\slk} + m) {\slvarepsilon}  \over (p-k)^2 -m^2} +
{{\slvarepsilon} \;({\slq} + {\slk} + m)\gamma^- \over (q+k)^2 -m^2}
\bigg]u( p)\nonumber\\
&&\int d^2 r
e^{i({\un q} + {\un k} - {\un p}) \cdot {\un r}}
\Big(V({\un r}) - 1\Big).
\label{eq:finalamp_dilep}
\end{eqnarray}
The two terms of the amplitude are shown in Fig. \ref{fig:dilep}
where again, as in the real photon production case, the virtual photon
can be radiated either before or after the multiple scattering from
the target.
%%%%%%%%%%%%%%%%%%%%%%%%%%%%%
%%%%%%%%%%%%%%%%%%%%%%%%%%%%%%
Note that this amplitude looks exactly as the amplitude for photon
production in (\ref{eq:finalamp_pho}). The difference is that, here
the photon $4$-momentum $k$ is off shell ($k^2 \neq 0$). Squaring the
amplitude, including the flux factors and averaging (summing) over the
initial (final) state degrees of freedom, the cross section is given
by
\begin{eqnarray}
&&\!\!{d\sigma^{q\,A\rightarrow q\,l^+l^-\,X}
\over d^2 b \,d^2 k \,d\ln M^2 \, dz}=
{{2 e_q^2 \, \alpha^2_{\rm em}} \over{3\pi}} \int 
{{d^2 q}\over{(2\pi)^4}} \sigma_{dipole}^F ({\un q} + {\un k})\nonumber\\
&&
\left\{
\left[{{1+(1-z)^2}\over{z}}\right]
{z^2 l_T^2 \over [k_T^2+M^2(1-z)] [({\un k} -z \, {\un l})^2 + M^2 (1-z)]}
\right.\nonumber\\ &&\left.  - z(1-z)\,M^2\left[{1\over
[k_T^2+M^2(1-z)]} - {1\over [({\un k} - z \, {\un l})^2+M^2(1-z)]}\right]^2
\right\}
\label{eq:cs_dilep}
\end{eqnarray}
where $M^2$ is the dilepton invariant mass squared, ${\un l} = {\un q}
+ {\un k}$ and we have ignored quark masses. As before, $z$ is the
fraction of the incoming quark light cone energy carried away by the
photon.

Since the dilepton production cross section is proportional to the electromagnetic
coupling constant $\alpha^2_{em}$, the production rate is quite small, specially 
at higher transverse momenta. One way of getting around this problem is to consider
transverse momentum integrated dilepton production cross section. One can then
investigate the behavior of this cross section with varying dilepton mass. To do this,
we write the dipole cross section in transverse coordinate space, and 
integrate over the transverse momentum of the dilepton pair $k_T$ in (\ref{eq:cs_dilep}).
We get  
\begin{eqnarray}
z {d\sigma^{q\,A\rightarrow q\,l^+l^-\,X}
\over d^2 b \,d M^2 \, dz}=&&
{{\alpha^2_{em}}\over{3\pi^2}}\, 
 {1-z \over z^2} \int \, dr_T^2 \, 
\sigma_{dipole}^F (x_g, {\un b}, r_T)
\nonumber\\
&&
\nonumber \\
&&
\Bigg[[1 + (1-z)^2] \,K_1^2[{\sqrt{1-z}\over z} M r_T] + 
2 (1-z) \, K_0^2[{\sqrt{1-z}\over z} M r_T]\Bigg]
\label{eq:cs_dilep_M}
\end{eqnarray}

To relate this to proton-nucleus scattering, we need to
convolute (\ref{eq:cs_dilep_M}) with the quark (and anti-quark) distributions
$q(x,M^2)$ in a deuteron (proton). As shown in 
\cite{Kopeliovich:2001hf, Johnson:2001xf, boris}, this can be
written in terms of the proton structure function $F_2$
\begin{eqnarray}
{d\sigma^{p\,A\rightarrow l^+l^-\,X}
\over d^2b \,d M^2\, dx_F}= &&
{{\alpha^2_{em}}\over{6\pi^2}}{1 \over x_q + x_g}\, 
\int_{x_q}^1 \,dz \,\int dr_T^2 \, {1-z \over z^2} \, F_2^{p} (x_q/z) \, 
\sigma_{dipole}^F (x_g,{\un b}, r_T)
\nonumber\\
&&
\nonumber \\
&&
\Bigg[[1 + (1-z)^2] \,K_1^2[{\sqrt{1-z}\over z} M r_T] + 
2 (1-z) \, K_0^2[{\sqrt{1-z}\over z} M r_T]\Bigg]
\label{eq:dpA}
\end{eqnarray}
where 
\begin{eqnarray}
x_q= {1\over 2} \bigg[\sqrt{x_F^2 + 4 {M^2 \over s}} + x_F\bigg ]\nonumber \\
x_g=  {1\over 2} \bigg[\sqrt{x_F^2 + 4 {M^2 \over s}} - x_F\bigg ]
\end{eqnarray}
with $x_F\equiv {M\over \sqrt{s}}[e^y - e^{-y}]$ and 
\[ F_2^{p}\equiv \sum_f x \, [q_f(x,M^2) + \bar{q}_f (x,M^2)]\] is the proton 
structure function.

Inclusion of small $x$ evolution of the target wave function is again
straightforward. As in the case of photons and elastic quark and gluon
production, the dilepton production cross section depends on the
dipole cross section which satisfies the JIMWLK or BK evolution
equations.  Using the solution of JIMWLK or BK for the dipole cross
section $\sigma_{dipole}^F$ in (\ref{eq:cs_dilep}) and
(\ref{eq:cs_dilep_M}) would automatically incorporate small $x$
evolution of the target nucleus wave function.  Finally, it is worth
noting that there is a deep relation between dilepton production in
proton-nucleus collisions and in DIS of virtual photons on a
hadron/nucleus due to the approximate crossing symmetry of the
amplitudes in the Color Glass Condensate \cite{Gelis:2002nn,boris}
(the crossing symmetry of the amplitude is approximate due to the
eikonal approximations made).

%%%%%%%%%%%%%%%%%%%%%%%%%%%%%%%%%%%%%%%%%%%%%%%%%%%%%%%%%%%%%%%%%%%%%%%%%%%%%%%%%%%%%%%%

\subsection{Hadronic Two-Particle Correlations}
\label{2pi}

In this Section we will discuss two-particle inclusive production in
$pA$ collisions. We will review production of two gluons, of a valence
quark and a gluon at forward rapidity and of a $q\bar q$ pair at
mid-rapidity.

%%%%%%%%%%%%%%%%%%%%%%%%%%%%%%%%%%%%%%%%%%%%%%%%%%%%%%%%%%%%%%%%%%%%%%

\subsubsection{Two-Gluon Production}

\begin{figure}[ht]
\begin{center}
\epsfxsize=10cm
\leavevmode
\hbox{\epsffile{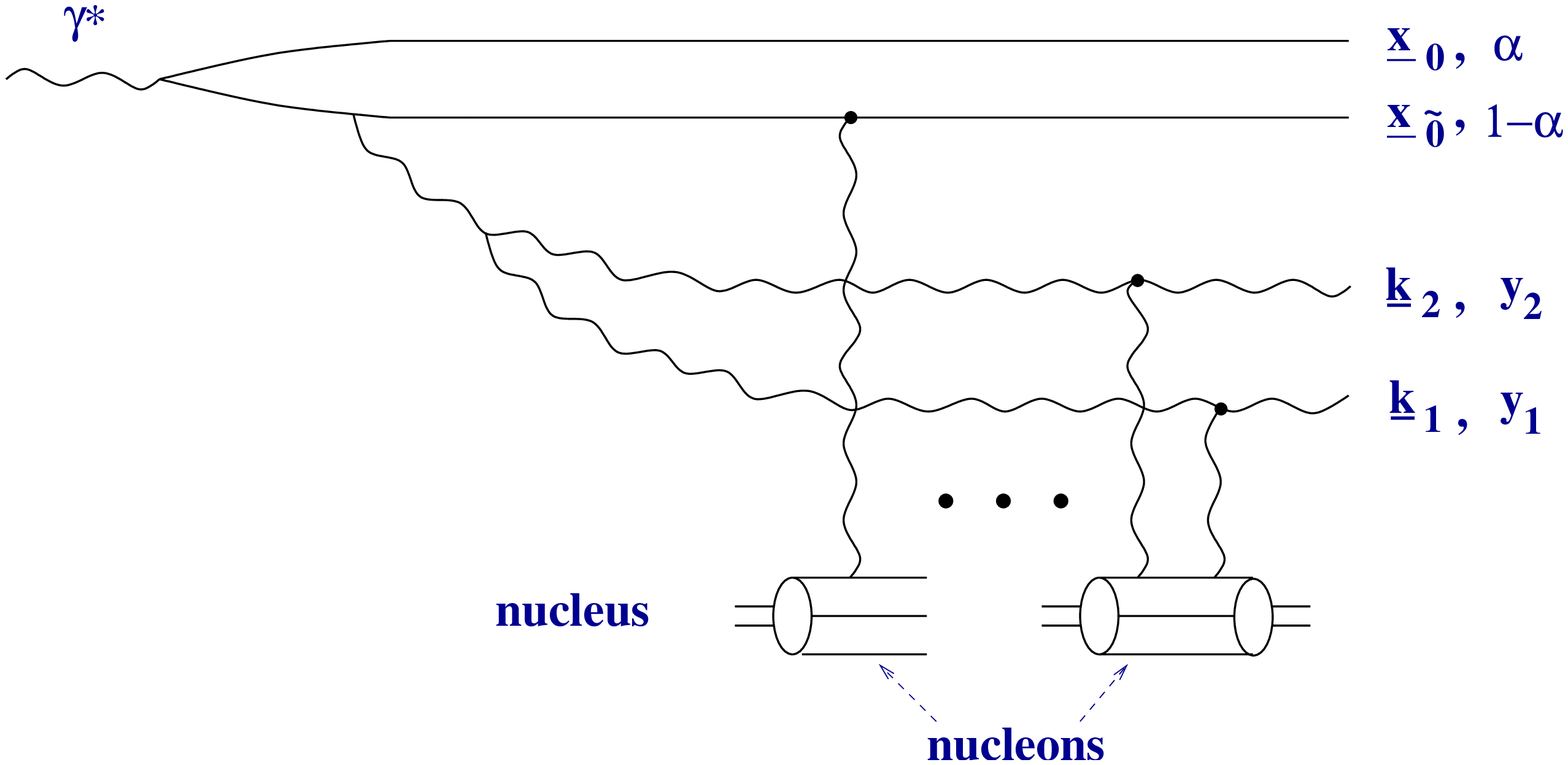}}
\end{center}
\caption{Two-gluon production in DIS on a nucleus including multiple 
rescatterings. }
\label{lodis}
\end{figure}

Here we are interested in calculating an inclusive production cross
section for two gluons with transverse momenta ${\un k}_1$, ${\un
k}_2$ and rapidities $y_1$, $y_2$ in $pA$ collisions or in deep
inelastic scattering. A diagram contributing to two-gluon production
in DIS is shown in \fig{lodis}. Again the target nucleus has rapidity
$0$ and the projectile (proton) has rapidity $Y$. The gluons are
produced at central rapidities $0 \ll y_1, y_2 \ll Y$. For simplicity
we assume that $y_2 \gg y_1$: generalization of the results we will
present below to $y_1 \sim y_2$ appears to be conceptually
straightforward but extremely cumbersome technically.
%%%%%%%%%%%%%%%%%%%%%%%%%%%%%
%%%%%%%%%%%%%%%%%%%%%%%%%%%%%%

Without going through details of the calculation, which was worked out
in \cite{JMK}, we will just quote the answer for two-gluon production
cross section at mid-rapidity for $q\bar q$--nucleus scattering
including small-$x$ evolution. After a tedious calculation one obtains
\cite{JMK}
\ben
\frac{d \sigma^{q {\bar q} \, A \ \rightarrow \ q {\bar q} \, G_1 G_2 X}}
{d^2 k_1 \, dy_1 \, d^2 k_2 \, dy_2} ({\un x}_{0{\tilde 0}})
\bigg|_{y_2 \gg y_1}\, = \,
 \int \ d^2 B \ \bigg\{ n_2 ({\un x}_0, {\un x}_{\tilde 0}, Y; {\un
 x}_1 , {\un x}_{\tilde 1}, y_1, {\un x}_2, {\un x}_{\tilde 2}, y_2)
\een
\ben
\times \, d^2 x_1 \, d^2 x_{\tilde 1} \, d^2 x_2 \, d^2 x_{\tilde 2} \, s
 ({\un x}_1, {\un x}_{\tilde 1}, {\un k}_1, y_1) \, s ({\un x}_{2},
 {\un x}_{\tilde 2}, {\un k}_2, y_2)
+ n_1 ({\un x}_0, {\un x}_{\tilde 0}, Y; {\un x}_1 , {\un
x}_{\tilde 1}, y_2) \, d^2 x_1 \, d^2 x_{\tilde 1} 
\een
\ben
\times \, \frac{\bas}{(2 \pi)^3}
\, \int \, d^2 x_2 \, d^2 x_{2'} \, 
e^{- i {\underline k}_2 \cdot {\underline x}_{22'}} \,
\bigg[ \left( \frac{{\un x}_{21}}{x_{21}^2} - 
\frac{{\un x}_{2{\tilde 1}}}{x_{2{\tilde 1}}^2} \right) \cdot 
\left( \frac{{\un x}_{2'1}}{x_{2'1}^2} - 
\frac{{\un x}_{2'{\tilde 1}}}{x_{2'{\tilde 1}}^2} \right) \, 
M ({\un x}_2, {\un x}_{2'}, {\un x}_{\tilde 1}, y_2; {\un k}_1,
y_1) \, S({\un x}_2, {\un x}_{2'}, y_2) 
\een
\ben
- \left( \frac{{\un x}_{21}}{x_{21}^2} - 
\frac{{\un x}_{2{\tilde 1}}}{x_{2{\tilde 1}}^2} \right) \cdot 
\frac{{\un x}_{2'1}}{x_{2'1}^2} \, 
M ({\un x}_2, {\un x}_1, {\un x}_{\tilde 1}, y_2; {\un k}_1,
y_1) \, S({\un x}_2, {\un x}_1, y_2)
- \left( \frac{{\un x}_{2'1}}{x_{2'1}^2} - 
\frac{{\un x}_{2'{\tilde 1}}}{x_{2'{\tilde 1}}^2} \right) \cdot 
\frac{{\un x}_{21}}{x_{21}^2} 
\een
\be\label{2Gincl}
\times \, M ({\un x}_1, {\un x}_{2'}, {\un x}_{\tilde 1}, y_2; {\un k}_1,
y_1) \, S({\un x}_1, {\un x}_{2'}, y_2) 
+ \frac{{\un x}_{21}}{x_{21}^2} \cdot \frac{{\un x}_{2'1}}{x_{2'1}^2}
\, M ({\un x}_1, {\un x}_1, {\un x}_{\tilde 1}, y_2; {\un k}_1, y_1) \,
+ \, (1 \leftrightarrow {\tilde 1}) \bigg] \bigg\}.
\ee
\eq{2Gincl} gives us the two-gluon inclusive production cross section for 
a scattering of a $q\bar q$ dipole on a nucleus. (Generalization to
$pA$ is discussed in \cite{JMK}.) The transverse coordinates of the
quark and anti-quark in the dipole are ${\un x}_0$ and ${\un
x}_{\tilde 0}$, with the impact parameter ${\un B} = ({\un x}_0 + {\un
x}_{\tilde 0})/2$. $n_1 ({\un x}_0, {\un x}_{\tilde 0}, Y; {\un x}_1 ,
{\un x}_{\tilde 1}, y_2)$ is given by \eq{eqn}: to convert it to the
notations of \eq{eqn} we write 
\be
n_1 ({\un x}_0, {\un x}_{\tilde 0}, Y;
{\un x}_1 , {\un x}_{\tilde 1}, y_2) \, \leftrightarrow \, n_1 ({\un
x}_{0\tilde 0}, {\un x}_{1\tilde 1}, {\un b} = \frac{1}{2} ({\un x}_0
+ {\un x}_{\tilde 0} - {\un x}_1 - {\un x}_{\tilde 1}), Y-y_2). 
\ee
$n_2 ({\un x}_0, {\un x}_{\tilde 0}, Y; {\un x}_1 , {\un x}_{\tilde
1}, y_1, {\un x}_2, {\un x}_{\tilde 2}, y_2)$ is the number of pairs
of dipoles formed by quark and anti-quark at ${\un x}_1$ and ${\un
x}_{\tilde 1}$ with rapidity $y_1$ and by quark and anti-quark at
${\un x}_2$ and ${\un x}_{\tilde 2}$ with rapidity $y_2$ in the
original dipole ${\un x}_{0\tilde 0}$ having rapidity $Y$. $n_2$ obeys
the following equation \cite{dip2,dip3}
\ben
n_2 ({\un x}_0, {\un x}_{\tilde 0}, Y; {\un x}_1 , {\un x}_{\tilde 1},
y_1, {\un x}_2, {\un x}_{\tilde 2}, y_2) \, = \, \frac{\bas}{2 \, \pi} \,
\int_{\mbox{max} \{ y_1, y_2 \}}^Y d y \ e^{ - 2 \, \bas \, \ln \left(
  \frac{x_{0{\tilde 0}}}{\rho} \right) (Y - y)} \, \int_\rho d^2 x_3
  \frac{x_{0{\tilde 0}}^2}{x_{30}^2 x_{3{\tilde 0}}^2}
\een
\ben
\times  \, \left[ n_1 ({\un
  x}_0, {\un x}_3, y; {\un x}_1 , {\un x}_{\tilde 1}, y_1) \, n_1
  ({\un x}_3, {\un x}_{\tilde 0}, y; {\un x}_2 , {\un x}_{\tilde 2},
  y_2) + n_1 ({\un
  x}_0, {\un x}_3, y; {\un x}_2 , {\un x}_{\tilde 2}, y_2) \, n_1
  ({\un x}_3, {\un x}_{\tilde 0}, y; {\un x}_1 , {\un x}_{\tilde 1},
  y_1) +  \right.
\een
\be\label{n2}
+ \left. n_2 ({\un x}_0, {\un x}_3, y; {\un x}_1 , {\un x}_{\tilde 1},
y_1, {\un x}_2 , {\un x}_{\tilde 2}, y_2) + n_2 ({\un x}_3, {\un
x}_{\tilde 0}, y; {\un x}_1 , {\un x}_{\tilde 1}, y_1, {\un x}_2 ,
{\un x}_{\tilde 2}, y_2)\right]
\ee
which is linear and can be solved after one finds $n_1$ from
\eq{eqn}.
%%%%%%%%%%%%%%%%%%%%%%%%%%%%%
\begin{figure}[ht]
\begin{center}
\epsfxsize=12cm
\leavevmode
\hbox{\epsffile{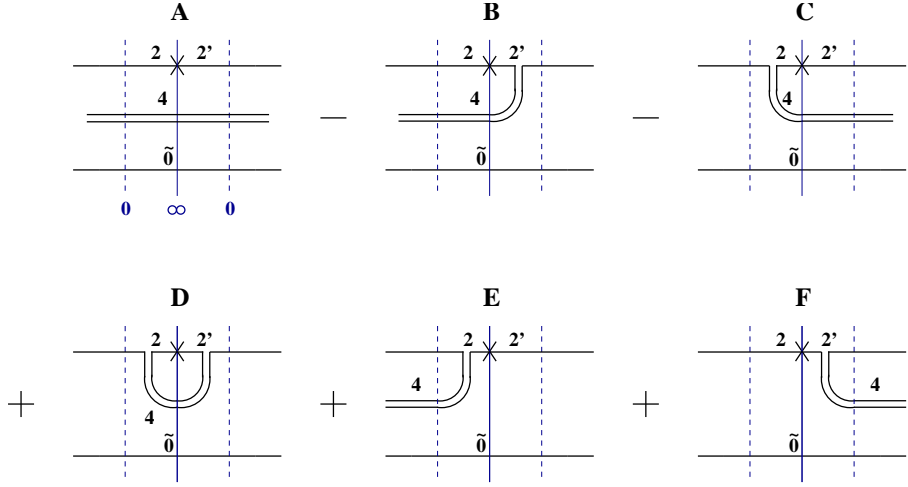}}
\end{center}
\caption{Diagrams describing one step of evolution for $M$.}
\label{mev}
\end{figure}
%%%%%%%%%%%%%%%%%%%%%%%%%%%%%%

The quantity $s$ in \eq{2Gincl} is defined in terms of gluon dipole
amplitudes as
\ben
s ({\un x}_1, {\un x}_2, {\un k}, y) \, \equiv \, \frac{\bas}{(2
\pi)^3} \, \int \, d^2 z_1 \, d^2 z_2 \,
e^{- i {\underline k} \cdot ({\underline z}_1 - {\underline z}_2 )} \,
\sum_{i,j = 1}^2 (-1)^{i+j} \,
\frac{{\underline z}_1 - {\underline x}_i}{|{\underline z}_1 - 
{\underline x}_i|^2} \cdot \frac{{\underline z}_2 - {\underline
x}_j}{|{\underline z}_2 - {\underline x}_j|^2} 
\een
\be\label{sdef}
\times \, \left[ N_G \left({\underline z}_1, {\underline x}_j, y \right) 
+ N_G \left({\underline z}_2 , {\underline x}_i, y \right) 
- N_G \left({\underline z}_1 , {\underline z}_2, y \right)
- N_G \left({\underline x}_i , {\underline x}_j, y \right) \right]
\ee
and is an integral component of single inclusive production cross
section from \eq{paclg1}. $S$ in \eq{2Gincl} is the $S$-matrix of a
quark dipole scattering on the target nucleus, related to the quark
dipole amplitude $N$ by
\be\label{S}
S ({\un x}_{0}, {\un x}_{\tilde 0}, Y) = 1 - N ({\un x}_{0}, {\un
x}_{\tilde 0}, Y),
\ee
where $N$, in turn, can be found from the nonlinear evolution equation
(\ref{eqN}). Note that the gluon dipole amplitude $N_G$ is defined by
\eq{ggqq}.

\begin{figure}[t]
\begin{center}
\epsfxsize=18.5cm
\leavevmode
\hbox{\epsffile{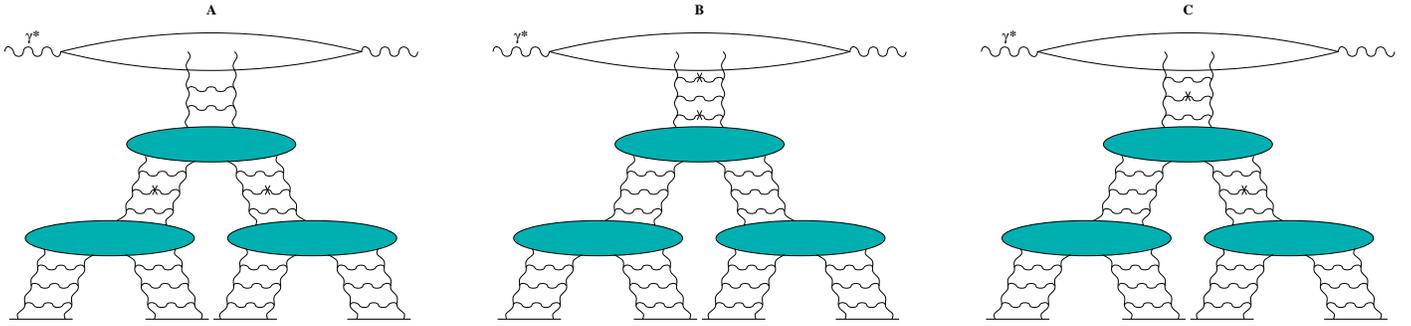}}
\end{center}
\caption{Feynman diagrams corresponding to double gluon production 
cross section given by \eq{2Gincl}. Emitted gluons are denoted by
crosses.  }
\label{fan2}
\end{figure}

Finally, the quantity $M ({\un x}_2, {\un x}_{2'}, {\un x}_{\tilde 1},
y_2; {\un k}_1, y_1)$ in \eq{2Gincl} has the physical meaning of
production amplitude of a gluon with transverse momentum ${\un k}_1$
and rapidity $y_1$ in an off-forward scattering of dipole $2{\tilde
1}$ into dipole $2'{\tilde 1}$ on the target nucleus. The evolution
equation for $M$ is \cite{JMK}
\ben
M ({\un x}_2, {\un x}_{2'}, {\un x}_{\tilde 0}, Y; {\un k}_1, y_1) \,
= \, e^{- \bas \, \ln \left( \frac{x_{2{\tilde 0}} x_{2'{\tilde 0}}
x_{22'}}{\rho^3} \right) \, (Y-y_1)} \, d({\un x}_2, {\un x}_{2'},
{\un x}_{\tilde 0}, {\un k}_1, y_1) +
\een
\ben
+ \frac{\bas}{2 \pi} \, \int d^2 x_4 \int_{y_1}^Y dy \, e^{- \bas \,
\ln \left( \frac{x_{2{\tilde 0}} x_{2'{\tilde 0}} x_{22'}}{\rho^3}
\right) \, (Y-y)} \, \bigg\{
\left( \frac{{\un x}_{42}}{x_{42}^2} - 
\frac{{\un x}_{4{\tilde 0}}}{x_{4{\tilde 0}}^2} \right) \cdot 
\left( \frac{{\un x}_{42'}}{x_{42'}^2} - 
\frac{{\un x}_{4{\tilde 0}}}{x_{4{\tilde 0}}^2} \right) 
\een
\ben
\times 
\bigg[ M ({\un x}_2, {\un x}_{2'}, {\un x}_4, y; {\un k}_1, y_1) 
+ \int d^2 x_a d^2 x_b \, n_1 ({\un x}_4, {\un x}_{\tilde 0}, y;
{\un x}_a, {\un x}_b, y_1) \, s ({\un x}_a, {\un x}_b, {\un k}_1, y_1)
\, [1 - N ({\un x}_2, {\un x}_{2'}, y)] \bigg] - 
\een
\ben
- \left( \frac{{\un x}_{42}}{x_{42}^2} - 
\frac{{\un x}_{4{\tilde 0}}}{x_{4{\tilde 0}}^2} \right) \cdot 
\left( \frac{{\un x}_{42'}}{x_{42'}^2} - 
\frac{{\un x}_{42}}{x_{42}^2} \right) \, 
M ({\un x}_4, {\un x}_{2'}, {\un x}_{\tilde 0}, y; {\un k}_1, y_1) 
\, [1 - N ({\un x}_2, {\un x}_4, y)] - 
\een
\be\label{Mev}
- \left( \frac{{\un x}_{42}}{x_{42}^2} - 
\frac{{\un x}_{42'}}{x_{42'}^2} \right) \cdot 
\left( \frac{{\un x}_{42'}}{x_{42'}^2} - 
\frac{{\un x}_{4{\tilde 0}}}{x_{4{\tilde 0}}^2} \right) \, 
M ({\un x}_2, {\un x}_4, {\un x}_{\tilde 0}, y; {\un k}_1, y_1) 
\, [1 - N ({\un x}_{2'}, {\un x}_4, y)] \bigg\} .
\ee
It includes the (nonlinear) evolution between the incoming dipole and
the produced gluon, which is shown in \fig{mev} using the same notation
as in \fig{eem}. The initial condition for the evolution in
\eq{Mev} given by $d({\un x}_2, {\un x}_{2'},{\un x}_{\tilde 0}, {\un
k}_1, y_1)$ has to be calculated separately by solving a linear
evolution equation (see Eqs. (27) and (28) in \cite{JMK}).

\eq{2Gincl} is represented graphically in \fig{fan2} in terms of 
traditional Feynman diagrams.  The first term in \eq{2Gincl}
corresponds to splitting of the original linear evolution in two,
which is described by $n_2$. Then each of the two ladders
independently produces a gluon with all the possible splittings
happening afterwards. This is illustrated in \fig{fan2}A.  The second
term in \eq{2Gincl} corresponds to nonlinear evolution successively
producing both gluons, after which all possible splittings are
allowed, as shown in Figs. \ref{fan2}B and \ref{fan2}C, where we have
divided the nonlinear evolution into the linear (Fig. \ref{fan2}B) and
nonlinear (Fig. \ref{fan2}C) parts. The linear evolution in this
second term in \eq{2Gincl} is given by $n_1$ and by $M$ from the
linear part of \eq{Mev}. This linear evolution leads to production of
both gluons $\# 2$ and $\# 1$ and is illustrated in
\fig{fan2}B. The initial conditions for \eq{Mev} are nonlinear. They 
include ladder splittings and are pictured by the fan diagram in the
lower part of \fig{fan2}B. One should note, however, that \eq{Mev},
while being linear in $M$, has extra factors of $1-N$ on its right
hand side. That means that evolution of $M$ includes ladder splittings
between the gluons $\# 2$ and $\# 1$, one of which is shown in
\fig{fan2}C. There the evolution leading to creation of gluon $\# 2$ 
is still linear since it is still given by $n_1$ in the second term on
the right hand side of \eq{2Gincl}. However, since the evolution in
the rapidity interval between the emitted gluons (evolution of $M$) is
nonlinear, splittings are allowed between the gluons $\# 2$ and $\#
1$, as depicted in \fig{fan2}C.
%%%%%%%%%%%%%%%%%%%%%%%%%%%%%
%%%%%%%%%%%%%%%%%%%%%%%%%%%%%%

The diagrams A and B in \fig{fan2} are the same as would have been
expected from AGK cutting rules \cite{AGK} (see also \cite{KT} for
similar correspondence between the dipole model results and AGK rules
expectations). However, the diagram C in \fig{fan2}, while being
included in \eq{2Gincl}, is prohibited by AGK cutting
rules. Therefore, we seem to observe a direct violation of the AGK
rules in QCD. Since AGK rules have never been proven for QCD, one
should not be too surprised that they do not work here. It is
interesting to note that AGK violation sets in at the level of the
2-gluon production: single gluon inclusive production cross section
calculated in \cite{KT} adheres to AGK rules and so does the
diffractive DIS cross section calculated in \cite{KL}.

The violation of AGK cutting rules in \eq{2Gincl} is due to non-linear
terms in \eq{Mev}, which are in turn due to late time (after the
interaction) gluon emissions at light cone times $\tau > 0$. These
terms were not important for the calculation of the total cross
section in the dipole model \cite{dip1,dip2}: there they were found to
cancel \cite{CM}. Thus if one would try to construct an analogy
between the fan diagrams \cite{GLR} and dipole calculations
\cite{yuri_bk} based on the correspondence of total cross sections,
one would omit such terms. Since the fan diagrams seem to adhere to
AGK rules, this omission would lead to the erroneous conclusion that
AGK rules should work for the two-gluon production cross
section. However, as we have seen above, these late time emissions are
important for single \cite{KT} and double inclusive gluon production,
violating the AGK rules for the latter. What appears to fail here is
the one-to-one correspondence between the fan diagrams and dipole
calculations.

Another difference between the result (\ref{2Gincl}) and the direct
application of AGK rules to calculating inclusive cross section done
in \cite{Braun2} is that nonlinear splittings may start exactly at the
point in rapidity when the softer of the produced gluons is emitted in
the diagram B or exactly at the point of emission of both gluons $\#
1$ and $\# 2$ in the diagram A. Similar discrepancy was already
observed when comparing the single gluon inclusive cross section
calculated in \cite{KT} to the results of \cite{Braun2}.

It is interesting to note that \eq{2Gincl} can not be cast in the form
adhering to $k_T$-factorization, as it was done for \eq{incl}
transforming it into \eq{ktdip} \cite{JMK}. This indicated a breakdown
of $k_T$-factorization for two-particle production. Below we will
observe the same failure of $k_T$-factorization for the case of quark
production \cite{BGV,NSZZ}.

\eq{2Gincl} gives a complete result for two-gluon production cross 
section in saturation/Color Glass Condensate formalism. It can be used
to describe two-particle correlations in $dAu$ collisions at RHIC and
in $pA$ collisions at LHC. Unfortunately \eq{2Gincl} is rather
complicated. In order to make \eq{2Gincl} easier to implement, it is
highly desirable to find some way of simplifying it. At this time we
are not aware of any simplification of \eq{2Gincl} in the general
case. Nevertheless, in certain kinematic regimes \eq{2Gincl} may be
simplified. For instance, if the center of mass energy of the
collision is not too high or if the transverse momenta of the produced
gluons are sufficiently large ($|{\un k}_1|, |{\un k}_2| \gsim Q_s$),
the nonlinear saturation effects, such as ladder splittings, could be
neglected. This implies that the diagrams in Figs. \ref{fan2} A and C
are small with the linear part of the diagram B dominating the cross
section. This is the well-known leading twist result
\cite{lr} (see also \cite{ktflow}).

In \cite{KLM2}, using $k_T$-factorization formula, Kharzeev, Levin and
McLerran suggested that back-to-back particle correlations would be
suppressed in $p(d)A$ collisions, when the rapidity interval between
the jets is large, e.g., when one of the jets ($y_2$) is going in the
proton fragmentation region and the other one ($y_1$) is at
mid-rapidity at RHIC (the so-called Mueller-Navelet jets
\cite{MN}). This prediction will be discussed in more detail in Sect. 
\ref{fet}. While preliminary data from STAR collaboration appears to
confirm this prediction \cite{Ogawa}, it would be important and
interesting to generalize the approach of \cite{KLM2} beyond the
$k_T$-factorization formula, which is valid only at high $k_T \gg
Q_s$. \eq{2Gincl} provides us with an opportunity to carry out such a
generalization.

%%%%%%%%%%%%%%%%%%%%%%%%%%%%%%%%%%%%%%%%%%%%%%%%%%%%%%%%%%%%%%%%%%%%%%

\subsubsection{Gluon--Valence Quark Production}

Here we consider production of a high $x$ (valence) quark and a gluon
from scattering of a quark in the proton wave function on the target
nucleus, treated as a Color Glass Condensate. This is expected to be
the dominant process in the projectile proton fragmentation region
where one probes the high $x$ degrees of the freedom in the proton and
the small $x$ degrees of freedom in the target nucleus. This process
is the $\alpha_s$ correction to the scattering of the high $x$ quark
on the target considered in the previous sections.

\begin{figure}[htb]
\begin{center}
\epsfxsize=12cm
\leavevmode
\hbox{\epsffile{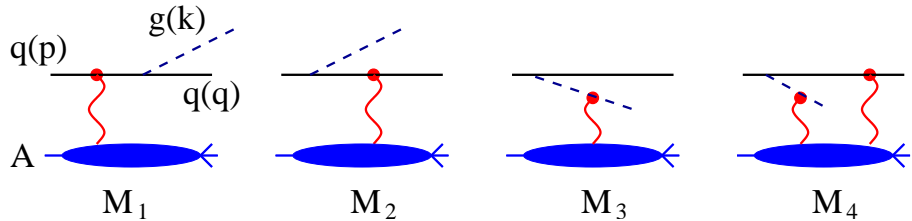}}
\end{center}
\caption{Production of a quark and a gluon in quark-nucleus scattering.}
\label{fig:qgx_1234}
\end{figure}

Production of a valence quark and a gluon was considered in
\cite{JMK}. Here we go through the main part of the derivation and
refer the reader to \cite{JMK} for more detail. The starting point is
the amplitude for scattering of a incoming massless on-shell quark on
the target, producing an on-shell quark and gluon shown in
% Fig. \ref{fig:qA-qgx}
 \begin{eqnarray}
q(p)\,A \rightarrow q(q)\, g(k)\, X
\label{eqn:qAtoqgX}
\end{eqnarray}
which is given by 
\begin{eqnarray}
{\cal M} \!\!\!\!& = \!\!\!\!&g\! \int d^4x \,d^4y \,d^4z \,d^4r \, 
d^4\bar{r} \,e^{i(q\cdot z + k\cdot r - p\cdot y)} 
\bar{u}(q) [i\stackrel{\rightarrow}{\slpartial}_z ] S_F (z,x) \gamma^{\nu} 
t^c S_F(x,y)  [i\stackrel{\leftarrow}{\slpartial}_y ] u(p) \nonumber \\
&&G_{\nu\rho}^{cb}(x,\bar{r}) D^{\rho\mu}_{ba}(\bar{r},r) \, \epsilon_{\mu}(k)
\label{eq:amp_qg}
\end{eqnarray}
where $S_F$, $G_{\nu\rho}$ are the quark and gluon propagators in the
classical field background and $D^{\rho\mu}$ is the "inverse" propagator (which
amputates the external gluon line), defined by
\begin{eqnarray}
\int d^4 r \,G^{0cb}_{\nu\rho} (x,r)\, D^{\rho\mu}_{ba} (r,y) \equiv 
\delta^c_a \, \delta_{\nu}^{\mu} \, \delta^4 (x-y)
\end{eqnarray}
and 
$G_{\nu\rho}^0$ is the free gluon propagator and  the quark and gluon lines 
with a thick circle represent the propagators in the background field as illustrated
in Fig. \ref{fig:v_quark}. The quark and gluon propagators in the classical background
field are known \cite{hw,ayala}. Here we write them as an interacting part and a free
part as before
\begin{eqnarray}
S_F(q,p)&\equiv &(2\pi)^4 \delta^4(p-q)\, S^0_F
(p) + S_F^0 (q) \,\tau_f (q,p) \, S_F^0 (p) \nonumber \\ 
G^{\mu\nu}(q,p)&\equiv&
(2\pi)^4 \delta^4(p-q)\, G^{0\mu\nu} (p) + G^{0\mu}_{\rho}(q) \,
\tau_g (q,p)\, G^{0\rho\nu}(p)
\label{eq:prop_decomp}
\end{eqnarray}
where the free propagators are 
\begin{eqnarray}
S_F^0 (p) &=& i{\slp \over p^2} \,\,\,\,\,\,\,\,\!\!\! \mbox{and}
\,\,\,\,\,\,\,\,\,\,\!\!\!  G^0_{\mu\nu}(k) = {i\over k^2}
\bigg[-g_{\mu\nu} + {\eta_\mu k_\nu + \eta_\nu k_\mu \over \eta \cdot
k}\bigg]
\end{eqnarray}
and $\eta_\mu$ is the light cone gauge vector so that $\eta \cdot A
\equiv A^- = 0$ defines the gauge we choose to use. One of the 
advantages of using this gauge is that the interaction part of the gluon 
propagator in \eq{eq:prop_decomp}, denoted here by $\tau_g (q,p)$, is 
diagonal in Lorentz indices (proportional to $g_{\mu\nu}$)  which 
is why we can write it as in \eq{eq:prop_decomp}.  The interacting parts
of the quark and gluon propagator $\tau_f$ and $\tau_g$ are defined
as 
\begin{eqnarray}
\tau_f (q,p)\equiv (2\pi)\delta(p^- - q^-) \,\gamma^-\, \int d^2x\, 
e^{i ({\un q} - {\un p})\cdot {\un x}}\, [V({\un x}) - 1]\\
\tau_g (q,p) \equiv 2p^- \, (2\pi)\delta(p^- - q^-) \, \int d^2x\, 
e^{i ({\un q} - {\un p}) \cdot {\un x}}\, [U({\un x}) - 1]
\label{eq:tau_def}
\end{eqnarray}
where $V$ is the multiple scattering matrix in the fundamental representation
defined in (\ref{eq:v_fun}) and $U$ is the analogous matrix in the adjoint representation.
The scattering amplitude then can be written as
\begin{eqnarray}
{\cal M}(q,\lambda ,k;p)\equiv  
\epsilon_\mu^{(\lambda)} (k) \,[M_1^\mu + M_2^\mu + M_3^\mu + M_4^\mu]
\label{eq:amp_mu}
\end{eqnarray}
where $\epsilon_\mu^{(\lambda)} (k) $ is the polarization vector of the 
produced gluon and 
\begin{eqnarray}
\!\!\!\!M_1 \!\!\!\!&=&\!\!\!\! -g \,\bar{u}(q)\, \slepsilon \,t^a\,S_F^0(q+k)\, 
\tau_f(q+k,p)\,u(p) \label{eq:amp_12341} \\
M_2 \!\!\!\!&=&\!\!\!\! -g\, \bar{u}(q)\, \tau_f(q,p-k)\, S_F^0(p-k)\, \slepsilon \, 
t^a\, u(p) \\
M_3 \!\!\!\!&=&\!\!\!\! -g \,\bar{u}(q) \, \gamma_{\nu} t^b\,\tau_g^{ba}(k,p-q)\,u(p)\, 
G_0^{\nu\mu}(p-q)\, \epsilon_\mu (k) \\
M_4 \!\!\!\!&=&\!\!\!\! -g \! \int \!d^4l \,\bar{u}(q) \tau_f(q,p-l) S_F^0(p-l)\gamma_{\nu}
t^b \tau_g^{ba}(k,l) u(p) G_0^{\nu\mu}(l)\epsilon_\mu (k). 
\label{eq:amp_1234}
\end{eqnarray}
The different terms in Eqs. (\ref{eq:amp_12341})--(\ref{eq:amp_1234})
are depicted in \fig{fig:qgx_1234} where the solid line denotes
the incoming quark, the thick wavy line attached to a full circle
denotes multiple scattering from the target and the dashed line
represents the radiated gluon.
%%%%%%%%%%%%%%%%%%%%%%%%%%%%%
%%%%%%%%%%%%%%%%%%%%%%%%%%%%%%

$M_1$ corresponds to radiation of a gluon before the multiple
scattering before the quark and the target while $M_2$ corresponds to
the incoming quark scattering on the target and then radiating a
gluon. $M_3$ depicts the case when the radiated gluon multiply
scatters from the target, while in $M_4$ both the radiated gluon and
the final state quark multiply scatter from the target.  To get the
production cross section, we need to square the amplitude. This is
straightforward but tedious. First, one can do the $l^-$ integration
in $M_4$ using the delta function $\delta (l^- - k^-)$ and the $l^+$
integration via contour methods using the $p-l$ pole which sets $l^- =
k^-$ and $l^+ = - l_\perp^2/2 q^-$.

Contribution of $M_1$ and $M_2$ terms is identical (up to color
matrices) to the photon production case except for the extra
contribution coming from the pieces of the gluon propagator
proportional to the gauge vector $\eta_{\rho}$ which is not present in
the case of the inclusive photon production. This warrants some
clarification. In single photon production, photons are emitted from
the incoming quark and can not be emitted from the target (CGC) which,
to leading order in $\alpha_s$ is comprised of gluons only. Radiation
of a photon from the target is suppressed for two reasons; first it is
a higher order in $\alpha_s$ correction since one would need a gluon
in the target wave function to split into a quark anti-quark pair
which would then radiate a photon and second, it is suppressed in the
soft photon limit (it is proportional to $z$ where $z$ is the fraction
of the parent quark the radiated photon carries away). Therefore, in
the case of photon production, only the $g_{\mu\nu}$ part of the
propagator contributes. In case of gluon production, one needs to
consider radiation of gluons from the target as well as from the
projectile quark. One can avoid these contributions by working in the
light cone gauge as shown by \cite{KM}. This however necessitates
inclusion of the contribution of the extra terms, proportional to the
gauge vector, in the propagator. The contribution of $M_1$ and $M_2$
is therefore
\begin{eqnarray}
&&(M_1^\mu + M_2^\mu) \left[-g_{\mu\nu} + 
{(k_\mu \eta_\nu + \eta_\mu k_\nu)\over  \eta \cdot k}\right] (M_1^\nu + M_2^\nu)  
 = 16 p^-p^- \nonumber\\
&&\Bigg\{ {z(1-z)^2 \over [z {\un k} - (1-z){\un q}]^2}
Tr [V^{\dagger}({\un q} + {\un k}) - (2\pi)^2 \delta^2({\un q} + {\un k})]  t^a \, t^a 
[V({\un q} + {\un k}) - (2\pi)^2 \delta^2({\un q} + {\un k})] \nonumber\\
&+& {z(1 - z)^2 \over k_T^2} 
Tr [V({\un q} + {\un k}) - (2\pi)^2 \delta^2({\un q} + {\un k})]  t^a \, t^a 
[V^{\dagger}({\un q} + {\un k}) - (2\pi)^2 \delta^2({\un q} + {\un k})]  \nonumber\\
&+& 2 z^2  \bigg[V^\dagger ({\un q} + {\un k}) t^a - 
t^a V^\dagger ({\un q} + {\un k}) \bigg] 
\bigg[ {1 \over [z {\un k} -(1-z) {\un q}]^2} t^a V({\un q} + {\un k}) - 
{1\over k_T^2} V({\un q} + {\un k}) t^a \bigg]
\nonumber \\
&+&  \Bigg[
(1-z)^2 (1 + z^2) {q_T^2 \over k_T^2  [z {\un k} - (1-z) {\un q}]^2} + {z^2(1-z^2) 
\over  [z{\un k} - (1-z) {\un q}]^2}
- {1-z^2 \over k_T^2}\Bigg] \nonumber\\ 
&&Tr \, t^a [V^{\dagger}({\un q} + {\un k}) - (2\pi)^2 \delta^2({\un q} + {\un k})] 
t^a [V({\un q} + {\un k}) - (2\pi)^2
\delta^2({\un q} + {\un k})]\Bigg\}
\label{eq:amp_12}
\end{eqnarray}
where $Tr$ denotes trace of color matrices. A tedious but straightforward calculation 
of the rest of the diagrams leads to 

\begin{eqnarray}
|M_3^{\dagger}M_1| &=& 16 \,p^-p^- \, z(1 + z^2) 
{q_T^2 -z {\un q} \cdot ({\un q} + {\un k}) \over q_T^2 [z{\un k} - (1-z){\un q}]^2}
[U^{\dagger ab}({\un q} + {\un k}) - \delta^{ab}(2\pi)^2 \delta^2({\un q} + {\un k})] 
\nonumber\\
&&Tr \, t^b\, t^a [V({\un q} + {\un k}) - (2\pi)^2 \delta^2({\un q} + {\un k})] \nonumber\\
|M_3^{\dagger} M_2|&=& 16 \,p^-p^- z (1 + z^2) 
{{\un q} \cdot {\un k} \over q_T^2 k_T^2} 
 [U^{\dagger ab}({\un q} + {\un k}) - \delta^{ab}(2\pi)^2 \delta^2({\un q} + {\un k})] 
 Tr\,t^b [V({\un q} + {\un k})\nonumber\\ &&- (2\pi)^2 \delta^2({\un q} + {\un k})]\,t^a \nonumber\\
|M_3|^2 &=& 16 p^- p^- {z(1+ z^2) \over q_t^2} 
[U^{\dagger ab}({\un q} + {\un k}) - \delta^{ab}(2\pi)^2 \delta^2({\un q} + {\un k})] 
[U^{ca}({\un q} + {\un k}) - \delta^{ca} (2\pi)^2 \delta^2({\un q} + {\un k})] \nonumber\\
&&Tr\,  t^b \, t^c \nonumber\\ 
|M^{\dagger}_3 M_4| &=&- 16\, p^-p^- z (1 + z^2) \int {d^2l \over (2\pi)^2} 
{{\un q}\cdot {\un l}  \over q_T^2 l_T^2} \,
[U^{\dagger ab}({\un q} + {\un k}) - \delta^{ab}(2\pi)^2 \delta^2({\un q} + {\un k})]  
\nonumber\\
&&[U^{ca}({\un k} - {\un l}) - \delta^{ca}(2\pi)^2 \delta^2({\un k} - {\un l})] \,
Tr\,t^b [V({\un q} + {\un l}) - (2\pi)^2 \delta^2({\un q} + {\un l})] t^c 
\nonumber\\
|M_4^{\dagger} M_1| &=& - 16 p^-p^- z (1 + z^2) \int {d^2l \over (2\pi)^2}
{(1 - z) {\un q} \cdot {\un l} - z {\un k} \cdot {\un l} \over l_T^2 [z{\un k} - (1-z){\un q}]^2} 
[U^{\dagger ab}({\un k} - {\un l}) - \delta^{ab}(2\pi)^2  \delta^2({\un k} - {\un l})]
\nonumber\\  
&&Tr\, t^b [V^{\dagger }({\un q} + {\un l}) - (2\pi)^2 \delta^2({\un q} + {\un l})]  t^a 
[V({\un q} + {\un k}) - (2\pi)^2 \delta^2({\un q} + {\un k})] \nonumber\\
|M_4^{\dagger} M_2| &=& -16 p^-p^- z (1 + z^2) \int {d^2l \over (2\pi)^2} 
{{\un k} \cdot {\un l} \over l_T^2 k_T^2} 
[U^{\dagger ac}({\un k} - {\un l}) - \delta^{ac}(2\pi)^2  \delta^2({\un k} - {\un l})] 
 \nonumber\\
&&Tr \,t^c  [V^{\dagger }({\un q} + {\un l}) - (2\pi)^2 \delta^2({\un q} + {\un l})]  
[V({\un q} + {\un k}) - (2\pi)^2 \delta^2({\un q} + {\un k})] t^a \nonumber \\
|M_4|^2 &=& 16 p^-p^- z (1 + z^2) 
 \int {d^2l \over (2\pi)^2} {d^2\bar{l} \over (2\pi)^2} 
{{\un l} \cdot     \bar{ {\un l}}  
\over l_T^2 \bar{l}_T^2} 
[U^{\dagger ac}({\un k} - \bar{ {\un l}}) - \delta^{ac}(2\pi)^2 
\delta^2({\un k} - \bar{{\un l}})] 
\nonumber\\
&&[U^{ab}({\un k} - {\un l}) - \delta^{ab}(2\pi)^2  \delta^2({\un k} - {\un l})] 
 Tr  \, t^c t^b 
[V^{\dagger}({\un q} + {\un l}) - (2\pi)^2 \delta^2({\un q} + {\un l})]\nonumber \\
&&[V({\un q} +  \bar{{\un l}}) - (2\pi)^2 \delta^2 ({\un q} +  \bar{{\un l}})]
\label{eq:amp_34}
\end{eqnarray}
where for any interference term, there is an equal conjugate contribution. To get
the cross section, one needs to average (sum) over the initial (final) state degrees
of freedom and include the phase space and flux factors. The invariant cross 
section for production of a quark with transverse momentum $q_T$ and energy 
fraction $z$ and a gluon with transverse momentum $k_T$ (and energy fraction 
$1-z$) is given by
\begin{eqnarray}
z \, (1-z) \, {d\sigma^{qA\rightarrow qgX} \over dz \,d^2 q\,d^2k } = 
{\alpha_s \over 128 \, \pi^4 \, p^- p^-}  \,|M|^2 
\label{eq:cs_qg}
\end{eqnarray}
where $|M|^2$ is defined in (\ref{eq:amp_mu}) and given in
(\ref{eq:amp_12},\ref{eq:amp_34}).  This cross section looks quite
more complicated than single inclusive quark or gluon
production. Nevertheless, the degrees of freedom are again Wilson
lines in fundamental or adjoint representation as before. Here, one
has correlators of more than two Wilson lines (which were not present
in single inclusive production of quarks and gluons).  To do a
quantitative study of two hadron production in proton-nucleus cross
section, one needs to convolute this cross section with the
distribution function of the quark in a proton and the two hadron
fragmentation function.

One can use the two particle cross section derived here (after the
convolution with distribution and fragmentation functions) to
investigate two hadron correlations in proton-nucleus collisions at
RHIC and LHC. The expectation is that due to classical multiple
scattering (no quantum evolution), the correlation function will
become wider. Inclusion of quantum corrections can be achieved by
solving the JIMWLK equations for the higher point functions of the
Wilson lines and using the solution in the above cross section. This
should be valid as long as the produced quark and gluon have the same
(or similar) rapidities. One expects that inclusion of quantum
evolution would lead to reduced correlation between the produced
hadrons.  The effects of quantum evolution on the two hadron
correlation should be most dramatic when the produced hadrons are
widely separated in rapidity, as considered in the previous section.

%%%%%%%%%%%%%%%%%%%%%%%%%%%%%%%%%%%%%%%%%%%%%%%%%%%%%%%%%%%%%%%%%%%%%%

\subsubsection{Quark--Anti-quark Pair Production}

We finally consider production of a quark anti-quark pair in proton-nucleus
collisions. This has been investigated in detail in 
\cite{Kharzeev:2003sk, Tuchin:2004rb,Gelis:2003vh,Blaizot:2004wu, Blaizot:2004wv}. 
Here, we follow the derivation of the \cite{Blaizot:2004wu,
Blaizot:2004wv} since it is most closely related to the Color Glass
Condensate formalism. For technical details and more references, we
refer the reader to \cite{Blaizot:2004wu}. The starting point is the
classical background field of a single nucleus in terms of the nuclear
color charge density given by
\begin{eqnarray}
A^\mu_A (x)=-g\delta^{\mu-}\delta(x^+)\frac{1}{\partial_T^2}
\rho_A ({\un x}) 
\end{eqnarray}
in the covariant gauge and its Fourier transform in momentum space 
\begin{eqnarray}
A^\mu_A (q)=2\pi g\delta^{\mu-}\delta(q^+)
\frac{\rho_A ({\un q})}{q_T^2}\,\, .
\label{eq:A_mom}
\end{eqnarray}
Note that (\ref{eq:A_mom}) is linear in the color charge density even
though it does describe the classical field created by a dense
nucleus. This is a property of the covariant gauge chosen here and is
not true in a general gauge.

Since proton is treated as a dilute object while the nucleus taken to
be a Color Glass Condensate, one needs to keep only the linear term in
the color charge density of the proton while in the nucleus, the color
charge density is kept to all orders. The general solution of the
classical equations of motion for a proton-nucleus collision was found
in \cite{Blaizot:2004wu,DM} and is given by
\begin{eqnarray}
A^\mu(q)= A_{reg}^\mu(q)
+\delta^{\mu -} A_{sing}^-(q)\; ,
\end{eqnarray}
where $ A_{reg}^\mu$ is given by  
 \begin{eqnarray}
 A_{reg}^\mu(q)&=& A_p^\mu(q) +
\frac{ig}{q^2 + i \, \epsilon \, q^+} \int\frac{d^2 k_1}{(2\pi)^2}
\bigg\{
C_{u}^\mu (q, {\un k_1})\, 
\left[U( {\un k_2})-(2\pi)^2\delta( {\un k_2})\right]
\nonumber\\
&&
+ C_{v\, , reg}^\mu(q)\, 
\left[\tilde{U}({\un k_2}) - (2\pi)^2 \delta ({k_2})\right]
\bigg\}
\frac{{\rho_p}({\un k_1} )} 
{k_{1\, T}^2}\; 
\label{eq:field_1}
\end{eqnarray}
and the singular term is given by 
\begin{eqnarray}
A_{sing}^-(q)\equiv
-\frac{ig}{q^+}\int \frac{d^2 k_1}{(2\pi)^2}
\left[\tilde{U}({\un k_2}) - (2\pi)^2\delta({\un k_2})\right]
\frac{{\rho_p}({\un k_1})}{k_{1\, T}^2}
\end{eqnarray}
where $U$ and $\tilde{U}$ (note the factor of $1/2$ in the definition
of $\tilde{U}$) are matrices in the adjoint representation and
\begin{eqnarray}
U({\un x}) &\equiv& {\cal P} \exp\left[i g\int_{-\infty}^{+\infty}
dz^+ A_A^- (z^+,{\un x}) \cdot T
\right]\; ,\nonumber\\
\tilde{U}({\un x}) &\equiv &{\cal P} \exp\left[i \frac{g}{2}\int_{-\infty}^{+\infty}
dz^+ A_A^-(z^+,{\un x})\cdot T
\right]\; ,
\end{eqnarray}
and $T^a$ are the generators of the adjoint representation of $SU(N)$
and ${\un k_1}$ is the momentum transfer from the proton and 
${\un k_2} \equiv {\un q} - {\un k_1}$ is the momentum transfer from the nucleus. The
coefficients $C_{u}^\mu (q,{\un k_1})$ and $C_{v, reg}^\mu (q)$ are
defined by
\begin{eqnarray}
&& 
C_{u}^+(q,k_{1\, T})\equiv -\frac{k_{1\, T}^2}{q^- + i\epsilon}\;,\; 
C_{u}^-(q,k_{1\, T})\equiv \frac{k_{2\, T}^2 - q_T^2}{q^+}\;,\;
C_{u}^i(q,{\un k_1})\equiv -2 k_1^i\; \;
\nonumber\\
&& 
C_{v, reg}^+ (q) \equiv 2q^+ \;\;
C_{v, reg}^- (q) \equiv 2q^-- \frac{q^2}{q^+}\;\;
C_{v, reg}^i (q) \equiv 2 q^i
\end{eqnarray}
and $C_{v, reg}^\mu (q)$ and $C_{v}^\mu (q)$ are related through
$C_{v, reg}^\mu (q)=C_v^\mu (q) + \delta^{\mu -}\frac{q^2}{q^+} $. To
get the amplitude, one again uses the quark propagator in the
background field of the nucleus which includes multiple scattering
from the target nucleus and includes multiple scatterings from the
quark, anti-quark and gluon lines. The final amplitude is given by
\begin{eqnarray}
&&\!\!\!\!\!\!\!\!\!\!\!\!
{\cal M}(q,p)\!=\!g^2\!\int\!\frac{d^2 k_1 } {(2\pi)^2}
\frac{d^2 k} {(2\pi)^2}
\frac{\rho_{p,a}( {\un k_1})}{k_{1,\, T}^2}
\!\int\!\! d^2 x d^2 y
e^{i {\un k} \cdot {\un x}}
e^{i( {\un p}\!+\! {\un q}\!-\!{\un k}\!-\!{\un k_1})\cdot {\un y}}
\nonumber\\
&&\times\Bigg\{
\frac
{\overline{u}(q)\gamma^+(\slq-\slk+m)\gamma^-
(\slq-\slk-\slk_1+m)\gamma^+
[V ({\un x}) t^a \, V^\dagger ({\un y})]
v(p)}
{2p^+[( {\un q} - {\un k})^2+m^2]+2q^+[({\un q} - {\un k} - {\un k_1})^2+m^2]}
\nonumber\\
&& + 
\overline{u}(q)\left[
\frac {\slC_u  (p+q, {\un k_1})}  {(p+q)^2}
-
\frac{\gamma^+}{p^++q^+}
\right]t^b v(p)\, U^{ba}({\un x})
\Bigg\}\; .
\label{eq:M_qqbar}
\end{eqnarray}
where $V$ is in the fundamental representation and $p$ and $q$ are the
momenta of the quark and anti-quark and $m$ denotes the quark
mass. Note that the Wilson line $\tilde{U}$ (with the $1/2$ in the
exponent) has dropped out of the expression for the amplitude due to
cancellation between the regular and singular contributions.To get the
cross section, we need to square this amplitude, sum (average) over
the final (initial) state degrees of freedom and color average over
the sources which is tedious but straightforward. Here we just give
the final result \cite{Blaizot:2004wv}
\begin{eqnarray}
&&\frac{d\sigma}{d^2 p d^2 q dy_p dy_q}=
\frac{\alpha_s^2}{(2\pi)^4 C_F}
\int d^2 k_1 d^2 k_2 
\frac{\delta^2 ( {\un p}  + {\un q} - {\un k_1} - {\un k_2})}
{k_{1,\, T}^2\, k_{2,\, T}^2}
\nonumber\\
&&
\times\Bigg\{
\int d^2k d^2 k^{\prime} \, 
{\rm tr}_{\rm d}
\Big[(\slq\!+\!m)T_{q\bar{q}}(\slp\!-\!m)
\gamma^0 T_{q\bar{q}}^{\prime\dagger}\gamma^0\Big]
\varphi_{A}^{q\bar{q},q\bar{q}}
({\un k}, {\un k_2} \! - \! {\un k};
{\un k}^\prime, {\un k_2}\! - \! {\un k}^\prime)
\nonumber\\
&&\;\;
+\int d^2 k \, 
{\rm tr}_{\rm d}
\Big[(\slq\!+\!m)T_{q\bar{q}}(\slp\!-\!m)
\gamma^0 T_{g}^{\dagger}\gamma^0\Big]
\varphi_{A}^{q\bar{q},g}
({\un k}, {\un k_2}\!-\! {\un k};{\un k_2})
\nonumber\\
&&\;\;
+\int d^2 k \, 
\!{\rm tr}_{\rm d}
\Big[(\slq\!+\!m)T_{g}(\slp\!-\!m)\gamma^0 T_{q\bar{q}}^{\dagger}\gamma^0\Big]
\varphi_{A}^{q\bar{q},g}
({\un k},{\un k_2}\!-\! {\un k};{\un k_2})
\nonumber\\
&&
+ \,  {\rm tr}_{\rm d} \, 
\Big[(\slq\!+\!m)T_{g}(\slp\!-\!m)\gamma^0 T_{g}^{\dagger}\gamma^0\Big]
\varphi_{A}^{g,g}({\un k_2})
\Bigg\}
\varphi_p({\un k_1})
\label{eq:cs_qqbar}
\end{eqnarray}
where 
\begin{eqnarray}
&&T_{q\bar{q}}( {\un k_1},{\un k})\equiv 
\frac{\gamma^+(\slq-\slk+m)\gamma^-(\slq-\slk-\slk_1+m)\gamma^+}
{2p^+[( {\un q}\!-\!{\un k})^2+m^2]+2q^+[({\un q}\!-\!{\un k}\!-\!{\un k_1})^2+m^2]}
\nonumber\\
&&T_{g}({\un k_1})\equiv 
\frac{\slC_{L}(p+q,{\un k_1})} {(p+q)^2}
\label{eq:Tqqbar-Tg}
\end{eqnarray}
and  $T_{q\bar{q}}, T_{q\bar{q}}^\prime, T_g$ are related to Eqs.~(\ref{eq:Tqqbar-Tg}) via 
\begin{eqnarray}
T_{q\bar{q}}\equiv T_{q\bar{q}}({\un k_1},{\un k})\;,\;
T_{q\bar{q}}^\prime
\equiv T_{q\bar{q}}({\un k_1},{\un k_t}^\prime)\; ,\; 
T_g\equiv T_{g}({\un k_1})
\end{eqnarray}
and $\varphi$'s are the Fourier transforms of two, three and four
point functions of Wilson lines \cite{Blaizot:2004wv}. For example,
\begin{eqnarray}
\varphi_A^{q\bar{q},g}({\un l};{\un k})\equiv
\frac{2\pi^2 R_A^2 l_T^2}{g^2 N_c}
\int d^2 x\, d^2 y\, 
e^{i({\un k} \cdot {\un x} + ({\un l} - {\un k})\cdot {\un y})}\; 
{\rm tr} \, \left<
V ({\un x}) t^a V^\dagger ({\un y}) t^b U_{ba}(0)
\right>
\end{eqnarray}

There are a few remarks in order here. First, the quark anti-quark
production cross section is again expressed in terms of the Wilson
lines just like the other cross sections in Color Glass Condensate
approach to high energy proton-nucleus collisions. Furthermore, the
standard $k_T$-factorization does not hold in general, at least in its
most common form, even though one does recover the $k_T$-factorized
form of the cross section in the leading twist region. For a
quantitative study of the breakdown of $k_T$ factorization in this
context, we refer the reader to \cite{Fujii:2005vj}.  We also note
that one can integrate over the momenta of the final state anti-quark
(or quark) in order to get the cross section for single inclusive
quark production. Finally, to include quantum corrections of the form
$\alpha_s \, \ln 1/x$ in the cross section above one needs to solve
the JIMWLK equation for the $n$ point function of Wilson line and use
them in the cross section in (\ref{eq:cs_qqbar}). This may then be
used to quantitatively investigate production of heavy (or light)
quark anti-quark pairs in proton-nucleus collisions at RHIC and LHC.

%%%%%%%%%%%%%%%%%%%%%%%%%%%%%%%%%%%%%%%%%%%%%%%%%%%%%%%%%%%%%%%%%%%%%%%%%%%%%%%%%%%%%%%%

\section{Results from dAu Collisions at RHIC}
\label{data}

Now we are going to review the experimental data obtained from $d+Au$
experiments at RHIC. We will show that at mid-rapidity $d+Au$
collisions provide a control experiment giving us evidence of jet
quenching due to strong final state interactions indicating formation
of quark-gluon plasma. We will then proceed by reviewing the results
from forward rapidity in $d+Au$ collisions: they confirm the
qualitative expectations of saturation/Color Glass Condensate physics,
summarized in \fig{toy}. Therefore we argue that RHIC experiments have
seen the first experimental confirmation of Color Glass Condensate in
nuclear collisions.

%%%%%%%%%%%%%%%%%%%%%%%%%%%%%%%%%%%%%%%%%%%%%%%%%%%%%%%%%%%%%%%%%%%%%%%%%%%%%%%%%%%%%%%%

\subsection{Mid-Rapidity}

\subsubsection{Cronin Effect in dAu}

%%%%%%%%%%%%%%%%%%%%%%%%%%%%%
\begin{figure}
\vspace*{1.5cm}
\begin{center}
\begin{tabular}[t]{cc} 
\epsfig{file=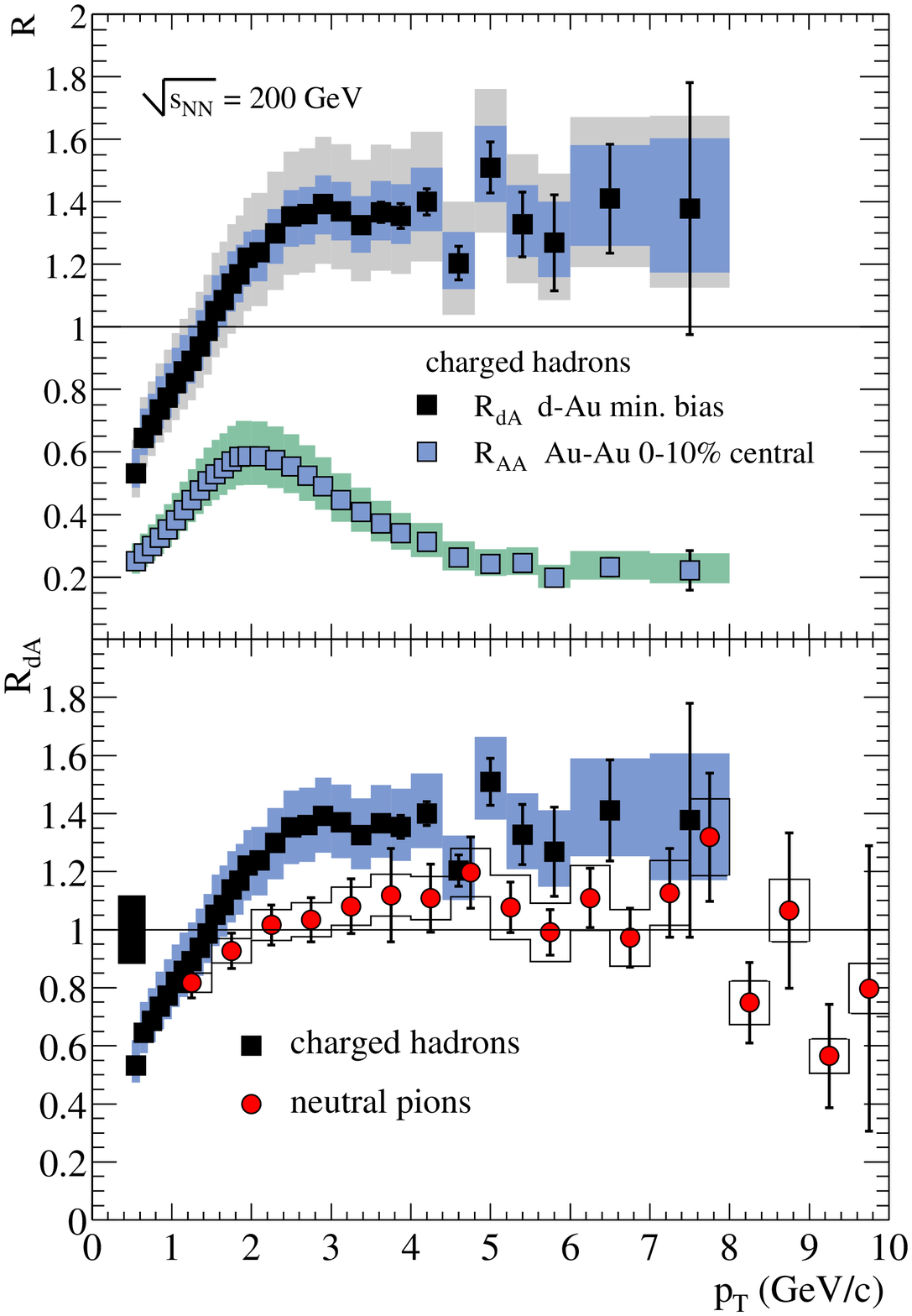,width=8cm} & \vspace*{-12cm} \epsfig{file=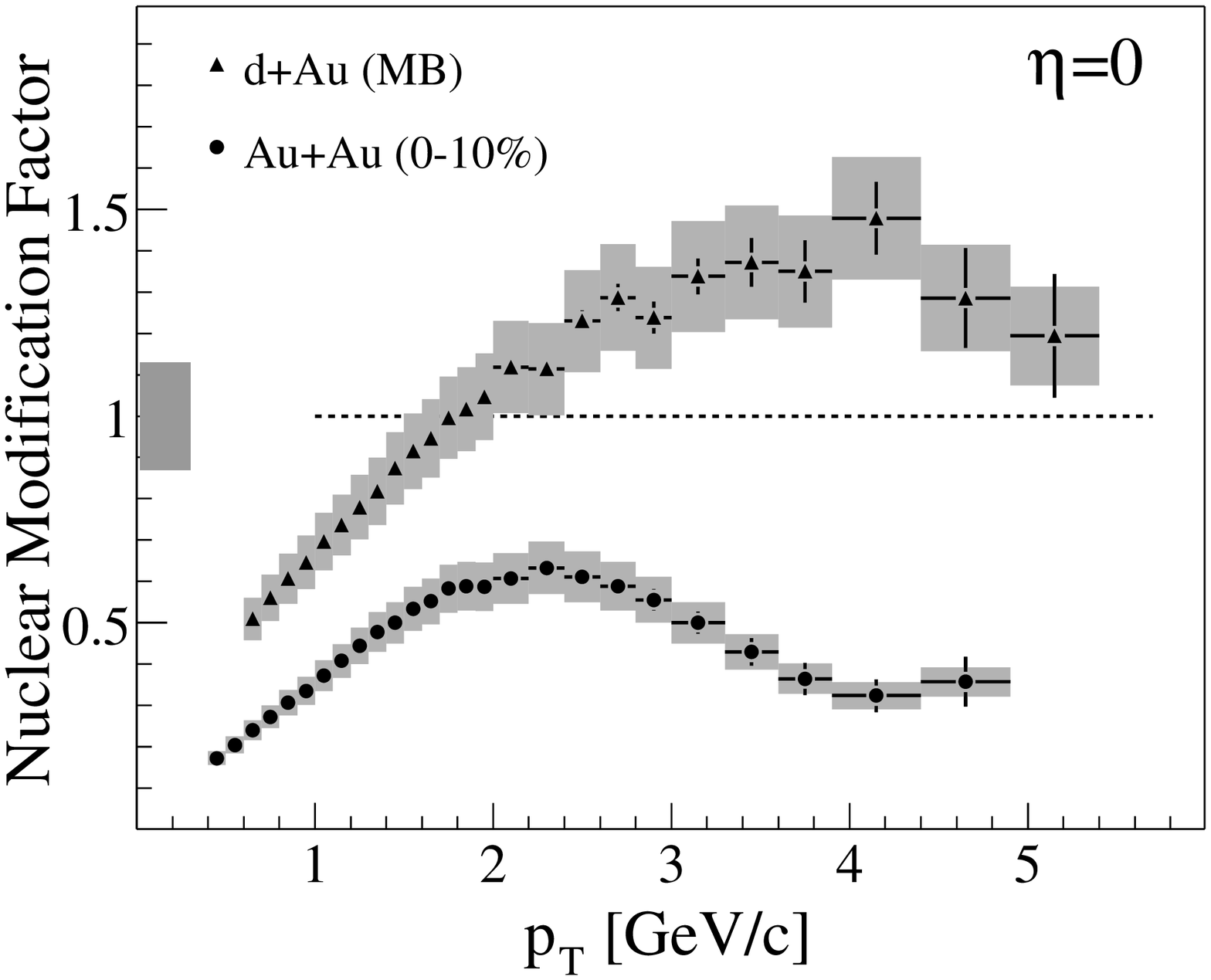,width=7cm} \\ 
\hspace*{8cm} & \vspace*{-5cm} \epsfig{file=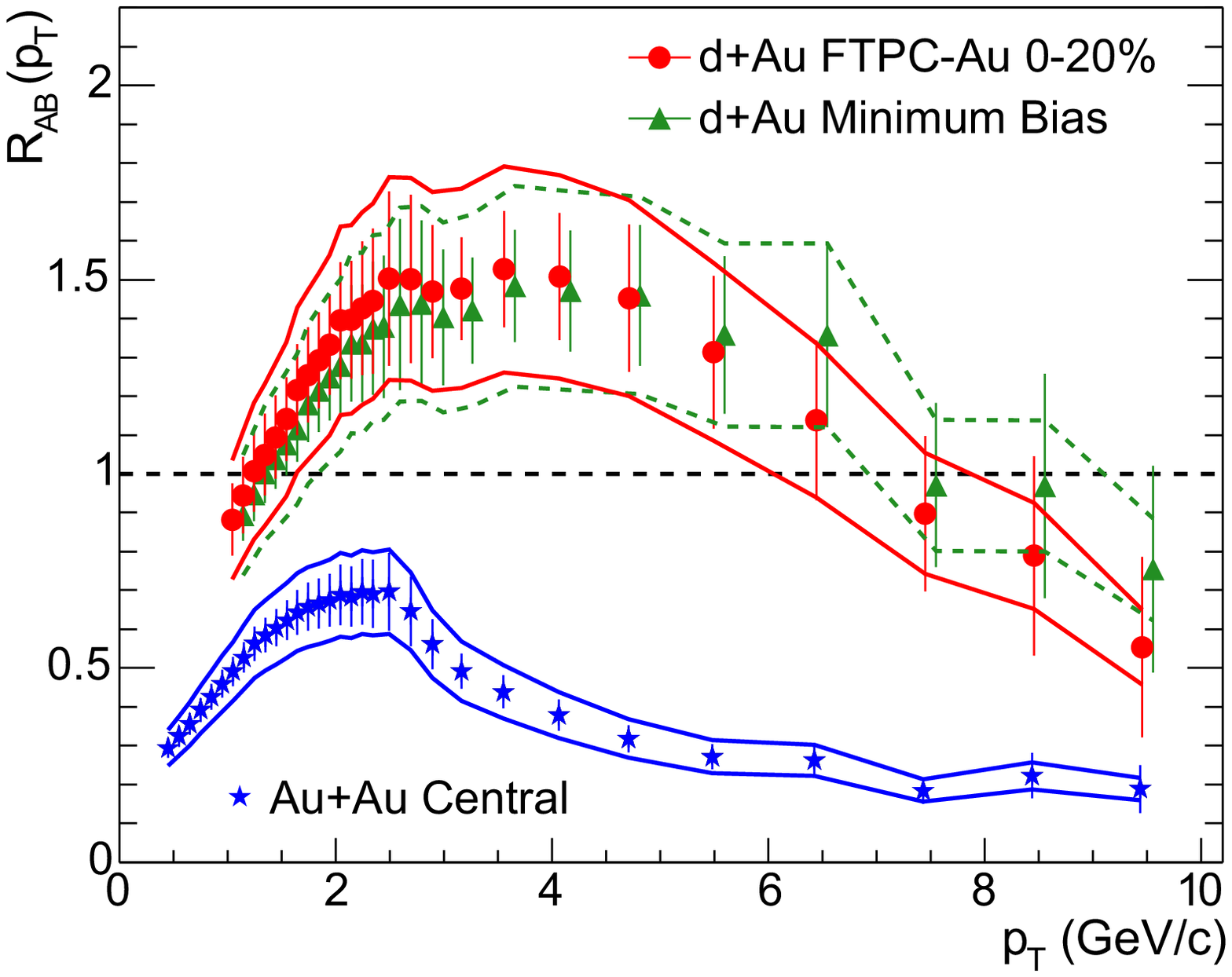,width=7cm}
\end{tabular}
\end{center}
\caption{Nuclear modification factor for charged hadrons and neutral pions 
produced in $d+Au$ and $Au+Au$ collisions at mid-rapidity, as reported
by (clockwise from left) PHENIX \protect\cite{phenix}, STAR \protect\cite{star}
and BRAHMS \protect\cite{brahmsaa} collaborations.}
\label{pbs}
\end{figure}
%%%%%%%%%%%%%%%%%%%%%%%%%%%%%%

The experimental program at RHIC involved $d+Au$ collisions at the
center of mass energy $\sqrt{s} = 200$~GeV per nucleon. While the
above analysis of particle production was, strictly speaking, designed
for $pA$ and DIS, we believe that its results can be applied to $d+Au$
collisions as well.  $d+Au$ scattering also involves a scattering of a
small projectile on a large nucleus, similar to DIS and $pA$. The only
assumption we have made in our above analysis was that the saturation
scale of the projectile is much smaller than transverse momentum of
produced particles, which allowed us to neglect saturation effects in
the projectile's wave function.  The fact that deuteron has the atomic
number of $A=2$ increases its saturation scale compared to that of a
proton by a small factor of $A^{1/3} \approx 1.26$. Therefore, the
region of transverse momentum above the deuteron saturation scale,
where our analysis is applicable, would still be very wide in $d+Au$
collisions. It is, therefore, a good approximation to apply our above
conclusions to $d+Au$ collisions.

\begin{figure}[hbt]
\begin{center}
\epsfxsize=8cm
\leavevmode
\hbox{\epsffile{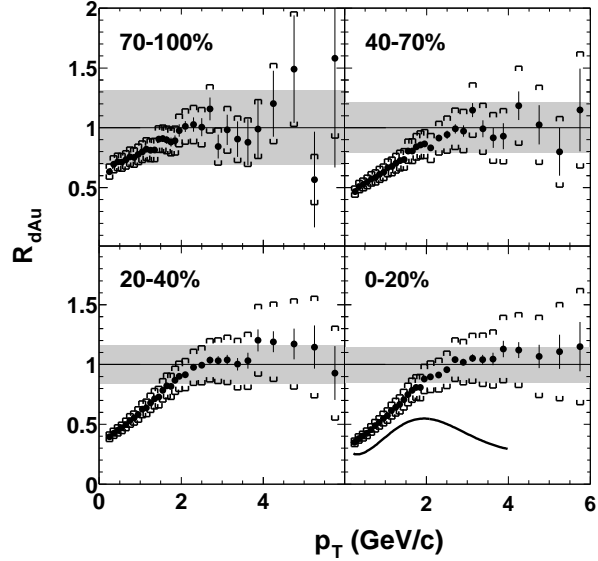}}
\end{center}
\caption{Nuclear modification factor for charged hadrons produced in 
$d+Au$ collisions reported by PHOBOS collaboration
\protect\cite{phobos} in pseudo-rapidity range of $0.2 < \eta < 1.4$
for four bins of centrality.}
\label{phobos_mide}
\end{figure}

Nuclear modification factor for charged hadrons produced in $d+Au$
collisions at mid-rapidity was measured by all four experimental
collaborations at RHIC \cite{phenix,phobos,star,brahmsaa}. The data
from BRAHMS \cite{brahmsaa}, PHENIX \cite{phenix} and STAR \cite{star}
collaborations is shown in Fig. \ref{pbs} and the data from PHOBOS
collaboration \cite{phobos} is shown in \fig{phobos_mide}. One can see
that $R^{dAu}$ at mid-rapidity at RHIC exhibits Cronin enhancement
\cite{Cronin}, in agreement with theoretical expectations of multiple 
rescattering models, shown in \fig{cron}
\cite{KNST,KM,ktbroadening1,ktbroadening2,Vitev03,ktbroadening3,BKW,KKT}.
%%%%%%%%%%%%%%%%%%%%%%%%%%%%%
%%%%%%%%%%%%%%%%%%%%%%%%%%%%%%

The PHOBOS data, presented in \fig{phobos_mide} allows one to study
$R^{dAu}$ as a function of centrality. Analyzing \eq{rpa1}, along with
its asymptotic high-$p_T$ limit of Eq. (\ref{rpaghkt}), we see that
high-$p_T$ $R^{dAu}$ in the quasi-classical multiple rescattering
approach is an increasing function of $A$, or, equivalently,
centrality. These theoretical expectations are confirmed by the data
in \fig{phobos_mide}, where $R^{dAu}$ is clearly an increasing
function of the collision centrality. 

%%%%%%%%%%%%%%%%%%%%%%%%%%%%%
\begin{figure}[hbt]
\begin{center}
\epsfxsize=6.cm
\leavevmode
\hbox{\epsffile{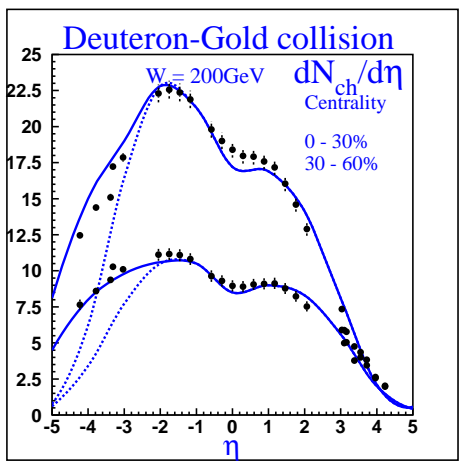}}
\end{center}
\caption{Predictions of saturation model from \cite{KLN} (solid and dashed lines) 
for the total charged particle multiplicity $dN_{ch}/d\eta$ in $d+Au$
collisions compared to the experimental data from
\protect\cite{brahms_mult} as a function of pseudo-rapidity. }
\label{kln}
\end{figure}
%%%%%%%%%%%%%%%%%%%%%%%%%%%%%%
An intriguing property of the PHOBOS data from \fig{phobos_mide}
\cite{phobos} is that the height of the Cronin maximum in it is much 
lower than in the data from BRAHMS, PHENIX and STAR collaborations
shown in \fig{pbs} \cite{brahmsaa,phenix,star}. The origin of this
disagreement appear to be in the fact that PHOBOS was collecting data
in the pseudo-rapidity interval of $0.2 < \eta < 1.4$, whereas the
other three collaborations presented the data at $\eta =0$. This
decrease of Cronin maximum with increasing rapidity can already be
interpreted as a precursor of the suppression predicted by saturation
physics shown above in \fig{toy}. However, we will postpone this
discussion till Sect. \ref{cgcsuppr} where the data at more forward
rapidities will be presented.

Another interesting feature of the $d+Au$ data at mid-rapidity is that
other observables, such as integrated charged hadron multiplicity as a
function of rapidity \cite{phobos_mult,brahms_mult}, appear to be in
agreement with saturation/CGC models. In \fig{kln} we show a
prediction of saturation model from \cite{KLN} for the total charged
particle multiplicity $dN_{ch}/d\eta$, compared to the data from
\cite{brahms_mult}. The model in \cite{KLN} uses $k_T$-factorization 
formula for gluon production from \eq{ktpa} integrated over all $k_T$
with the unintegrated gluon distribution on the nucleus ``frozen'' at
a constant at transverse momenta below $Q_s$ and falling off as $\sim
1/k_T^2$ for $k_T > Q_s$ (for details see \cite{KLN}).  As one can see
the agreement between the prediction from \cite{KLN} and the data is
very good.

%%%%%%%%%%%%%%%%%%%%%%%%%%%%%%%%%%%%%%%%%%%%%%%%%%%%%%%%%%%%%%%%%%%%%%%%%%%%%%%%%

%\subsubsection{Evidence for Quark-Gluon Plasma: High-$p_T$ Suppression 
% and Disappearance of Back-to-back Correlations in AuAu Collisions}

%\subsubsection{Back-to-back Correlations}

%%%%%%%%%%%%%%%%%%%%%%%%%%%%%%%%%%%%%%%%%%%%%%%%%%%%%%%%%%%%%%%%%%%%%%%%%%%%%%%%%

\subsubsection{$dAu$ as a Control Experiment for Quark-Gluon Plasma Production}

$dAu$ collisions can be used as a control experiment for formation of
quark-gluon plasma (QGP) in $AuAu$ collisions at RHIC. As was
originally suggested in \cite{Bj,BDMPS1,BDMPS,EL,EL2}, creation of a
dense hot medium, such as QGP, in high energy heavy ion collisions
should lead to a depletion of high-$p_T$ particles produced in the
collisions. This phenomenon is known as {\sl jet quenching}
\cite{EL}. The underlying physics is rather straightforward: high-$p_T$ 
particles produced in the medium would loose energy in interactions
with the medium. As a result there would be less high-$p_T$ particles
produced in the heavy ion collisions than one would naively expect by
scaling the number of high-$p_T$ particles produced in a $pp$
collisions by the number of elementary nucleon-nucleon collisions
$N_{coll}$.

One usually measures this suppression by analyzing the nuclear
modification factor for $AA$ collisions, which is defined as
\be
R^{AA} \, = \, \frac{\frac{dN^{AA}}{d^2 k_T \, dy}}{N_{coll} \,
\frac{dN^{pp}}{d^2 k_T \, dy}}.
\ee
Jet quenching due to {\sl energy loss} of particles in the medium
would result in $R^{AA}$ being suppressed. The calculations of such
suppression were carried out in \cite{BDMPS1,BDMPS,EL}, where both
multiple rescatterings in the medium and medium-induced particle
emissions were taken into account.

The data on $R^{AA}$ collected at RHIC was reported in
\cite{phenixAA,phobosaa,starAA,brahmsaa}. Here those data are shown in 
\fig{pbs}, along with the nuclear modification factor for $d+Au$ collisions. 
As on can see from \fig{pbs}, jet quenching was observed in $Au+Au$
collisions at RHIC with $R^{AuAu} < 1$ for all measured $p_T$. 

For jet quenching to become a signal of creation of quark-gluon plasma
one has to prove that the observed suppression is really due to energy
loss of jets in the produced matter, which would be a {\sl final}
state effect. In principle, jet quenching can be accounted for by the
small-$x$ evolution effects in the nuclear wave functions, similar to
those described for $pA$ collisions in Sect. \ref{qepa} leading to
suppression shown in \fig{toy}. Such suppression would be the {\sl
initial} state effect. In fact, it has even been suggested in
\cite{KLM} that the suppression observed in 
\cite{phenixAA,phobosaa,starAA,brahmsaa} could be due to initial state 
effects induced by small-$x$ evolution. The initial state suppression
should be observable already in $d+Au$ collisions, while the final
state energy loss is typical only for $AA$ collisions and should not
manifest itself in $d+Au$. In that sense $d+Au$ scattering is a
control experiment for QGP formation \cite{GM1}.

The data presented in \cite{phenix,phobos,star,brahmsaa} and shown
here in Figs. \ref{pbs} and \ref{phobos_mide} clearly indicate absence
of suppression in $R^{dAu}$. The Cronin enhancement observed in these
data is in agreement with the classical gluon field dynamics for $pA$
collisions of Sect. \ref{clpa}: the same initial state classical
fields would lead to Cronin-like enhancement of $R^{AA}$ in $AA$
collisions \cite{JMNV}. One can therefore conclude that jet quenching
observed for $AA$ collisions in
\cite{phenixAA,phobosaa,starAA,brahmsaa} is due to strong {\sl final}
state interactions leading to energy loss of partons in the medium.

%%%%%%%%%%%%%%%%%%%%%%%%%%%%%
\begin{figure}[htb]
\begin{center}
\epsfxsize=8cm
\leavevmode
\hbox{\epsffile{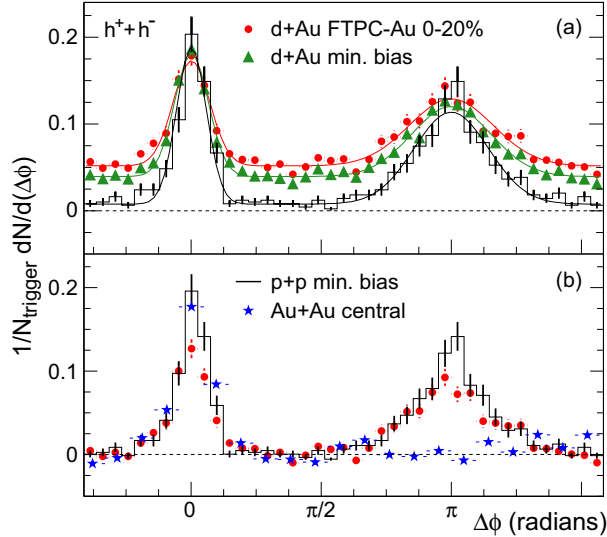}}
\end{center}
\caption{Two-particle correlation functions for $pp$, $d+Au$ and central $Au+Au$ 
collisions measured by STAR collaboration at mid-rapidity
\protect\cite{starb2b}.}
\label{star_corr}
\end{figure}
%%%%%%%%%%%%%%%%%%%%%%%%%%%%%%

The case for production of hot and dense medium is strengthened by
observation of two-particle correlations. In $pp$ collisions a
high-$p_T$ jet is usually accompanied by another high- or
intermediate-$p_T$ jet balancing its transverse momentum. This is
known as {\sl back-to-back} jet correlations. In the presence of dense
medium absorbing many of the produced jets, such back-to-back
correlations should be reduced, and possibly wiped out completely. To
see whether this takes place, and to eliminate possible initial state
nuclear wave functions effects, one has to study two-particle
correlation functions in $Au+Au$, $d+Au$ and $pp$ collisions. 

The data on such correlations was reported by the STAR collaboration
in \cite{starb2b} and is shown here in
\fig{star_corr}. \fig{star_corr} depicts the correlation function
(with constant background subtracted) between a trigger particle
(hadron) with $4$~GeV~$<p_T (trig) < 6$~GeV and and associated
particle with $2$~GeV~$< p_T < p_T (trig)$ for $pp$, $d+Au$ and
central $Au+Au$ collisions as a function of the azimuthal angle
$\Delta \phi$ between the two particles. First of all one should note
that back-to-back correlations in $Au+Au$ collisions (with $\Delta
\phi = \pi$) are completely wiped out in \fig{star_corr}. Secondly, 
the back-to-back correlations in $pp$ and $d+Au$ are, approximately,
the same. Therefore one concludes that disappearance of back-to-back
correlations in $Au+Au$ collisions is due to {\sl final} state
effects.

Final state interactions have to be quite strong to eliminate more
than half of the jets (see \fig{pbs}) and completely wipe out their
back-to-back correlations (see \fig{star_corr}). Such strong
interactions are likely to lead to thermalization of the produced
dense medium: if partons interact strongly with high-$p_T$ jet
particles, they would interact strongly with each other reaching
thermal equilibrium. It is therefore quite plausible that $d+Au$
scattering experiments indicate the formation of QGP in $Au+Au$
collisions.

%%%%%%%%%%%%%%%%%%%%%%%%%%%%%%%%%%%%%%%%%%%%%%%%%%%%%%%%%%%%%%%%%%%%%%%%%%%%%%%%%%%%%%%%

\subsection{Forward Rapidity}

\subsubsection{Suppression at all $p_T$: Evidence for the Color Glass Condensate}
\label{cgcsuppr}

\begin{figure}[htb]
\begin{center}
\epsfxsize=18.5cm
\leavevmode
\hbox{\epsffile{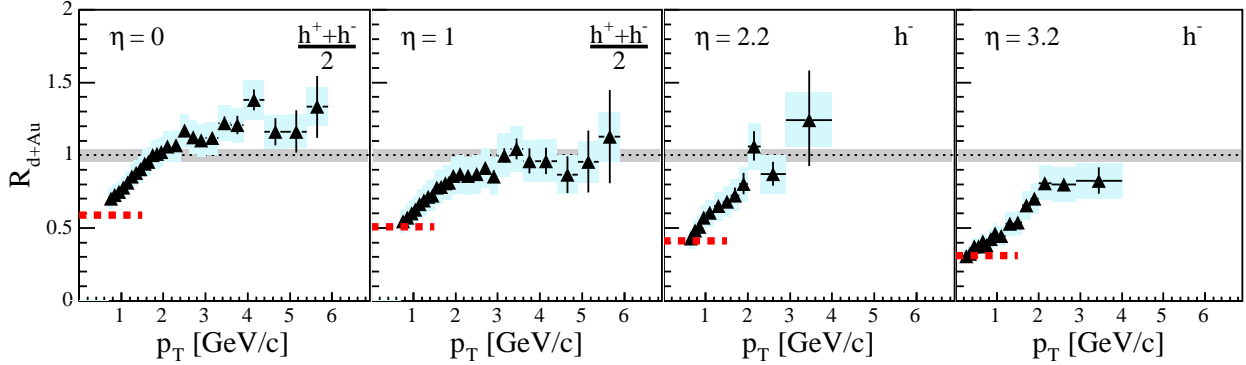}}
\end{center}
\caption{Nuclear modification factor for charged hadrons at rapidities 
$\eta = 0, 1$ and for negatively charged hadrons at rapidities $\eta =
2.2, 3.2$ reported by BRAHMS collaboration in
\protect\cite{brahms-2}.}
\label{bdata}
\end{figure}

The results on nuclear modification factor $R^{dAu}$ at forward
rapidity were first presented by BRAHMS collaboration in
\cite{brahms-1}. The final version of the data published in 
\cite{brahms-2} is shown in here in \fig{bdata}. There the nuclear 
modification factor $R^{d+Au}$ is plotted as a function of transverse
momentum $p_T$ for charged hadrons at rapidities $\eta = 0, 1$ and for
negatively charged hadrons at rapidities $\eta = 2.2, 3.2$. The data
of \fig{bdata} demonstrates suppression of hadron production as one
goes from central towards forward rapidity. We can clearly see that
Cronin maximum gradually disappears and {\sl suppression} sets in as
rapidity becomes larger: this clearly {\sl confirms} the qualitative
prediction of \fig{toy} based on Color Glass Condensate/saturation
physics \cite{KLM,KKT,AAKSW}.
%%%%%%%%%%%%%%%%%%%%%%%%%%%%%
%%%%%%%%%%%%%%%%%%%%%%%%%%%%%%

\begin{figure}[ht]
\begin{center}
\epsfxsize=10cm
\leavevmode
\hbox{\epsffile{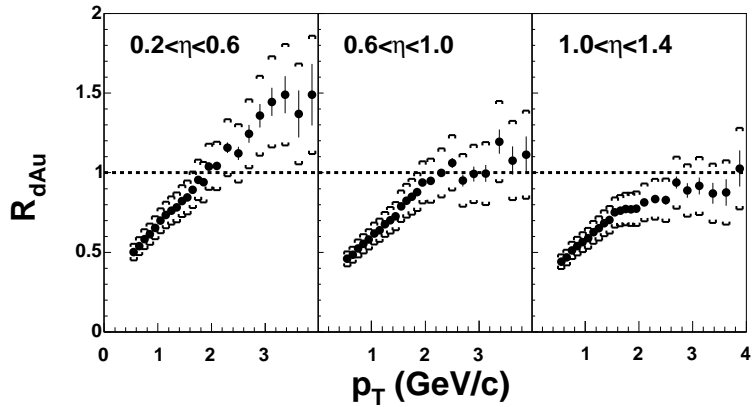}}
\end{center}
\caption{Data on nuclear modification factor $R^{d+Au}$ for hadrons 
presented by PHOBOS collaboration in \protect\cite{phobosdA}.}
\label{F1}
\end{figure}

BRAHMS data on forward suppression in $d+Au$ from \cite{brahms-2} is
independently confirmed by other RHIC experiments. In \fig{F1} we show
the data on hadronic nuclear modification factor $R^{d+Au}$ reported
by PHOBOS collaboration in \cite{phobosdA}. While PHOBOS collaboration
does no have such a wide rapidity acceptance as BRAHMS, one can see
the onset of suppression of $R^{d+Au}$ at high $p_T$ already at
pseudo-rapidities of $\eta = 1.0 \div 1.4$.

%%%%%%%%%%%%%%%%%%%%%%%%%%%%%
%%%%%%%%%%%%%%%%%%%%%%%%%%%%%%

%%%%%%%%%%%%%%%%%%%%%%%%%%%%%
\begin{figure}[ht]
\begin{center}
\epsfxsize=10cm
\leavevmode
\hbox{\epsffile{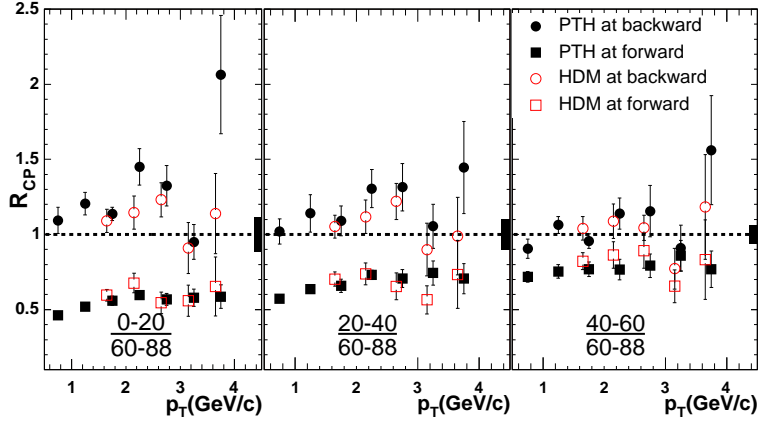}}
\end{center}
\caption{Data on $R^{CP}$ for hadrons presented by PHENIX collaboration 
in \protect\cite{phenixdA}.}
\label{hrcp}
\end{figure}
%%%%%%%%%%%%%%%%%%%%%%%%%%%%%%
%%%%%%%%%%%%%%%%%%%%%%%%%%%%%
\begin{figure}[hb]
\begin{center}
\epsfxsize=8cm
\leavevmode
\hbox{\epsffile{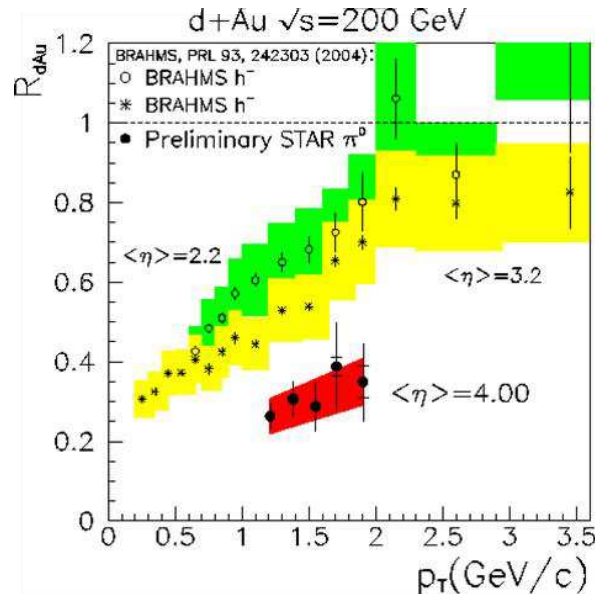}}
\end{center}
\caption{Data on $R^{d+Au}$ for neutral pions presented by STAR collaboration 
in \protect\cite{stardA} superimposed with the BRAHMS collaboration
data for negatively charged hadrons from \protect\cite{brahms-2}.}
\label{Frcp}
\end{figure}
%%%%%%%%%%%%%%%%%%%%%%%%%%%%%%

In Figs. \ref{hrcp} and \ref{Frcp} we show the data presented by
PHENIX \cite{phenixdA} and STAR \cite{stardA} collaborations. Instead
of $R^{pA}$ from \eq{rpaex} one sometimes is interested in ratio of
particle production rates in central over peripheral collisions, which
is defined by
\be\label{rcp}
R^{CP} (k_T, y) \, = \, \frac{\frac{1}{N_{coll}} \, \frac{d
N^{pA}}{d^2 k \, dy} \mbox{(central)}}{\frac{1}{N_{coll}} \, \frac{d
N^{pA}}{d^2 k \, dy} \mbox{(peripheral)}}.
\ee 
As once can see by comparing \eq{rcp} to \eq{rpaex}, $R^{CP}$ is very
similar to $R^{pA}$, with the difference of using the peripheral $pA$
collisions instead of $pp$ as a reference in the denominator. The
conclusions of the above analysis of $R^{pA}$ summarized in \fig{toy}
would also apply to $R^{CP}$: we would also expect a transition from
Cronin enhancement to suppression in $R^{CP}$ with increasing
rapidity.

%%%%%%%%%%%%%%%%%%%%%%%%%%%%%%%%%%%%%%%%%%%%%%%%%%%%%%%%%%%%%%%%%%%%%%%
\begin{figure}[hb]
\begin{center}
\begin{tabular}{cccc}
\epsfig{file=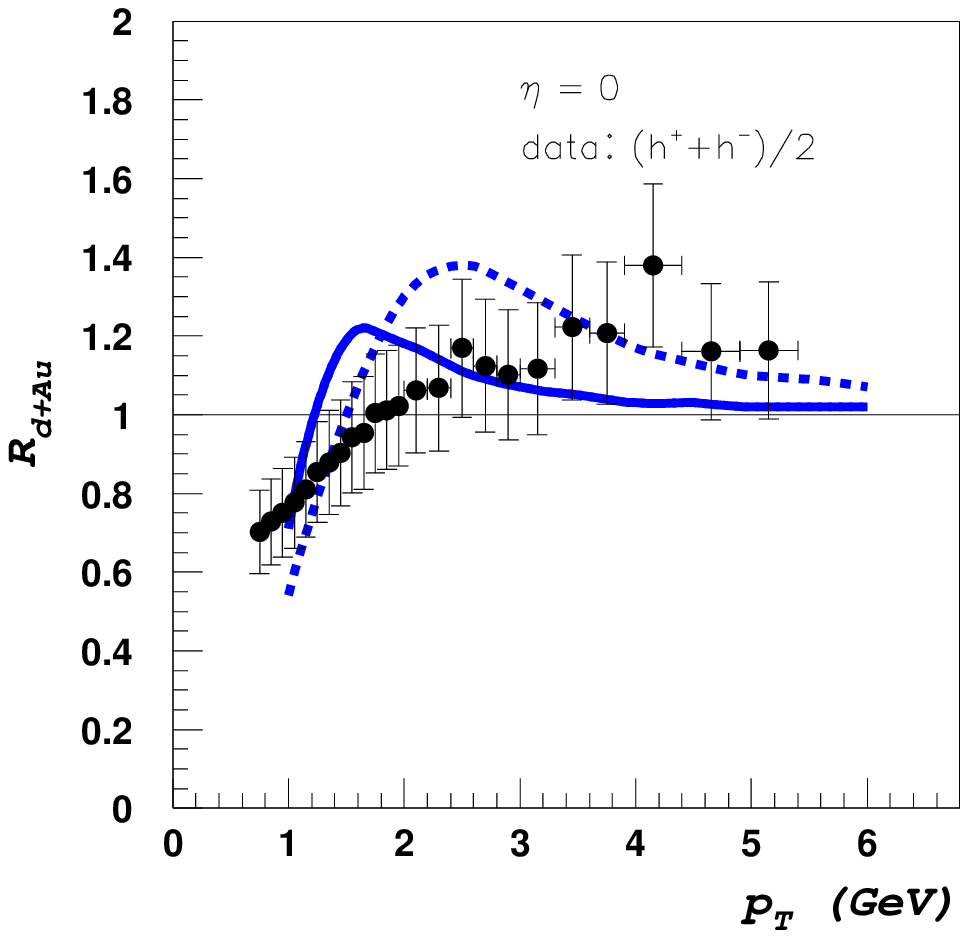,width=5.2cm} \hspace*{-1.5cm}& 
\epsfig{file=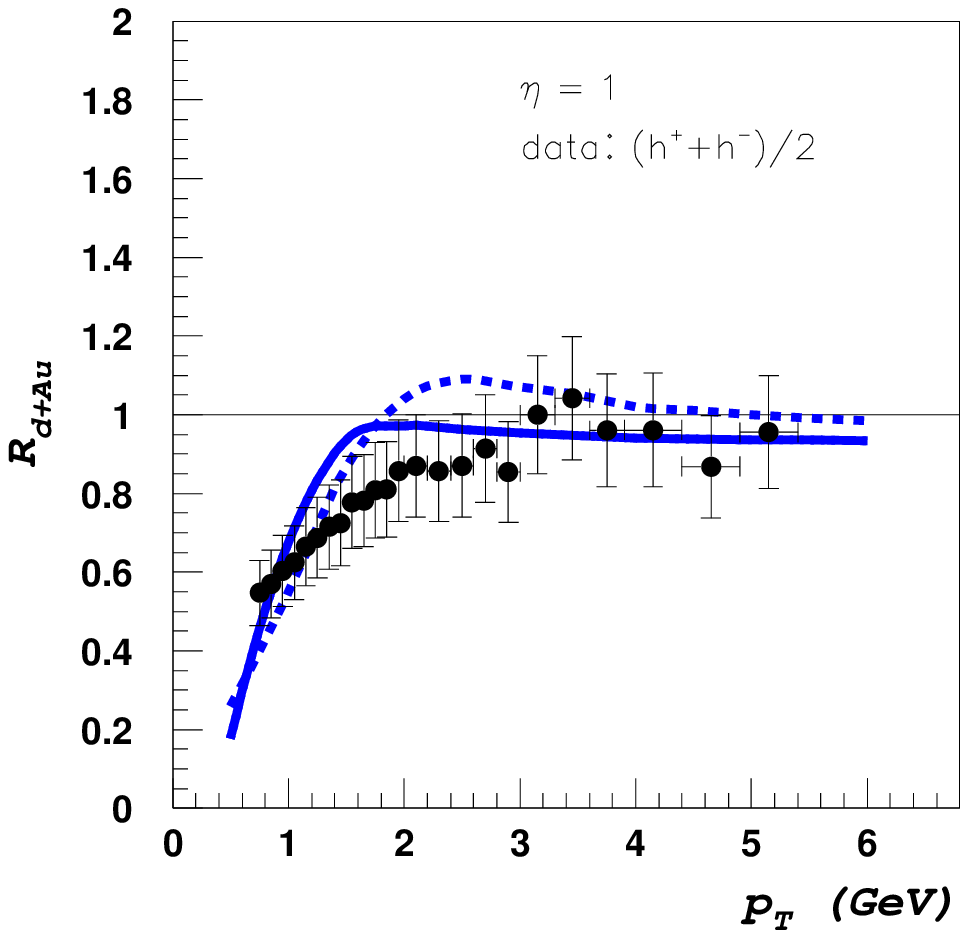,width=5.2cm} \hspace*{-1.5cm}& 
\epsfig{file=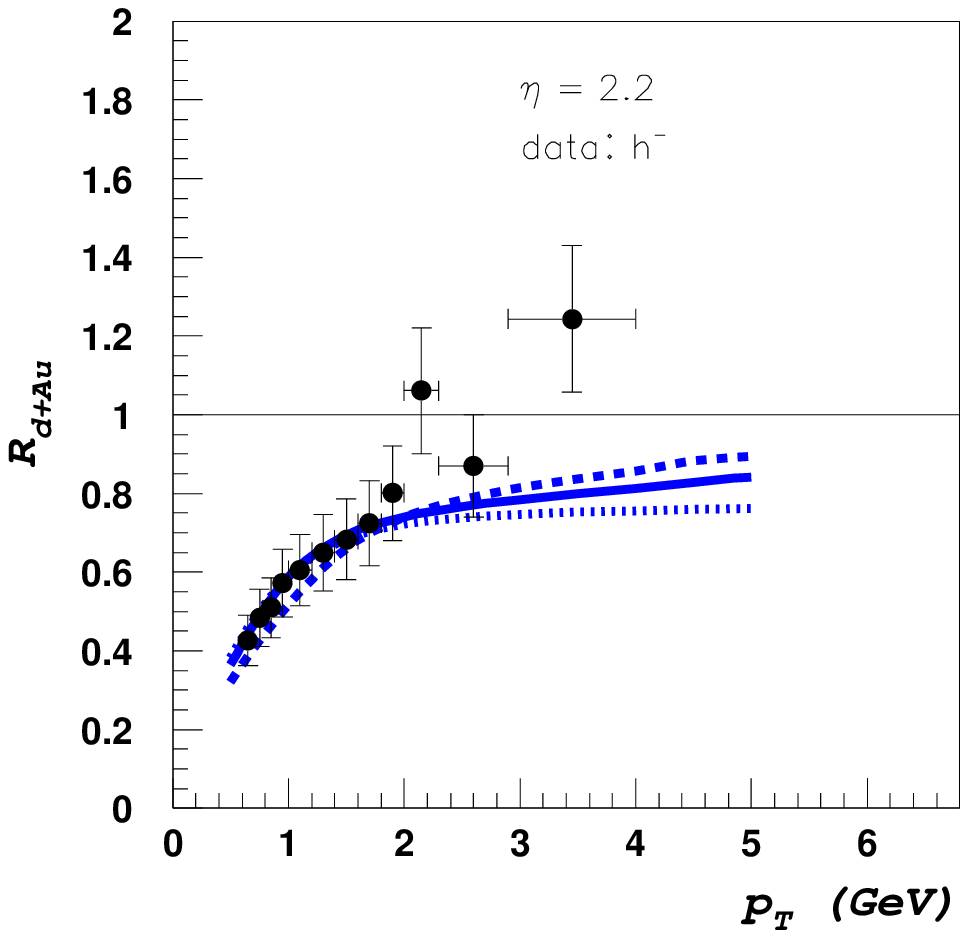,width=5.2cm} \hspace*{-1.5cm}&  
\epsfig{file=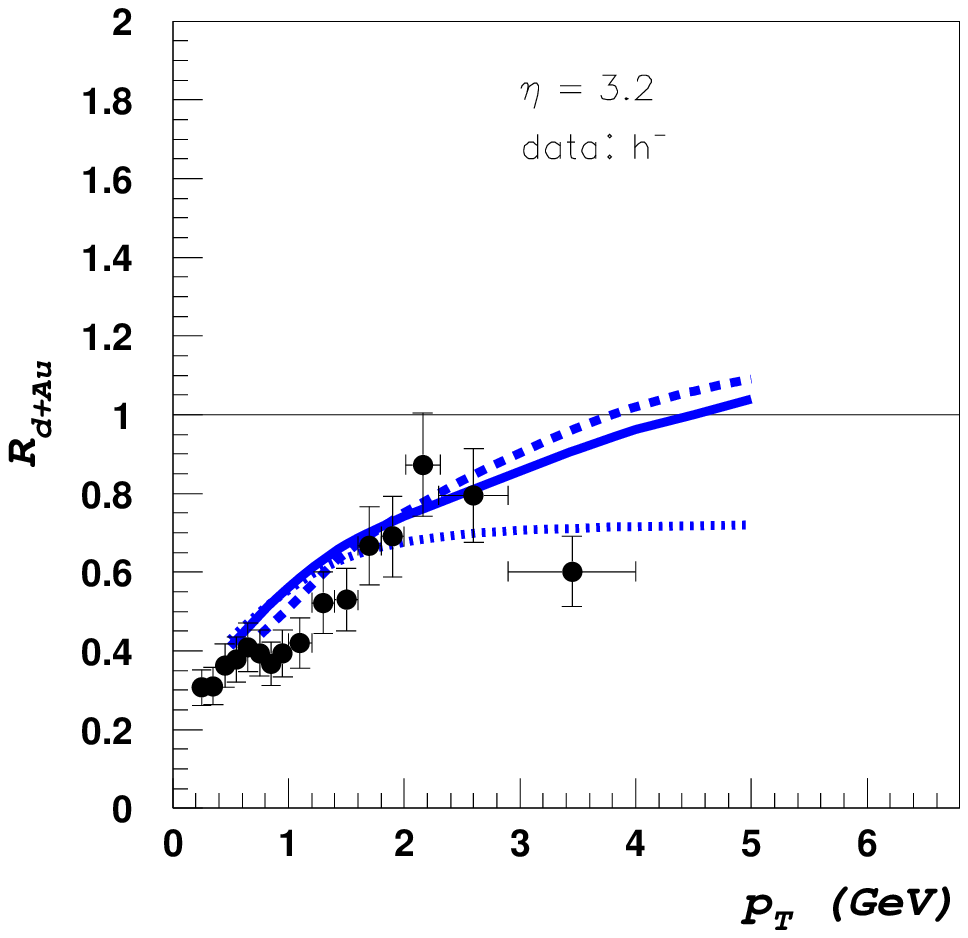,width=5.2cm}
\end{tabular}
\end{center}
\caption{Fit of nuclear modification factor $R_{dAu}$ of charged particles 
for different rapidities from \protect\cite{KKT2}. In the top two
figures, corresponding to $\eta=0,1$, the solid line corresponds to
$(h^-+h^+)/2$ contribution calculated with $\kappa=0$ in the
isospin-independent approximation, while the dashed line gives the
same $(h^-+h^+)/2$ contribution but with $\kappa=1$~GeV. In the lower
two plots, corresponding to $\eta=2.2,3.2$, the solid line gives the
$h^-$ contribution calculated in the constituent quark model with
$\kappa=0$, the dashed line gives the same $h^-$ contribution for
$\kappa=1$~GeV, while the dotted line at $\eta=2.2,3.2$ gives the
$(h^++h^-)/2$ contribution with $\kappa=0$ (see text).  Data is from
\cite{brahms-2}.}
\label{fig:rda}
\end{figure}
%%%%%%%%%%%%%%%%%%%%%%%%%%%%%%%%%%%%%%%%%%%%%%%%%%%%%%%%%%%%%%%%%%%%%%%

\fig{hrcp} plots PHENIX data on $R^{CP}$ for hadrons as a function of $p_T$
for three different centrality bins both for forward rapidity ($1.4 <
\eta < 2.2$, denoted by squares) and for backward rapidity 
($-2.2 < \eta < -1.4$, denoted by circles) \cite{phenixdA}. We see
that $R^{CP}$ is suppressed at forward pseudo-rapidities $1.4 < \eta <
2.2$ in agreement with \fig{bdata}. \fig{Frcp} presents preliminary
STAR data on $R^{d+Au}$ for $\pi^0$'s as a function of transverse
momentum $p_T$ at rapidity $\eta = 4$ \cite{stardA} superimposed with
BRAHMS data on negatively charge hadrons at lower rapidities. One can
see that suppression gets progressively stronger at forward
rapidities, in agreement with \fig{bdata}, leading to a very strong
suppression of $\pi^0$'s at $\eta =4$.

Of course the qualitative confirmation of the CGC-based predictions is
a very important experimental evidence for saturation/CGC
physics. However it is important to demonstrate the quantitative
agreement of BRAHMS data \cite{brahms-2} with the CGC expectations. A
saturation/CGC-based model was constructed in \cite{KKT2}, which
involved a new parameterization of the dipole-nucleus scattering
amplitude $N_G$ (see \eq{eq:kkt} below and discussion around it), with
the saturation scale matching that of the Golec-Biernat--W\"{u}thoff
model of DIS on a proton \cite{GBW}. Together with a simple model for
$n_G$, the resulting $N_G$ from \eq{eq:kkt} was used in \eq{paevc1} to
give the gluon production cross section. At forward rapidities
high-$p_T$ particle production is close to the kinematic limit, which
is obtained by demanding that projectile's Bjorken $x$ is less than
$1$, $x_p = (k_T /\sqrt{s}) \, e^\eta \le 1$, leading to $k_T \le
\sqrt{s} \, e^\eta$. The effective $x_1$ is rather large and valence
quark contribution becomes important. The latter is given by
Eqs. (\ref{eq:cs_v_quark}) and (\ref{eq:cs_dipole_F}) and was also
taken into account in
\cite{KKT2}. To compare the quark and gluon production cross sections 
with the data one indeed needs to convolute them with fragmentation
functions.  The resulting fit of BRAHMS data from \fig{bdata} is shown
in \fig{fig:rda} (see also \cite{JamalH}). Since BRAHMS data at
rapidities $\eta = 0,1$ is for total charged hadrons, while the data
at rapidities $\eta = 2.2, 3.2$ is for negatively charged hadrons, the
charge asymmetry issues had to be taken into account \cite{GSV}. One
may also worry that in the given kinematics the saturation-based
description of high-$p_T$ hadron production in $pp$ collisions may not be
well-justified: in that sense the success of the saturation model from
\cite{KKT2} in describing particle spectra in $pp$ collisions can be
considered as just a parameterization of the $pp$ data. What is
important is that saturation correctly describes particle production
in $d+Au$ collisions, i.e., the numerator in \eq{rpa}. To model
non-perturbative effects a shift of the saturation scale $\kappa$ was
introduced in \cite{KKT2}: however, this shift can be put to zero
without significantly impacting the quality of the fit, as one can see
from \fig{fig:rda}. (The non-perturbative shift appears to improve the
agreement with the data at mid-rapidity (the dashed line on the left
panel in \fig{fig:rda}) indicating that non-perturbative effects may
still be important at corresponding values of Bjorken $x$.) We,
therefore, conclude that BRAHMS data from
\fig{bdata}
\cite{brahms-2} is in both qualitative and quantitative agreement with
the saturation/CGC physics.

The model from \cite{KKT2} is based on \eq{ktpa} with the unintegrated
gluon distribution functions calculated in the saturation/CGC
approach. In that sense it is consistent with the model of \cite{KLN}
used to describe total charged particle multiplicity in $d+Au$
collisions, as shown in \fig{kln}. Both models are inspired by
saturation physics, but reflect different aspects of it: the model
from \cite{KKT2} deals with the high-$p_T$ hadronic spectra, and is,
therefore, more accurate at high-$p_T$. At the same time the model in
\cite{KLN} concentrates on total particle multiplicity, which is most 
sensitive to low-$p_T$ hadron production: therefore, that model is
more precise at low-$p_T$.

%%%%%%%%%%%%%%%%%%%%%%%%%%%%%%%%%%%%%%%%%%%%%%%%%%%%%%%%%%%%%%%%%%%%%%%
\begin{figure}[htb]
\begin{center}
\begin{tabular}{cccc}
\epsfig{file=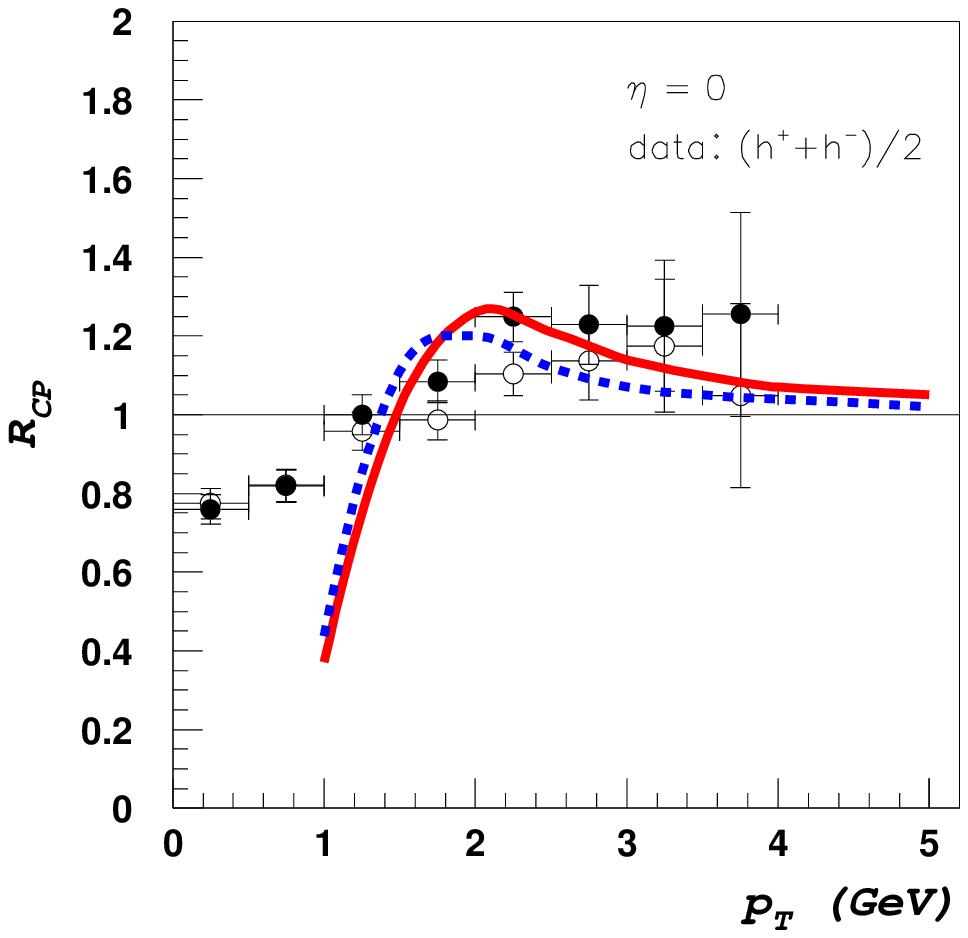,width=5.2cm} \hspace*{-1.5cm}& 
\epsfig{file=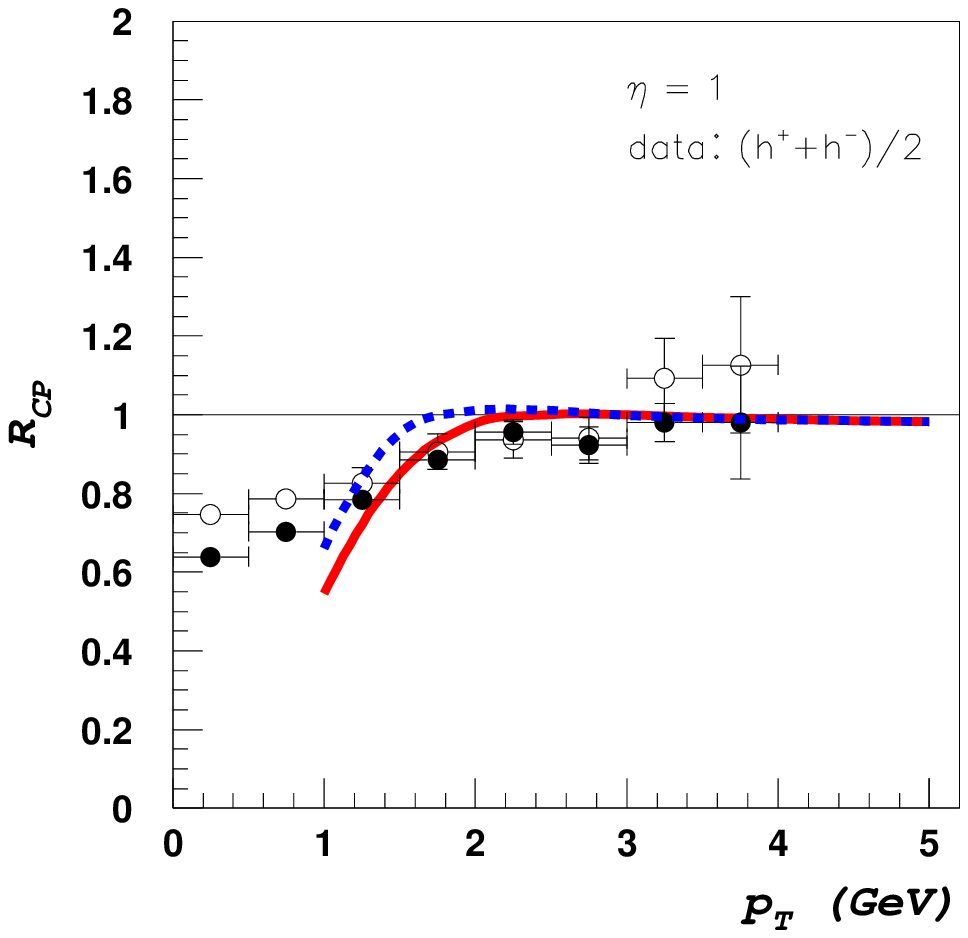,width=5.2cm} \hspace*{-1.5cm}& 
\epsfig{file=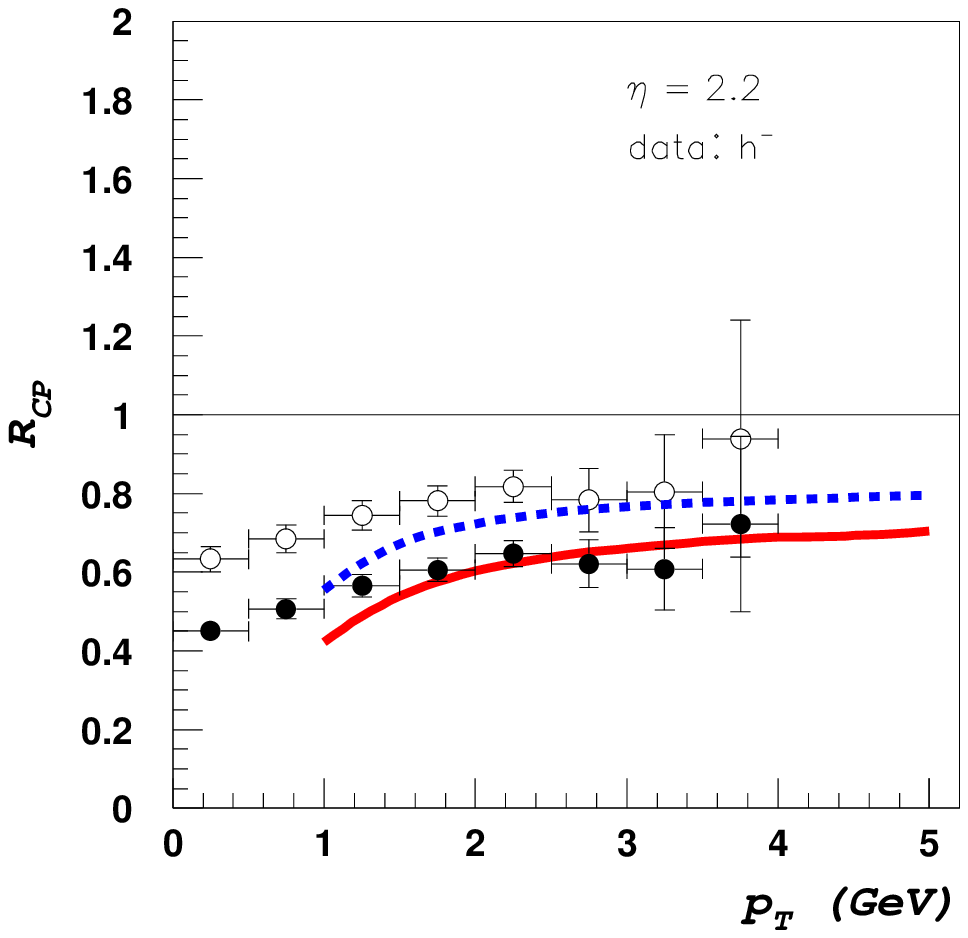,width=5.2cm} \hspace*{-1.5cm}&  
\epsfig{file=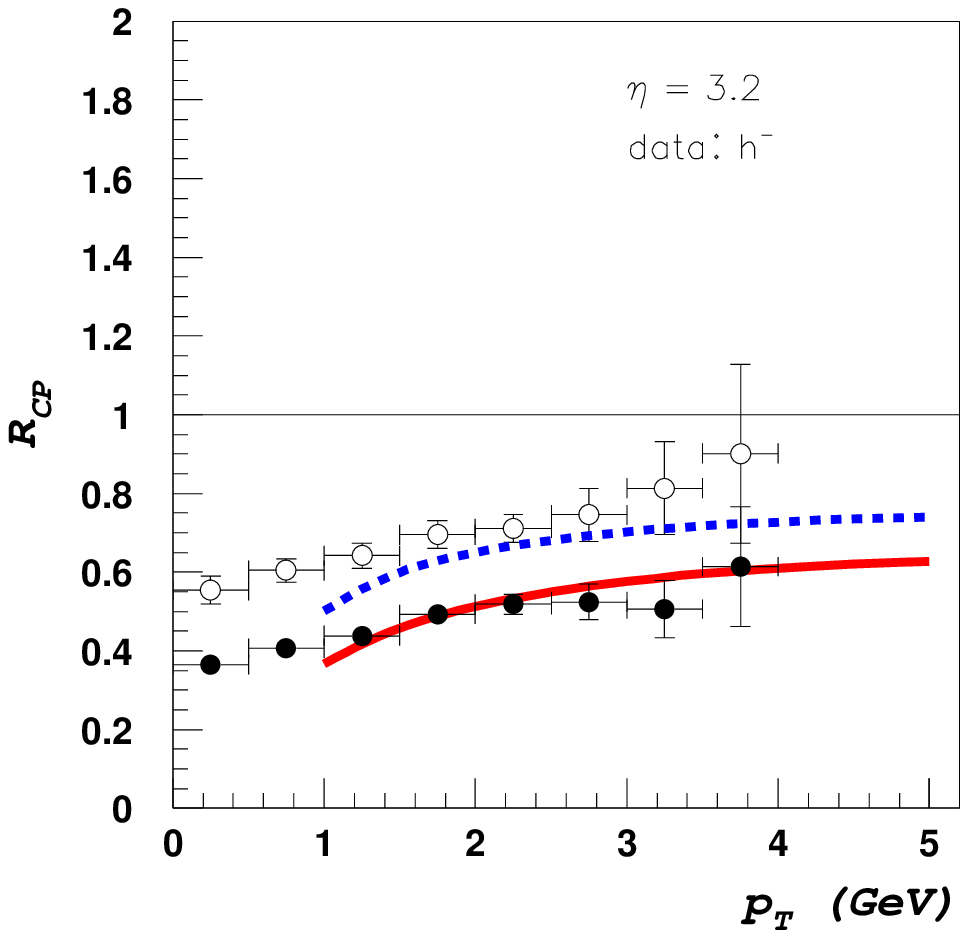,width=5.2cm}
\end{tabular}
\end{center}
\caption{Data from \cite{brahms-2} on nuclear modification factor $R_{CP}$ of charged 
particles $(h^++h^-)/2$ for rapidities $\eta=0,1$ and of negatively
charged particles $h^-$ for $\eta=2.2,3.2$ superimposed with the
results of the model from \cite{KKT2}. Full and open dots, described
by the solid and dashed lines correspondingly, give the ratio of
particle yields in 0-20$\%$ and 30-50$\%$ centrality events
correspondingly divided by the yields from 60-80$\%$ centrality events
scaled by the mean number of binary collisions \cite{brahms-2}.}
\label{fig:rcp}
\end{figure}
%%%%%%%%%%%%%%%%%%%%%%%%%%%%%%%%%%%%%%%%%%%%%%%%%%%%%%%%%%%%%%%%%%%%%%%

In \fig{fig:rcp} we show BRAHMS data on $R^{CP}$ for the same
rapidities as in \fig{fig:rda}. Full dots correspond to more central
collisions, while open dots correspond to more peripheral collisions
(see the caption of \fig{fig:rcp}). One can see that, in agreement
with the expectation of saturation/CGC (see Sect. \ref{clpa}), the
Cronin maximum at $\eta =0$ is an increasing function of
centrality. At forward rapidities the centrality dependence changes:
$R^{CP}$ becomes a decreasing function of centrality, again in
agreement with expectations of saturation/CGC approach (see
Sect. \ref{qepa}). On top of this qualitative agreement between the
data and the theory of CGC, in \fig{fig:rcp} we observe quantitative
correspondence of the data and the results of the saturation-inspired
model from \cite{KKT2}.

There have been several attempts to explain the suppression of
particle production from \fig{bdata} in the frameworks alternative or
parallel to CGC. In \cite{GSV,Ramona} the suppression of
\cite{brahms-2} was analyzed in the framework of standard shadowing
models. The conclusion of \cite{GSV} was that the observed suppression
is much stronger than one would anticipate using the common
parameterizations of nuclear shadowing.\footnote{A successful fit of
the data from \cite{brahms-2} by shadowing models would only produce
new parameterizations of quark and gluon shadowing, without
calculating them from first principles, as we have done here in the
CGC framework.} In \cite{QV} it was argued that multiple rescattering
in the nuclei would lead to observed suppression: however, this
conclusion appears to be in contradiction with our Sect. \ref{clpa},
where we prove that multiple rescatterings lead to enhancement of the
nuclear modification factor. Recently a new model was proposed in
\cite{KNPJS} in which the origin of suppression in \fig{bdata} was
attributed to energy conservation which was imposed by Sudakov
from-factors. The model of \cite{KNPJS} includes both multiple
rescatterings and small-$x$ evolution: hence at the moment the
conceptual differences between the CGC approach and the model from
\cite{KNPJS} are not clear. Parton recombination models also appear 
to be able to accommodate the data by completely redefining the
hadronization mechanism \cite{HYF}.

%%%%%%%%%%%%%%%%%%%%%%%%%%%%%%%%%%%%%%%%%%%%%%%%%%%%%%%%%%%%%%%%%%%%%

\subsubsection{Future Experimental Tests}
\label{fet}

Despite the successes of the Color Glass Condensate in predicting the
suppression of the hadron production cross section in proton
(deuteron)-nucleus collisions as compared to proton-proton collisions
in the forward rapidity region at RHIC and the change of centrality
dependence as one goes from mid to forward rapidity, one would like to
quantitatively understand to which degree CGC is the dominant physics
at forward rapidity RHIC and LHC collisions and possibly at
mid-rapidity LHC collisions. So far, the main evidence for CGC at RHIC
comes from the $p_T$ spectrum of single inclusive hadron production in
deuteron-gold collisions in the forward rapidity region.

A difficulty with a quantitative understanding of the physics behind
the observed suppression of the hadron spectra in deuteron-nucleus
collisions at RHIC is that hadronization by its nature is
non-perturbative and is not well-understood from first principles. To
avoid non-perturbative physics or to, at least, minimize its effects,
one needs to consider high $p_T$ processes. Exactly at what $p_T$
non-perturbative physics can be safely neglected can not be addressed
by pQCD and one needs experimental input. The RHIC data on transverse
momentum spectra of pions produced in proton-proton collisions in
forward rapidity suggest that NLO pQCD works well down to $p_T \sim
1.5 -2$ GeV \cite{sofer}. Therefore one may expect the weak coupling
methods employed by CGC to work at similar $p_T$'s for differential
cross sections, even though perturbative approaches usually work much
better for ratios of cross sections due to possible (partial)
cancellations of large corrections in both the numerator and the
denominator.

The experimental situation is complicated by the fact the forward
rapidity data published by BRAHMS is for negatively charged hadrons,
production of which is suppressed, as compared to neutral or charge
averaged hadrons, in proton-proton collisions. This can bias the
nuclear modification factor as pointed out in \cite{GSV}. However, the
preliminary data on neutral pion production from STAR at rapidity
$\eta = 4$ shown above in \fig{Frcp} demonstrates clear suppression up
to $p_T \approx 2$~GeV. If this preliminary STAR data is confirmed, it
will help clarify the role of isospin effects in the observed
suppression of the hadron spectra at forward rapidity.

The situation will improve drastically at LHC where due to an order of
magnitude higher center of mass energy, the saturation scale is
expected to be higher so that the weak coupling methods of CGC will be
much more reliable and the available phase space in transverse momenta
will be much larger even at rapidities as high as $5-6$ which should
eliminate or minimize phase space edge effects even for moderately
large transverse momenta.

Nevertheless, one can further test applicability of the Color Glass
Condensate at RHIC by considering its electromagnetic signatures, for
example in photon and dilepton production considered earlier. As noted
above, the CGC building blocks for single particle production cross
sections in proton-nucleus collisions are the fundamental and adjoint
dipole cross sections, describing the scattering of quarks and gluons
from the target. Therefore, one can "measure" or (more precisely)
constrain the dipole cross section in one process and then use it to
make predictions for other processes.  This is in direct analogy with
particle production cross sections in pQCD where one has universal
(process independent) ingredients in the form of distribution and
fragmentation functions, which are measured in one process and then
used to make predictions for other processes, which give rise to the
predictive power of the theory.

Here we consider models of the dipole cross section (we consider those
which include the small $x$ evolution governed by JIMWLK or BK
equations only) which have been used to fit the data, either in DIS at
HERA or pA at RHIC. We then use these models to make predictions for
photon and dilepton production cross sections and their modification
factors at RHIC and LHC. We show that measuring the photon production
cross section at RHIC can impose significant constraints on the models
of the dipole cross section and further clarify the role of the Color
Glass Condensate at RHIC.

The first model of the dipole cross section we consider is that of
Iancu, Itakura and Munier \cite{IIM} which was used to fit the proton
structure functions at HERA. It has the correct behavior in the
saturation region and in the geometric scaling region but does not
have the correct double logarithmic limit built in. Furthermore, it is
fit to the data on proton targets and not on nuclear targets. The IIM
model of the dipole cross section is
\begin{eqnarray}
\int d^2 b \, N (r_T, b_T, \ln 1/x_g ) \, \equiv 
\, \pi \, R^2  \,\,{\cal N} \,(x_g, r_T Q_s)
\label{eq:cs_iim}
\end{eqnarray}
where 
\begin{eqnarray}
{\cal N} \,(x_g, r_T Q_s) = 1 - e^{-a \ln^2 (b\,r_T Q_s)},   \,\,\,\,\,\,\,\,\,\,
\mbox{for} \hspace*{5mm} r_T \, Q_s > 2 
\nonumber
\end{eqnarray}
and 
\begin{eqnarray}
{\cal N} \,(x_g, r_T Q_s) = {\cal N}_0 \exp \Bigg\{ 2 \ln ({r_T Q_s \over 2})
\bigg [\gamma_s + {\ln 2/r_T Q_s \over \kappa \lambda \ln 1/x_g} \bigg] \Bigg\},
\,\,\,\,\,\,\,\,\,\, \mbox{for} \hspace*{5mm}
r_T \, Q_s \le 2 
\label{eq:cs_param}
\end{eqnarray}
The constants $a, b$ are determined by matching the solutions at $r_T
\, Q_s =2$ while $\gamma_s = 0.63$ and $\kappa = 9.9$ are determined
from LO BFKL.  The form of the saturation scale
$Q_s^2$ is taken to be $Q_s^2 \equiv (x_0/x)^{\lambda}$~GeV$^2$ with
$x_0, \lambda, R, {\cal N}_0$ determined from fitting the HERA data on
proton structure function $F_2$ \cite{IIM}.

\vspace*{.5cm}
%%%%%%%%%%%%%%%%%%%%%%%%%%%%%
\begin{figure}[ht]
\begin{center}
\epsfxsize=7cm
\leavevmode
\hbox{\epsffile{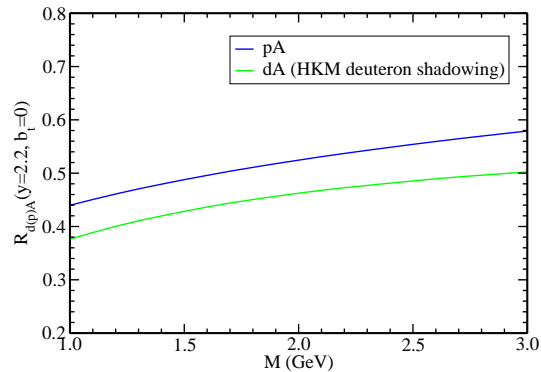}}
\end{center}
\caption{Invariant mass dependence of the nuclear modification factor for 
dilepton production in proton(deuteron)-nucleus scattering at RHIC.}
\label{fig:R_dilep_rhic}
\end{figure}
%%%%%%%%%%%%%%%%%%%%%%%%%%%%%%

This model of the dipole cross section was used in
\cite{Jalilian-Marian:2004er} to make predictions for the transverse
momentum integrated nuclear modification factor $R_{pA}$ for dilepton
production cross section in deuteron (proton)--nucleus collisions at
RHIC. (For the $p_T$-dependence of dilepton $R_{pA}$ see
\cite{BGD}.)  Here we show the main result of \cite{Jalilian-Marian:2004er} 
for the nuclear modification factor for central collisions and at
$y=2.2$ in Fig. \ref{fig:R_dilep_rhic} which shows a clear suppression
in analogy with the nuclear modification factor in hadron production
at RHIC that can be understood as being due to the quantum evolution
of the dipole-nucleus cross section, present in both hadron and
dilepton production cross sections. In the case of a deuteron
projectile, nuclear shadowing (anti-shadowing) of the deuteron wave
function is included by using the HKM parameterization \cite{HKM}.

\begin{figure}[htb]
\begin{center}
\epsfxsize=7cm
\leavevmode
\hbox{\epsffile{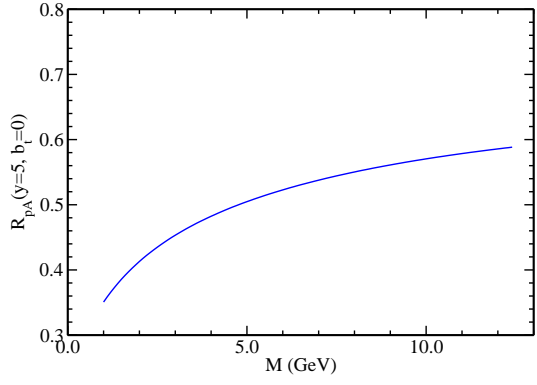}}
\end{center}
\caption{Invariant mass dependence of the nuclear modification factor for 
dilepton production in proton-nucleus scattering at LHC.}
\label{fig:R_dilep_lhc}
\end{figure}

We also show the nuclear modification factor for central collisions
and at rapidity $y=5$ for the integrated over transverse momentum
dilepton nuclear modification factor at LHC as a function of the
dilepton invariant mass $M$ in Fig. \ref{fig:R_dilep_lhc}, where again
a similar suppression is seen. Here one can extend the calculations to
much higher dilepton masses since the saturation scale is very large
due to the large forward rapidity and the order of magnitude increase
in the center of mass energy at LHC as compared to RHIC.

%%%%%%%%%%%%%%%%%%%%%%%%%%%%%
%%%%%%%%%%%%%%%%%%%%%%%%%%%%%%

\begin{figure}[ht]
\begin{center}
\epsfxsize=7cm
\leavevmode
\hbox{\epsffile{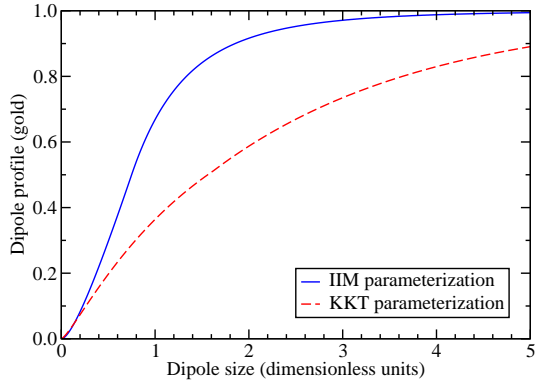}}
\end{center}
\caption{Quark anti-quark dipole profile for a nuclear target.}
\label{fig:dipole_a}
\end{figure}

We note that the IIM parameterization of the dipole cross section is
originally done for a proton target. Here we have assumed a simple
scaling of the dipole profile function by $A^{1/3}$ in order to use it
for nuclear targets. There is another parameterization of the dipole
cross section, due to Kharzeev, Tuchin and one of the authors
\cite{KKT2}, which has been used to fit the hadron production data in
deuteron-nucleus collisions at RHIC. This model has the advantage that
it has the correct high $p_t$ limit built in, unlike the IIM
parameterization. It is given by
\begin{eqnarray}
N (r_T, x_g) \, = \, 1- \exp\left[-\frac{1}{4} \left(r_T^2
\frac{C_F}{N_c} \, Q_s^2\right)^{\gamma(x_g,r_T^2)}\right].
\label{eq:kkt}
\end{eqnarray}
where the anomalous dimension $\gamma(y, r_T^2)$ is parameterized as 
\begin{eqnarray}
\gamma(y, r_T^2) \, = \, \frac{1}{2}\left(1+\frac{\xi 
(y, r_T^2)}{\xi (y, r_T^2) + \sqrt{2 \,\xi (y, r_T^2)}+  7
\zeta(3)\, c} \right),
\label{eq:andim}
\end{eqnarray}
with $y= \ln 1/x_g$ and  
\begin{eqnarray}
\xi (y, r_T^2) \, = \, \frac{\ln\left[1/( r_T^2 \, Q_{s0}^2 ) 
\right]}{(\lambda/2)(y-y_0)}\,,
\label{eq:xi}
\end{eqnarray}
and $c$ is a constant fitted to the data. This parameterization has
the correct high-$p_T$ behavior where $\gamma \rightarrow 1$. In
Fig. \ref{fig:dipole_a} we show the profile function for the two
dipole parameterizations as a function of the dimensionless variable
$r_T Q_s$ for a nuclear target for $x_g = 1.6 \,\times \, 10^{-4}$
(which corresponds to $p_T =1.5$ GeV at $y=3.8$ at RHIC). Clearly, the
two dipole profiles are quite different.

%\vspace{8mm}
%%%%%%%%%%%%%%%%%%%%%%%%%%%%%
%%%%%%%%%%%%%%%%%%%%%%%%%%%%%%

We now use the two parameterizations of the dipole cross section to
calculate the photon production cross section in deuteron-gold
collisions at RHIC. To get the cross section, we need to convolute the
partonic production cross section in (\ref{eq:frag_pho}) with the
quark (and anti-quark) distribution function in a deuteron. We
therefore use
\cite{Jalilian-Marian:2005zw}
\begin{eqnarray}
{d\sigma^{d(A)\, A \rightarrow \gamma(k_T, y_{\gamma})\, X}\over 
d^2b \, d^2k \, dy_{\gamma}}\!\!\! &=& \!\!\!
{1\over (2\pi)^2} \sum_f  \int dx_q \,
[q_f(x_q,k_T^2) + \bar{q}_f(x_q,k_T^2)]\, {D_{\gamma/q}(z,k_T^2) \over z}
\sigma_{dipole}^F (x_g,{k_T\over z}, {\un b})
\label{eq:cs_pho_da}
\end{eqnarray}
to calculate the nuclear modification factor for photon production in deuteron-gold
collisions at RHIC at $y=3.8$ and for the most central collisions. This will be 
eventually measured at RHIC by the STAR detector.

\begin{figure}[hb]
\begin{center}
\epsfig{file=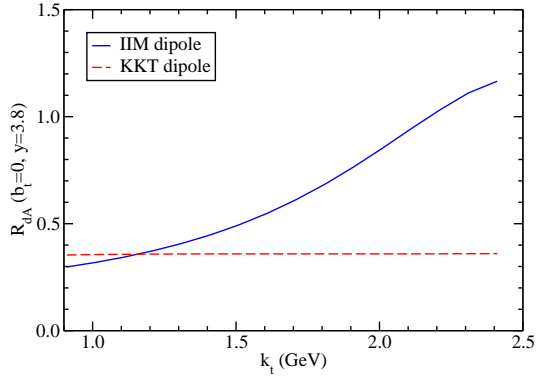,width=7cm}
\caption{Nuclear modification factor for photon production.}
\label{fig:R_pho_rhic}
\end{center}
\end{figure}

It is clear that the two parameterizations lead to quite different
results shown in \fig{fig:R_pho_rhic}. This may be understood to be
due to the different dipole shapes in the two approaches, so that a
future measurement of this ratio can help constrain the models of the
dipole cross section. Each parameterization of the dipole cross
section has its advantages and disadvantages. For example, the IIM
parameterization is a fit to the proton structure function, which
involves a convolution of the dipole cross section with the photon
wave function squared. Therefore, the dipole cross section is weighed
preferentially in the physical cross section and some dipole sizes
play a bigger role than others. The KKT parameterization on the other
hand is a fit to particle production data which is a less inclusive
observable than a fully inclusive structure function and therefore is
more constraining on the dipole cross section. However, it is partly
complicated by hadronization and the need for a convolution with the
hadron fragmentation functions. Electromagnetic signatures of the
Color Glass Condensate therefore offer a more precise means by which
the quantitative aspects of CGC may be probed.

~\hspace*{5mm}
%%%%%%%%%%%%%%%%%%%%%%%%%%%%%
%%%%%%%%%%%%%%%%%%%%%%%%%%%%%%

%%%%%%%%%%%%%%%%%%%%%%%%%%%%%%%%%
%%%%%%%%%%%%%%%%%%%%%%%%%%%%%%%%%%%%%%%%%%%%%%%%%%%%
\begin{figure}[ht]
\begin{center}
\epsfig{file=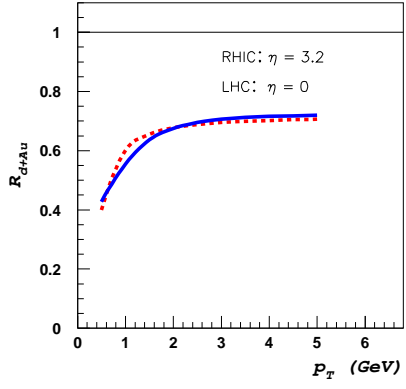,width=6.5cm}
\caption{Nuclear modification factor $R^{pA}$ of charged particles 
$(h^++h^-)/2$ at LHC energies $\sqrt{s}=5500$~GeV at mid-rapidity
$\eta=0$ as predicted in \protect\cite{KKT2} (dashed line) versus
$R^{dAu}$ of $(h^++h^-)/2$ for RHIC energies $\sqrt{s}=200$~GeV at
$\eta=3.2$ (solid line).}
\label{lhc}
\end{center}
\end{figure}
%%%%%%%%%%%%%%%%%%%%%%%%%%%%%%%%%%%%%%%%%%%%%%%%%%%

Having a parameterization of the dipole-nucleus scattering amplitude
$N$ should allow one to make predictions for particle production at
LHC. In \fig{lhc} we present a prediction from \cite{KKT2} for the
nuclear modification factor $R^{pA}$ of charged particles at
mid-rapidity at LHC (dashed line). It appears that the amount of
suppression to be seen at mid-rapidity $pA$ collisions at LHC is
roughly equal to the currently observed at forward rapidity $d+A$
collisions at RHIC (solid line in \fig{lhc}). It would be very
interesting to verify this prediction. If suppression at mid-rapidity
$pA$ collisions at LHC is observed, then one would be able to
appreciate how special RHIC energy range is: while the center of mass
collisions energy at RHIC is high, it is still low enough to have
Cronin enhancement at mid-rapidity in $d+Au$ collisions, allowing to
use $d+Au$ as a control experiment for the suppression of particle
production observed in $Au+Au$ indicating creation of quark-gluon
plasma.
%%%%%%%%%%%%%%%%%%%%%%%%%%%%%
\begin{figure}[b]
\begin{center}
\epsfxsize=6cm
\leavevmode
\hbox{\epsffile{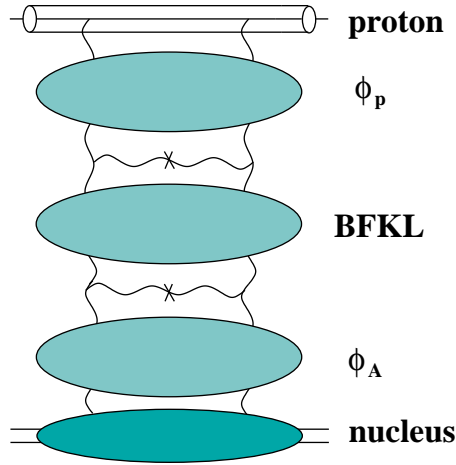}}
\end{center}
\caption{Two-gluon production in $pA$ collision in $k_T$-factorization 
approximation, which includes linear BFKL evolution only.}
\label{2Gkt_fig}
\end{figure}
%%%%%%%%%%%%%%%%%%%%%%%%%%%%%%

Another interesting experimental test of Color Glass Condensate
physics is in measuring two-particle correlations. While the exact
numerical analysis of \eq{2Gincl} leading to numerical predictions to
be verified experimentally has not been done yet, one can use the fact
that for $k_{1T}, k_{2T} \gg Q_s$ all the higher-twist multiple
rescattering effects are not important and the two-gluon production
cross section (\ref{2Gincl}) (generalized to $pA$) reduces to the
following expression for gluon multiplicity \cite{lr}
\ben
\frac{d N^{p \, A}}
{d^2 k_1 \, dy_1 \, d^2 k_2 \, dy_2 \, d^2 B} ({\un x}_{0{\tilde 0}})
\bigg|_{y_2 \gg y_1}\, = \, \frac{\as^2 \, N_c}{\pi^2 \, C_F} \, 
\frac{1}{{\un k}_1^2 \, {\un k}_2^2} \, \int d^2 q_1 \, d^2 q_2 \, \phi_p ({\un q}_2^2, Y-y) 
\, f ({\un k}_2 - {\un q}_2, {\un q}_1, y_2 - y_1)
\een
\be\label{2Gkt}
\times \, \phi_A (({\un k}_1 - {\un q}_1)^2, y_1),
\ee
where $f ({\un k}_2 - {\un q}_2, {\un q}_1, y_2 - y_1)$ is given by
the BFKL equation (\ref{bfkleq}) with the initial conditions from
\eq{bfkl_init} and both $\phi_p$ and $\phi_A$ have linear evolution only. 
\eq{2Gkt} is the expression for two-gluon production 
in $k_T$-factorization formalism \cite{lr} applied to $pA$
collisions. It is illustrated in \fig{2Gkt_fig}.

Following Kharzeev, Levin, and McLerran \cite{KLM2}, we apply
\eq{2Gkt} to the case when the rapidity interval between the two produced 
gluons is very large. For instance, in RHIC $d+Au$ kinematics, we are
interested in the case when one of the gluons is produced at
mid-rapidity and the other one is produced at forward deuteron
rapidity. This is the case of Mueller-Navelet jets, which were
originally proposed in \cite{MN} by Mueller and Navelet as an
observable allowing to study experimentally the BFKL dynamics.
%%%%%%%%%%%%%%%%%%%%%%%%%%%%%
\begin{figure}[hbt]
\begin{center}
\epsfxsize=8cm
\leavevmode
\hbox{\epsffile{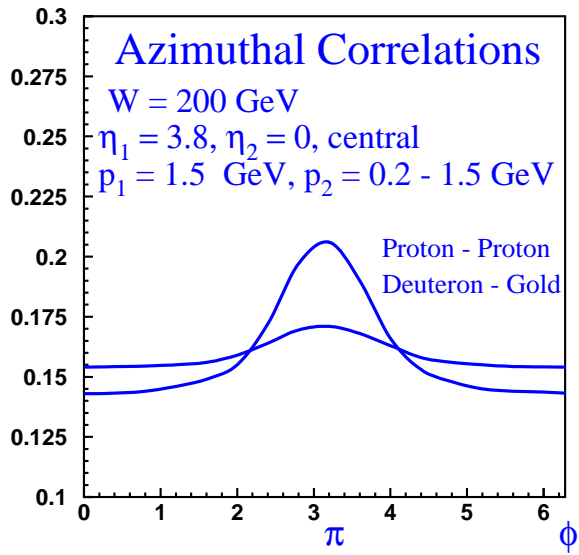}}
\end{center}
\caption{Prediction from \protect\cite{KLM2} of suppression of back-to-back 
correlations of high-$p_T$ hadrons produced in $d+Au$ collisions as
compared to $pp$ collisions. }
\label{dazim}
\end{figure}
%%%%%%%%%%%%%%%%%%%%%%%%%%%%%%

Since \eq{2Gkt} also applies to gluon production in $pp$ collisions,
one can compare the strength of back-to-back correlations in $pA$ and
$pp$ collisions with one gluon at mid-rapidity and the other one at
forward rapidity. The authors of \cite{KLM2} argued that BFKL
evolution between the two produced gluons would deplete back-to-back
correlations of produced gluons due to onset of anomalous dimension
(\ref{nu0val}) in $pA$ scattering. The effect is similar to
suppression in $R^{pA}$ discussed above in Sect. \ref{qepa}. The
resulting prediction for suppression of back-to-back correlations from
\cite{KLM2} is shown in \fig{dazim}, where the two-particle correlator
is plotted as a function of the azimuthal angle between the particles
for $d+Au$ and $pp$ collisions (with a constant background
subtracted). Once can see that the strength of back-to-back
correlations should be reduced due to the anomalous dimension effects
in $d+Au$ collisions. Experimental verification of the qualitative
behavior predicted in \fig{dazim} by the authors of \cite{KLM2} would
be an interesting and important test of CGC. Preliminary results from
STAR collaboration \cite{Ogawa} appear to confirm the qualitative
expectation of suppression of back-to-back correlations for particles
produced with a large rapidity interval between them, but more
experimental tests are needed to reach a conclusion.

%%%%%%%%%%%%%%%%%%%%%%%%%%%%%%%%%%%%%%%%%%%%%%%%%%%%%%%%%%%%%%%%%%%%%%%%%%%%%%%%%%%%%%

\section{Conclusions}
\label{conc}

In writing a review of a rapidly developing field, one is bound to
omit recent developments, which happen in parallel to the writing of
the article. The topic of pomeron loop corrections to JIMWLK and BK
evolution equations has received a lot of attention in the recent
literature \cite{loop1,loop2,loop3,loop4,loop5}. Pomeron loops are
corrections to the non-linear evolution equations which are not
enhanced by powers of color charge density, or, equivalently, powers
of $A$. A typical pomeron loop brings in an extra (parametric) factor
of $\as^2 \, \exp \{ (\alpha_P - 1) \, Y\}$
\cite{dip3,yuri_bk}. Such corrections become important when 
$\as^2 \, \exp \{ (\alpha_P - 1) \, Y\} \, \sim \, 1$. Therefore,
JIMWLK and BK evolution equations are valid up to rapidities
\cite{dip3,yuri_bk}
\be\label{qetop}
Y \, \le \, Y_{loop} \, = \, \frac{1}{\alpha_P -1} \, \ln \frac{1}{\as^2}.
\ee
\eq{qetop}, along with \eq{qcond}, describes the applicability region
of JIMWLK and BK evolution equations. For rapidities higher than
$Y_{loop}$ pomeron loop corrections become important and have to be
included. In the recent papers \cite{loop1,loop2,loop3,loop4,loop5}
corrections to JIMWLK evolution were discussed, which would take into
account pomeron loop contributions. The consequences of pomeron loop
corrections for observables in $pA$ collisions are not clear at
present.

Another important class of processes which we have omitted here are
the exclusive and semi-exclusive processes. In the quasi-classical
approximation the diffractive DIS structure functions were calculated
in \cite{Hebecker,KMc}. In the framework of dipole model one can
construct an evolution equation governing single diffractive
dissociation amplitudes for DIS, as it was done in \cite{KL}. The
results of both \cite{Hebecker,KMc} and \cite{KL} can easily be
generalized to $pA$ collisions. Diffractive particle production was
calculated for both DIS and $pA$ in the quasi-classical approximation
in \cite{yuridiff}. Unfortunately, due to limited detector
capabilities, it appears to be very difficult, if not impossible, to
measure these exclusive processes at RHIC. Since our review here is
dedicated to RHIC physics we had to omit such processes.

We conclude by pointing out once again that we have reviewed the state
of the field at the moment of writing. Future developments are likely
to bring in many new exciting results, possibly modifying some of the
discussed conclusions and adding on to the material presented here.

%%%%%%%%%%%%%%%%%%%%%%%%%%%%%%%%%%%%%%%%%%%%%%%%%%%%%%%%%%%%%%%%%%%%%%%%%%%%%%%%%%%%%%

\addcontentsline{toc}{section}{Acknowledgments}
\section*{Acknowledgments}

We would like to express out gratitude to Adrian Dumitru, Dima
Kharzeev, Alex Kovner, Francois Gelis, Kazu Itakura, Genya Levin,
Larry McLerran, Al Mueller, Dirk Rischke, Mark Strikman, Derek Teaney,
Kirill Tuchin, Raju Venugopalan, and Heribert Weigert for many
productive and enjoyable collaborations on the subject. We would like
to thank Alberto Accardi, Rolf Baier, Ian Balitsky, Jean-Paul Blaizot,
Misha Braun, Miklos Gyulassy, Ulrich Heinz, Edmond Iancu, Boris
Kopeliovich, Xin-Nian Wang, and Urs Wiedemann for many stimulating and
informative discussions.

The work of J.J-M. is supported in part by the U.S. Department of
Energy under Grant No. DE-FG02-00ER41132. The research of Yu. K. is
supported in part by the U.S. Department of Energy under Grant
No. DE-FG02-05ER41377.

%%%%%%%%%%%%%%%%%%%%%%%%%%%%%%%%%%%%%%%%%%%%%%%%%%%%%%%%%%%%%%%%%%%%%%%%%%%%%%%%%%%%%%%%%

\end{document}